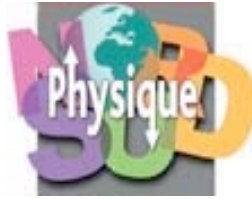

# 1$^{er}$ Congrès Nord – Sud de Physique

## La recherche et l'enseignement de la Physique

### 9 au 13 avril 2007

## Université Mohamed 1$^{er}$
## Faculté des Sciences
## Oujda - Maroc

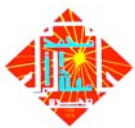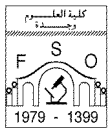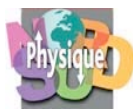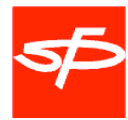

# Table des Matières

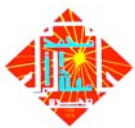 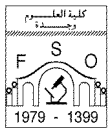 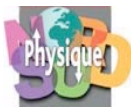 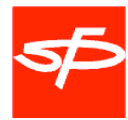

# Le mot de la co-présidente du Congrès
## Faïrouz MALEK / CNRS-IN2P3 et SFP

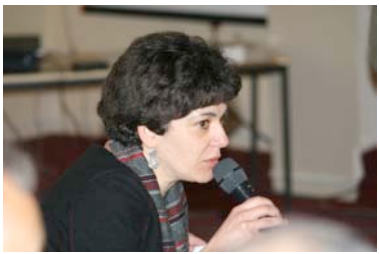

L'impact des actions de l'Année Mondiale de la Physique a été très sensible dans nos relations avec les pays du Maghreb. Le réseau qui s'est structuré grâce à ces actions nous a permis d'envisager la réalisation d'un congrès scientifique au Maroc qui associerait des enseignants chercheurs des trois pays du Maghreb mais qui pourrait être ouvert à des universitaires francophones du pourtour méditerranéen. Après une année et demi d'efforts pour *coordonner* les partenariats, affiner le contenu scientifique et surtout de surmonter les difficultés de communication dues essentiellement à cette maudite fracture numérique et technologique, *ce projet* se réalise aujourd'hui.

C'est grâce à la motivation des promoteurs, les membres des sociétés savantes de physique, les collègues universitaires du Maghreb que cette réunion voit le jour. Le soutien des différentes autorités politiques a été déterminant : qu'ils en soient remerciés. Nous souhaiterions pérenniser cette coopération qui va se mettre en place lors de ce congrès de Oujda. Une formule qui nous semble bien appropriée à ce projet serait celle d'un Fonds de Solidarité Prioritaire « physique et développement » auprès du ministère des affaires étrangères français. Il s'agit d'un projet ambitieux, dédié à la Zone de Solidarité Prioritaire, dont font partie le Maghreb et l'Afrique Subsaharienne. Il comporterait deux objectifs : les aspects fondamentaux de la physique, sous l'angle de la diffusion et de l'appropriation des connaissances avec des outils aidant à combler le fossé Nord-Sud et aussi, bien sûr, ses applications, fortement déterminées par les exigences du développement durable. Il permettrait de mieux structurer la coopération scientifique sur la physique et ses applications entre la France et les pays africains, en lui apportant une meilleure visibilité internationale.

Ce congrès, comme vous pouvez le constater dans son contenu scientifique est très éclectique dans les sujets qu'il propose de traiter. Les ateliers dont la pédagogie et les outils sont présentés sont modernes et tout à fait adaptés aux pays du Maghreb. Ce congrès a aussi une ambition de mettre en place une école doctorale en physique au Maghreb, de rassembler les sociétés de physique déjà existantes ou d'en créer dans les pays où elles n'existent pas. C'est un projet ambitieux, mais nous sommes tous très motivés. Ce sont, nous l'espérons, les premières pierres d'une coopération fructueuse et équitable !

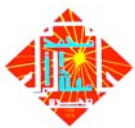 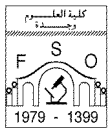 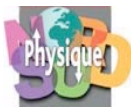 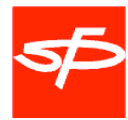

# Présentation du congrès et de ses objectifs

Les structures universitaires pédagogiques connaissent une mutation importante tant en Europe que dans les pays maghrébins. Les contenus et les formes pédagogiques doivent s'ajuster en fonction des données nouvelles de la science, des besoins d'ouvertures entre disciplines et des moyens pédagogiques disponibles pour l'enseignement qui se diversifie de plus en plus. Il nous semble intéressant et opportun, d'organiser une rencontre pour échanger et confronter nos expériences avec l'objectif élaborer des collaborations universitaires nouvelles sur le plan de l'enseignement et de la recherche.

Il s'agit de renouveler des enseignements traditionnels en intégrant les données scientifiques nouvelles et en tenant compte des développements actuels et des progrès réalisés dans les techniques d'investigation du monde micro et nanoscopique.

Les frontières des enseignements de la physique se sont déplacées et les sujets d'étude prennent plus naturellement en compte les ouvertures trans-disciplinaires (telles que mécanique, physico chimie, utilisation des moyens et réseaux informatiques) et des sujets touchant aux problèmes économiques et sociaux actuels même dans des cursus de base (énergie, environnement, eau et sols, instrumentation médicale).

La préoccupation de penser à des débouchés concerne toutes les filières, et les enseignements doivent prendre plus en compte le passage des problèmes de base aux
applications. En particulier les procédés peuvent être introduits dans des enseignements de base.
Les moyens d'enseignement en ligne et sur DVD en particulier se sont développés et il est intéressant de les identifier et d'en analyser les possibilités. Les expositions interactives et leur apport seront évoqués comme outils de sensibilisation des jeunes aux orientations scientifiques professionnalisantes. De même nous examinerons le renouveau des pratiques expérimentales (T.P. ; Stages en laboratoire…).

Il faudra aussi explorer les diverses pistes pour la réalisation et la diffusion de livres d'enseignement pour le plus grand nombre d'étudiants ainsi que d' améliorer la gestion des bibliothèques universitaires en tenant compte des possibilités des informations en ligne.

Le congrès aura plusieurs orientations et thématiques :

- Faire une revue des <u>temps forts de la recherche dans les 3 pays du Maghreb</u> ;
- Passer en revue <u>les développements réalisés dans divers domaines</u> ainsi que <u>les grandes orientations de recherche</u> susceptibles de fournir les bases d'un programme d'enseignement universitaire mieux adapté aux réalités nouvelles. La parole sera donnée à des chercheurs de renommée internationale qui présenteront des exemples pris dans leurs domaines de spécialité (la physique en recherche fondamentale comme la compréhension de la matière et l'univers, les matériaux et nanostructures, les procédés, la rhéologie et la tribologie, la modélisation et la simulation, la physique des ondes et l'imagerie etc …) ;
- l'échange d'expériences relatives à la mise en place du système LMD. Ceci se fera par l'intervention de représentants des universités participantes, par des témoignages des représentants des secteurs socio-économiques et par des forums ciblés.

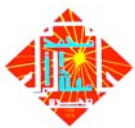 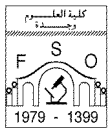 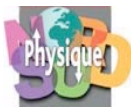 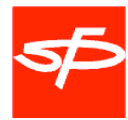

- **La relation de la physique avec la société** sera vue sous deux aspects :
  à travers la contribution des développements de la physique à la solution des problèmes de la société moderne d'une part (santé, eau, risques naturels, énergies renouvelables, informatique, télécommunications, microélectronique, etc.) et à travers la contribution de l'enseignement de la physique dans les formations professionalisantes et qualifiantes d'autre part.

Le congrès visera à mettre en place une **école pré doctorale de physique au Maghreb** qui pourrait étendre et prolonger l'action du centre de Trieste (ICTP)[1] en s'inspirant aussi du travail fait en mathématiques autour du CIMPA[2].

L'un des objectifs est que cette manifestation soit dans l'avenir régulièrement organisée dans une relation qui devrait utiliser des regroupements scientifiques, dans le pourtour méditerranéen, comparables à la SFP en France alors que les contacts actuels sont souvent très morcelés.

Ce sera de même l'occasion de susciter la mise en place de **sociétés savantes sœurs, au Maroc en particulier, et pourquoi pas une société de physique du Maghreb.**

# Les détails du Congrès

→ En amont : de l'enseignement de la physique et comment il s'enrichit de l'apport de la recherche.
→ En aval : de l'interaction avec les besoins de développement des pays du "SUD"

**Lieu de la manifestation :** Faculté des Sciences, Université Mohamed 1er Oujda, Maroc
**Période :** 9 avril 2007 – 13 avril 2007

**Présidence du Congrès :**  F. Malek / SFP et CNRS France : fmalek@lpsc.in2p3.fr
                et J. Derkaoui/Doyen Univ. Oujda Maroc : derkaoui@sciences.univ-oujda.ac.ma

**Haut Comité de parrainage :**
**Mr Claude Cohen-Tannoudji** (Prix Nobel de Physique),
**Mr Jean Coudray** (CIRUISEF[3]),
**Mr Robert Klapisch** (Président de la fondation « Partager le Savoir»),

**Promoteurs du projet :**

**SBP** (Société Belge de Physique), **SFP** (Société Française de Physique), **STP** (Société Tunisienne de Physique), Université d'Oujda, Maroc**.**

---

[1] International Centre for Theoretical Physics: http://www.ictp.it/
[2] International Centre for Pure and Applied Mathematics : http://www.cimpa-icpam.org/index.php
[3] Conférence Internationale des Responsables d'Universités et Institutions à dominante Scientifique et technique d'Expression Française.

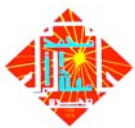 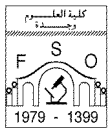 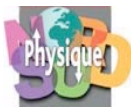 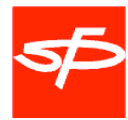

**Soutiens et Partenaires académiques et institutionnels :**

**AUF** (Agence Universitaire de la Francophonie), **CFP** (Comité Français de Physique), **EPS** (European Physical Society)**, CNRS** (Centre National de la Recherche Scientifique)-France, **CNRST** (Centre National de la Recherche Scientifique et Technique)-Maroc, **DRIC** (Direction des Relations Internationales et de la Coopération)-France, **IFO** (Institut Français de l'Oriental), **MAE** (Ministère des Affaires Etrangères)-France, **Ambassades de France**, **MENESFCRS**(Ministère de l'Education nationale, de l'enseignement supérieur, de Formation des Cadres et de Recherche Scientifique)-Maroc, **Académie Hassan II** des Sciences et Techniques, **Ministère de l'emploi**-Maroc, **Agence pour le Développement des Provinces de la Région de l'Oriental** .

**Localement au Maroc :**

- Wilaya de la Région Orientale
- Conseil de la Région Orientale
- Commune Urbaine d'Oujda

**Sponsors Locaux et Régionaux au Maroc :** Holcim, Sonacid, Maroc Telecom, RAM, IRCOD , CRI , Lafarge, Lyonnaise des eaux, St Gobin, Onyx (Vivendi), Midi

# L'organisation du Congrès

## Comité scientifique international

**D. Bideau** (Univ. Rennes et SFP, France), **H. Bouchriha** (Tunisie), **J. Derkaoui** (Doyen Univ. Oujda, Maroc), **M. Darche** (CCST Orléans et SFP, France), **E. Guyon** (ENS Paris et SFP, France), **A. Kamili** (Doyen Univ. Tetouan, Maroc), **M. Leduc** (SFP et LKB-Paris, France), **F. Malek** (SFP et CNRS, France), **R. Maynard** (Président SFP, France), **MD. Mitiche** (AAP et UMMTO, Algérie), **V. Pierrard** (Présidente SBP, Belgique), **A. Ould Haouba** (Mauritanie et AUF), **A. Sadok** (Doyen Univ. Casa, Maroc), **E. Tahri** (Univ. Oujda, Maroc), **M. Telmini** (Président STP, Tunisie), **F. Terki** (SFP et Univ. Montpellier, France), **A. Thalal** (Univ. Marrakech, Maroc), **J. Treiner** (Paris, France), **A. Weiner** (Univ. Orsay et SFP, France).

## Comité d'organisation local

**R. Baddi** (département de Physique), **N. Benazzi** (EST), **M. Ben El mostafa** (département de Physique), **H. Chatei** (département de Physique), **M. Dahmani** (département de Physique), **J. Derkaoui** (département de Physique), **H. Dekhissi** (département de Physique), **H. El Boudouti** (département de Physique), **F. Fethi** (département de Physique), **F. Maaroufi** (département de Physique), **A. Mezghab** (département de Physique), **T. Ouali** (département de Physique), **L. Roubi** (ENSAO), **E.H. Tahri** (département de Physique), **Y. Tayalati** (Faculté Pluridisciplinaire Nador), **A. Ziyyat** (département de Physique).

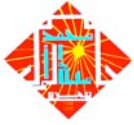 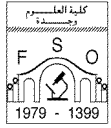 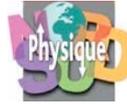 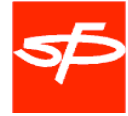

# Promoteurs du Congrès

**Société Française de Physique** 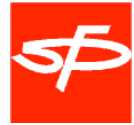

**Université d'Oujda, Faculté des Sciences** 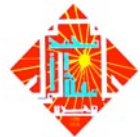 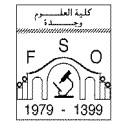

# Partenaires

**Société Tunisienne de Physique** 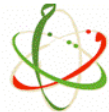

**Société Belge de Physique** 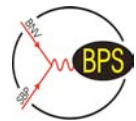

**Institut Français de l'Oriental** 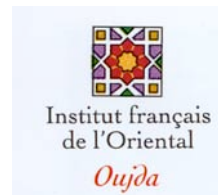

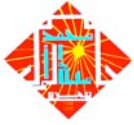 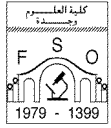 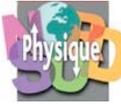 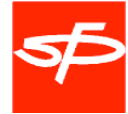

# Sponsors

Centre National de Recherche Scientifique et Technique

Académie Hassan II des Sciences et 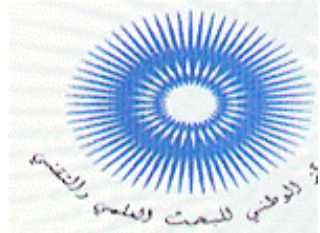 Techniques

Agence de Développement de l'Oriental

Agence Universitaire de la Francophonie 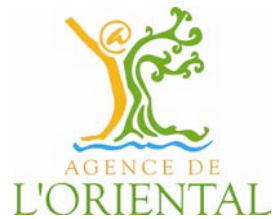

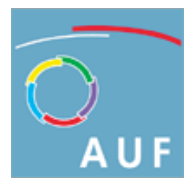

Ministère des Affaires étrangères Français (MAE) et Ministère de l'Éducation Nationale et de l'Enseignement supérieur et de la Recherche Français (DREIC)

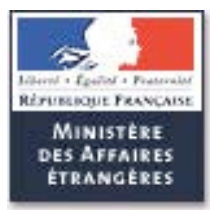 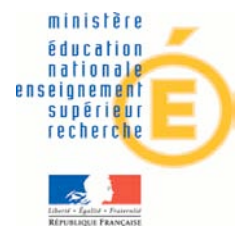

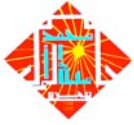 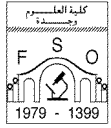 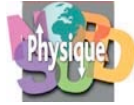 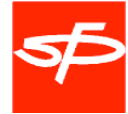

Le Centre National de la Recherche Scientifique (CNRS)

et l'Institut National de Physique Nucléaire et de Physique des Particules (IN2P3)

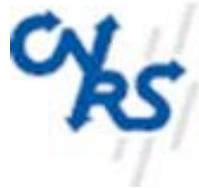 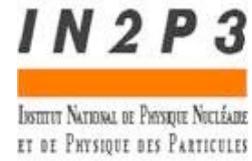

Comité Français de Physique

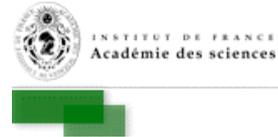

Fondation « Partage du Savoir »

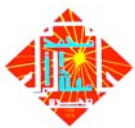 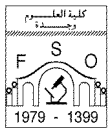 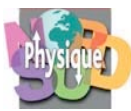 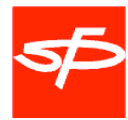

# Liste des Participants

| Nom | Prénom | Email |
|---|---|---|
| Abbaoui | Elbekay | abdou@sciences.univ-oujda.ac.m |
| ABbbes | Oukacha | okimabbes@yahoo.fr |
| Aghanim | Nabila | Nabila.Aghanim@ias.u-psud.fr |
| Ahmed | Al Hattab | |
| Ait moussa | Abdelaziz | a_aitmoussa@yahoo.fr |
| Amor | Nadji | amor.nadji@synchrotron-soleil. |
| Amraqui | Samir | Samir.nova@gmail.com |
| Aynaou | Hassan | hassan_aynaou@yahoo.fr |
| Badaoui | Mohammed | nouralhakim@yahoo.com |
| Baddi | Ramdane | rambaddi@yahoo.fr |
| Bedira | Rachida | Rachida.Bedira@fst.rnu.tn |
| Bekkouch | Kamal | kamihab@yahoo.fr |
| Benelmostafa | M'hammed | moustafa_55@yahoo.fr |
| Benlahsen | mohamed | mohamed.benlahsen@sc.u-picardi |
| Bideau | Daniel | daniel.bideau@univ-rennes1.fr |
| Blizak | Djanette | bdmeriem@yahoo.fr |
| Bouhali | Othmane | obouhali@ulb.ac.be |
| Boujrhal | Fatima-Zahra | boujrhal@yahoo.fr |
| Boukadir | Said | boukadirsaid@yahoo.fr |
| Bourahla | Ahmed | abourahla@hotmail.com |
| Boustela | Jamal | j_boustela@yahoo.fr |
| Boutigny | Dominique | boutigny@in2p3.fr |
| Brohan | Luc | luc.brohan@cnrs-imn.fr |
| Chaabelasri | Elmiloud | chaabelasri@gmail.com |
| Chadli | Elhassan | chadli@sciences.univ-oujda.ac. |
| Chafi | Noureddine | chafinouredine@yahoo.fr |
| Chasseriaux | Jean-Michel | jean-michel-chasseriaux@clora. |
| Chatei | Hassan | chateikariat@yahoo.fr |
| Cheikh | Monia | monia.cheikh@fst.rnu.tn |
| Chérigier-Kovacic | Laurence | Laurence.Kovacic@univ-provence |
| Chetouani | Abdelaziz | abd.chetouani@menara.ma |
| Dahmani | Mohammed | mdahmani2004@yahoo.fr |
| Dannoun | Saifaoui | |
| Darche | Michel | mldarche@free.fr |
| David | Sylvain | sdavid@ipno.in2p3.fr |
| Dekhissi | Hassane | hdekhissi@yahoo.fr |
| Dekhissi | Bouchra | dekhissi@cppm.in2p3.fr |
| Derkaoui | Jamal Eddine | derkaoui@fso.ump.ma |
| Diaf | EL Yamani | e_diaf@hotmail.com |
| Dihmani | Nadia | nadia.dihmani@hotmail.com |
| Diouri | Mohamed | diouri@sciences.univ-oujda.ac. |
| Dokhane | Nahed | nahed_dokhane@yahoo.fr |
| Douady | Stéphane | douady@lps.ens.fr |
| Ech-Chadi | Saïd | saidech@gmail.com |
| El Bojaddaini | Mohamed | boj2med@yahoo.fr |
| El Boudouti | El Houssaine | elboudouti@yahoo.fr |
| El Hassouani | Youssef | hassouani@yahoo.fr |
| El Khayati | Naima | elkhayat@fsr.ac.ma |
| El Marssi | Mimoun | mimoun.elmarssi@u-picardie.fr |
| El Moussaouy | Abdelaziz | azize10@yahoo.fr |
| Elgaied | Mohamed Moncef | moncef.elgaied@mes.rnu.tn |
| Elhafayani | Mohamed Larbi | Elhafyani@est.uni-oujda.ac.ma |
| Elkharrim | Abderrahman | elkharrim@yahoo.fr |
| Errahmani | Ahmed | ahmederrahmani1@yahoo.fr |
| Fethi | Fouad | fethi@sciences.univ-oujda.ac.ma |

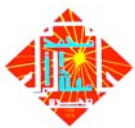 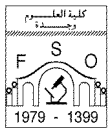 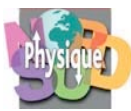 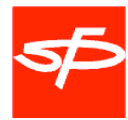

| | | |
|---|---|---|
| **Guellati** | **Saïda** | guellati@spectro.jussieu.fr |
| **Guyon** | **Etienne** | guyon@pmmh.espci.fr |
| **Hamal** | **Mohammed** | mo_hamal@yahoo.fr |
| **Hammaoui** | **Dahmane** | hammaouidahmane@yahoo.fr |
| **Hammouch** | **Malika** | hammouchmalika@yahoo.fr |
| **Hanchi** | **samir** | samir_hanchi@yahoo.fr |
| **Haouari** | **Souad** | souadmn@yahoo.fr |
| **Inchaouh** | **jamal** | inchaouh@mailcity.com |
| **Jaziri** | **sihem** | sihem.jaziri@fsb.rnu.tn |
| **Jenffer** | **Patrice** | patrice.jenffer@espci.fr |
| **Kamal** | **Hirech** | just4phy@hotmail.com |
| **Kamar** | **Zakaria** | kaka_zaka06@yahoo.fr |
| **khadraoui** | **mohamed** | khamohamed@hotmail.com |
| **Ladj** | **Rabah** | r_ladj@hotmail.com |
| **Lahlou** | **Fouad** | lahloufouad@hotmail.com |
| **Leduc** | **Michele** | leduc@lkb.ens.fr |
| **Maaroufi** | **Fatiha** | mazinema@yahoo.fr |
| **Maaza** | **Malik** | Maaza@tlabs.ac.za |
| **Maitte** | **Bernard** | bernard.maitte@univ-lille1.fr |
| **Mahani Zouhir** | | |
| **Mahy Rachid** | | |
| **Malek** | **Fairouz** | fmalek@lpsc.in2p3.fr |
| **Mariano-Goulart** | **Denis** | d-mariano_goulart@chu-montpell |
| **Maynard** | **Roger** | roger.maynard@grenoble.cnrs.fr |
| **Medari** | **leila** | l_medari@yahoo.fr |
| **Melville** | **Peter** | peter.melville@iop.org |
| **Mezzane** | **Daoud** | chef.physique@fstg-marrakech.a |
| **Mimouni** | **Jamal** | jamalmimouni@yahoo.com |
| **Mohsine** | **Youssef** | mohsine7@hotmail.com |
| **Mokrani** | **Arezki** | Arezki.Mokrani@cnrs-imn.fr |
| **Elouariachi** | **Mustafa** | mostafa14600@gmail.com |
| **Moussa** | **Abdelilah** | a.moussa@sciences.univ-oujda.a |
| **Moussaid** | **Driss** | Dr_moussaid@yahoo.fr |
| **Moussaoui** | **Mohammed Amine** | mouamine1@yahoo.fr |
| **Oldache** | **mustapha** | oldachemustapha@hotmail.com |
| **Omont** | **Alain** | omont@iap.fr |
| **Ouchrif** | **Mohamed** | ouchrif@cern.ch |
| **Oukouiss** | **Abdelkarim** | aoukouiss@yahoo.fr |
| **Ooould El Bah** | **Menny** | menny@yahoo.fr |
| **Pierrard** | **Viviane** | viviane.pierrard@oma.be |
| **Rabhi** | **Mohammed** | mohammedrabhi@yahoo.fr |
| **Ray** | **El-Mokhtar** | ray@univ-paris12.fr |
| **Robert** | **Jean-Louis** | jean-louis .robert@ges.univ-mo |
| **Saifoui** | **Dennoun** | d.saifaoui@fsac.ac.ma |
| **Seccia** | **Francis** | francis.seccia@univ-provence.f |
| **Seddouki** | **Abdeladim** | abdou_fso81@yahoo.fr |
| **Suzor-Weiner** | **Annick** | annick.suzor-weiner@u-psud.fr |
| **Tahani** | **Abdelouahad** | wtahani@yahoo.fr |
| **Tahri** | **El Hassan** | hassanfa@yahoo.com |
| **Tayalati** | **Yahya** | tayalati@cern.ch |
| **Telmini** | **Mourad** | mourad.telmini@fst.rnu.tn |
| **Terki** | **Férial** | terki@ges.univ-montp2.fr |
| **Terki** | **Hassaine Mounir** | terki_mounir@yahoo.fr |
| **Tsouli** | **Najib** | tsouli@hotmail.com |
| **Van Damme** | **Henri** | henri.vandamme@espci.fr |
| **Yahla** | **Houari** | hyahla@yahoo.fr |
| **Zineb** | **Elmouridi** | zinebelmouridi@gmail.com |
| **Smail** | **Bel Aaouja** | smail.bel@gmail.com |
| **Al Marrakchi** | | almarrakchi1@yahoo.fr |

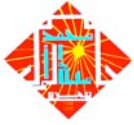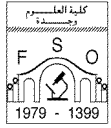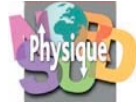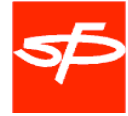

# Quelques Photos du Congrès

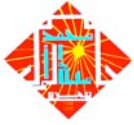 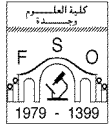 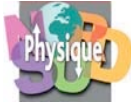 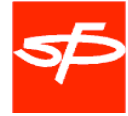

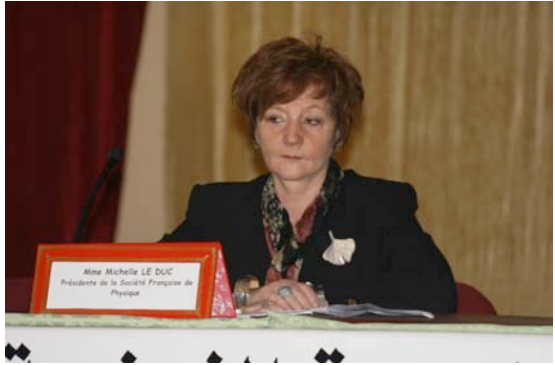 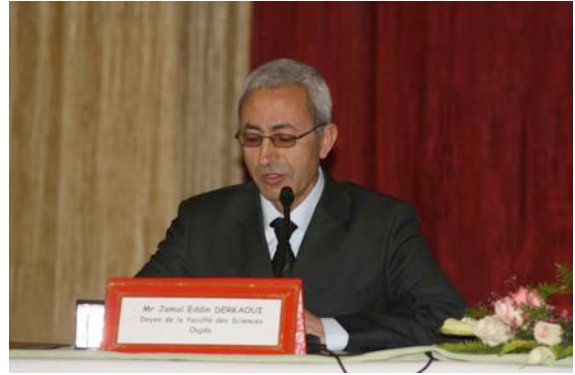

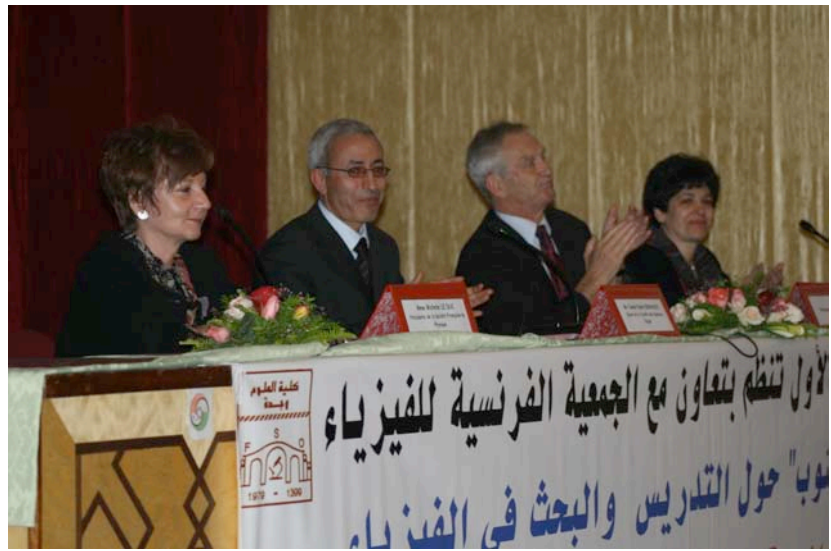

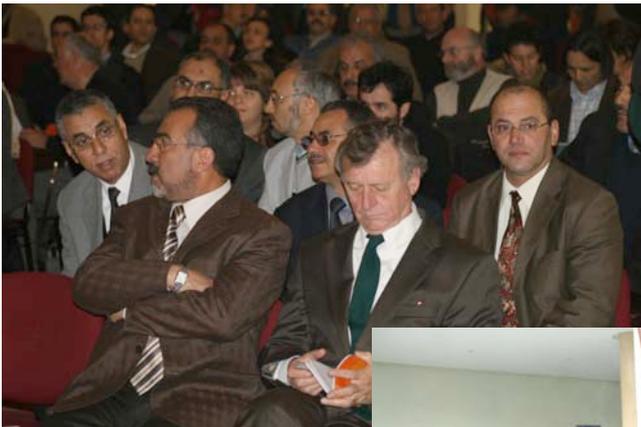 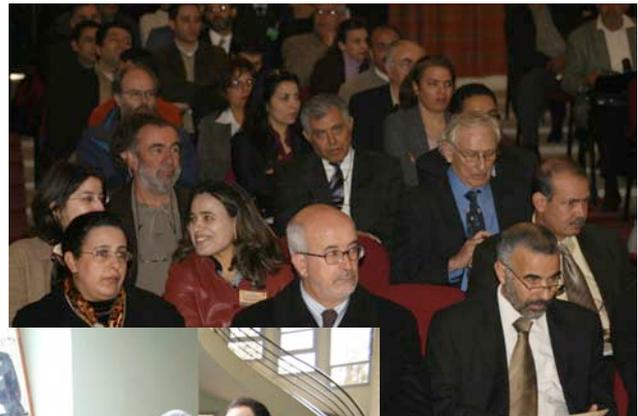

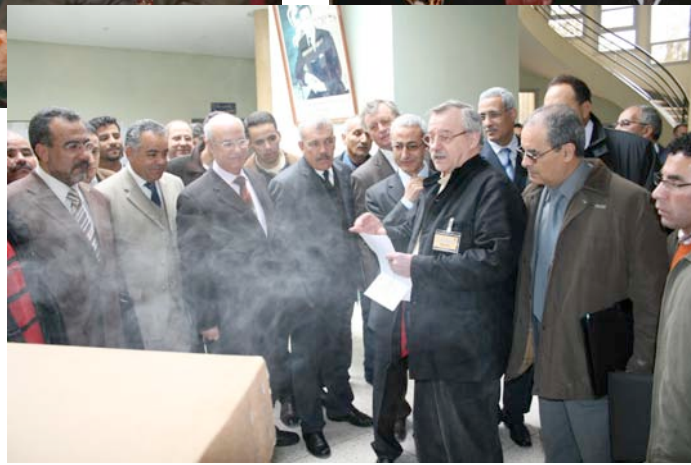

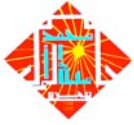 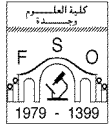 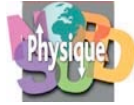 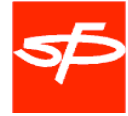
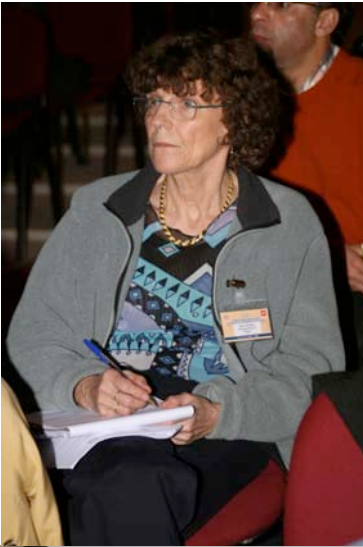 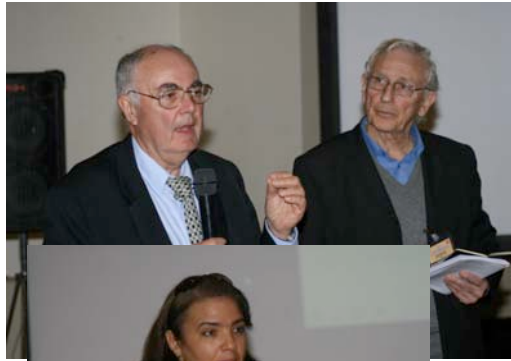 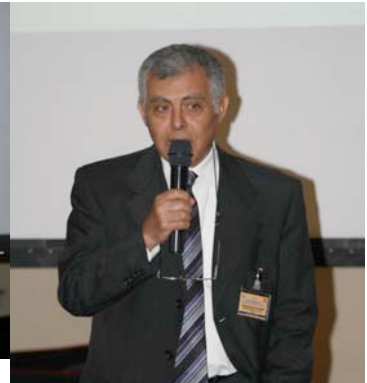
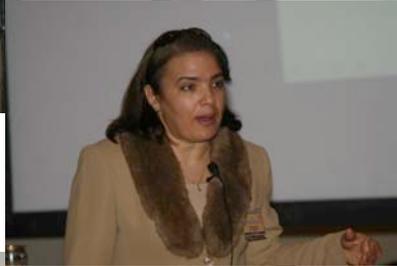 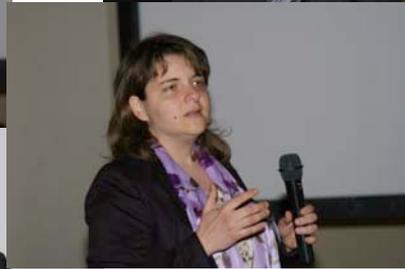
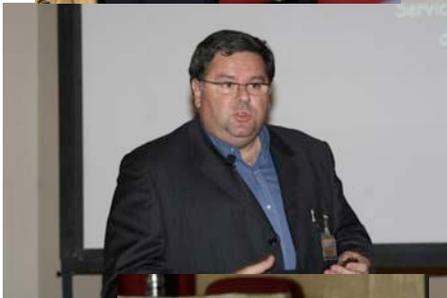 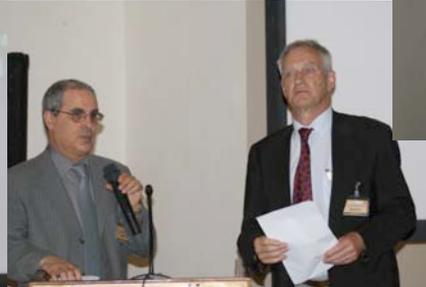 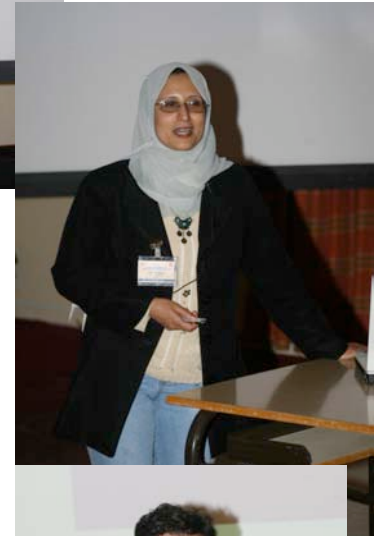
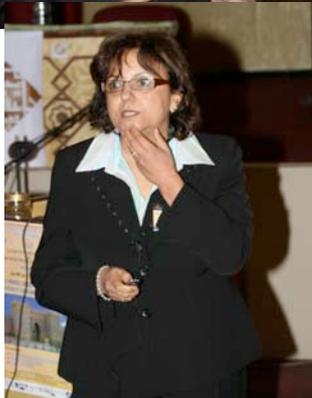 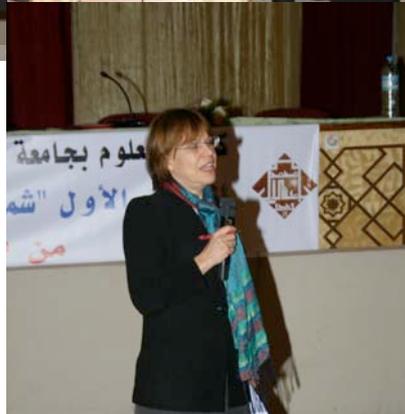
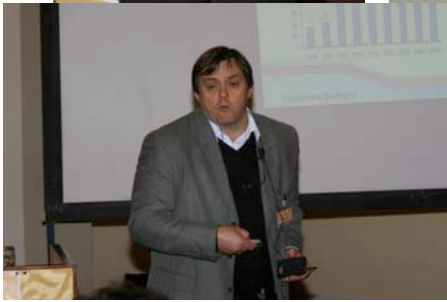 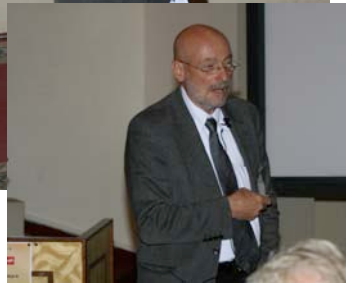 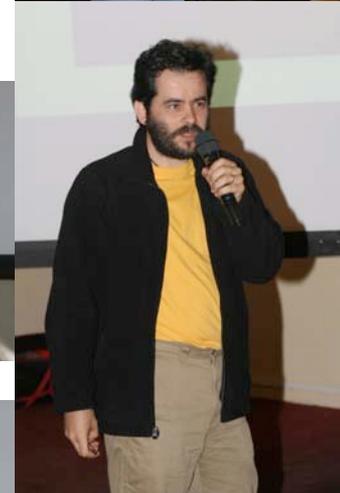
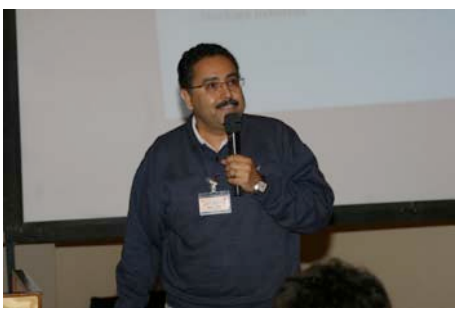 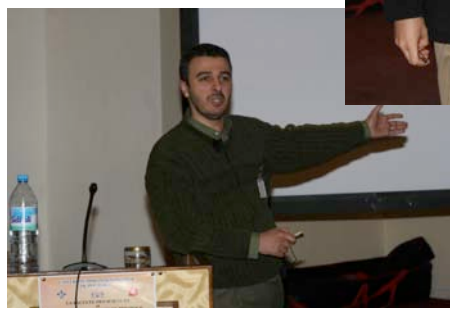

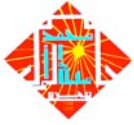
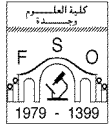
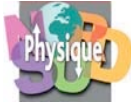
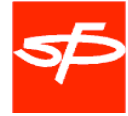
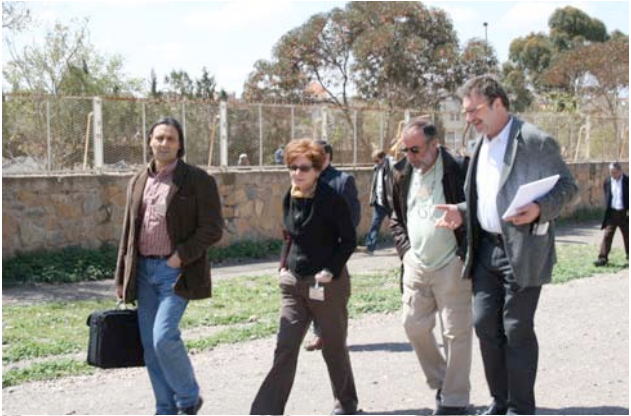
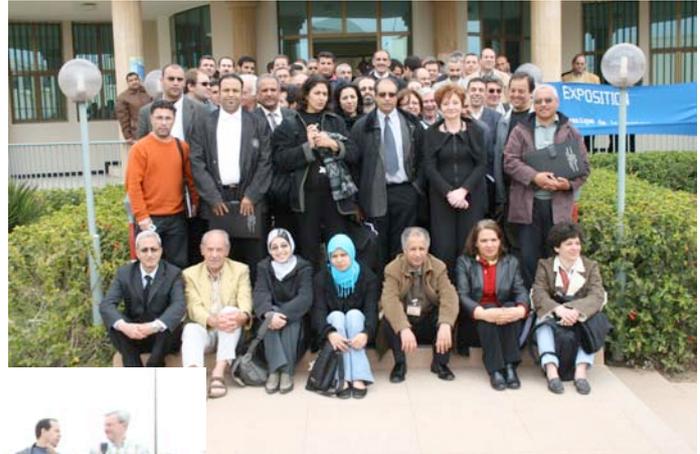
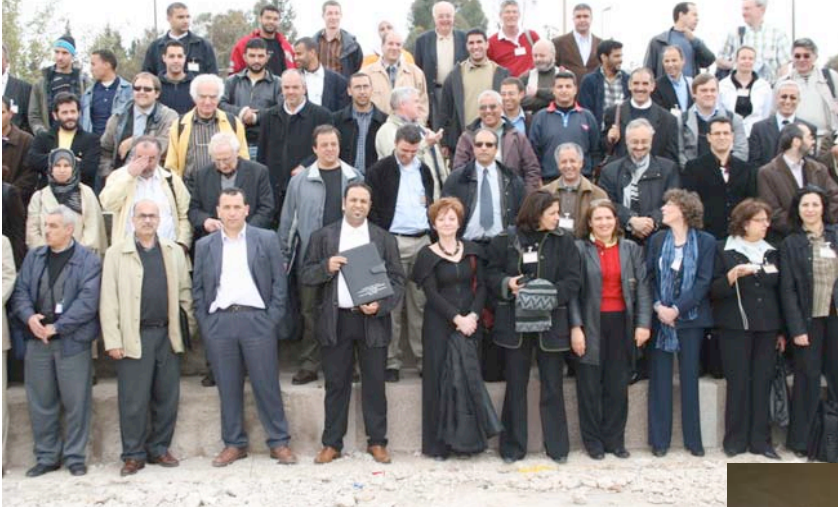
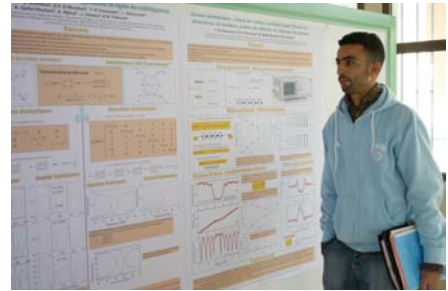
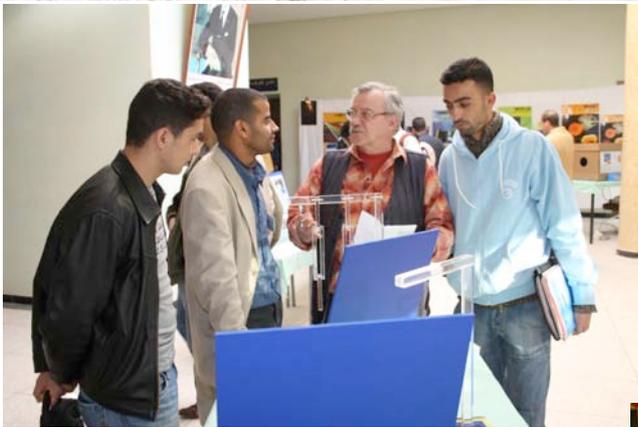
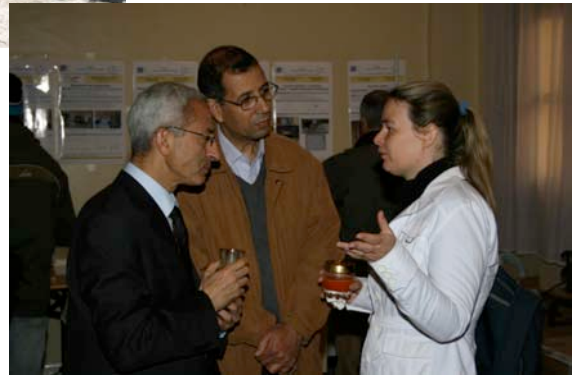
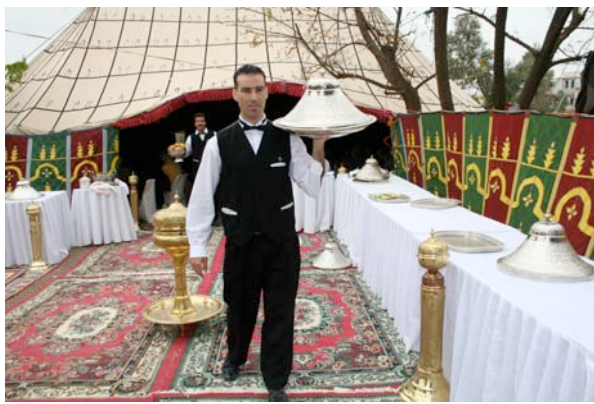
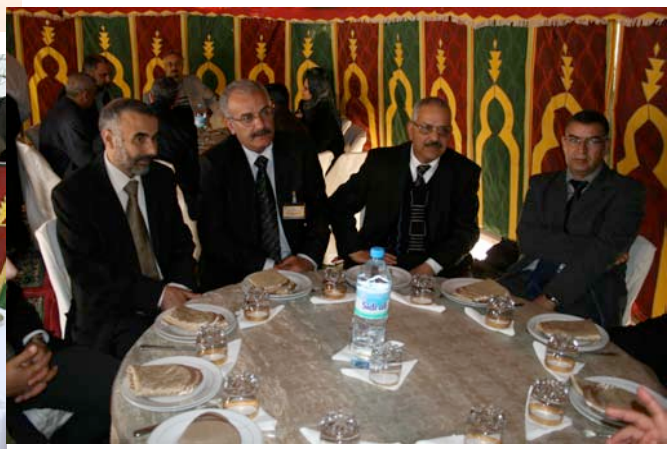

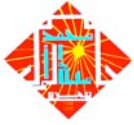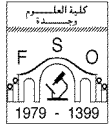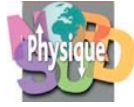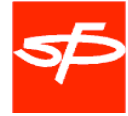

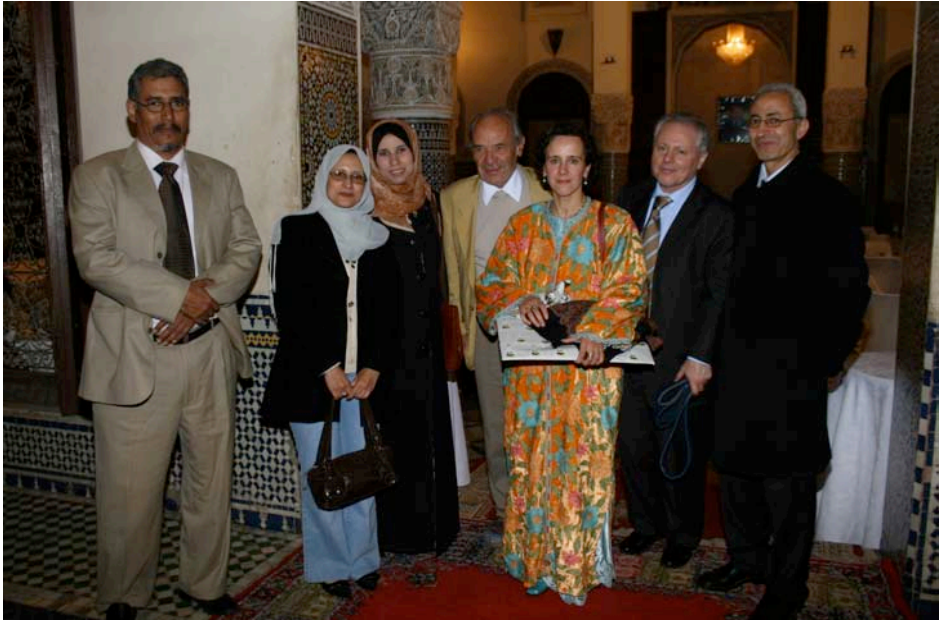

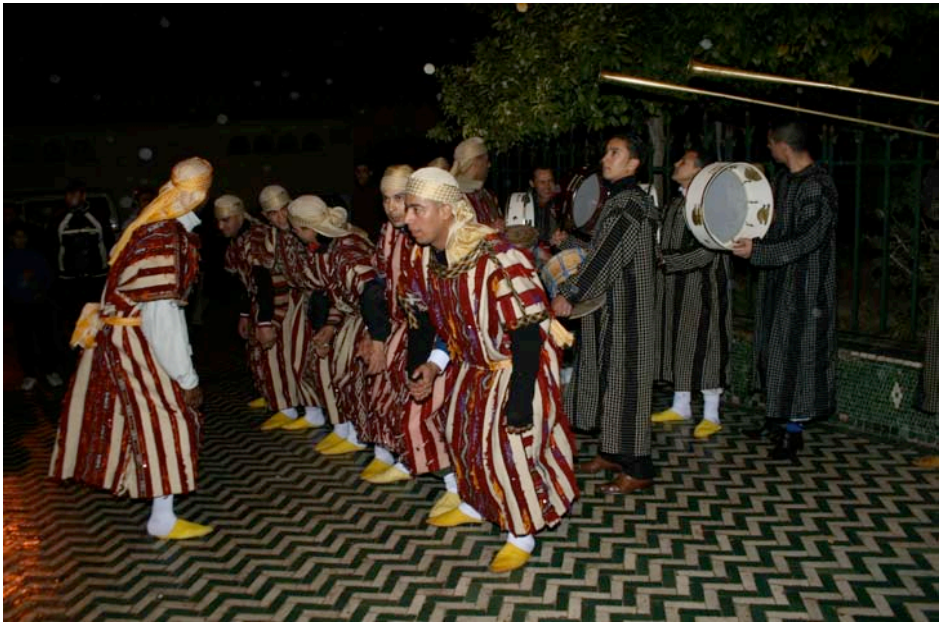

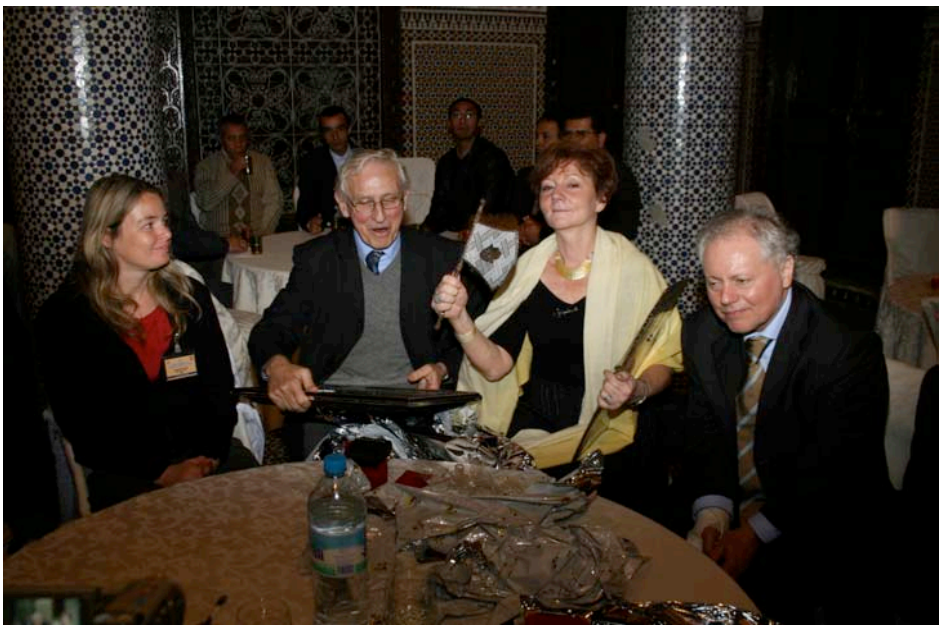

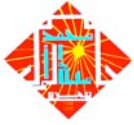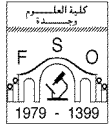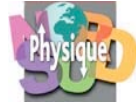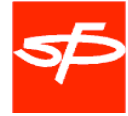

# Le programme scientifique

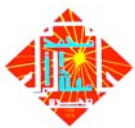 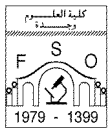 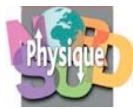 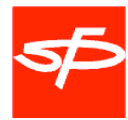

## LUNDI

*8h30:* Accueil
**Ouverture :   Session 1**
Présidence de session : Jamal Eddine Derkaoui

*9h:*     **Allocutions d'ouverture**
Président Université Oujda, Présidente SFP, co-présidents du Congrès
*9h30:*  **Demain la physique**
Édouard Brézin (Académicien, ancien président de l'académie des Sciences
et ancien président de la SFP, France)

*10h30: Pause café/thé*

Inauguration de l'exposition
« **Fousaifissa Al Physia** »

**De la recherche dans sa relation avec l'enseignement :   Session 2**
Présidence de session : Édouard Brézin

*11h-12h30*
- Revue des temps forts de la recherche au Maroc
Ahmed El Hattab (Directeur des Sciences, Rabat, M)
- Revue des temps forts de la recherche en Tunisie
M. Moncef El Gaied (DG RSR, Tunisie)

***12h30 : Déjeuner***

*15h30-17h30*
**De la recherche dans sa relation avec l'enseignement : Session 4**
Présidence de session : Michèle Leduc

- Le problème de l'énergie : Sylvain David (CNRS Orsay, F)
- L'exemple du LHC et la contribution marocaine : Abdeslam Hoummada  (Université de Casablanca, M)
- Les projets SESAME et SOLEIL, Nadji Amor (Synchrotron Soleil)

**Intervention Spéciale de Maroc Télécom**

*17h30-19h*
**La démarche expérimentale en physique : Atelier 1**
Modérateur : Etienne Guyon
Pourquoi une démarche expérimentale en Licence ? Exemples de pratiques et d'outils

*19h: Fin de journée*

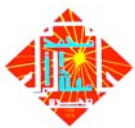 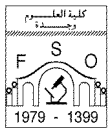 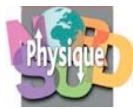 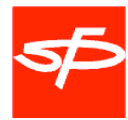

## **MARDI**

**De la recherche dans sa relation avec l'enseignement :   Session 5**
Présidence de session : Roger Maynard

*9h-10h30*
- Le monde des nanosciences : Jean-Louis Robert (Prof Univ. Montpellier, F)
- Nouveaux nano-matériaux pour la conversion et stockage de l'énergie solaire : Luc Brohan (CNRS Nantes, F)
- Approche numérique des matériaux : Arezki Mokrani (CNRS Nantes, F)

*10h30:* Pause café/thé

**De la recherche dans sa relation avec l'enseignement :   Session 6**
Présidence de session : Fairouz Malek

*11h-12h*
- Constantes fondamentales et métrologie laser : Saida Guellati (CNAM, F)
- Avancées récentes et questions ouvertes en cosmologie : Nabila Aghanim (IAS Orsay, F)

*12h : Déjeuner*

**De la recherche dans sa relation avec l'enseignement :   Session 7**
Présidence de session : Larbi Roubi

*14h-16h :*      Témoignages
- Hymne à la physique fondamentale ; Vicissitudes et promesses algériennes : Jamal Mimouni (Univ. Constantine, A)
- Le cas de l'Afrique du Sud : Malik Maaza (Afrique du Sud)

*16h:* Pause café/thé

*16h30-18h*
**La démarche expérimentale en physique : Atelier 2**
Modératrice : Férial Terki

• Des démarches expérimentales innovantes : option science en seconde

*18h*
Visite guidée de l'exposition « **Fousaifissa Al Physia** »

*19h:  Fin de journée*

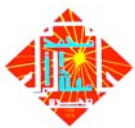 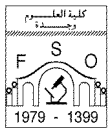 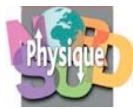 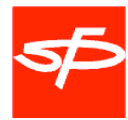

## MERCREDI

**De l'enseignement de la physique : Session 8**
Présidence de session : Daniel Bideau

*8h30-10h*
• De l'enseignement de la physique : Jacques Treiner (Prof Paris, F)

• La réforme pédagogique de l'université Marocaine : Jamal Eddine Derkaoui (Univ. Oujda, M)

• Le système LMD en France : Luce Abouaf (Univ. Paris, F)

*10h: Pause café/thé*

**De l'enseignement de la physique : Session 9**
Présidence de session : Fatiha Maaroufi

*10h30-12h30*
• Le système LMD en Tunisie : Sihem Jaziri (Tunisie)

• Recherche et enseignement à la FST Univ; Nouakchott : Menny Ould El Bah (Mauritanie)

• Le système LMD à l'Université de Constantine : Jamal Mimouni (U. Constantine, A)

• Le problème de la formation doctorale – création d'école doctorale au Maghreb ?
Dominique Salin (MSTP, F)

*13h30 : Déjeuner*

*14h30-15h*
Visite Ecole nationale des Sciences Appliquées de Oujda

demi-journée **DECOUVERTE CULTURELLE : Session 10**

*20h30: Dîner du congrès (Dar Es Sebti)*

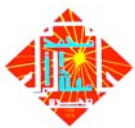 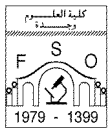 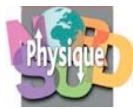 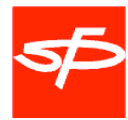

## **JEUDI**

**De la recherche dans sa relation avec l'enseignement : Session 11**
Présidence de session : F. Malek

*8h30-10h30*
- Les fluides et la rhéologie : Etienne Guyon (ESPCI, F)
- Le sol, l'eau, les dunes : Stéphane Douady (LPS-ENS, F)
- Matériaux de construction des bétons aux constructions de terre :
  Henry Van Damme (ESPCI, F)
- La recherche à l'Université Mohamed 1er : A. Anane (Oujda)

*10h30: Pause café/thé*

**Physique et Société : Session 12**
Présidence de session : Ramdane Baddi

*11h-13h15*
- Du bon usage des grands instruments : Alain Fontaine (CNRS, F)
- La Molécule d'hydrogène dans tous ses états : Mourad Telmini (STP,T)
- Une illustration des liens entre physique théorique et imagerie médicale :
  Denis Mariano-Goulart (CHU Montpellier, F)
- Coopération Amiens-Marrakech : Mimoun El Marsi

*13h15: Déjeuner*

**Physique et Société : Session 13**
**La société de l'information et la fracture numérique Nord-Sud**
Présidence de session : Annick Weiner

*14h30-17h*
- La fracture numérique et les évolutions technologiques :
  Dominique Boutigny (CNRS Lyon, F)
- Magrid et Eumed Grid : Othmane Bouhali (Univ. Bruxelles, B)
- La formation des journalistes scientifiques : Bernard Maitte (Univ. Lille, F)

*17h : Pause café/thé*

**Physique et Société : Session 14**
Présidence de session : Sihem Jaziri

*17h30-18h30*
- L'approche du genre et les problèmes spécifiques aux jeunes femmes en recherche et en enseignement supérieur : Monia Cheikh (Univ. Tunis, T)
- Enseignement, recherche et développement économique :
  Nahed Dokane (Univ. Boumerdes, A)
- La physique, art de l'homme moderne : M. Diouri (Fac. Sciences, Oujda)

*19h30 : Fin de journée*

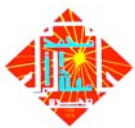 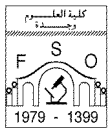 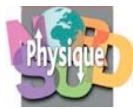 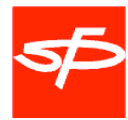

## VENDREDI

**Physique et Société : Session 15**
**Projets de coopération scientifique**
Présidence de session : Mourad Telmini (STP)

*8h30-10h20*
- Le fonds de solidarité prioritaire :
Jean-Claude Topin (MAE, F), Roger Maynard (SFP)
- Engagement de l'Europe pour la coopération scientifique :
Jean-Michel Chasseriaux, UE
- La commission « physique et développement » de l'IUPAP
et le Groupe inter-divisions Physique pour le Développement de l'EPS :
Annick Weiner (Univ. Paris-Sud Orsay, F)

*10h45: Pause café/thé*

**Physique et Société : Session 16**
Présidence de session : Roger Maynard

*11h10-12h40*
- Le problème de la formation doctorale : Sihem Jaziri (Univ. Tunis,T)
- Encadrement et co-tutelle : Annick Weiner (Univ. Paris-Sud Orsay, F)
- Ecoles doctorales du type "Les Houches" : Michèle Leduc (LKB Paris, F)

*12h40: Déjeuner*

**La démarche expérimentale en physique : Atelier 3**
Modérateur : Michel Darche

*14h30-16h30*
De l'enseignement de la physique à la communication grand public
- La caravane de la physique en Algérie : J. Mimouni (Univ. Mentouri , Constantine)
- La formation de médiateurs à l'Université : F. Lahlou (Univ. Kenitra, M)
- Des expositions à l'association "Avenir pour la Science et le savoir" : E.H. Tahri (Univ. Oujda)
- Des TP de labos à la communication grand public : P. Jenffer (Univ. Paris XI)
- La circulation des expositions de physique depuis 3 ans : M. Darche (Centre•Sciences - Orléans)

*16h30: Pause café/thé*

**Physique et Société : Session 17**
Présidence de session : Michèle Leduc (France)
**Rôle des Sociétés Savantes dans la coopération**

*17h-19h*
- Exemple de la Société Tunisienne de Physique : Mourad Telmini, STP
- Exemple de la Société Belge de Physique : Viviane Pierrard, BE
- Réunion de discussion sur le projet de « Société Marocaine de Physique »
- Conclusions et perspectives : Michèle Leduc, présidente SFP

*19h30 : clôture*

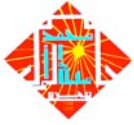 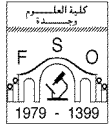 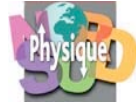 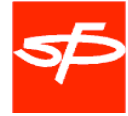

# Une Recherche de Pointe

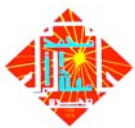 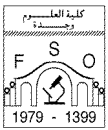 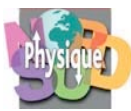 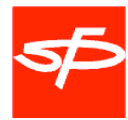

# La problématique des sources d'énergie du futur

Sylvain DAVID , Chargé de recherche CNRS
Institut de Physique nucléaire, Orsay, France

## 1 Introduction

La production d'énergie mondiale atteint 10 milliards de tonnes équivalent pétrole (tep) chaque année. Elle est assurée essentiellement par du pétrole, du gaz et du charbon, de façon très inégalitaire au niveau de la planète. Si les pays riches gaspillent, de nombreux pays en voie de développement et très peuplés tendent légitimement à augmenter massivement leur consommation dans les décennies à venir. Les scénarios énergétiques prévoient une augmentation de 50 à 300% de la production mondiale d'énergie d'ici 2050. A cela s'ajoutent des contraintes imporantes de réserves et de risque climatique. Il est d'ores et déjà évident qu'une telle augmentation ne pourra se faire sur le modèle actuel, basé sur les énergies fossiles, dont les réserves sont limitées, et dont l'utilisation conduit à des émissions massives de $CO_2$ responsable d'un changement climatique de grande ampleur.

Si tout doit être fait pour limiter nos consommations, ou au moins la rendre la plus efficace possible, cela ne sera pas suffisant, du fait de l'augmentation de la population mondiale, et du souhait légitime des pays pauvres à se développer. De nouvelles sources d'énergie doivent donc impérativement se déployer rapidement, et la recherche a un rôle primordial à jouer pour les rendre efficaces et économiques dans les années qui viennent.

## 2 Limiter l'utilisation des fossiles

Les derniers travaux du GIEC (Groupement International d'Experts du Climat) nous disent qu'il faut réduire par 4 les émissions de gaz à effet de serre au niveau mondial afin de limiter l'augmentation de température à +2°C par rapport à l'ère pré-industrielle. En imaginant une émission par habitant partagée entre 6 milliards d'habitant, cela signifie qu'un français devrait réduire ses émissions par 8, un allemand par 12 et un américain du nord par 20, et compte tenu de l'augmentation annoncée de la population mondiale, les taux de réduction demandée sont encore plus grand pour 2050. L'enjeu est donc énorme, car il faut fournir au monde l'énergie dont il a besoin tout en réduisant massivement l'utilisation des combustibles fossiles, émetteurs de $CO_2$, qui sont la base de notre approvisionnement énergétique.

Les combustibles fossiles représentent plus de 75% de la production mondiale d'énergie. Le pétrole devrait voir sa production décroître d'ici 2020 et le gaz d'ici 2040. Le charbon représente quant à lui des réserves pour deux siècles environ, et il existe d'immenses quantités de combustibles fossiles de mauvaise qualité (lignite, schistes bitumineux, hydrates de méthanes, …), difficiles à extraire, et dont l'impact sur l'environnement pourrait s'avérer catastrophique. Le captage et le stockage du $CO_2$ (sous terre) peut être envisagé dans certains cas, notamment pour les production centralisées d'électricité par des centrales à charbon, mais ne pourra de toute façon pas s'appliquer au transport ni au chauffage individuel. Mais rappelons que la Chine par exemple installe chaque année une capacité électrique supplémentaire de 50 GWe, soit presque un parc électrique français par an , mais à base de charbon ; ceci devrait nous pousser à imposer rapidement le captage et le stockage du $CO_2$.

L'enjeu semble donc de mettre en place des alternatives au pétrole conventionnel et au gaz, afin d'éviter une utilisation durable du charbon et des fuels lourds, dont l'impact climqtique serait catastrophique. On doit donc être prêt à mettre en place ce changement radical dans notre production d'énergie bien avant 2050.

## 3 Mettre en place des sources alternatives

Le développement de nouvelles sources d'énergie qui ne produisent pas de $CO_2$ apparaît donc comme incontournable, quelques soient les efforts que les pays riches pourront faire dans la maîtrise de la demande. Ces sources alternatives sont bien connues et relativement bien quantifiées.

Le nucléaire apparaît comme la seule source disponible rapidement à grande échelle, mais nécessite une mobilisation importante de capitaux et une acceptation publique. Aavant la fin du siècle, il faudra également développer une technologie innovante (surgénérateurs) pour assurer une production d'énergie nucléaire massive sur plusieurs siècles.

L'énergie solaire est un gisement très important, mais sa mise en œuvre reste encore extrêmement chère et complexe. Elle est cependant déjà compétitive dans des zones dépourvues de réseaux électriques, notamment les pays du Sud, pour qui le photovoltaïque est une voie intéressante pour accélérer le développement énergétique. Le chauffage solaire direct mériterait d'être développé partout dans le monde, il s'agit d'une technologie simple et quasiment compétitive du point de vue économique, et pourrait couvrir 10% des besoins en chauffage d'un pays comme la France. Enfin, une production centralisée d'électricité solaire pourrait être assurée par des centrales solaires thermodynamiques, qui consistent à chauffer un fluide (en concentrant les rayons avec des miroirs) destiné à faire tourner une turbine.

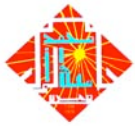 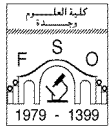 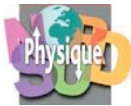 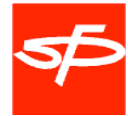

L'énergie éolienne représente un gisement limité et ne pourra sans doute dépasser 10% de la production électrique, et toujours de façon intermittente et aléatoire.

La biomasse est une voie intéressante, rapidement utilisable pour le chauffage (déjà 10% de la consommation d'énergie dans le monde). Les biocarburants semblent capables de remplacer 5 à 10% du pétrole utilisé dans les transports, moyennant une augmentation importante de la surface des terres cultivées.

Le stockage de l'énergie est loin d'être maîtrisé. Il représente un défi technologique important, et pourrait rendre les énergies intermittentes plus intéressantes dans l'avenir. La voie de l'hydrogène est intéressante pour le futur, mais demande encore beaucoup d'effort pour la rendre efficace. En effet, le rendement global de la filière est encore très faible (environ 5%), car il faut produire l'hydrogène (ce n'est pas une source d'énergie mais un vecteur !), le conditionner, le transporter, et l'utiliser dans des moteurs ou des piles à combustibles.

Enfin, la fusion thermonucléaire représente une source massive, mais risque de ne pas être disponible avant la fin du siècle, et ne peut être considérée comme une façon de « régler » le problème énergétique et climatique qui concerne les décennies à venir.

## 4 Le rôle majeur de l'électricité dans le futur

On s'aperçoit vite qu'une grande partie des sources du futur qui n'émettent pas de $CO_2$ sont dédiées à la production d'électricité. C'est le cas du nucléaire, de l'éolien, de l'hydraulique, du photovoltaïque, du solaire thermodynamique, et du charbon avec captage et stockage du $CO_2$.

Cela signifie qu'il va falloir nécessairement reporter sur l'électricité une grande partie de notre consommation énergétique actuelle, et cela nécessite de mettre en place de nouvelles technologies. C'est le cas par exemple des transports, et le déploiement massif de voitures électriques, semble une voie essentielle à soutenir. Les technologies sont dès aujourd'hui disponibles pour un usage urbain (véhicule de masse faible, distance journalière limitée), et ne demandent qu'un soutien politique et industriel pour se déployer.

Rappelons également que le chauffage électrique permet, quand l'électricité est produite sans $CO_2$, de limiter considérablement le recours au fuel domestique ou au gaz, grands émetteurs de $CO_2$, c'est le cas dela France par exemple.

## 5 En conclusion

La fin des combustibles fossiles conventionnels doit être sérieusement envisagé dans les décennies à venir, pour des raisons de réserves limitées et d'impact catastrophique sur le climat. Aucune politique ne semble aujourd'hui à la mesure de l'enjeu immense qui nous attend. La mise en place de sources d'énergie alternatives est incontournable, mais extrêmement difficile à mener : il s'agit en général de technologies complexes et très chères, et de sources d'énergie dont le potentiel est physiquement limité. Toutes doivent être développées en parallèle.

Dire que la recherche a un rôle majeur à jouer, et que la collaboration entre les différentes régions de la planète, est une grande évidence. Le problème des sources d'énergie et du climat est bie entendu un problème mondial, et le thème de l'énergie devrait devenir très rapidement le principal thème de collaboration entre les pays du Nord et du Sud, au niveau de l'industrie, de la recehrche et de l'enseignement supérieur, et ceci dans l'intérêt de tous !

**Références**
*L'énergie de demain, Techniques, Environnement, Economie*, JL Bobin, E. Huffer, H. Nifenecker, EDP Sciences (2005).
*L'énergie, Ressources, Technologies et Environnement*, C. Ngo, Dunod, 2004
**Collectif « Sauvons le Climat »**
http://gasnnt.free.fr/sauvonsleclimat/prologue.html
**Ecole Energies et Recherches**
http://eer.in2p3.fr/
**Société Française de Physique (rubrique énergie)**
http://sfp.in2p3.fr/

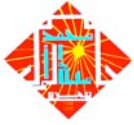 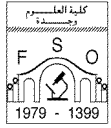 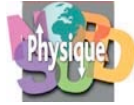 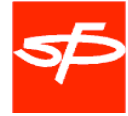

# Le LHC et la participation Marocaine à l'expérience ATLAS

Abdeslam HOUMMADA

Faculté des Sciences, Université HASSAN II Ain Chock

**Résumé**

Depuis Septembre 1996, le Maroc a été officiellement admis à participer à l'expérience ATLAS au CERN, après trois années de participation, à titre officieux au programme RD3, de quelques physiciens Marocains. Dans cette présentation sont données outre une brève présentation du CERN [1] et du LHC, les différentes phases qu'à connu la contribution Marocaine à l'expérience ATLAS.

## 1 Introduction : Le CERN

Le Centre Européen de Recherche Nucléaire (CERN) fondé il y a une cinquantaine d'années a quatre missions essentielles :
- La recherche scientifique
- L'innovation technologique
- Le développement de ressources humaines
- Les collaborations Internationales

Il est considéré à ce jour comme le plus grand laboratoire de physique au monde, avec plus de 6000 utilisateurs de toutes les nationalités.

Après les succès enregistrés auprès du collisionneur électrons-positrons (LEP) par la découverte des bosons intermédiaires $W^+$, $W^-$ et $Z°$ ainsi que la détermination du nombre de familles de neutrinos légers. Ces découvertes prédites par le modèle standard de Weinberg, Salam et Glashow ont ouvert la voie à la construction du Collisionneur à protons, le Large Hadron Collider (LHC), avec quatre expériences (ATLAS, CMS, LHCb, ALICE).

Cependant les succès du LEP, ont laissé entrouverte plusieurs questions du modèle standard. Parmis les principales questions nous pouvons citer :

- l'asymétrie matière-antimatière ;

- l'unification des quatres interactions : electrofaible, forte et gravitionnelle ;

- le nombre de familles de particules ;

- l'origine de la masse, dont le modèle standard la lie à un nouveau type de particules, le boson de Higgs.

L'origine de la masse des particules peut être considérée comme la pierre angulaire du modèle standard. Selon la théorie de Higgs, elle serait due à un champ imprégnant tout l'espace, non encore observé, qui est responsable des masses des particules. A ce champ il serait associé une particule, le boson de Higgs.

## 2 Le Large Hadron Collider

Le Large Hardon Collider (LHC) est un collisionneur en construction au CERN, qui fera entrer en collision des paires de protons, chacun avec une énergie d'à peu près 7000 fois la masse du proton (7 TeV) créant environ un milliard de collisions par seconde. Le LHC est construit à l'intérieur d'un tunnel de 27 km de circonférences à une profondeur moyenne de 100 m. L'énergie au centre de masse de 14 TeV permettra de recréer les conditions de l'Univers à un âge de $10^{-11}$ à $10^{-14}$ secondes.

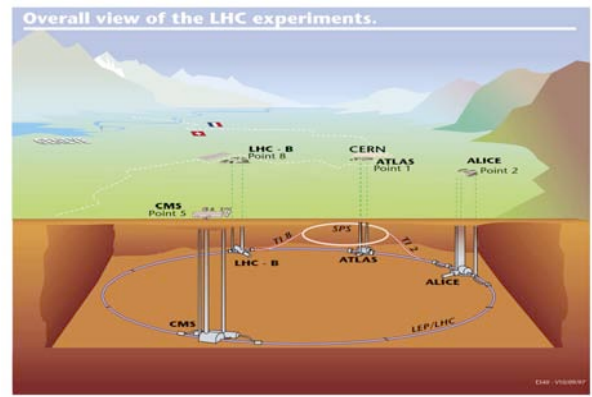

FIG. 1: Vue du LHC et des sites des expériences

D'autre part les expériences auprès du LHC en plus de la recherche du boson de Higgs, pourront éclaircir la nature de la matière noire et les propriétés du plasma primordial. Aux énergies du LHC les expériences telles que ATLAS ou CMS pourront aussi explorer de nouvelles dimensions en énergie et en espace.

A ces expériences planétaires regroupant plusieurs centaines de physiciens de plusieurs dizaines de pays, participent activement d'une manière visible quelques pays du Sud tels que le Maroc, l'Inde, Israel et le Pakistan. Par ailleurs plusieurs pays ont signé récemment des conventions de coopération avec le CERN ; les Emirats Arabes Unies, L'Egypte, la Jordanie et l'Arabie Saoudite.

A côté de ses programmes de recherche le CERN organisent annuellement plusieurs écoles ouvertes à tous les jeunes scientifiques y compris des pays du Sud. Parmi ces écoles on peut citer : l'école de Physique et l'école d'informatique et de calcul.

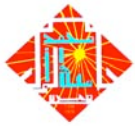 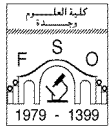 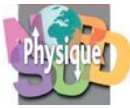 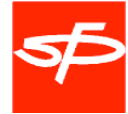

## 3 Le Maroc dans ATLAS

Depuis 1993, la participation Marocaine dans ATLAS [2] a couvert plusieurs activités. Au tout début d'ATLAS, le souci majeur était la conception et le choix des matériaux non polluants. Concernant la conception nous avons participé au programme RD3 dédié à la mise au point d'un calorimètre à argon liquide et particulièrement à son détecteur frontal ; le preshower. Cette période a été caractérisé par une intense activité de simulation et de construction de prototypes. Enfin de compte le choix de la collaboration s'est porté sur un détecteur plus simple, le Presampler, qui permet de corriger l'énergie du calorimètre électromagnétique pour tenir compte des pertes dans les matériaux du détecteur interne.

Concernant le choix des matériaux la construction d'une station d'irradiation neutrons, auprès de l'accélérateur SARA [3] à Grenoble, a permis de sélectionner les matériaux à même de supporter les fortes fluences attendues et qui seront de l'ordre de $10^{13}$ à $10^{14}$ neutrons par cm$^2$ et par an.

### 3.2 Contribution à la construction du pré-échantillonneur central d'ATLAS

Le pré-échantillonneur est placé sur la face avant du calorimètre électromagnétique, il est constitué de 64 secteurs identiques répartis sur deux tonneaux. Chaque secteur est constitué de 8 modules de différentes longueurs, chaque module est une succession d'anodes et de cathodes séparées par une couche active de 2 mm d'argon liquide. Le nombre total d'électrodes est de 50000 anodes et 50000 cathodes. Le Maroc a pris en charge les tests mécaniques et électriques des 50000 anodes du pré-échantilloneur.

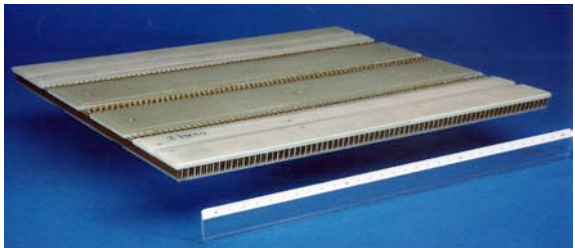

FiG. 2 : Module du presampler

Les tests mécaniques avaient pour but de ne sélectionner que les anodes ayant une épaisseur de 320 à 370 µm. Aux anodes sélectionnées sont soudées 4 résistances de 1 MΩ. L'anode est alors soumise à un test de haute tension de 3 KV pendant 24 heures afin de tester la tenue de la haute tension pendant une longue durée. A la fin l'anode est soumise à un banc de test électronique qui permet de mesurer les courants de fuite et la forme du signal électrique correspondant à une tension de charge et de décharge à 2 KV. Cette période de tests et de qualification des anodes a duré de 1997 à 2003.

La qualification des cathodes par contre a été réalisée au KTH de Stockholm en Suède.

L'assemblage des modules en secteur et l'installation des cartes mères électroniques ainsi que les tests cryogéniques ont été effectuées au LPSC à Grenoble.

L'ensemble des 64 secteurs ont été installés au CERN au cours de l'été 2004.

### 3.3 Simulation physique et analyse des données

Le groupe Marocain a participé à plusieurs campagnes de faisceaux tests et de prises de données, ainsi qu'à l'analyse des données. Ces données ont permis d'étudier les performances du calorimètre électromagnétique [4].

En parallèle plusieurs canaux physiques ont été étudiés et constitueront une base d'analyse des données au démarrage du LHC
- La recherche de bosons additionnels neutres Z'.
- La recherche de résonances décroissant en paires de quarks tops, $H/A \to t\bar{t}$ par exemple.
- L'étude de la violation de CP dans le canal :
  $B_s \to J/\psi + \phi \to \mu^+\mu^-K^+K^-$

L'accès aux données du LHC passera nécessairement par la grille de calcul LCG. Un effort est déjà fourni dans ce sens au niveau du Maroc et devra aboutir à une liaison avec la grille LCG.

## 4 Conclusion :

La participation Marocaine à l'expérience ATLAS est un exemple réussit de la coopération Nord-Sud. Les retombées scientifiques technologiques et aussi d'ordre humain sont très importantes. Nombreux sont les jeunes physiciens Marocains qui ont été formés, au plus haut niveau, au sein de la collaboration en alternant les séjours entre différents laboratoires Français, Suédois et le CERN. Nous espérons que les efforts déployés pour la construction du détecteur seront poursuivis pour l'exploitation des données, afin d'assurer une contribution Marocaine visible et performante.

## Références

[1] http://www.cern.ch
[2] ATLAS collaboration, ATLAS Technical Proposal, CERN/LHCC/94-33
[3] Belymam A. et al., Nucl. Inst. And Meth. In Phys. Res. B, 1998, 134, 217-223
[4] Akhmadaliev A. et al., Nucl. Inst. And Meth. In Phys. Res. A, 2002, 480, Issues 2-3, 508.

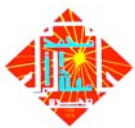 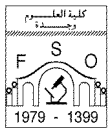 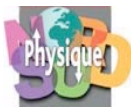 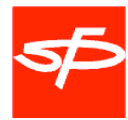

# Manipulation d'atomes par laser et métrologie des constantes fondamentales

Saïda GUELLATI-KHELIFA
Maître de Conférences
Conservatoire National des Arts et Métiers
Laboratoire Kastler Brossel CNRS-ENS-UPMC
4, place Jussieu 75252 Paris Cedex 05
guellati@spectro.jussieu.fr

## 1 Introduction

Depuis quelques années, un débat sur une révision du système international d'unités, s'est instauré au sein de notre communauté, celle de la « métrologie fondamentale » [1,2] . Cette réflexion est suscitée par le fossé qui se creuse depuis une dizaine d'années entre la réalisation des unités du système international SI et les nouvelles méthodes expérimentales générées par les récentes avancées scientifiques et technologiques.

Dans ce contexte, le contrôle et le refroidissement des degrés de libertés externes et internes des atomes par le rayonnement laser, ont eu une part considérable. Il serait très ambitieux de ma part de faire l'inventaire précis de ces techniques expérimentales tellement le domaine des applications est vaste. Je présente dans ce papier les principaux axes de recherches pour mieux définir les perspectives d'une collaboration avec les acteurs de la recherches dans les pays du bord de la méditérannée.

## 2 Mesures à haute exactitude avec les atomes froids : état de l'art en 2007

Les premières applications des techniques de manipulations d'atomes par laser ont porté sur la spectroscopie à très haute résolution dans le domaine micro-onde et le domaine optique. Les deux expériences pionnières [3 , 4] qui ont consacré la faisabilité de la fontaine atomique à l'aide d'une source d'atomes froids ont mis en évidence des potentialités extraordinaires en termes de stabilité et d'exactitude. Ces résultats ont suscité rapidement une dynamique de recherche particulière pour satisfaire des exigences strictes en termes de techniques d'interrogation, de détection et du contrôle de l'environnement des atomes. Il a fallu quelques années pour aboutir à la réalisation d'étalons primaires de césium avec une stabilité de $1,6 \times 10^{-14} \tau^{-1/2}$ et une exactitude de $6,5 \times 10^{-16}$. En stabilité, ces performances sont actuellement limitées par le bruit de projection quantique (à court terme) et la stabilité en fréquence de l'oscillateur local (à long terme) [5]. La limite ultime est quant à elle manifestement imposée par l'accélération de la gravité. Sur terre, un gain sur la largeur de la résonance se fera fatalement au détriment de l'exactitude. A ce titre, des projets d'horloge atomique en orbite *(Pharao: Projet d'Horloge Atomique par Refroidissement d'Atomes en Orbite.)* et dans l'espace *(ACES: Atomic Clock Ensemble in Space.)* devraient pousser plus loin les limites intrinsèques.

Le développement des horloges optiques a également trouvé un renouveau, au sein de ce domaine de recherche. Le facteur de qualité des étalons optiques est $10^4$ fois plus important que celui d'une horloge à césium et permettrait un gain égal en stabilité. Ce gain a été longtemps compromis par la perte en exactitude conférée à la sensibilité aux déplacements lumineux. Je dirais que cette difficulté est tombée à la suite de la proposition de H.Katori, de développer des horloges à réseaux optiques. L'idée consiste à piéger les atomes dans un réseau lumineux en régime de Lamb-Dicke. Ce confinement permet d'augmenter le temps d'interrogation et de s'affranchir de l'effet de recul. L'étude de Katori a surtout montré qu'il est possible d'annuler au premier ordre le déplacement lumineux en ajustant la fréquence des lasers produisant le réseau optique [6]. Ce type d'horloge permettrait d'atteindre une exactitude dans la gamme $10^{-17}$-$10^{-18}$ [7]. De plus, les peignes de fréquences optiques générés par un laser femtoseconde [8] assurent une intercomparaison souple et surtout très fiable entre les horloges optiques favorisant ainsi leur développement.

La seconde voie a porté sur l'exploitation les propriétés quantiques des atomes froids pour concevoir des interféromètres atomiques. Ces interféromètres permettent de lire "la phase" de l'onde atomique. Cette phase est très sensible à l'environnement extérieur de l'atome et peut ainsi fournir des mesures fines des perturbations perçues au sein de cet environnement (champs inertiels, forces extérieures...) [9] . Deux particularités méritent d'être soulignées à ce niveau. La première est liée au fait que, dans les interféromètres atomiques utilisés actuellement, la séparation et la recombinaison des ondes atomiques sont réalisées grâce à l'impulsion du photon de l'onde électromagnétique qui induit le changement de l'état interne. Ce lien entre les degrés de libertés internes et externes "accroît" considérablement la sensibilité de ce type d'interféromètres. La seconde particularité, est que la mesure de la phase est souvent ramenée à une mesure de temps ou de fréquence. Aujourd'hui, ces deux grandeurs physiques sont mesurées avec un excellent niveau d'incertitude. Plusieurs expériences d'interférométrie atomique fleurissent dans le monde, notamment celles dédiées à la mesure de champs de rotation ou d'accélération (gyromètre et gravimètre à atomes). Je ne citerai dans ce registre que l'expérience pionnière du groupe de S. Chu qui a permis d'une part une mesure de l'accélération de la gravité [10] et d'autre part, celle de la fréquence associée au recul de l'atome suite à l'absorption ou l'émission d'un photon [11].

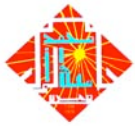 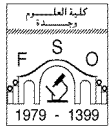 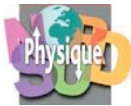 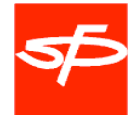

Par ailleurs, la dynamique particulière des atomes, soumis à une force extérieure F dans un réseau optique, décrite par les oscillations de Bloch [12], a défini un autre concept expérimental particulièrement élégant et puissant. Dans ce point de vue, le paquet d'onde atomique oscille localement, avec une fréquence caractéristique appelée fréquence de Bloch et définie par $\nu_B = F\lambda/2h$ où $\lambda$ est la longueur d'onde optique et h la constante de Planck. Pour avoir une idée des ordres de grandeurs, une force extérieure comprise entre $10^{-4}$ x Mg 100 x Mg (M étant la masse atomique et g l'accélération de la gravité), correspondrait à une fréquence de Bloch $\nu_B$ dans la gamme [1 Hz, 1 MHz]. Plusieurs projets se dessinent autour de cette idée, notamment pour la mesure de faibles forces au voisinage de surfaces avec une resolution spatiale à l'échelle du micromètre [13].

Dans notre groupe au laboratoire Kastler Brossel à Paris, nous avons exploité le mécanisme des oscillations de Bloch pour mesurer la vitesse de recul d'un atome qui absorbe ou émet un photon. Chaque oscillation de Bloch correspond à un transfert cohérent et précis d'une impulsion égale à deux fois l'impulsion du photon absorbé. Cette mesure nous a permis de réaliser une nouvelle détermination de la constante de structure fine $\alpha$ avec une insertitude relative de $10^{-9}$ [14].

## 6. Conclusion et perspectives

Au délà de leur grand impact sur la métrologie fondementale, les techniques de manipulations d'atomes par laser sont, aujourdhui incontournables dans les expériences récentes de physique atomique. Elles offrent un moyen très efficace pour la formation par la recherche (technologie laser, ultravide, électronique, traitement de donnée,…). Pour promovoir la physique atomique moderne dans les pays du sud, il me semble tout a fait opportun de démarrer une expérience permettant de réaliser une source d'atomes froids. Ce type de dispositif expérimental est aujourdhui tout à fait accessible avec un budget de recherches raisonnable. Ces expériences peuvent à terme aboutir sur des projets plus ambitieux comme la réalisation de la première horloge à atomes froids du Maghreb.

## Remerciements

## Références

[1] C. J. Bordé, Phil. trans. R. Soc. A 363 (2005) 2177-2201.
[2] I. M. Mills, P.J Mohr, T.J Quinn, B. N. Taylor and E.R. Williams, Metrologia **43**, 227-246 (2006
[3] M. Kasevich, E. Riis, and S. Chu, Phys. Rev. Lett. **63**, 126 (1989).
[4] A. Clairon, C. Salomon, S. Guellati and W.D. Phillips, Europhys. Lett. **16**, 165 (1991).
[5] C.Vian, P. Rosenbusch, H. Marion, S. Bize, L. Cacciapuoti, S. Zhang, M. Abgrall, D. Champbon, I. Maksimovic, P. Laurent, G. Stantarelli, A. Clairon, A. Luiten, M. Tobar and C. Salomon, IEEE. Trans. Inst. Meas, **54** (2005) 833.
[6] H. Katori, T. Ido and M. Kuwata-Gonokami, J. Phys. Soc. Jnp. **68**, 2479 (1999).
[7] H. Katori, M. Takamoto, V. G. Pal'chikov and V. D. Ovsiannikov, Phys. Rev. Lett. **91**, 173005-1 (2003).
[8] R.Holzwarth, Th. Udem, and T. W. Hänsch, Phys. Rev. Lett. 85, 2264 (2000). - Th.Udem, R.Holzwarth, T.W. Hänsch, Nature 416, 233 (2002).
[9] T.L. Gustavson, P. Bouyer, and M. A. Kasevich, Phys. Rev. lett **78**, 2046 (1997).
[10] M. Kasevich and S. Chu, Appl. Phys. **B54**, 321 (1992).
[11] A. Wicht, J. M. Hensley E. Sarajlic and S. ChuA, Physica Scripta T102, 82 (2002).
[12] ] M. Ben Dahan, E. Peik, J. Reichel, Y. Castin, and C. Salomon, Phys. Rev. Lett **76**, 4508 (1996).
[13] I. Carusotto, L. Pitaevskii, S.Stringari, G. Mudungno and M. Ingusio, Phys. Rev. Lett. **95**, 093202 (2005).
[14] P.Cladé, E. de. Mirandes, M. Cadoret, S. Guellati-Khélifa, C. Schwob, F. Nez, L. Julien and F. Biraben, Phys. Rev. Lett. **96**, 033001-1 (2006)

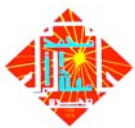 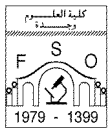 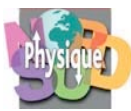 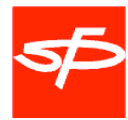

# La molécule d'hydrogène dans quelques uns de ses états

**Mourad TELMINI, Maître de Conférences**

LSAMA, Département de Physique, Faculté des Sciences de Tunis, Tunisie

**Résumé**

La molécule d'hydrogène H$_2$ est la plus petite molécule. Pourtant, son étude quantique est loin d'être triviale, notamment lorsqu'elle est dans un état de Rydberg où dans un état doublement excité. Dans cet article, nous décrivons une méthode théorique originale qui permet de décrire ces états exotiques de la molécule d'hydrogène ; le modèle halfium.

## 1 La molécule d'hydrogène

La molécule d'hydrogène est la molécule la plus petite et la plus légère. Formée de deux protons et de deux électrons, c'est un édifice lié dont les constituants sont maintenus en cohésion par le biais de l'interaction électromagnétique. Malgré les progrès impressionnants de la mécanique quantique et de la physique atomique et moléculaire, plusieurs propriétés de cette molécule restent énigmatiques. Les approches théoriques basées sur les interactions de configuration et les méthodes ab initio de chimie quantique restent confinées à l'état fondamental et aux premiers états excités.

Par ailleurs, cette molécule a connu un regain d'intérêt ces dernières années, suite la crise annoncée des énergies fossiles, et à l'intérêt croissant pour les énergies renouvelables, dont l'hydrogène sous sa forme moléculaire est l'un des vecteurs les plus prometteurs. L'hydrogène est sorti du laboratoire pour être utilisé de manière de plus en plus croissante comme carburant. Les techniques de production, de transport et de stockage sont opérationnelles et plusieurs équipes de recherche et de développement à travers le monde travaillent sur le sujet. L'Islande, pays insulaire de l'Atlantique nord, peu doté en énergies fossiles, a décidé d'être à l'horizon 2050 la première écomonie du monde, dont l' énergie est basée exclusivement sur l'hydrogène.

La molécule d'hydrogène est également d'une importance capitale en astrophysique. Sa présence a été mise en évidence dans les planètes géantes gazeuses du système solaire (Jupiter et Saturne), mais aussi dans l'espace interstellaire. Les observations des sondes et télescopes spatiaux, notamment ISO, Hubble et Spitzer (voir Figure 1), ont montré à quel point la détection de H$_2$ peut aider à comprendre la composition et l'évolution de notre univers.

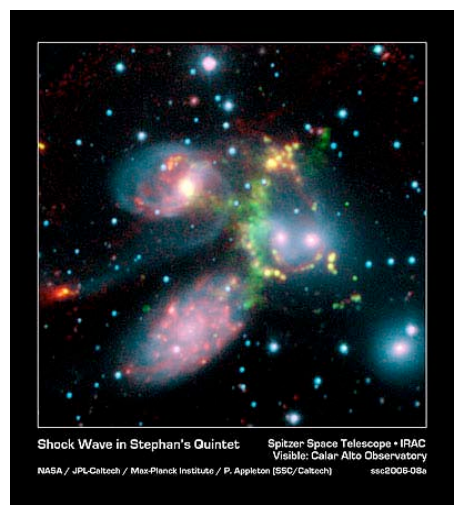

FIG. 1: Photo prise par le télescope stapial infrarouge Spitzer (Août 2006), où on voit un groupe de galaxies siège d'une onde de choc d'une puissance inouïe, grâce à l'observation de l'émission de l'hydrogène moléculaire

## 2 Le modèle halfium

Le modèle halfium [1] étudie la molécule d'hydrogène dans le cadre de l'approximation de Born-Oppenheimer, qui consiste à séparer le mouvement lent des noyaux, du mouvement très rapide des électrons. Le modèle combine la méthode variationnelle de la matrice R avec la théorie généralisée du défaut quantique multi-voies, en coordonnées sphéroïdales. L'équation de Schrödinger électronique est résolue pour une distance internucléaire donnée. L'espace des configurations est divisé en deux parties ; *(i)* un volume de réaction de forme ellipsoïdale dont les foyers correspondent aux positions des noyaux fixes et *(ii)* le reste de l'espace qui est la zone asymptotique. Dans le volume de réaction, « toutes » les interactions sont prises en compte, notamment l'interaction bi-électroniqque coulombienne, et les interactions non locales et d'échange. En revanche, dans la zone asymptotique, un seul électron est supposé pouvoir évoluer. C'est une première limitation du modèle, qui ne permet pas l'étude de l'ionisation double, mais qui permet néanmoins d'étudier les états de Rydberg et les états doublement excités. De plus, l'électron externe est supposé être soumis à un potentiel purement coulombien à deux centres dont les charges nettes valent chacune la moitié d'une charge élémentaire positive. Cette hypothèse, décrit de manière effective l'écrantage partiel des protons par l'électron interne. Le système ainsi décrit (deux charges positives demi-entières et un électron) est un système moléculaire fictif que nous avons baptisé halfium, et qui a donné son nom au modèle. Dans le volume de réaction, nous cherchons des solutions particulières de l'équation de Shcrödinger qui satisfiont le principe variationnel de Kohn et qui ont donc une dérivée logarithmique stationnaire sur la surface de réaction. Dans la région asymptotique, la molécule d'hydroène est approximée par la particule halfium, dont l'équation de Schrödinger est exactement soluble en coordonnées sphéroïdales. La continuité de la fonction d'onde et de sa dévivée sur la surface de réaction achève de déterminer la fonction d'onde dans tout l'espace à chaque énergie. Toute l'information est condensée dans une matrice compacte dont les éléments diagonaux sont trivialement liés aux défauts quantiques. L'application ultérieure des conditions aux limites (à l'infini) permet de quantifier l'énergie de la molécule dans le spectre lié, et de déterminier les positions et les largeurs des résonances asociées aux états doublement exciés dans le contiumm électronique

## 3 Applications

Le modèle halfium a été appliqué avec succès aux principales symétries de la molécule $H_2$[1], aussi bien pour déterminer les courbes de potentiel que les propriétés des résonances. Son point fort est la description ab initio unifiée du spectre discret et du continuum électronique. Les défauts quantiques sont obtenus sous la forme d'une fonction à deux variables ; l'énergie $E$ et la distance internucléaire $R$, et doivent avoir une variation lisse avec ces deux variables (voir figure 2). C'est une condition nécessaire pour pouvoir effectuer des interpolations efficacement et aussi pour pouvoir appliquer la technique de la transformation de repère, qui dans le cadre de la théorie du défaut quantique, permet d'inclure les degrés de liberté de vibration et de rotation des noyaux.

Pour vérifier la validité des calculs effectués avec le modèle halfium, nous avons comparé nos résultats avec ceux des méthodes ab initio de chimie quantique pour les états bas, les seuls étudiés par ces méthodes. L'accord est raisonable, étant entendu que notre but n'est pas de rivaliser avec ces méthodes puissantes pour les basses énergies, mais plutôt d'étudier les états très excités où ces méthodes échouent. Nous avons aussi fait des prédictions sur les résonances associées aux états doublement excités autoionisants jusqu'à une énergie de 45 eV au dessus du fondamental, où le calcul fait intervenir une dizaine de voies en interaction.

Les premièrs calculs de niveaux ro-vibroniques incluant des interactions de spin en été faits très récemment en collaboration avec l'équipe expérimentale de F. Merkt (ETH, Zurich). Les résultats seront publiés prochainement.

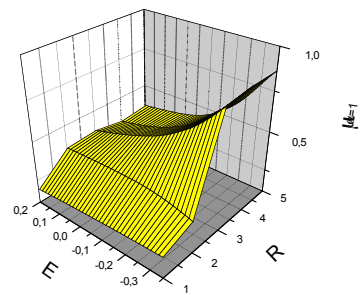

FIG. 2: Variation du défaut quantique propre de l'une des voies de la symétrie $^1\Sigma_g^+$ de $H_2$ en fonction de l'énergie $E$ et de la distance internucléaire $R$ (pour plus de détails voir réf. [1])

## 3 Conclusion et perspectives

Le modèle halfium est une approche originale qui a permis pour la première fois de décrire les spectres liés et le continuum électronique de la molécule $H_2$ avec un traitement ab initio efficace. Le modèle est potentiellement généralisable à toute molécule diatomique à deux électrons actifs. Les dimères d'alcalins sont des candidats parfaits pour ce projet, d'autant plus, qu'il interviennent dans les collisions ultra-froides et les processus de photo-association, où les expériences sont en plein essor. Ce dernier point est à mettre en lien avec d'autres travaux effectués récemment en collaboration avec E. Charron, A. Suzor-Weiner et L. Pruvost (Orsay), sur la manipulation laser cohérente d'atome froids et la description de dispositifs d'optique atomique utilisant des atomes de rubidium froids [2].

**Références**
[1]  M. Telmini and Ch. Jungen, Phys. Rev. A **68**, (2003) 062704;
   M. Telmini, S. Bezzaouia and Ch. Jungen, Int. J. Quant. Chem. **104**, (2005) 530;
   S. Bezzaouia, M. Telmini and Ch. Jungen, Phys. Rev. A **70**, (2004) 012713;
   M. Telmini, Ch. Jungen, S. Bezzaouia and H. Oueslati, J. Phys. Conf. Ser. **4** (2005) 256;
   H. Oueslati, M. Telmini and Ch. Jungen, Mol. Phys. **104**, (2006) 187.
[2]  N. Gaaloul, A. Suzor-Weiner, L. Pruvost, M. Telmini and E. Charron, Phys. Rev. A 74 (2006) 023620.

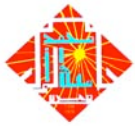 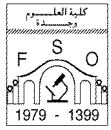 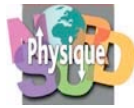 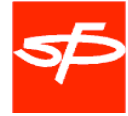

# Apports récents de la physique à la tomographie médicale

Denis MARIANO-GOULART, MCU-PH

Service de Médecine Nucléaire. CHU et Faculté de Médecine, Montpellier, France

**Résumé**

Physiciens et mathématiciens ont un rôle des plus en plus reconnu au sein des structures hospitalières d'imagerie médicale. Ils contribuent à la radioprotection, au choix puis au contrôle de qualité des appareils d'imagerie, et au développement de méthodes innovantes en traitement du signal. En survolant les récents progrès obtenus en tomographie par émission de positons, cette brève contribution vise seulement à donner quelques exemples du rôle que peuvent jouer des physiciens au sein de telles structures de soin. On insistera sur le fait que ces recherches ne nécessitent que des mises à disposition occasionnelles de tomographes utilisés par ailleurs dans un cadre hospitalier. Elles ne nécessitent en conséquence aucun investissement coûteux et sont susceptibles d'être rapidement commercialisées ou utilisées en routine clinique, participant ainsi à l'essor économique d'une région.

## 1 Introduction

Les modalités radiologiques d'imagerie médicale (radiographie, échographie, IRM) fournissent au médecin une information utile sur l'anatomie de différents organes. Ces techniques peuvent parfois être mises à profit pour exploiter une information fonctionnelle (IRM fonctionnelle par exemple), mais c'est surtout la médecine nucléaire qui donne accès à une véritable imagerie métabolique. En évaluant la distribution volumique de l'activité d'un radio-isotope (émetteur $\gamma$ ou $\beta^+$) lié une molécule vectrice d'intérêt biologique, ces techniques scintigraphiques permettent de quantifier une fonction d'organe ou un métabolisme. Inventé au milieu du XX° siècle, la tomographie par émission de positons (TEP) a connu depuis 20 ans un développement remarquable au sein des structures hospitalières. Celui-ci s'explique d'une part par la possibilité de marquer la plupart des molécules de base de la biochimie au moyen de radio-isotopes émetteurs $\beta^+$, et d'autre part par l'apport de la détection en coïncidence des deux photons d'annihilation du positon émis par le radio-traceur.

Cette paire de photons permet de s'affranchir d'une collimation physique et rend possible l'exploitation complète de données tri-dimensionnelles de projection lors de l'étape de reconstruction tomographique [1]. Il s'ensuit un gain important en qualité d'image, par rapport aux techniques d'imagerie scintigraphiques par émission de photons uniques (gamma). Ces progrès technologiques ont conduit les physiciens à tenter d'optimiser le signal de TEP et s'attachant en particulier à y corriger les artefacts d'acquisition inhérents à la détection de coïncidences fortuites ou aux interactions (Compton, photoélectrique) subies par les photons d'annihilation avant leur détection. Ces dernières ont pu être efficacement prises en compte en couplant les TEP à des tomodensitomètres (scanner X). Au moyen des ces mesures de transmission d'un faisceau polychromatique de rayons X, il est en effet possible d'accéder à la distribution volumique des tissus osseux, musculo-graisseux ou aérien traversés par chaque paire de photons d'annihilation émise lors d'un examen TEP réalisé dans les mêmes conditions que la tomodensitométrie. Les travaux des physiciens ont alors permis de développer des techniques directes pour la correction de l'effet photoélectrique et des méthodes de simulation du diffusé Compton (méthodes de Monte-Carlo).

L'amélioration apportée par ces travaux lors de la dernière décennie a été telle que la plupart des TEP installés désormais sont couplés à un tomodensitomètre X. L'étape suivante consiste à optimiser la distribution volumique de radioactivité reconstruite en exploitant de façon adéquate la redondance des données de projection 3D en TEP. Dans ce qui suit, nous allons, à titre d'illustration, montrer comment une petite équipe de physiciens peut assez facilement participer aux progrès en la matière en ne disposant que de quelques accès aux TEP installés dans les structures hospitalières et des capacités de calcul désormais accessibles sur tout ordinateur personnel du commerce.

## 2 Reconstruction tomographique 3D directe.

Les progrès récents en reconstruction 3D directe dans le domaine de Fourier se fondent sur un résultat fondamental du à J. Radon (1917). Etant donné la connaissance des projections orthogonales $p_\omega$ d'une distribution volumique inconnue de radioactivité f selon une direction $\omega \in \Omega$, Radon montra que la transformée de Fourier de f dans le plan $\omega^\perp$ s'identifie à celle de ses projections.

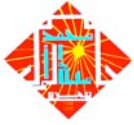 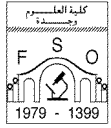 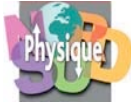 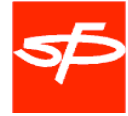

Pour qu'une reconstruction de la distribution f soit possible, il faut donc que le domaine Ω contienne au moins un arc de grand cercle (condition d'Orlov qui assure une affectation à tout voxel de l'espace 3D) et que les données de projection ne soient pas tronquées dans les plans $\omega^\perp$ (invariance des projections en translation).

La condition d'Orlov est vérifiée pour la géométrie cylindrique des TEP cliniques. En revanche, l'invariance des projections n'est pas vérifiée pour les projections obliques. Il est donc nécessaire d'estimer les données de projections obliques manquantes avant toute utilisation de la formule de Radon (ou de l'algorithme de rétroprojection filtrée qui en découle). Enfin, l'implémentation du théorème de Radon conduit à de délicates opérations d'interpolation dans les trois dimensions de l'espace de Fourier qu'il convient de maîtriser.

## 3 Contexte International

Une solution à ces difficultés a été proposée en 1997 par l'équipe de M. Defrise à Bruxelles [2]. Cette équipe a montré qu'au prix d'une approximation acceptable en pratique, les données de projection transverse pouvaient être optimisées en moyennant celles-ci avec des données de projection obliques recueillies à d'autres niveaux de coupes. Cette technique (« FORE ») permet de réorganiser les données de projections 3D en un jeu de projections 2D dont le rapport signal/bruit est amélioré. La reconstruction tomographique est alors réalisée coupe par coupe (2D). Cette méthode a immédiatement été implémentée dans les caméras TEP hospitalières, avec un succès certain.

Peu de temps après, en 2001 et 2006, l'équipe de S. Matej à l'université de Pennsylvanie a ravivé l'intérêt pour l'inversion directe au moyen du théorème de Radon 3D en proposant des techniques sophistiquées d'interpolation dans l'espace de Fourier [3], à condition de disposer de données de projection obliques non tronquées.

## 4 Méthode, analyse, résultats

Au sein de la faculté de médecine de Montpellier, une petite équipe composée d'un physicien, d'un mathématicien et d'un médecin nucléaire proposa de ré-écrire les équations qui avaient permis à M. Defrise de développer l'algorithme FORE et montra qu'il était possible d'estimer n'importe donnée de projection oblique manquante à partir des projections effectivement acquises, permettant ainsi de disposer d'un jeu de projections 3D complet et optimisé en terme de rapport signal/bruit sur les TEP cylindriques de routine clinique [4]. Par ailleurs, il fut possible de prouver que la réorganisation des données au sein d'un ou de deux plans de projection fixes permettait de diminuer le bruit de calcul lié à l'interpolation des données dans le domaine de Fourier 3D [5]. L'ensemble de ces résultats permet, à qualité d'image et à dose injectée au patient égales, de réduire d'un facteur 2 le temps d'acquisition des examens de TEP réalisés au sein des services de médecine nucléaire.

## 5 Pertinence dans le contexte maghrébin, africain et propositions de collaborations

Les facultés de médecine et les services médico-techniques des centres hospitaliers sont de plus en plus souvent amenés à utiliser des appareils de haute technologie que ce soit à des fins diagnostiques ou thérapeutiques. La définition de protocoles d'utilisation de ces appareils, leur maintenance tout comme le développement de méthodes innovantes en traitement ou analyse du signal sont autant d'occasions pour les physiciens de collaborer à l'amélioration des connaissances en la matière, d'améliorer la prise en charge des patients, voire d'exploiter commercialement leurs innovations.

Profitant de ressources technologiques acquises et exploitées par des centres de soins, ces travaux ne nécessitent aucun dispositif expérimental coûteux supplémentaire. Ils sont accessibles à tout scientifique disposant d'un simple ordinateur personnel et de ressources bibliographiques. La principale difficulté pour développer ce type de projet de recherche reste la nécessaire multi-disciplinarité des membres de l'équipe qui nécessite que chacun fasse un pas vers le domaine de l'autre de manière à définir avec précision des thèmes de recherche pertinents, c'est-à-dire utiles pour la prise en charge des patients, intéressants d'un point de vue scientifique et réalisables à court ou moyen terme.

## 6 Conclusion

Compte tenu de l'équipement déjà significatif des pays du Maghreb en appareils médicaux de haute technologie, de tels projets de recherche réalisés en collaboration entre physiciens et médecins pourraient être développés sans grande difficulté en Afrique du nord. Initiés au sein de structures universitaires, ces projets permettent relativement facilement de financer des étudiants en thèse de sciences, d'initier la création d'entreprises émergentes et de produire des outils directement utilisables au sein des services hospitaliers partenaires. Ils peuvent ainsi contribuer à l'essor économique des pays impliqués.

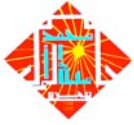 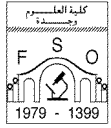 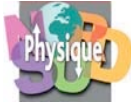 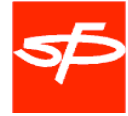

## Références

[1] P.E. Valk, D.L. Bailey, D.W. Towsend, M.N. Maisey. Positron Emission Tomography. Basic Sciences and Clinical Practice. 2003. Springer.

[2] M. Defrise, P.E. Kinahan, W. Townsend, C. Michel, M. Sibomana, D.F. Newport, "Exact and approximate rebinning algorithms for 3-D PET data" IEEE Trans Med Imaging, vol. 16, no. 10, pp.145-158, 1997.

[3] S. Matej, I.G Kazantsev. "Fourier-based reconstruction for fully 3-D PET: optimization of interpolation parameters" IEEE Trans. Med. Imag., vol.25, no. 7, pp.845-854, 2006.

[4] F. Ben Bouallègue, J.F. Crouzet, C. Comtat, M. Fourcade, B. Mohammadi, D. Mariano-Goulart. "Exact and Approximate Fourier Rebinning Algorithms for the Solution of the Data Truncation Problem in 3D PET"
Sous presse dans IEEE Trans Med Imaging 2007.

[5] D. Mariano-Goulart, J.F. Crouzet, "A new reconstruction method in the Fourier domain for 3D positron emission tomography," CR. *Physique*, vol. 6, pp. 133-137, 2005

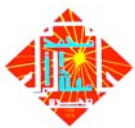 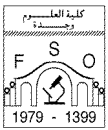 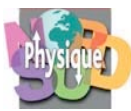 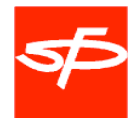

# « Le monde des nanosciences »

## Jean-Louis ROBERT

GES UMR 5650 CNRS- Université de Montpellier 2
34095 Montpellier cedex 5

Les nanosciences et les nanotechnologies s'intéressent aux objets et aux phénomènes à l'échelle du millionième de millimètre et révolutionnent des domaines aussi variés que la conception de matériaux, la microélectronique, les technologies de l'information, la biologie, la médecine et la pharmacologie ; elles sont susceptibles d'apporter de nouvelles solutions, par exemple dans le traitement de problèmes concernant l'environnement et l'énergie.

À l'échelle mondiale, les nanosciences et les nanotechnologies sont devenues un secteur stratégique en croissance rapide, avec un potentiel de développement économique considérable. Cependant, si les nanotechnologies offrent des perspectives nouvelles dans de nombreuses applications, leur développement soulève de nombreuses questions tant au plan éthique que sociétal.

Dans cette présentation, nous rappelons que les nanosciences portent sur l'étude et la compréhension des phénomènes observés à l'échelle du milliardième de mètre dans des structures, systèmes ou objets dont la taille est de quelques nanomètres et dont les propriétés découlent spécifiquement de cette taille nanométrique.

Nous montrons comment s'est développé ce domaine de recherche, selon deux approches: une approche descendante et une approche ascendante. Dans l'approche descendante, on utilise les procédés de nano-fabrication issus de la microélectronique pour réaliser des nano-objets ou nano-systèmes.

Dans l'approche ascendante, on fabrique ces nano-objets ou nano-systèmes « atome par atome » ou "molécule par molécule ».

La taille d'un nano-objet étant mille fois plus petite que la longueur d'onde de la lumière visible, il ne peut être observé au microscope optique. Inventé par Binning et Rohrer, prix Nobel de physique en 1985, le microscope à effet tunnel permet à la fois de voir et de manipuler les atomes : cette invention est sans aucun doute à l'origine de l'engouement des chercheurs pour les nanosciences.

Au travers de divers exemples de réalisation, nous insistons sur le caractère fortement interdisciplinaire de ce domaine de recherche : à l'échelle du nanomètre, l'analyse et la compréhension des phénomènes observés, qu'ils soient physiques, chimiques ou biologiques reposent sur la confrontation d'approches scientifiques et technologiques étroitement mêlées, En ce sens, nanosciences et nanotechnologies sont mariées de façon durable.

Nous soulignons également l'enjeu de la formation sous tous ses aspects (fondamental, technologique, éthique…) qui est considérable puisqu'elle conditionne le développement même de ce domaine d'activité.

En particulier, la création d'équipes interdisciplinaires suppose qu'au cours de leur formation les chercheurs aient acquis à la fois une grande maitrise des concepts de base et une large ouverture d'esprit vers les autres disciplines. Il s'agit là d'un défi à relever visant à décloisonner les disciplines.

Enfin, on doit noter que la métrologie, à l'échelle du nanomètre, est encore à inventer. La définition des protocoles d'élaboration et de caractérisation des nano-objets s'accompagne de la mise en place de nouveaux concepts et de nouveaux outils de mesure. Leur développement est primordial pour définir les normes nécessaires à la fabrication de produits issus des nanotechnologies et pour évaluer valablement les risques associés, en particulier, à leur utilisation.

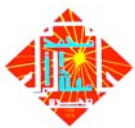 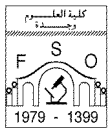 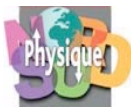 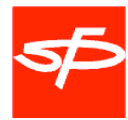

# Approche numérique des matériaux

Arezki MOKRANI

Institut des Matériaux Jean Rouxel UMR 6502
2, rue de la Houssinière, BP 32229
44322 Nantes cedex 3, France

**Résumé**

La modélisation et la simulation dans le domaine de la physique et chimie des matériaux a connu une forte évolution ces dernières années. Des codes de calculs de plus en plus complexes sont développés pour des simulations portant à la fois sur une grande gamme d'échelles, allant de l'électron à la nanostructure de plusieurs atomes avec les différents types de matériaux, solides organiques, inorganiques et sous différentes formes, amas libres ou déposés, surfaces, polymères, nanotubes, etc... Dans cette contribution, nous donnerons un exemple de calculs ab initio de structures électroniques basés sur la DFT, de matériaux magnétiques et semi-conducteurs.

## 1   Introduction

La modélisation numérique en physique, d'une façon générale et en science des matériaux en particulier a pris un essor considérable ces dernières années. L'approche numérique a pour but l'interprétation des observations expérimentales et la prédiction de nouvelles propriétés des matériaux. La compréhension des phénomènes physiques nécessite une description à l'échelle atomique où la taille et l'organisation géométrique jouent des rôles importants. Le défi majeur est de modéliser des systèmes aussi proches que possible de ceux élaborés au sein du laboratoire.

La complexité tant au niveau de la structure géométrique que sur la composition chimique font que la modélisation de ces systèmes nécessite tout un panel d'approches allant des méthodes semi-empiriques aux méthodes *ab initio*. Il est alors nécessaire de faire cohabiter les méthodes dites semi-empiriques pouvant traiter des systèmes très complexes et les méthodes a*b initio* plus précises mais applicables à des systèmes plus simples préservant un minimum de périodicité et présentant un nombre d'atomes non équivalents réduit.

Plusieurs équipes de chercheurs travaillant dans le domaine de la modélisation et la simulation ont été créées ces dernières années dans les Universités africaines et notamment en Afrique du Nord. Deux facteurs essentiels ont favorisé le développement de la « computer science » dans ces pays émergents. Il y a d'abord l'explosion des moyens de communication sur Internet qui facilitent l'accès aux codes de calculs et leurs documentations, ainsi que l'utilisation de centre de calculs à distance. Par ailleurs, les moyens de calculs de plus en plus performants deviennent maintenant accessibles aux chercheurs moyennant des budgets relativement modestes.

Dans ce qui suit nous présenterons quelques résultats de calculs de structure électronique faits en collaboration avec des chercheurs des Universités de Tizi-Ouzou et d'"Oran, Algérie.

## 2   Magnétisme des métaux de transition 3d dans le germanium

L'association de matériaux magnétiques et de matériaux semi-conducteurs a conduit à l'émergence d'une nouvelle génération de composés électroniques, où les porteurs se distinguent non seulement par leur charge mais aussi par leur spin qui constitue un degré de liberté supplémentaire. Rendre un semi-conducteur ferromagnétique à température ambiante devient alors un enjeu très important pour la communauté travaillant dans cette nouvelle discipline qu'est la spintronic.

Dans ce travail de modélisation mené en collaboration avec les Universités d'Oran et de Tizi-Ouzou, on s'intéresse au magnétisme des éléments magnétiques (Mn,Fe,Co,Ni) dans une matrice semi-conducteur de germanium. Des calculs de structure électronique *ab initio* dans le cadre de la DFT, utilisant le code de calcul TB-LMTO [1] ont été effectués dans le cas des sytèmes $XGe_2$ (X=Mn,Fe,Co) en massif et sous forme de films ultra-minces. La structure cristalline de ces composés est tétragonale de type $Al_2Cu$ (C16). Les résultats de l'optimisation de la géométrie par les calculs donnent des paramètres de mailles en parfait accord avec les mesures expérimentales dans le cas du composé $FeGe_2$ [2]. Dans tous les systèmes considérés, on cherche toutes les solutions correspondantes aux différentes configurations magnétiques ainsi que leur énergie totale permettant de préciser l'état fondamental.

Sur la figure 1 on présente les résultats obtenus dans le cas des composés MnGe2, FeGe2 et CoGe2 en volume. Les courbes (a), (b) et (c) présentent la variation de l'énergie totale en fonction du paramètre de maille. L'état fondamental correspond à la configuration ferromagnétique dans le cas du composé MnGe2 et anti-ferromagnétique pour le composé FeGe2, tandis que le CoGe2 et non magnétique.

Sur les courbes (d), (e) et (f) présentant les densités d'états locales des électrons d du métal de transition, on observe un déplacement du niveau de Fermi de gauche à droite du pic quand on passe du Mn, qui se trouve à gauche du Fe dans le tableau périodique, au Co qui se trouve à droite.

Dans le cas du composé NiGe2, le niveau de Fermi se déplaçant encore plus à droite, la densité d'états est très faible au niveau de Fermi et seule la solution non magnétique existe. Ces observations sont en parfait accord avec le critère de Stoner.

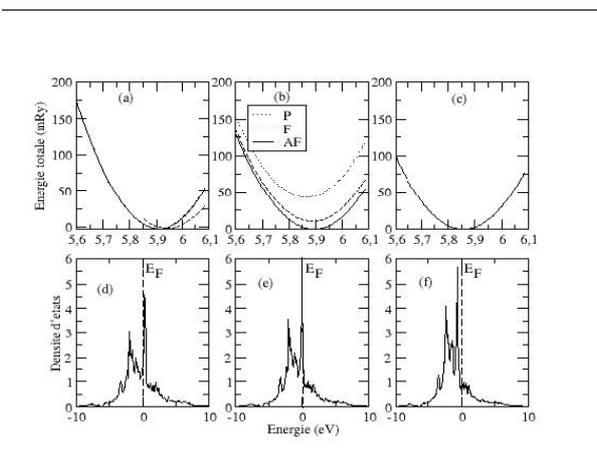

*a* pour les différentes configurations magnétiques obtenues pour les composés MnGe2 (a), FeGe2 (b) et CoGe2 (c). Les courbes (d), (e) et (f) présentent les densités d'états non magnétiques des électrons d du Mn, Fe et Co dans leur composé respectif.

Dans le cas de films ultra-minces (quelques plans atomiques), l'ordre magnétique est différent de celui du volume. On observe un ordre ferromagnétique dans le cas des films de FeGe2 jusqu'à une épaisseur de 7 plans atomiques de Fe et 6 plans de Ge pour lequel on retrouve la configuration anti-ferromagnétique du volume. Dans le cas des composés CoGe2 et NiGe2 non magnétiques en volume, on observe une apparition d'un ordre magnétique dans les cas des films ultraminces de CoGe2, tandis que les films de NiGe2 ne présentent aucun magnétisme [3-4].

Afin de tenir compte des effets de la relaxation du réseau dans le cas des films, une autre approche de type DFT avec des pseudo potentiels [5] a été utilisée. Les résultats de ces calculs montrent que seules les valeurs des moments magnétiques sont sensibles à la relaxation, les couplages magnétiques restant inchangés.

## Conclusion

Dans cette communication nous avons présenté un exemple de travail de modélisation numérique de matériaux fait en collaboration avec des chercheurs des universités d'Oran et de Tizi-Ouzou en Algérie. Ces collaborations s'articulent essentiellement sur des co-encadrements de thèses avec les universités de Nantes et de Strasbourg (ULP). Bien que les échanges de codes de modélisation et l'accès aux machines de calculs par les réseaux Internet soient très importants, les rencontres entre chercheurs restent indispensables. L'organisation d'écoles thématiques est un moyen de favoriser ces rencontres et permettre la transmission des connaissances.

En effet, les rencontres telle que celles organisées par la Société Française de Physique lors du congrès d'Oujda, favorisent non seulement les coopérations Nord-Sud mais créent aussi de nouvelles passerelles pour les coopérations Sud-Sud.

## Références

[1] O.K. Andersen, O. Jepsen, Phys. Rev. Lett.53(1984) 2571

[2] A. Siad, A. Mokrani, C. Demangeat, Surface Science 576 (2005) 158

[3] A. Mokrani, A. Siad, A. Lounis, M. Benakki and C. Demangeat, INTAS workshop, Hierarchy of Scales in Magnetic Nanostructures, Bochum (2007)

[4] A. Siad, Thèse de Doctorat, Université d'Oran (2007)

[5] Ab initio pseudopotential code DACAPO developed at CAMPOS (www.camp.dtu.dk/campos)

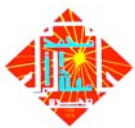 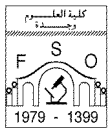 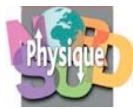 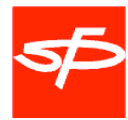

# Nano OxTi pour la conversion et le stockage de l'énergie solaire

Luc BROHAN, chercheur CNRS

Institut des Matériaux de Nantes, (IMN) CNRS - UMR 6502, 44322 Nantes

**Résumé**

Les oxydes de titane, structurés à l'échelle du nanomètre possèdent des propriétés remarquables qui doivent permettre entre autres la réalisation de cellules photovoltaïques de IIIème génération à haut rendement et le développement de photobatteries rechargeables. L'approche pluridisciplinaire du projet nécessite la mise en place d'un réseau de compétences international impliquant les pays du Maghreb.

## 1 Introduction

L'objectif général de cette communication est de favoriser la mise en place d'un réseau interdisciplinaire de compétence dans le domaine des nanomatériaux pour la conversion et le stockage de l'énergie solaire impliquant : la synthèse des nanomatériaux et leur intégration dans les dispositifs, incluant les approches théoriques (établissement d'une plateforme théorique de connaissances fondamentales) et analytiques (projet autour des grands instruments). Nous présentons quelques résultats récents ainsi que les avantages potentiels de ces nouveaux nanomatériaux photosensibles pour la réalisation de cellules photoélectrochimiques à haut rendement (III) et le développement de photobatteries.

## 2 Description du projet de recherche

Ce projet vise à créer une rupture technologique dans le domaine des cellules photovoltaïques et des photobatteries par le développement d'un système de conversion et de stockage de l'énergie solaire qui soit efficace, de faible coût, et qui réponde aux normes environnementales : cellules photovoltaïques de $III^{ème}$ génération à haut rendement (de type « metallic intermediate band solar cell » (MIBCELL)) et photobatterie rechargeable.

Outre la réalisation des photo-dispositifs dont les rendements sont susceptibles de dépasser les limites actuelles, les retombées scientifiques doivent permettre un accroissement considérable des performances des composés à base d'oxyde de titane dans les applications, photo-électrolyse de l'eau et photocatalyse (dépollution de l'eau et de purification de l'air).

## 3 Contexte International

### 3.1 Le photovoltaïque

Le secteur énergétique, au niveau mondial, doit accompagner une demande croissante avec une raréfaction des ressources tout en répondant aux menaces du changement climatique induites par l'émission de gaz à effet de serre[1] Les changements dans les technologies utilisées, le basculement des sources conventionnelles vers les énergies renouvelables, dépendront évidemment du prix relatif du kWh produit. Dans ce contexte, l'énergie solaire possède de nombreux atouts tant sur le plan environnemental qu'économique. En 2005, le marché mondial des installations photovoltaïques a atteint un record avec une capacité installée de 1460 mégawatts[2], correspondant à une croissance moyenne annuelle de 34% au cours des 10 dernières années[3].

---

[1] Report on the Basic Energy Sciences Workshop on Solar Energy Utilization April 18-21 2005 http://www.sc.doe.gov/bes/reports/files/SEU_rpt.pdf;
European Environment Agency, 2005. The European environment - State and outlook 2005. Copenhagen ; Greenhouse gas emission trends and projections in Europe 2006, EEA report No 9/2006 ;
Sciences Physiques Etats Unis, Nanosciences, Microélectronique, Matériaux Juin 2006.
[2] Marketbuzz 2006 report – http//www.seia.org/
[3] Lawrence L. Kazmerski, Solar photovoltaics R&D at the tipping point: A 2005 technology overview, Journal of Electron Spectroscopy and Related Phenomena 150 (2006) 105–135

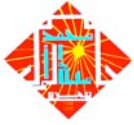 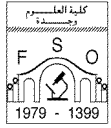 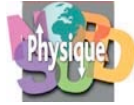 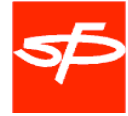

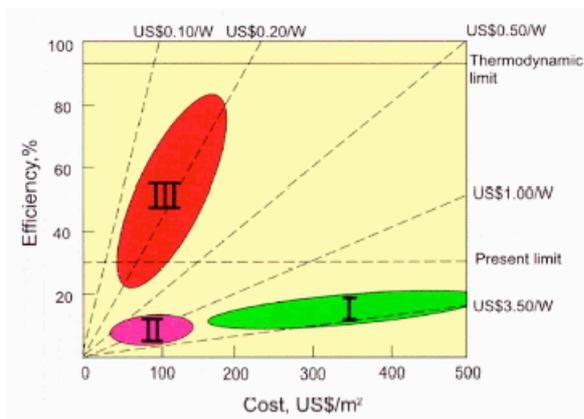

*Figure 1 : Comparaison de l'efficacité des cellules PV et de leur coût par unité de surface : la ligne diagonale indique le prix du module installé (2001) par Watt-crête. La limite théorique de 31% pour les cellules Shockley-Queisser (S-Q) est quasiment atteinte avec les cellules solaires à base de silicium mono-cristallin. Les cellules PV de III$^{ème}$ génération (cellules solaires multijonction ou semi-conducteurs présentant des effets de taille quantique) peuvent dépasser ces limites.*

Le rapport entre l'efficacité et le coût reste un critère déterminant aussi bien dans le choix des matériaux que dans la technologie à développer (fig. 1)[4]. Le paramètre pertinent à prendre en considération concerne le coût du système installé par Watt-crête (ligne diagonale en pointillés) qui dépend de l'efficacité de conversion, du prix des modules, de leur interconnexion et de la maintenance.

Dans un futur proche, le coût des matériaux et leur disponibilité orienteront l'évolution de la technologie photovoltaïque vers des cellules présentant les efficacités les plus élevées possibles. De nouveaux concepts sont nécessaires pour produire des dispositifs photovoltaïques de troisième génération (**III**) à faible coût et haute performance. Le photovoltaïque utilise les procédés de l'industrie électronique, et comme cette dernière, il va profondément évoluer avec l'émergence des nanotechnologies. Selon de récents rapports internationaux, « les nanosciences… permettront de dépasser les limitations actuelles et de générer les ruptures technologiques attendues pour le développement d'une économie basée sur les énergies renouvelables ».

Au-delà du module qui porte les cellules exposées au soleil, le stockage constitue le point faible de la filière énergétique et la recherche dans le domaine des photobatteries rechargeables reste encore à un stade exploratoire.

## 3.2 TiO$_2$ et Energies Renouvelables

Durant cette dernière décennie, les propriétés photochimiques de TiO$_2$ suscitent un engouement croissant auprès de la communauté scientifique, aussi bien dans les aspects fondamentaux qu'appliqués[5][6][7][8]. Les applications potentielles de TiO$_2$ concernent deux domaines environnementaux majeurs de ce siècle : la protection de l'environnement à travers la photocatalyse d'oxydation (PCO) (purification de l'air, dépollution de l'eau), la photohydrophylicité (PSH) (matériaux autonettoyants…)[9][10] et la production d'énergie avec le photovoltaïque (cellule photoélectrochimique à colorant : Grätzel)[11] et la photoélectrolyse de l'eau pour la production d'hydrogène. La création d'une paire électron trou, sous irradiation UV, à la fois sur le TiO$_2$, lui-même et les réactions chimiques ou de transfert d'électron qui s'en suivent sont au cœur des photo-dispositifs.

La compréhension des mécanismes photoélectrochimiques aux interfaces constitue, très probablement, une des clefs qui permettra d'optimiser la conversion de l'énergie solaire en énergie chimique[12][13]. Dans les réactions photocatalytiques, si la plupart des auteurs s'accordent sur la participation du catalyseur dans les processus de photodégradation des composés organiques, puisque les atomes de sa surface sont incorporés dans les produits de la réaction, des avancées majeures restent nécessaires. Parmi celles-ci, définir la nature chimique et électronique des états de surface et identifier la structure des sites actifs constituent actuellement les principaux verrous.

[4] M.A. Green, K. Emery, D.L. King, S. Igari, W. Warta, Prog. Photovolt: Res. Appl. 11 39 (2003).
[5] O. Carp, C.L. Huisman and A. Reller, Progress in Solid State Chemistry 2004, 32, 33–177
[6] D. M. Blake, NREL/TP-570-26797 (1999), NREL TP-510-31319 (2001) Report ; www.osti.gov/bridge
[7] Ulrike Dielbold, "The surface science of titanium dioxide", Surface Science Reports 48 (2003) 53-229.
[8] H.A. Al-Abadleh, V.H. Grassian / Surface Science Reports 52 (2003) 63–161.
[9] A. Mills, S.-K. Lee, Journal of Photochemistry and Photobiology A: Chemistry 152 (2002) 233–247.
[10] R. Wang, K. Hashimoto, A. Fujishima, M. Chikumi, E. Kojima, A. Kitamura, M. Shimohigoshi, T. Watanabe, Nature 388 (1997) 431.
[11] B. O'Regan and M. Grätzel, Nature, 353, (1991).
[12] Basic Research Needs for the Hydrogen Economy, Office of Science, US dep. of Energy May 13-15 (2003).
[13] Annual Energy Outlook 2004 with projection to 2025, DOE/EIA-0383, January (2004).

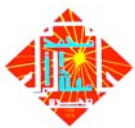 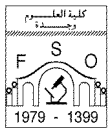 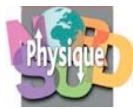 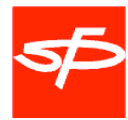

# 4 Méthode, analyse, résultats

**De la polycondensation des oxydes de titane…**

Afin d'améliorer les performances de l'oxyde de titane, il importe de maîtriser la morphologie et la nature des faces cristallines développées. La mise en œuvre de méthodes de chimie douce (T<200°C), acido-basique, permet de contrôler, non seulement, la nature de la variété allotropique de $TiO_2$ et la dimensionnalité du réseau iono-covalent (Ti-O) : 0D, 1D, 2D ou 3D, mais aussi la croissance des plans cristallographiques : nanofibres, nanofeuillets, demi-nanotubes ou nanorubans. Certains de ces nanomatériaux possèdent des propriétés remarquables dans le domaine de la photocatalyse d'oxydation[14] et du stockage de l'énergie. La cyclabilité des supercapacités, en charge - décharge et leur capacité de stockage croissent considérablement lorsque la dimension du matériau cathodiques diminue. En contrôlant le facteur de forme (rapport de la longueur sur le diamètre) des nanorubans de $TiO_2(B)$, la capacité de $Li_xTiO_2(B)$, pour x=0.5, augmente de 140 mAh.g$^{-1}$ à 175 mAh.g$^{-1}$. De plus, l'amélioration des vitesses de cyclage et l'accroissement de la réversibilité indiquent que la cinétique d'intercalation n'est plus limitée par la diffusion des ions $Li^+$ dans le solide.

**… à la génération d'une photobatterie**

Notre approche consiste à analyser, modéliser, comprendre et exploiter les propriétés originales des nanomatériaux récemment synthétisés à l'IMN, précurse[15] et sols-gels photosensibles[16] à base d'oxyde de titane, afin de développer de nouveaux photo-dispositifs et d'optimiser leur fonctionnement.

Les sols et gels d'oxyde de titane, structurés à l'échelle du nanomètre, possède une forte densité de sites actifs en raison d'un rapport surface/volume élevé et par suite une structure électronique dans laquelle les états de surface jouent un rôle déterminant. La présence de tels états est théoriquement susceptible d'accroître l'efficacité des cellules solaires photoélectrochimiques. Comparativement aux composés massifs, les sols et gels contenant des nanoparticules semi-conductrices, en interaction forte avec le solvant, possèdent des propriétés électroniques et optiques inusuelles, liées à des transferts de charges aux interfaces semiconducteur/adsorbat/solvant. Le changement réversible de coloration des échantillons correspondant à la réduction de $Ti^{IV+}$ et à l'oxydation de $Ti^{III+}$ et donc à la charge et à la décharge des électrons a permis la réalisation de dispositifs prototypes capables de convertir (cellule Photovoltaïque) et de stocker l'énergie solaire (Photobatterie rechargeable).

L'étudie des modifications structurales induites par l'irradiation, et des propriétés photo-électrochimiques aux interfaces (nature chimique et électronique des états de surface (MIBs), structure des sites actifs…) doivent permettre d'optimiser la conversion de l'énergie solaire en énergie chimique.
Nos orientations de recherche se déclinent selon sept axes:

- Aquérir une meilleure compréhension de la photo-électrochimie de l'interface oxyde/contre-ion, notamment en étudiant *in*-situ (EXAFS, XANES, XPS, RX…) les modifications structurales et électroniques photo-induites dans les nouveaux nanomatériaux, précurseurs et sols-gels photosensibles d'oxyde de titane.
- Apprendre à exploiter les nouveaux phénomènes physiques associés aux particules semi-conductrices de taille quantique (MIBs) présentes dans les sols-gels (absorbeur) et développer de nouvelles méthodes de synthèse et de dépôt adaptées.
- Comprendre et exploiter les phénomènes d'auto-assemblage des précurseurs d'oxyde de titane (semi-conducteur de type n) pour réaliser des films minces présentant des réseaux mésoporeux ordonnés et multi-échelles.
- Caractériser la structure des dépôts (Réflectivité des rayons X, Diffusion des rayons X en incidence rasante GISAXS).
- Développer les outils analytiques (Diffraction de rayons X et Spectroscopie XANES résolues en temps et sous irradiation) et les méthodes théoriques nécessaires à la compréhension des processus photo-électrochimiques aux interfaces. La plupart des outils théoriques utilisés dans ce projet impliquent de nouvelles approches fondamentales qui seront validées par l'expérience. L'image microscopique fournie par la modélisation de ces oxydes et de l'interface oxyde/électrolyte permettra de corréler l'effet de l'irradiation aux modifications des propriétés structurales et électroniques de ces systèmes.
- Réaliser les cellules PV III$^{ème}$ génération et photo-batterie, mesurer leurs caractéristiques et optimiser leur fonctionnement afin de dépasser de manière significative les limites actuelles.
- Intégrer les cellules dans des dispositifs pour exploitation.

---

[14] C.-W. Peng, M. Richard-Plouet, T.-Y. Ke, C.-Y. Lee, H.-T. Chiu, C. Marhic, E. Puzenat and L. Brohan, Advanced functional materials, soumise (2007).

[15] L. Brohan, H. Sutrisno, E. Puzenat, A. Rouet, H. Terrisse. Aquo-oxo chlorure de titane, procédé pour sa préparation. French CNRS patent priority N° 0305619 (09/05/2003), L. Brohan, H. Sutrisno, E. Puzenat, A. Rouet., H. Terrisse. Titanium aquo-oxochloride and preparation method thereof. International Publication n° WO 2004/101436 A2 ( 25/11/2004). European patent (EP) n° 04 742 604.4 (24/11/2005). Japan (JP) CNRS patent n°2006-530327 (16/10/2006). United States (US) CNRS patent n° 018344/0578 (04/02/2006).

[16] L. Brohan, H. Sutrisno, Y. Piffard, M. Caldes, O. Joubert. Polymère sol-gel à base de $TiO_2$, French CNRS patent priority N° 0201055 (29/01/2002). L. Brohan, H. Sutrisno, Y. Piffard, M. Caldes, O. Joubert, E. Puzenat, A. Rouet. Polymère sol-gel à base d'oxyde de titane. International Publication N° WO 03/064324 A3 (07/08/2003). European (EP) patent n° 03 734 737.4 (14.01.2003). United States (US) patent n° (in progress). Japan (JP) patent n° 2003-563956 (03/08/2004).

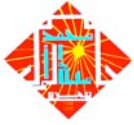 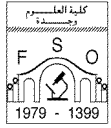 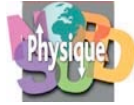 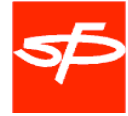

### Collaborations

Ce programme s'inscrit dans le cadre du projet régional (Pays de la Loire) PERLE (Pôle Emergent pour la Recherche Ligérienne en Energie) et bénéficie des soutiens financiers suivants :
- ACI Nano (2004-2007) (coordinateur L. Brohan).
- ANR PV (2006-2009) (coordinateur L. Brohan).
- Région Pays de la Loire (Bourse Post-doc 2007)
- Total, Energies Renouvelables (BDI 2006-2009),
- Vaillant - Saunier-Duval (CIFRE 2006-2009).

L'approche pluridisciplinaire du projet (**ANR-06-PSPV-015**) allie les compétences de chimistes (IMN/CS, UMR 6502), physiciens (LPEC, UMR 6087 Le Mans, GMCM, UMR 6626 Rennes) et théoriciens (CTMM, UMR 5636, Montpellier) à celles de spécialistes des couches minces et des interfaces (IMN/PCM, UMR 6502, LPEC, UMR 6087), du photovoltaïque (POMA, UMR 6136, Angers), des dispositifs (LAAS, UPR 8001, Toulouse) et de deux industriels fortement impliqués dans ce domaine, TOTAL (Energies Renouvelables) et Vaillant (Saunier-Duval).

Le groupe de chimistes du solide a acquis, de longue date, une reconnaissance internationale dans le domaine de la chimie douce et des oxydes de titane. Celle-ci s'est étendue récemment à la photocatalyse à travers les réseaux COST 540 WG1 ("new nanoarchitectures for better photocatalysis in Europe") et EJIPAC (« European-Japanese Initiative on Photocatalytic Applications and Commercialization »), aux nanomatériaux (SLONANO-06) et au photovoltaïque (IPS-International Conference on Photochemical Conversion and Storage of Solar Energy). Un partenariat avec Taïwan a été mis en place (2 thèses soutenues (2001, 2006) dont une en co-tutelle) et une extension internationale est en cours avec la Slovénie.

## 5  Pertinence dans le contexte Maghrébin, Africain et propositions de collaborations

### Contexte

La part des énergies renouvelables dans la consommation énergétique mondiale, évaluée aujourd'hui à 6%, devrait se situer à hauteur de 12% d'ici à 2020. La promotion des énergies renouvelables constitue un des grands axes de la politique énergétique et environnementale de l'Afrique et des Pays du Maghreb en particulier[17]. Le Maghreb dispose de ressources suffisantes pour satisfaire tous ses besoins énergétiques en raison d'un potentiel certain en matière d'ensoleillement : entre 2000 et 3900 heures par an. L'énergie quotidienne reçue sur la majeure partie du territoire peut être estimée à 5 KWh/m$^2$, soit une puissance d'environ 1700 KWh/m$^2$/an dans le Nord et 2200 KWh/m$^2$/an dans les régions du Sud. Les secteurs pouvant tirer profit de ce potentiel énergétique sont multiples, allant de l'électrification (avec ou sans raccordement aux réseaux nationaux), jusqu'à l'alimentation en eau des populations en passant par l'irrigation agricole et la production d'eau chaude sanitaire. La promotion de l'énergie solaire qui vise à diversifier les sources énergies et protéger l'environnement (protocole de Kyoto), nécessite toutefois de réduire ses coûts d'utilisation (énergie solaire Photovoltaïque) par rapport aux énergies traditionnelles.

### Propositions de collaborations

La mise en place d'un partenariat de R&D sur le thème conversion et stockage de l'énergie solaire est en cours avec l'Algérie. Cette activité concerne principalement des mesures de caractéristiques I/V *in situ*, de cellules photovoltaïques de II$^{ème}$ et III$^{ème}$ générations au Centre International des Energies Renouvelables localisé à Ghardaïa au sud de l'Algérie.

Par ailleurs, nous proposons d'intégrer notre projet au réseau développé dans le cadre des programmes JOULE et INCO de la commission Européenne et la société algérienne Sonelgaz[18]. Des collaborations avec nos collègues d'Afrique du sud sont en cours de discussion et un projet ERA-net sera déposé en 2008.

---

[17] Eau Energie développement durable, http://europa.eu/scadplus/leg/fr/lvb/r12532.htm,
Revue Stratégie du réseau Energie N°40 (2006): "Les politiques de soutien à l'énergie solaire" - DGTPE http://www.missioneco.org/documents/303/115250.pdf,
[18] A. Maafi, Renewable Energy 20 (2000) 9-17

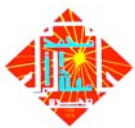
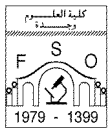
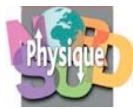
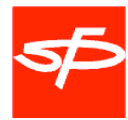

Des projets communs de recherche et/ou d'enseignement pourraient se développer dans les prochains mois avec :
- l'université de Tlemcen qui en partenariat avec L'École Nationale Polytechnique, l'Université de Constantine et le CDER[19] (Algérie) a récemment créé (2006-2007) une école Doctorale "Énergies Renouvelables".
- le laboratoire de Physico-chimie des Semiconducteurs (L.P.C.S) à l'université Mentouri de Constantine (Algérie) (Professeur Chari Abdelhamid) (Co-tutelle de thèse).

# 6 Conclusion

Cette première conférence a contribuée à la définition d'objectifs communs. Le projet présenté devrait contribuer à apporter des solutions durables aux difficultés de diffusion et d'utilisation des Energies Renouvelables. Il reste à coordonner nos actions, et à rendre plus cohérentes nos stratégies internes pour parvenir à un système énergétique mondial efficace et plus conforme aux objectifs du développement durable.

**Remerciements**
Je voudrais remercier les coordinateurs d'avoir pris l'initiative de cette 1ère rencontre internationale Nord-Sud, et féliciter les organisatrices (eurs) pour leur gentillesse et leur disponibilité.

---

[19] Programme d'énergies renouvelables en Algérie, journal El Watan, 22 février 2006

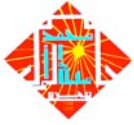 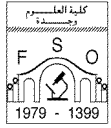 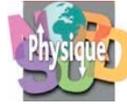 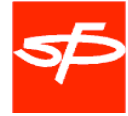

# Les fluides et la rhéologie

Etienne GUYON

Laboratoire de physique et mécanique des milieux hétérogènes (PMMH)
ESPCI 10 rue Vauquelin 75005 Paris, France

**Résumé**

La mécanique des fluides a longtemps été considérée en France comme un domaine privilégié des mathématiques appliquées. Une approche duale, mise en place par des physciens issus de la matière condensée, des matériaux et de la physique statistique vise à privilégier l'observation ainsi que le recours à l'image en faisant appel à des situations modèles. Elle commbine des expériences et des analyses simples mises en regard l'une de l'autre. Elle se prête à des enseignements au niveau L et au delà, et débouche sur des applications dans des domaines variés des sciences de l'ingénieur et de celles de l'environnement en particulier. Nous illustrerons ce propos sur des exemples de mécanique des fluides newtoniens et de rhéologie.

## 1 Introduction

Nous disposons aujourd'hui d'un matériel pédagogique important pour enseigner la mécanique des fluides. Le recours à l'image initié par un *album of fluid motions(1)* et une remarquable série de films édité par *a national committee for fluid mechanics(2)* dans les années 1970 a suscité l'interet d'une communauté de physiciens en France autour de P.G. de Gennes. Nous nous sommes efforcés de dégager la pédagogie qui pouvait être asssociée à l'étude de ces images dans *ce que disent les fluide(3)*. Par ailleurs un enseignement d'*hydrodynamique physique*
 *(4)* s'est mis en place à l'Université à coté d'un apppproche plus traditionelle en France proche des mathématiquees appliquées. Elle utilise des raisonnements simples basés sur l'analyse dimensionnelle, le recours à l'analogie et aux nombres sans dimensions ainsi qu'un aller et retour entre des propriétés physico chimiques des matériaux au niveau microscopique et des lois de comportement. Les exposés de S. Douady sur les milieux granulaires et la formation des dunes ainsi que celui de H. Vandamme sur les matériaux de construction sont des illustrations de ce type d'approche et montrent leur interet pour des applications

## 2 Les fluides newtoniens

### 2.1 forces et déformations

Les déformations d'un solide élastique peuvent être mis en regard avec les taux de déformation d'un fluide visqueux. Si les forces sont définies sous une forme tensorielle (tenseur des contraintes) il est possible de présenter la relation entre forces et déformations, et de proposer des applications les lois linéaires de la mécanique, à partir de situations simples (élongation, cisaillement). Le problème des grandes déformations rencontrées en particulier en Sciences de la Terre peut alors être abordé. Ainsi les coëfficients tels que le module d'Young le coefficient de Poisson et la viscosité deviennent des notions familières au même titre que la résistance électrique. Un travail de même type peut être fait sur les forces en surface et la capillarité. Nous disposons d'un ouvrage *goutttes, bulles, perleset ondes (5)* enrichi d'un CD présentant 150 expériences simples sur la capillarité. Des TP tels que ceux présentés à Oujda dans l'atelier 1 sur *l'armoire de la physique* (écoulement de Poiseuille, chute d'une bille dans un tube; phénomènes capillaires) font ainsi appel à des des mesures simples de mécanique des fluides.

### 2.2 viscosité dynamique et cinématique

En régime instationnaire, la viscosité cinématique est un coefficient de diffusivité de la quantité de mouvement d'un fluide. L'utilisation d'un récipient posé sur une platine de tourne disques offre une riche palette d'expériences (6) allant de la mise en évidence de cette viscosité cinématique, des effets d'écoulements secondaires omniprésents lorsque l'effet de la viscosité n'est pas assez efficace pour modifier le mouvement d'un fluide ( cas des fluides peu visqueux). Ele permet aussi de faire connaissance avec la physique riche et subtile des écoulements en rotation et de l'effet des forces de Coriolis qui se superpose à la force centrifuge, à l'échelle planétaire.

Le rapport entre le temps de la diffusion par la viscosité à celui des mouvements convectifs à la même échelle est le *nombre de Reynolds,* Re. Il caractérise les conditions de passage d'un régime laminaire à turbulent.

### 2.3 Le mélange

Une des secteurs d' applications les plus importantes de la mécanique des fluides concerne les problèmes de mélange, qu'il s'agisse de Génie des Procédés et de l'Environnement. Le mélange aux petites échelles (tel qu'il intervient en microfluidique) est généralement peu efficace d'autant que dans de nombreux liquides le *nombre de Prandtl*, rapport d'un temps de diffusion moléculaire à un temps de diffusion viqueuse, peut être élevé (huiles silicones). La réversibilité obtenue aux petits nombres de Reynolds (faux mélange) peut être contrariée par des stratégies de *mélange chaotique* (superbe application directe du chaos déterministe à la mécanique des fluides). Le mélange turbulent, et la trubulence aux grands Re ne sont toujours pas totalement compris. Se combinent des effets de turbulence statistique et de grandes structures cohérentes ( exemple en météorologie).

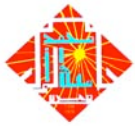 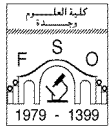 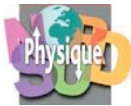 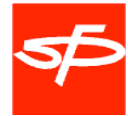

## 3    La rhéologie

### 3.1  viscoélasticité

Un temps intrinsèque des matériaux $T_i$ controle des comportements intermédaires entre ceux d'un solide élastique et d'un liquide, sa *viscoélasticité*. Dans le cas de l'eau ce temps est très court ($10^{-12}$ s) et seul son ocmportement visqueux est observé. Dans un polymère, des constantes de temps $T_i$ longues sont dues aux désenchevètrement des chaînes ou à leur étirement sous écoulement . Ce temps est mis en regard de celui de l'observateur $T_O$ ( temps d'observation, période d'une oscillation imposée ou, encore, inverse d'un taux de cisaillement). Le rapport $T_1/T_O$, ou *nombre de Deborah*, est petit pour un comportement de type liquide, et grand pour le solide. Des bons ouvrages d'enseignement de rhéologie (7) existent qui montrent le vaste spectre des applications et s'appuient sur des données complémentaires de chimie. Ce riche domaine de *la matière molle* connait un développement considérable avec des actions thématiques en France et une collaboration avec le Maghreb (colloque de Marrakech organisé par la société marocaine des polymères ) qui devrait pouvoir être entretenue dans le cas d'une grande société de physique marocaine.

### 3.2 comportements complexes

Les comportements non linéaires obtenus sous forte sollicitation tels que *rupture fragile ou ductile* sont d'une importance cruciale et sont omniprésents. Il peut sembler étonnant qu'ils ne soient pas introduits dans un enseignement scientifique généraliste. Les notions de constante de temps permettent de comprendre ces deux régimes et la façon dont on passe de l'un à l'autre en variant la température ou la composition. Ceci concerne aussi bien la métallurgie que les polymères et se prête aussi à une description multiéchelles qui va des dislocations ou de la fracture élémentaire à l'ouvrage d'art. Les matériaux de construction traités par H.Vandamme sont un domaine privilégié d'applications qui concerne prioritairement les pays du Sud.
Les effets de vieillissement et d'évolution lente au cours du temps sont aussi d'importance cruciale, en particulier dans le cas des matériaux de construction.
Un travail récente sur l'évolution sur la rhéologie des boues (8) a été présenté expérimentalement à Oujda. Il met en jeu des effets de *thixotropie* ( diminution de la viscosité lorsque le cisaillement augmente) , de *veillissement* ( durcissement au cours du temps), et de *fluide à seuil* (la boue ne s'écoule que au dessus d'une contrainte seuil). De tels effets sont importants pour comprendre les effets de glissement de terrain, par exemple. Ils sont exploités dans des applications vairées ( génie pétrolier ; peintures..).

## 4   Propositions  de collaboration

Le réseau MIAM ( MIlieux Aléatoires Macroscopiques; voir le bulletin *info-miam*) que nous créé et que nous animons (avec Daniel Bideau en particulier ) est très actif en France. Il a mis en place des Ecoles thématiques qui pourraient être prolongées dans le cadre d'un projet des "Houches du Maghreb" suggéré à Oujda. sur la thématique des matériaux désordonnés. La constitution d'un réseau thématique sur cette thématique à l'échelle d'un pays maghrébin ou inter-maghrébine serait une bonne chose. Ce reseau pourrait être la base d'un master "delocalisé" voire d'une ecole doctorale également multi sites. Ce permettrait de résoudre le problème de la masse critique scientifique, inhérent à la fois au niveau master et école doctorale. Ceci nécessite avant tout un état des lieux relativement précis des forces en présence.
Il existe aussi une action sables et désertification pilotée par D. Bideau dont la présentation de Stéphane Douady a rendu partiellement compte. Celle ci a donné lieu à des Ecoles entre autres en Tunisie et Mauritanie et est active dans les pays du Maghreb.
Soulignons aussi les actions de diffusion des savoirs. Les expositions *grains de sable* et *mozaïque de la physique*s ainsi qu'un projet en cours *architecture de terre* sont héritières des recherches actuelles en France et connaissent une diffusion en Afrique dont la présentation de Michel Darche (atelier 3) a bien rendu compte.

Enfin, ces sujets se prêtent à une formation par un apprentissage actif de type *main à la pate* à l'Ecole *(*voir les livres de la série *graines de science(9)*). Un numéro récent de l'excellente revue pédagogique *Textes et documents pour la classe* (TDC) traite de*s écoulements de matière (10)* et offre des séquences pédagogiques pour le collège et le lycée qui vont dans le sens du développement de l'apprentissage expérimental dont le congrès de Oujda a souligné la nécessité, sans avoir à recourir à des équipements chers ou sophistiqués.

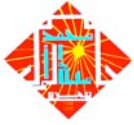 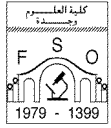 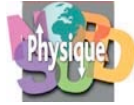 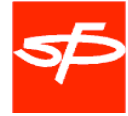

# Références

[1] *An album of fluid motion* M. Van Dyke (Parabolic press)

[2 *N.C.F.M. films*. De nombreuses séquences se trouvent dans le CD rom multilingues *multimedia fluid mechanics* C.U.P (2002)

[3] *Ce que disent les fluides* E.G, J.P. Hulin & L. Petit Belin (2006)

[4] *Hydrodynamique physique* E.G., J.P. Hulin & L.Petit - EDP sciences (2001)

[5] *Gouttes, bulles, perles et ondes* P.G. de Gennes, F. Brochard & D. Quéré Belin (2005)

[6] *exercices Hydrodynamique Physique* M. Fermigier Dunod (1999)

[7] *comprendre la rhéologie* P.Coussot & J.L. Grossiord EDP sciences (2002)

[8] P. Coussot & al Phys.Rev.lett. $\underline{88}$ 175201 (2002)

[9] *Graines de science 5 et 8* Le pommier (2003 et 2008)

[10] *les écoulements de la matière* Editions CNDP (15 avril 2007)

.

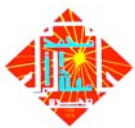 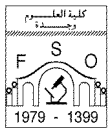 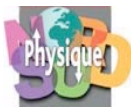 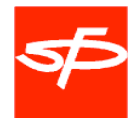

# Matériaux de construction et milieux granulaires consolidés

## Henri VAN DAMME

Physico-Chimie des Polymères et des Milieux Dispersés,
Ecole Supérieure de Physique et Chimie Industrielles (ESPCI), Paris, France

**Résumé**

En plus de sa beauté poignante, l'architecture vernaculaire recelle des trésors de sobriété et d'efficacité techniques qui, dans un contexte de boulversement climatique probable et de développement juste et durable, méritent une re-visite profonde. La physique et la physico-chimie de la matière en grains et de son espace dual – l'espace poreux, y compris les fluides qu'il peut contenir – fournissent le cadre naturel de cette démarche.

## 1 Habitat et urbanisme : les défis

Le terre portera probablement 9 milliards d'habitants dans quarante ans et sans doute 12 milliards à la fin du siècle. Cette augmentation brute de la population mondiale cache une autre évolution, tout aussi importante : c'est son urbanisation. En 2000, 26% de la population des pays du Sud vivaient en zone urbaine. En 2025, ce chiffre devrait passer à 37% en Asie, 42% en Afrique et 75% en Amérique latine. Un cinquième de la population urbaine vivra alors dans de très grandes villes, des mégalopoles de plus de 4 millions d'habitants. La pression sur l'environnement de telles concentrations humaines est énorme : consommation d'eau, énergie, alimentation, évacuation des déchets…

Malheureusement, l'urbanisation va aussi de pair avec l'augmentation des bidonvilles. La terre compte à ce jour près d'un milliard d'êtres humains entassés dans des bidonvilles, soit près d'un habitant de ville sur trois. Selon le Programme des Nations Unies pour les établissements humains (« ONU-Habitat »), cette situation ne devrait pas s'améliorer. On estime à 1,4 milliard le nombre des habitants de bidonvilles en 2020, c'est-à-dire autant que la population de toute la Chine.

Tout ceci pose l'urbanisme, au sens le plus large du terme, comme l'une des sciences-clés de notre futur presque immédiat. Nous pourrions sans doute continuer à nous développer en étendant à la terre entière la médecine et les moyens de communication informatiques que nous connaissons actuellement dans le monde développé, mais nous ne pourrons *pas* généraliser nos modes de construction et de vie collective sans courrir à la catastrophe. Le bâtiment est responsable d'environ la moitié de notre consommation d'énergie et de 25% de notre production de $CO_2$. Ceci concerne la construction, qui est assez brève, et, surtout, la *vie* du bâtiment, qui est beaucoup plus longue. Les transports représentent l'autre poste majeur. Les schémas de développement urbain, qui intègrent habitat et mobilité, sont donc en première ligne des améliorations possibles.

La structure des villes et leur dynamique sont incontestablement des points essentiels, mais leur matérialité l'est tout autant. Le ciment – le composant-clé du béton, le matériau le plus utilisé au monde avec 9 milliards de $m^3$ coulés par an – est à lui-seul responsable de 10% des émissions de $CO_2$. Et les « gisements » d'utilisation accrue de ce matériau que représentent les pays émergents sont incomparablement supérieurs à ce que nous connaissons pour le moment. Dans le même temps, le bâtiment est en passe de détrôner (et l'a peut-être déjà fait) l'emballage comme premier domaine d'utilisation des polymères de synthèse, issus du pétrole. Le verre, malgré son image environnementale excellente, reste un matériau extrêmement énergivore, même recyclé, tout comme l'acier. Tout ceci fait qu'il est impératif d'inventer de nouveaux matériaux de structure ou, à tout le moins, de faire évoluer ceux que nous connaissons.

## 2 Le projet : jeter un regard neuf sur la terre et établir des ponts avec d'autres matériaux

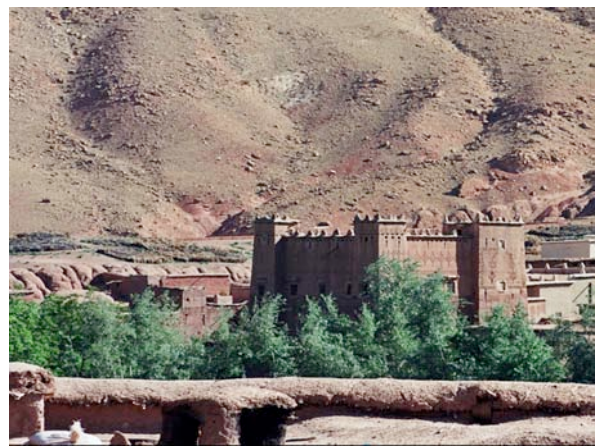

Fig. 1. Un des nombreux ksars en terre crue qui jalonnent les vallées de l'Atlas.

La re-visitation de certaines pratiques constructives anciennes comme la construction en terre crue est, face aux défis qui viennent d'être évoqués, une source puissante d'inspiration. On estime à 30% la fraction de la population mondiale vivant dans des habitations en terre : briques de boue séchée (adobes), terre compactée, torchis… Rien qu'en France, on estime à plus d'un million le nombre de maisons de ce type, certaines, encore (bien) habitées, vieilles de plus de trois siècles. En architecture, c'est près de 20% des édifices inscrits au patrimoine mondial de l'UNESCO qui sont en terre. On y trouve des forteresses, des édifices religieux des habitats collectifs, et même des villes denses d'immeubles allant juqu'à quatorze étages. L'Afrique Maghrebine – en particulier, la région de l'Atlas – et l'Afrique sub-saharienne recèlent de véritables trésors.

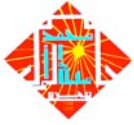 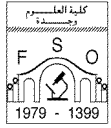 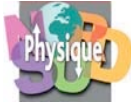 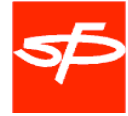

Ce savoir traditionnel qui a permis, avec des moyens frugaux, des réalisations remarquables, non seulement sur le plan architectural mais aussi sur le plan bioclimatique, mérite d'être revisité. Au premier chef, c'est la physique du matériau terre, véritable béton naturel, qu'il faut comprendre. Ceci concerne sa cohésion et, plus généralement, ses propriétés mécaniques mais aussi sa relation particulière avec l'eau, y compris les changements de phase de cette dernière en milieu confiné. Cette re-visite aura également des retombées dans la valorisation agricole de sols de faible qualité.

Il serait faux de croire qu'il s'agit d'une physique bien établie. Par exemple, aussi étonnant que cela paraisse, le simple phénomène de cohésion capillaire entre deux grains sphériques – plus exactement, sa dépendance par rapport à la quantité d'eau dans le pont capillaire – n'est à peu près correctement compris que depuis une dizaine d'années. C'est la prise en compte d'une propriété complètement négligée jusque là – la rugosité de surface – qui a donné la clé du problème [1-3].

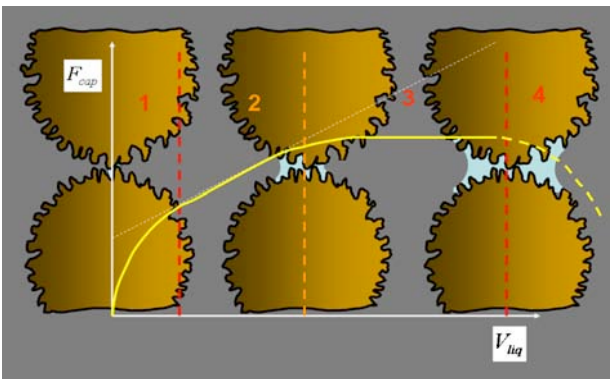

Fig.2. L'existence d'une rugosité introduit plusieurs régimes dans la cohésion capillaire entre grains sphéroïdaux. Le résultat est le passage par un optimum de cohésion pour un contenu en eau relié à la taille des grains et aux paramètres de cette rugosité.

Une autre question est le rôle de la polydispersité granulaire extrême du matériau terre, avec des tailles de grains allant de quelques centimètres au nanomètre. Les premiers résultats permettent de comprendre le rôle particulier que joue le tandem argile-eau dans cette cohésion.

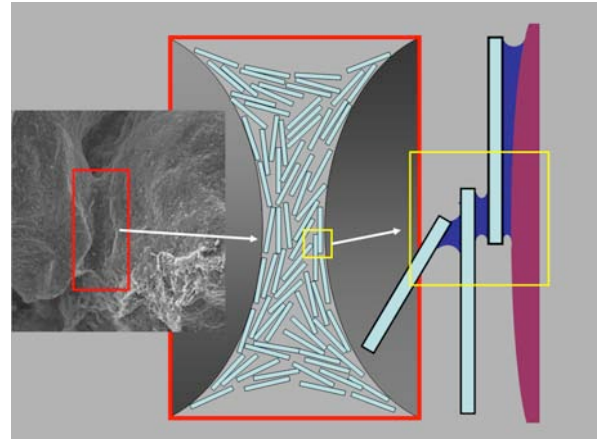

Fig. 3. Un modèle de terre, constitué de grains de sable et d'argile, permet d'étudier les bases de son comportement mécanique [David Gelard, thèse, UJF, Grenoble, 2005]

Bien des comportements typiques de la terre, comme les écoulements pâteux par exemple, recèlent encore des questions ouvertes. Que dire pour l'instant d'une propriété comme la plasticité, dont la phénoménologie est maîtrisée en mécanique des sols, mais dont la physique est balbutiante ? (sur ces deux points, voir la contribution d'Etienne Guyon).

Loin d'opposer des matériaux, cette re-visite de la terre, d'un point de vue de physicien, devrait permettre d'aboutir à une vision unifiée de nombreux matériaux de construction come le béton, la brique, le plâtre, les bétons de chaux, et même les bétons organiques. Loin de les opposer sur la base d'une logique commerciale ou « ecolo-dogmatique », la construction d'une base de connaissance physique devrait permettre d'effectuer les meilleurs choix.

## 3 Proposition de collaboration

Le projet de physique des **milieux granulaires consolidés** résumé ci-dessus est pertinent pour tous les partenaires d'une action nord-sud car les problèmes liés à l'évolution des matériaux de construction, de l'habitat et de l'urbanisme sont généraux. Il est complémentaire à la physique des dunes présentée par Stéphane Douady et à celle des fluides complexes présentée par Etienne Guyon. Ce projet d'ensemble peut se décliner en actions de recherches traditionnelles, en école du style des Houches, en actions d'éveil dans les écoles dans le style de *La Main à la Pâte*, ou encore en expositions itinérantes.

**Références**

[1] *What keeps sandcatles up?* D.J. Hornbaker et al., Nature, 387, 765 (1997)
[2] *How sandcastles fall,* Th. C. Halsey & A. J. Levine Phys. Rev. Lett., 80, 3141 (1998)
[3] *Moisture-induced ageing in granular media and the kinetics of capillary condensation,* L. Bocquet, E. Charlaix, S. Ciliberto & J. Crassous, Nature, 396, 735-737 (1998

# Mouvement et Chant des Dunes

Stéphane DOUADY

Laboratoire Matière et Systèmes Complexes (MSC),
CNRS-Université Paris-Diderot, Paris, France

**Résumé**

Les dunes, rares dans le Nord, sont une particularité du Sud, et plus particulièrement du Mahgreb. Leur mouvement, singulier ou collectif, leur réponse qu changement de vents, reste encore largement à étudier, d'autant plus au vue de l'impact pratique potentiel. Sans parler du surprenant chant des dunes. Leur étude est donc une priorité qui devrait être dévellopé dans le Sud en partenariat avec le Nord.

## 1 Introduction

Les dunes sont des "tas de sable formées par le vent". S'il se trouve du vent partout dans le monde, du sable libre de mouvement est plus rare, et se retrouve principalement dans les zones arides, comme au Sahara. Mais les dunes ne sont pas que des tas inertes, mais de véritables individus dont il faut comprendre la logique et le comportement.

Pour cela on peut s'appuyer sur deux méthodes, la première étant bien sûr l'observation sur le terrain. Une des premières étapes est de construire des méthodes de mesures à la foit simples, efficaces, et peu onéreuses, si l'on veut dévellopper un savoir faire local et une vraie autonomie. Mais un des désavantages inévitable est l'impossibilité d'expérimentation réelle, faute de contrôle et à cause de la soumission aux éléments naturels (le vent).

Pour cela il faut se tourner vers l'expérimentation en Laboratoire. Grâce au fait de remplacer l'air par de l'eau, on arrive à reproduire des dunes en laboratoires à l'échelle 1/1000eme (en fait au rapport de densité entre l'air et l'eau, voir figure 2). Cette technique qui commence a être bien contrôlée, reste peu onéreuse, et devrait être développée localement pour en assurer le contrôle et le dévellopement, en lein direct avec les besoins locaux.

Enfin le chant des dunes repose sur le mystère des sables musicaux, des sables bien particuliers qui ont en plus la fâcheuse tenance à s'user et à perdre leur capcités inhabituelles! Cela rends leur étude sur place nécessaire.

## 2 Mouvement des Dunes

### 2.1 L'Individu-Dune

La base de la dynamique des dunes a été posée depuis Bagnold[1]. La dune avance, le sable étant érodé de son dos par le vent, et redéposé plus en avant par avalanches. Comme la capacité du vent a entraîner le sable est a peu près constante, il en découle que la vitesse d'avancée des dunes est inversement proportionelle à leur taille.

Mais il reste encore à comprendre comment leur formes se décident, et comment leur individualité s'opère. Cela se comprends maintenant bien sur la bacrhane, la dune isolée en forme de croissant qui se forme dans un vent unidirectionnel : c'est le flux latéral de sable, qui, en suivant la plus grande pente, par du centre pour aller sur les côtés (figure 1) [2]. Ce faisant, ce flux de sable réduite l'érosion des bords, ce qui les ralentis, et ainsi un équilibre se forme, la différence de hauteur se calant sur une diférence de position (avancée) de telle sorte qu'avec les flux latéraux induit les vitesses soeint les mêmes. On, obtient ainsi un objet qui bouge de manière cohérente, un individu.

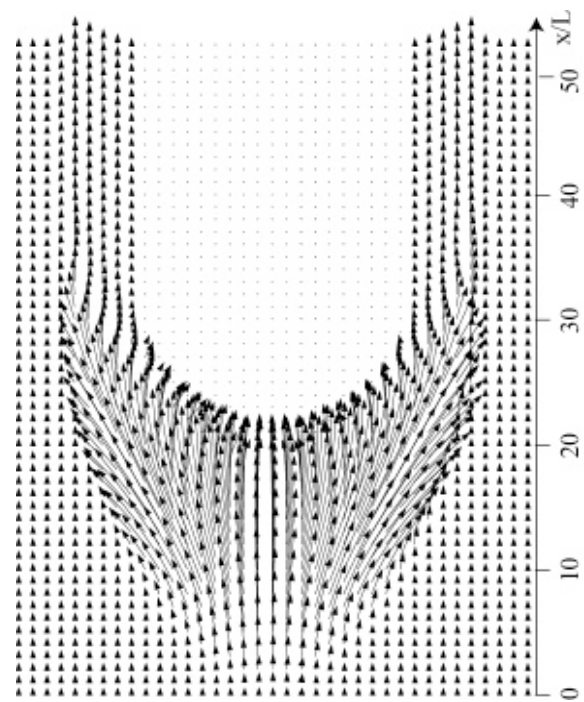

FIG. 1: Simulation numérique du flux de sable sur le dos d'un barchane. On voit le flux de sable latéral partant du centre vers les cornes.

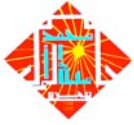 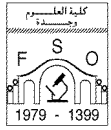 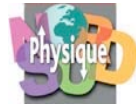 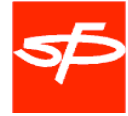

Ce raisonnement doit encore pouvoir être dévellopé sur des formes de dunes plus complexes, comme les dunes transversales, et leur transition avec les changemens de vents vers les dunes longitudinales, ou encore les dunes étoiles.

Mais il reste surtout un autre domaine d'intéraction essentiel à étudier: celui des intéractions entre ces individus-dunes (voir figure 2). En effet, les dunes perdent du sable, et un équilibre se fait par le sable qu'elles gagnes (provenant des autres dunes). Un équilibre peut s'établir, mais normalement cet équilibre est instable: une grosse dune aura tendance à gagner trop de sable et devenir de plus en plus grosse. Or on observe, comme sur la côte Occidentale du Sahara, des coridors de dunes barchanes qui conservent à peu près la même taille sur des centaines de kilomètres. L'explication de se phénomène ne peut provenir que des intéractions mutuelles des dunes, qui doit donc être étudié.

## 2.2 La Dune en Laboratoire

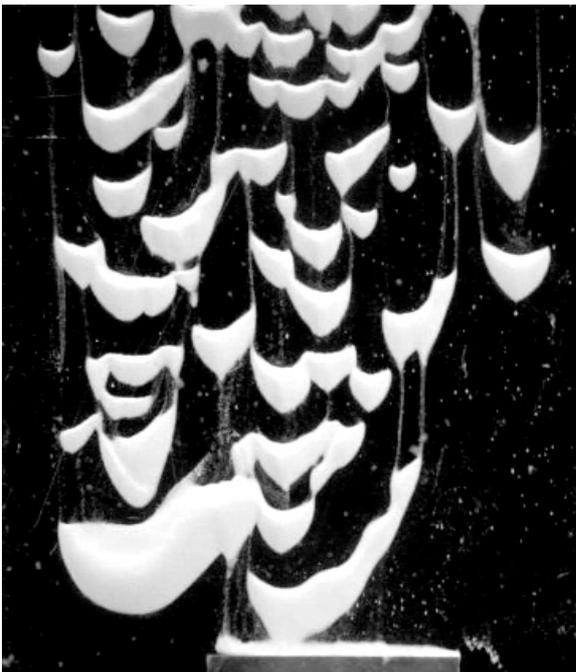

FIG. 2: Couloir de Barchanes reproduit en expérience à échelle réduite. On voit bien les flux de sables s'échapant des corne qui relient les différentes dunes.

Dans ce cadre la reproduction des dunes en laboratoire est un atout essentiel. Le principe est de mettre une petite quantité de sable dans un aquarium. Plutôt que de faire courrir l'eau (comme le vent) sur le fond fixe, il s'avère plus pratique de faire bouger rapidement le fond dans un aquarium immobile. Ce faisant on reproduit mieux les rafales de vent turbulent[3].

En faisant tourner la plaque, on peut aussi étudier de manière controler l'effet du changement de direction du vent, et ainsi chercher à reproduire les différent types de dunes, et comprendre les transitions entre elles.

## 2.3 La Dune Chantante

Certaines dunes ont la particularité d'émettre un son puissant (110 dB) lors qu'une avalanche assez importante interviens. Bien que l'on commence à comprendre que le son vient du mouvement synchronisé des grains de sable, ce phénomène conserve des mysères [4]. La première est de comprendre pourquoi ces grains de sables se synchronisent, au contraire d'autres grains de sable ordinaires. Le deuxième est de le relier aux particularités de ce sable, qui fait qu'il est justement "musical": en le triturant de différentes manières on arrive à en tirer des sons. Cela semble provenir non pas de la compositon, très variée, mais plus de la forme des grains, de leur tri, et surtout d'un vernis particulier, appelé "glaçure du désert". L'étude détaillée de ces phénomènes reste à faire, et rien ne semble plus aproprié que sur place, comme par exemple l'étude directe des conditions de dépôts de cette glacure.

# 3 Méthode, analyse, résultats, conclusion

Le suivi des dunes peut se faire avec des méhodes de triangulations simples, et des relevées de dépostions et d'érosion de sables au moyens de piques-repères. Le suivi du vent peut se faire avec des moyens courants. Enfin la reproduction en laboratoire ne demande pas de moyens très particuliers.

La compréhension du mouvement des dunes, individuels et collectifs, outre son aspect fondamental, et sa réapropriation par ses possesseurs, pourrait aussi conduire à des moyens originaux de lutte contre leur avancées.

### Remerciements
C'est un plaisir de remercier les coordinateurs des différentes sessions tout particulièrement les organisateurs et l'acceuil sur place!

### Références
[1] R. A. Bagnold. "The physics of blown sand and desert dunes", Chapman and Hall, London (1941).
[2] P. Hersen, "On the cresentic shape of Barchan Dunes", Eur Phys J B, (2004)
[3] P. Hersen, S. Douady et B. Andreotti "Relevant lengthscale of barchan Dunes", Phys. Rev. Lett. 89, 264301-4 (2002).
[4] S. Douady et al., "The song of the dunes as a self-synchronized instrument", PRL, 97, 018002 (2006)

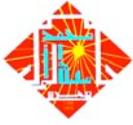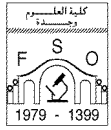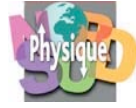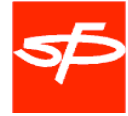

# Traitement
# de Grandes masses
# de Données

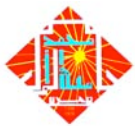 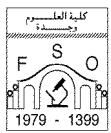 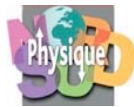 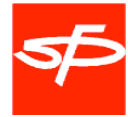

# La fracture numérique et les nouvelles technologies

## Dominique BOUTIGNY

Centre de Calcul de l'IN2P3 – Lyon / Villeurbanne - France

### Résumé

Dans cet exposé, nous donnons des éléments permettant d'appréhender le problème de l'inégalité d'accès aux technologies de l'information dans les pays émergeants. Nous présentons également quelques unes des techniques informatiques moderne sur lesquelles se basent certains domaines scientifiques. Enfin, nous proposons des pistes pour permettre aux pays du Sud d'aborder progressivement ces nouvelles technologies de l'information.

## 1 La fracture numérique

Selon l'encyclopédie collaborative en ligne, Wikipedia: *"La fracture numérique désigne le plus souvent l'inégalité d'accès aux technologies numériques, dont principalement l'ordinateur, et parfois le clivage entre les info-émetteurs et les info-récepteurs. Cette inégalité est fortement marquée entre les pays développés d'occident et les pays du Sud, dits en voie de développement"*.

Cette fracture numérique peut apparaître sous de nombreuses formes et n'est pas limitée aux pays en voie de développement. Une analyse détaillée du *"Centre de recherche pour l'étude et l'observation des conditions de vie"*[1], montre qu'il existe de multiples fractures numériques au sein de pays développés comme la France, liées à l'âge, à l'argent, à l'éducation etc…

### 1.1 Mesure de l'ampleur de la fracture numérique Nord / Sud

Créé à l'initiative de "L'International Committee on Future Accelerators (ICFA)", le SCIC ou "Standing Committee on Interregional Connectivity", étudie depuis plusieurs années l'évolution de la connectivité à l'Internet de nombreux sites scientifiques répartis sur la planète. Bien que non explicitement mentionné dans la mission du comité, le continent africain fait également l'objet de son étude.

Piloté par le Stanford Linear Accelerator Center (SLAC), le SCIC s'appuie sur un outil permettant de mesurer le temps de transit et le taux de perte de paquets d'informations transmises entre des sites serveurs et des sites receveurs. Ces deux grandeurs sont des estimateurs fiables de la qualité de la connectivité entre les sites.

Actuellement, 649 sites répartis dans 128 pays sont étudiés en permanence.

La figure suivante, extraite du rapport [2] montre l'évolution au cours du temps, de la bande passante disponible sur le réseau Internet entre l'Amérique du Nord et diverses régions du monde.

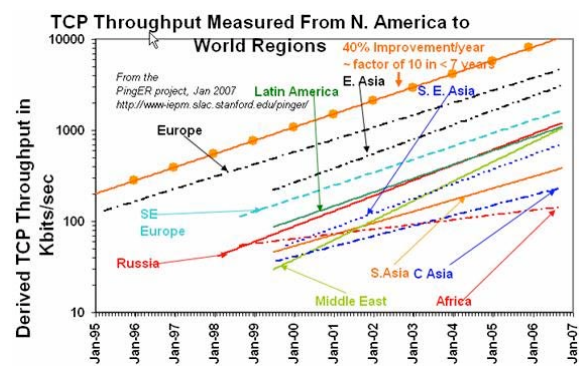

On peut constater que la plupart des régions voient leur connectivité augmenter d'environ 40% par an (soit un facteur 10 en un peu moins de 7 ans). Les exceptions sont le Moyen-Orient qui croît nettement plus vite et l'Afrique qui, à l'inverse voit son retard augmenter d'année en année.

Actuellement le continent Africain compte un retard moyen d'environ 12 ans par rapport à l'Europe.

### 1.2 Le cas particulier de l'Afrique du Nord

La mauvaise situation globale de l'Afrique est toutefois tempérée par le cas particulier de l'Afrique du Nord et, dans une certaine mesure, de l'Afrique du Sud. Au cours des dernières années, les pays du bord de la Méditerranée ont bénéficié du programme EUMEDconnect [3] financé à hauteur de 10 M€ par l'Union Européenne. Ce projet permet de relier les pays méditerranéens au grand réseau d'éducation et de recherche européen GEANT ainsi qu'au réseau sud-africain TENET. La bande passante disponible est de 155 Mbit/s pour l'Algérie et le Maroc, de 45 Mbit/s pour la Tunisie et de 32 Mbit/s pour l'Égypte.

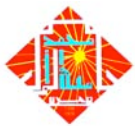 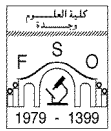 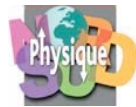 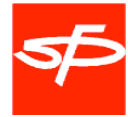

Ce type de projet européen n'est financé que pour un temps limité et devra déboucher sur des solutions pérennes, seules capables de garantir le développement de la connectivité sur le moyen et le long terme.

Il faut également noter que même quand la connectivité internationale est bonne, les universités et les laboratoires ont souvent du mal à en bénéficier pleinement en raison des difficultés de connexion sur l'épine dorsale du réseau. Ce phénomène universellement répandu est connu sous le nom de "problème du dernier kilomètre".

## 2 Les nouvelles technologies au service de la Science

Selon Wikipedia: "*Une grille informatique ou grid est une infrastructure virtuelle constituée d'un ensemble coordonné de ressources informatiques potentiellement partagées, distribuées, hétérogènes, externalisées et sans administration centralisée*"

La technologie des grilles informatiques devient un élément incontournable pour le traitement des données dans de nombreux domaines scientifiques.

### 2.1 La grille de calcul pour le LHC

Les expériences installées sur l'anneau de collision LHC vont fournir un flot de 15 Pétaoctets de données chaque année. Le stockage et le traitement d'une telle masse de données sont virtuellement impossibles dans un seul centre de calcul. Il est nécessaire d'utiliser toutes les ressources informatiques disponibles dans chacun des pays participant aux expériences. La solution retenue est une architecture de grille, coordonnée au niveau mondial au sein du projet W-LCG (Worldwide LHC Computing Grid).

L'implémentation et l'opération d'un tel système sont extrêmement complexes au niveau des grands centres formant l'ossature de la grille. En contrepartie, cette grille de calcul permet à toutes les équipes disposant de moyens informatiques même modestes, d'avoir accès à l'ensemble des données du LHC ainsi qu'à une puissance considérable de calcul distribué. La grille W-LCG constitue donc une opportunité unique pour les équipes de recherche des pays émergeants, qui pourront ainsi jouer un rôle de premier plan dans l'analyse des données du LHC.

### 2.1 La grille Européenne EGEE et les projets connexes

Une partie de la grille W-LCG s'appuie sur le projet européen EGEE (Enabling Grid for E-sciencE) qui vise à développer les logiciels et les procédures nécessaires à la mise en œuvre d'une architecture de grille pluridisciplinaire. Avec un financement de 32 M€ sur 2 ans et 90 partenaires répartis dans 32 pays, EGEE constitue la plus grande organisation de grille au monde. Initié par la physique des hautes énergies, EGEE est ouvert à de multiples champs disciplinaires.

Le projet EGEE possède plusieurs extensions qui visent à promouvoir les grilles de calcul dans les pays en voie de développement: EUMEDgrid pour les pays du pourtour méditerranéen, EUCHINAgrid en Asie et EELA en Amérique Latine. Cette dernière grille, forte de 700 CPU et de 60 Téraoctets de stockage sur disque est un véritable succès. De nombreux ateliers sont régulièrement organisés pour former les participants à la mise en œuvre et à l'utilisation des technologies des grilles de calcul dans le cadre de leurs projets de recherche.

Il faut souligner le caractère structurant des grilles de calcul qui permettent à des communautés entières de partager le même outil.

### 2.2 Les grilles légères

Toutes les applications ne nécessitent pas l'utilisation de grilles complexes. Il existe plusieurs logiciels de grilles qui permettent la mise en commun de ressources informatiques autour d'un projet scientifique. On peut citer BOINC[4], pour des applications très calculatoires, ou SRB[5] pour les échanges de données.

## 6 Conclusion

Bien que le Maghreb bénéficie d'une connectivité raisonnable, la situation de l'Afrique au niveau de la fracture numérique est particulièrement préoccupante. Une priorité politique doit absolument être fixée pour rattraper le retard en mettant l'accent non seulement sur le développement des liaisons internationales mais aussi sur l'interconnexion des universités.

Les technologies de grille offrent une opportunité unique de participer à de grands projets internationaux (LHC par exemple). Une mutualisation des ressources et la mise en place de projets collaboratifs inter-laboratoires permettent une utilisation optimale des nouvelles technologies de l'information au bénéfice du développement scientifique des pays émergeants.

**Références**
[1] *CREDOC – "Enquête sur les conditions de vie et les aspirations des Français" – juin 2006*
http://netissy.hautetfort.com/files/etude-credoc2006.2.pdf
[2] Rapport ICFA sur la connectivité Internet:
http://www.slac.stanford.edu/xorg/icfa/icfa-net-paper-jan07/
[3] Projet EUMEDconnect: http://www.eumedconnect.net/
[4] Projet BOINC: http://boinc.berkeley.edu/
[5] Projet SRB: http://www.sdsc.edu/srb/index.php/Main_Page

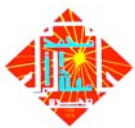 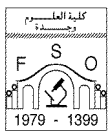 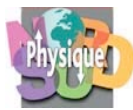 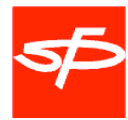

# La Grille de Calcul et la collaboration Belgique-Maroc

## Othmane BOUHALI[1], Chaker El AMRANI[2]

[1]Université Libre de Bruxelles, Bruxelles, Belgique
obouhali@ulb.ac.be
[2]Université Abdelmalek Essaadi, FST de Tanger, Maroc
celamrani@uae.ma

**Résumé**

Ce papier expose brièvement, la technologie de la Grille de Calcul et son apport essentiel à toute la communauté scientifique. La collaboration Belgo-Maorcaine à travers des associations de biologistes, de physiciens et de chercheurs en nouvelles technologies, sera abordée par la suite, et montrera le rôle bénéfique de ces structures dans le transfert technologique entre les deux pays. Ensuite, la collaboration réussie, dans le domaine de l'investigation en Grille de Calcul, entre la Belgique et le Maroc sera abordée. Elle est, en effet, le résultat d'un travail collectif et assidu, mené conjointement, depuis 2003, par l'Université Libre de Bruxelles (ULB), l'Université Abdelmalek Essaadi (UAE) et le Centre National pour la Recherche Scientifique et Technique (CNRST).

# 1 Introduction

Le Grid Computing ou Grille de Calcul est une approche novatrice visant à fédérer des moyens informatiques géographiquement dispersés et de permettre l'utilisation partagée et optimisée des ressources ainsi fédérée. Elle a pour objet de fournir, de manière transparente et sécurisée, à des communautés partageant les mêmes objectifs (Oganisations Virutelles), l'accès aux ressources de calcul et de données.

Les physiciens chercheurs sont, sans doute, parmi les plus demandeurs de puissance de calcul et de capacité de stockage.

Dans le cadre d'une collaboration entre des chercheurs Belges et Marocains, plusieurs activités ont été menées ces dernières années visant à promouvoir la technologie de la Grille de Calcul au sein des Universités Marocaines.

# 2 Collaboration Belgo-Marocaine

Pour des raisons linguistiques et sociales la Belgique est devenue l'une des destinations préférées des étudiants marocains désireux de poursuivre leurs études supérieures.

Conscients des défis auxquels ils sont affrontés, un bon nombre d'entre eux décident de s'organiser dans des structures à caractère scientifique. Citons deux exemples :

− BIOMATEC : association des biologistes marocains en Belgique créée au début des années 90.

− MASTeR/PHYMABEL : l'association PHYMABEL (Physiciens Marocains en Belgique) a été crée en 1995 par des étudiants marocains en Physique. En 2004 cette association s'est restructurée pour intégrer les cadres marocains travaillant dans les domaines des nouvelles technologies.

L'un des buts majeurs de ces associations est de participer au transfert de technologie entre les deux rives de la Méditerranée., à travers des initiatives de coopération et de collaboration. Plusieurs domaines sont concernés : la médecine, la biotechnologie, la physique et les nouvelles technologies. Récemment la Grille de Calcul a fait un autre sujet de collaboration que nous détaillerons dans la suite de ce papier.

# 3 Collaboration en matière de Grille de Calcul

La mise en place d'un projet de Grille de Calcul donnera accès à l'utilisation de moyens faramineux, en matière de puissance de calcul et de capacité de stockage, auxquels se heurte la communauté scientifique au Maroc. Afin de s'équiper d'une Grille de Calcul nationale qui couvre la plupart des établissements scientifiques, les Universités Marocaines devraient contribuer à ce projet par la conception et l'implémentation de Grilles locales.

L'Université Abdelmalek Essaadi est parmi les premiers établissements scientifiques Marocains à s'intéresser à cette nouvelle technologie.

## 3.1 Concept de la Grille de Calcul

Le concept de Grille a été mis au point pour répondre aux différents besoins dans la recherche scientifique et industrielle en optimisant au maximum l'utilisation des moyens de traitement et de stockage disponibles.

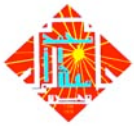 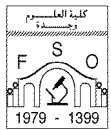 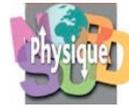 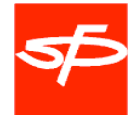

Un système de Grille repose sur les éléments suivants :
- des systèmes de traitement et de calcul, et des systèmes de stockage ;
- des mécanismes de communication par des réseaux haut-débit reliant les différents centres ;
- des services GRID réunis au sein d'une couche appelée intergiciel (Middleware en anglais);
- des logiciels d'applications spécifiques adaptés à l'architecture Grille.

*Cette nouvelle technologie est désormais opérationnelle en secteurs universitaires et industrielles, dans plusieurs pays comme aux Etats Unis, en Europe et dans des pays Asiatiques, dans le cadre dans de grands projets comme EGEE, EUMedGrid, SEE-Grid, EUChinaGrid, NAREGI, OSG, etc.*

### 3.2 Collaboration UAE (Maroc) et ULB (Belgique)

Le travail pour intégrer cette nouvelle technologie à l'UAE a débuté en 2004, grâce à la collaboration entre le groupe informatique du service de physique des particules élémentaires de l'ULB et le Département Génie Informatique de la FSTT de l'UAE.

Cette collaboration a abouti à plusieurs réalisations, on cite [1] :

1. Prototype d'une grappe d'ordinateurs (Cluster): qui a fait l'objet d'un projet de fin d'études en Maîtrise Génie Informatique. Il s'agit d'installer, de configurer et de gérer un ensemble d'ordinateurs (cluster) dédicacé au calcul intensif. Le système d'exploitation est basé sur Linux avec, comme gestionnaire de tâches, le logiciel CONDOR qui permet de créer un environnement HPC (High Throughtput Computing). Le groupe UAE a également développé, dans ce même contexte, une interface graphique aidant les utilisateurs à mieux utiliser le Cluster.

2. Prototype d'une autorité de certification nationale pour la Grille: l'autorité de certification (CA) est une entité qui délivre les certificats numériques pour les machines de la Grille ainsi que pour les utilisateurs. Une seule Autorité de certification doit exister par pays. Dans le cadre d'un projet de Diplôme d'Etudes Supérieures Appliquées (DESA) nous avons étudié la possibilité de la mise en place d'un CA au Maroc. Un prototype a été installé et des certificats ont été générés. Plusieurs aspects liés à la sécurité sur la Grille ont également été discutés.

3. Prototype d'une Grille de Calcul de test: dans le cadre d'un DESA, nous avons mis en place une station de test de Grille de Calcul. Grâce aux certificats délivrés par le prototype CA, des utilisateurs ont pu accéder à la Grille et y tourner des jobs de test.

4. Deux thèses de doctorat en cours de réalisation, dans ce même domaine d'intérêt.

5. Plusieurs publications et séminaires visant à promouvoir cette technologie au niveau national et international.

### 3.3 Projet MaGrid

La collaboration en termes de Grille de Calcul s'est accentuée avec l'implication du Centre Nationale pour la Recherche Scientifique et Technique (CNRST). En effet le CNRST joue un rôle primordial dans la promotion, le financement et la planification de la recherche scientifique au Maroc. Les trois groupes ont proposé l'implantation d'une plateforme nationale de Grille de Calcul : MaGrid. Une commission nationale s'est alors constituée pour développer une feuille de route basée sur deux phases : une phase pilote et une phase finale [2].

La phase pilote a pour but :
- D'acquérir l'expérience nécessaire quant à la gestion et l'administration d'un système basé sur la Grille ;
- D'identifier les éventuels problèmes et tester les solutions apportées.

La phase pilote concerne deux universités en plus du CNRST. Le groupe de l'UAE participe bien évidemment à cette phase. Son site est déjà opérationnel en plus de celui du CNRST. Une fois la phase pilote aura atteint ses objectifs, le projet MaGrid entamera sa phase finale qui consiste à étendre l'expérience à l'ensemble des universités et établissements de recherche.

## 4 Projet EUMedGrid

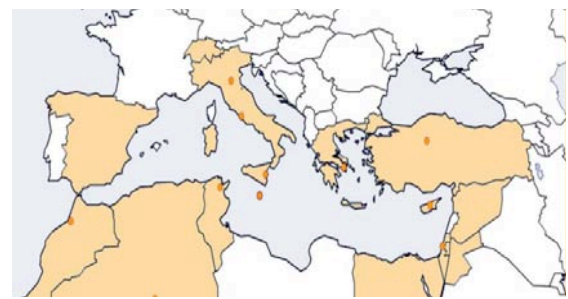

Fig. 1 : Carte des pays impliqués dans le projet EUMedGrid.

Le projet EUMedGrid (Empowering eSciences across the Mediterranean) [3] vise à accroître la conscience et les compétences dans le domaine de la Grille parmi les chercheurs de la région Méditerranéenne. Cette initiative permettra d'installer dans la région une infrastructure de Grille de Calcul pour la recherche [Fig. 1]. Cette plateforme peut intégrer la Grille Européenne EGEE [4] ainsi que d'autres initiatives analogues aux Balkans, en Europe du Nord, en Amérique Latine et en Asie. Le projet exploite les infrastructures télématiques existantes : le réseau Européen à haut débit GEANT2 et le réseau Méditerranéen EUMEDCONNECT. Le Maroc, représenté par le CNRST dans le projet EUMEDCONNECT, a pris part à cette initiative. Le CNRST et l'UAE sont déjà connectés à cette plateforme.

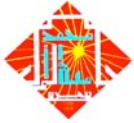 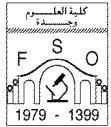 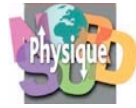 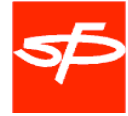

# 6 Conclusion

L'essentiel de la collaboration entre la Belgique et le Maroc en terme de Grille de Calcul a été exposé. Grâce à la coopération entre l'ULB, l'UAE et le CNRST, plusieurs objectifs ont été atteints, notamment, l'implémentation de la Grille Nationale MaGrid, et d'une autre locale installée à l'UAE.

Ces plateformes sont actuellement connectées à la structure méditerranéenne EmedGrid, offrant ainsi aux chercheurs Marocains des moyens faramineux de calcul et de stockage de données. En outre, cette collaboration a permis la publication de plusieurs travaux scientifiques de recherche dans ce même domaine d'intérêt.

## Références

[1] O. Bouhali et C. El Amrani, Proccedings of the Information and Communication Technologies International Symposium, Eds: M. Essaidi and N. Raissouni, ISBN: 9954-8577-0-2, June 2005, Tetouan, Morocco.

[2] O. Bouhali, C. El Amrani et R. Merrouch: to appear in Proccedings of the Information and Communication Technologies International Symposium, Fez, Morocco.

[3] EUMedGrid: http://www.eumedgrid.org

[4] EGEE: Enabling Grid for E-sciencE, http://public.eu-egee.org/

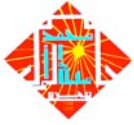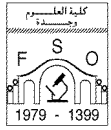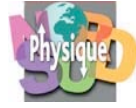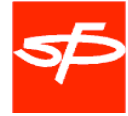

# Les réformes LMD

# au Maghreb et en France

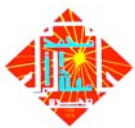 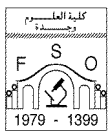 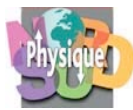 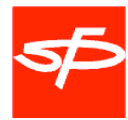

# De l'enseignement de la Physique

## Introduction par **Fatiha MAAROUFI**
Université d'Oujda, Maroc

La nouvelle réforme de l'enseignement aux universités; Licence, Mastère et Doctorat (LMD) est appliquée en Europe depuis quelques années (1999) et récemment au Maghreb:
- Septembre 2003 au Maroc
- Septembre 2004 en Algérie
- Septembre 2006 en Tunisie

Le système LMD, a été généralisé sur l'ensemble des universités au Maroc et partiellement en Algérie et en Tunisie, en concentration avec les enseignants sauf en Algérie.

Mme Siham Jaziri a donné un aperçu détaillé sur le système LMD en Tunisie; la structure de ce système est conforme à celle adoptée dans les pays européens mais tient compte de la réalité des institutions, des acquis des étudiants et des compétences des formateurs.

En Algérie, Mr Jamal Mimouni a présenté un témoignage sur l'université Mentouri de Canstantine en citant l'expérience de l'enseignement de deux modules qui s'inscrivent dans le cadre des modules de découverte; Physique et Applications et Histoire des Sciences.

Mr Menny OULD EL BAH, a fait un état de l'enseignement et de la recherche futurs en Mauritanie avec deux exemples de projet de recherche ayant donné de bons résultats; le premier sur la formation et la migration des dunes et le second relatif à la mise en place d'une base de données géoreférencées sur l'environnement de la ville de Nouakchott. Le système LMD sera appliqué En 2007 à l'université de Nouakchott.

La discussion qui a suivi chaque exposé s'est restreinte sur la comparaison du système dans les pays du Maghreb et l'évocation des problèmes rencontrés lors de l'enseignement des matières et l'évaluation des étudiants. En conclusion le système LMD, avec ses inconvénients et ses avantages, est le même aussi bien dans les pays du nord que ceux du sud.

Dans la dernière partie de la session, Mr Dominique Salin a expliqué pourquoi il y a nécessité de création d'Ecole Doctorale. Ces écoles, organisent la formation des docteurs et les préparent à l'insertion professionnelle d'une part et œuvre à la structuration de la recherche scientifique d'autre part. L'école doctorale n'est instauré dans aucun pays du Maghreb. Le ministère de l'enseignement supérieur marocain en concentration avec les enseignants prépare la charte de l'école doctorale marocaine pour la rentrée universitaire de 2007.

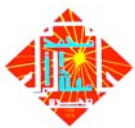 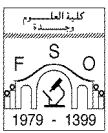 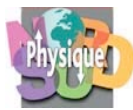 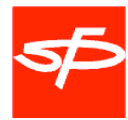

# La réforme pédagogique au Maroc : 4 ans après

Jamal Eddine DERKAOUI, Professeur

LPTPM, Faculté des Sciences, Université Mohamed 1$^{er}$, Oujda, Maroc

**Résumé**

La réforme pédagogique de l'université marocaine est en application depuis septembre 2003. Des changements qualitatifs peuvent être aperçus et un premier bilan peut être esquissé.

## 1 Introduction

Si l'université Al Qaraouyine de Fès rivalise d'ancienneté avec les plus vielles universités du monde, l'université au sens moderne du terme, date de 1957 avec l'Université Mohamed V de Rabat. Le Maroc compte aujourd'hui 14 universités, 320 000 étudiants et 15 000 enseignants.

L'université marocaine est appelée dans l'avenir à connaître un important accroissement du nombre d'étudiants qui devrait atteindre 1.000.000 à l'horizon 2015.

L'université marocaine est organisée en :
- établissements à accès ouvert: essentiellement les facultés qui recoivent 90% des effectifs des étudiants
- établissement à accès limité (avec sélection à l'entrée): telles les facultés des sciences et techniques, les écoles d'ingénieurs, les disciplines médicales, les écoles de gestion, etc.. qui représentent 10% des effectifs

## 2 Motivations de la réforme

L'ancien système avait certaines insuffisances :
- Absence d'autonomie des universités;
- Rigidité des programmes et des formations conduisant à une faible adaptation des cursus à l'évolution des besoins du marché de l'emploi;
- Importante déperdition dans les effectifs des étudiants ;

En moyenne, en sciences, on observait que 44% d'une promotion allaient au-delà du 1$^{er}$ cycle, 31% obtenaient leur licence (au bout de 5.6 ans en moyenne) et 9% avaient leur licence au bout de 4 ans.

La principale motivation interne de la mise en place du nouveau système est l'application des dispositions de la charte de l'éducation de la formation visant à :

- améliorer l'efficacité interne du système et réduire le taux de déperdition ;

- améliorer l'efficacité externe du système par un meilleur taux d'insertion des diplômés et meilleure adéquation des formations aux besoins;

- donner plus d'autonomie aux universités ;

Au niveau externe, la réforme est motivée par la facilité des échanges académiques qui découle de la similitude des parcours et des niveaux de diplômation : licence (bac+3), master (bac+5) et doctorat (bac+8).

## 3 Le système LMD marocain

Les principales caractéristiques du système LMD appliqué dans les universités marocaines sont :
- Les enseignements sont organisés en filières, semestres et modules;
- L'année universitaire est composée de deux semestres de 4 modules chacun;
- le semestre comporte 16 semaines d'enseignement et d'évaluation ;
- Le module (unité fondamentale du système de formation) équivaut à un horaire de 75 à 90 heures;

Ce système permet :
- une souplesse aux niveaux des méthodes d'enseignement et des modalités de contrôle des connaissances ;
- la possibilité de capitaliser les résultats acquis par l'étudiant ;
- la mise en place de passerelles entre différentes filières ;
- un processus d'orientation et de réorientation de l'étudiant plus facile.

Cependant le système appliqué ne prévoit pas de double service.

La procédure d'accréditation exige 3 niveaux de décision : le conseil de l'établissement concerné, le conseil d'université et la commission nationale de coordination de l'enseig,ement supérieur (CNCES) sur laquelle repose la césion ministérielle pour les diplômes nationaux et les deux premiers uniquement pour les diplômes d'université.

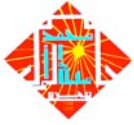 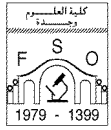 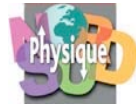 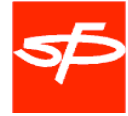

## 4 Mise en place de la réforme

La mise en place de la réforme s'est faite progressivement:

- Septembre 2003 : démarrage des licences fondamentales
- Septembre 2005 : démarrage des licences professionnelles ;
- Juin 2006: premiers licenciés ;
- Sept. 2006 : démarrage des masters et des masters spécialisés, des filières de techniciens et des filières de commerce et gestion ;

Les prochaines étapes concernent :

- le démarrage des filières d'ingénieurs en Septembre 2007 ;
- les premiers diplômés master en Juin 2008 ;
- le démarrage des nouveaux cycles doctoraux en Septembre 2008 ;

## 5 Un premier bilan

### Aspects qualitatifs

Les aspects qualitatifs accompagnant la réforme sont :

- Un plus grand choix de formation offert aux étudiants (plus de 400 filières accréditées pour 2006/07 au niveau du Maroc et 53 accrédités à ce jour à l'Université Mohamed 1$^{er}$ d'Oujda);
- Les possibilité de passerelles entre les filières et de réorientation des étudiants;
- Une intervention plus importante du milieu socio-économique dans la mise en place des filières (en particulier dans les filières professionnelles) ;
- Réponse plus rapide aux appels nationaux à offres de formations dans des domaines spécifiques répondant aux besoins particuliers des secteurs publics ou privés tels que pour les métiers de l'offshoring, les métiers de l'ingénieur, le travail social, etc..
- Émulation croissante entre les universités;

### Aspects quantitatifs

Bien que les statistiques ne soient pas assez significatives pour tirer des conclusions définitives, on peut constater un léger mieux au niveau de l'efficacité interne des formations. Le suivi de la première cohorte du nouveau système permet de voir que 17% ont quitté l'université au bout de la 1$^{ère}$ année, 12% des inscrits (17.5% des étudiants encore présents après 2 ans) ont eu leur licence en 3 ans.

## 6 Conclusion

En guise de conclusion, on peut noter que :

- la réforme de l'université marocaine était nécessaire;
- les moyens humains et matériels disponibles ne sont pas toujours à la hauteur des objectifs tracés;
- les aspects positifs engendrés par le nouveau système pédagogique sont perceptibles;
- le prochain défi est certainement celui de la réalisation des objectifs tant au niveau de la meilleure adéquation des formations aux besoins du marché qu'au niveau de la qualité des formations.

Il est par ailleurs très important, au moment où les autres pays maghrébins s'engagent dans le système LMD que des échanges réguliers d'expériences se mettent en place au niveau du Maghreb et avec la France.

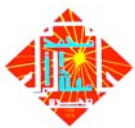 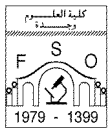 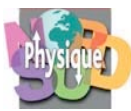 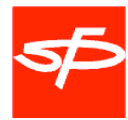

# La réforme LMD en Tunisie

**Habib BOUCHRIHA**
Faculté des Sciences de Bizerte, Tunisie

**Sihem JAZIRI**
Faculté des Sciences de Bizerte, Tunisie

*« … Nous poursuivons, par ailleurs, nos efforts pour la mise en œuvre des choix que nous avions décidés en vue de moderniser le système des diplômes universitaires dans notre pays, aux niveaux de la Licence, du Mastère et du Doctorat (L.M.D), en harmonie avec les normes et systèmes internationaux les plus évolués.*

*Nous appelons, à cet égard, à associer les divers membres du corps enseignant et de recherche universitaire à l'identification des meilleures voies pour la concrétisation de cette orientation et la promotion de ce système … ».*

<div style="text-align:right">Extrait du discours du Président de la République<br>Journée du savoir, 13 juillet 2005</div>

La Tunisie, à l'instar des autres pays développés a choisi de transformer progressivement les cursus universitaires à travers ce que l'on appelle la réforme LMD.

L'un des objectifs de cette réforme est d'assurer une véritable lisibilité de l'Enseignement Supérieur fondé sur les 3 grades LMD afin de faciliter la mobilité nationale et internationale. Et d'innover par l'organisation de nouvelles formations plus attractives et de qualité. Les parcours de formation sont construits sur des objectifs métiers et facilitant les réorientations par des passerelles. De nouveaux parcours seront construits pour répondre aux demandes du marché de l'emploi et s'appuyant sur des compétences locales.

## 1 – Mise en place : Réforme progressive basée sur la concertation

La démarche de la mise en place de la réforme est concertée et consensuelle. En effet les structures académiques et pédagogiques ont été associées. Les concertations ont été aux ordres de jours des Conseil des universités, dans la Conférence des doyens et des directeurs d'établissements, des Conseils scientifiques des établissements et des Conseils des départements. Plusieurs campagnes d'information auprès des enseignants et des représentants des étudiants ont été programmées. La Mise en place de la reforme va s'établir de façon progressive. La première vague a débuté en septembre 2006, une seconde vague suivra en septembre 2007. Finalement la dernière vague sera programmée en septembre 2008.

Après toutes ces réflexions et concertations, le ministère de l'enseignement supérieur a élaboré un mémoire de cadrage qui pose les objectifs et la structure de la réforme. La note de cadrage contient les dispositions relatives à l'inscription administrative et pédagogique, ainsi que l'organisation générale des examens. Elle prévoit des commissions de pilotages et détermine les règles de l'habilitation des parcours.

| Commissions sectorielles | |
|---|---|
| 1- Sciences Economiques | 12- Langues étrangères |
| 2- Sciences de gestion | 13- Arts |
| 3- Droit public | 14- Culture et Arts du spectacle |
| 4- Droit Privé | 15- Chimie |
| 5- Philosophie | 16- Physique |
| 6- Sciences humaines, sociales et religieuses | 17- Mathématiques |
| 7- Histoire | 18- Sciences géologiques |
| 8- Géographie | 19- Sciences biologiques |
| 9- Lettres et civilisation arabes | 20- Informatique |
| 10- Lettres et civilisation anglaises | 21- Sciences Appliquées et technologie |
| 11- Lettres et civilisation françaises | 22- Santé et technologies médicales |

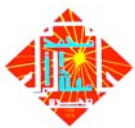
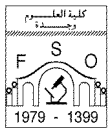
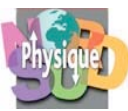
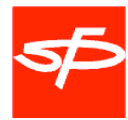

La structure est conforme à celle adoptée dans les pays européens mais tient compte de la réalité de nos institutions, des acquis de nos étudiants et des compétences de nos formateurs. Elle vise pour une meilleure employabilité et sera adaptée à l'évolution du progrès scientifique et technologique. Elle préserve le caractère national des diplômes.

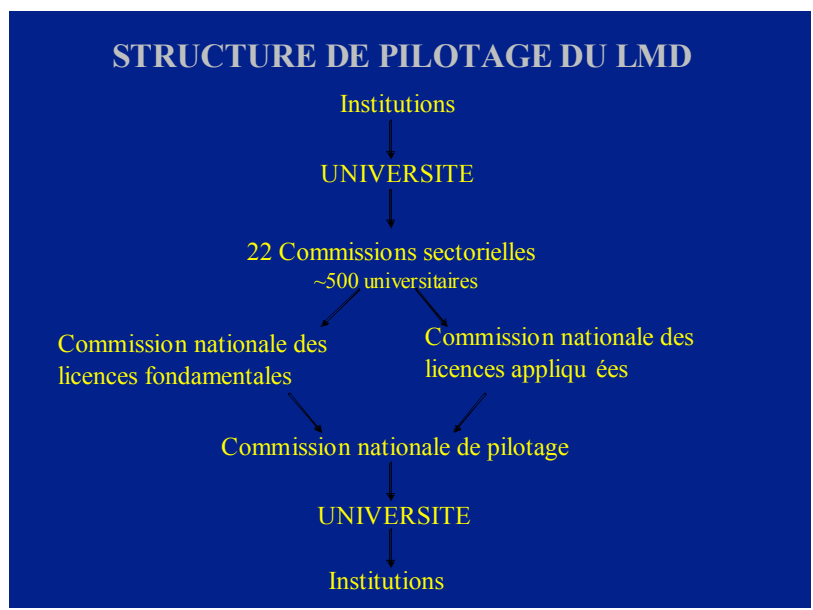

## 2 – L'étudiant au cœur de la réforme

Le ministère de tutelle a procédé à l'orientation progressive des étudiants. En se basant sur une formation flexible, un choix du parcours de formation et possibilités de sa modulation a été établi. La réforme prévoit la mobilité à l'échelle nationale et internationale avec une facilité de l'équivalence des diplômes et une meilleure intégration dans le marché de l'emploi. La formation est pluridisciplinaire avec la possibilité de passerelles entre les parcours. Concernant le protocole d'évaluation qui va être allégé tout en maintenant l'ancrage de la culture de l'effort.

Pour une meilleure lisibilité des dipôles Domaines de formation, Mentions de formation et Parcours de formation seront mentionnés dans la fiche d'orientation de chaque étudiant. Des exemples de domaines: Sciences et Technologie, Droit et Sciences Politiques , Sciences Economiques et Gestion  Lettres et Langues …..

Les parcours de formation conduisant aux grades de licence sont organisés en semestres.  Chaque semestre représente 30 crédits. Le nombre des crédits associés à chaque enseignement ou activité pédagogique est déterminé par rapport à cette norme.

L'inscription administrative est annuelle et à distance (e-inscription) et selon les procédures règlementaires en vigueur. Alors que l'inscription pédagogique est annuelle pour les unités obligatoires et semestrielles pour les unités optionnelles

L'organisation des examens est basée sur l'ancrage de la culture de l'effort et du savoir-faire et  garantit la valeur nationale du diplôme. On procedera à l'allégement du nombre des examens finals et la réduction de leur durée. On adoptera  le  principe du contrôle continu. L'évaluation semestrielle comprend i)régime mixte : 30% contrôle continu et 70% examen final et ii) régime continu : concerne les unités organisées en Travaux pratiques ou  en séminaires, ou tout autre unité à définir à cet effet.

Le passage est annuel. Une session principale est programmée à la fin de chaque semestre et une session de rattrapage à la fin  de l'année universitaire. En effet l'étudiant passe d'une année à l'autre de la licence :
i) S'il a obtenu la moyenne à toutes les UE (Unité d'Enseignement) de l'année universitaire.
ii) S'il a obtenu la moyenne annuelle générale par compensation entre les notes de toutes les UE de l'année.
iii) S'il a obtenu 75% des crédits alloués aux UE (45 crédits). Dans ce passage conditionné l'étudiant reste redevable des unités représentant les 15 crédits en instance et il doit les obtenir pour passer à l'année suivante.

Une UE est définitivement capitalisée lorsque l'étudiant y a obtenu la moyenne. L'étudiant peut capitaliser des éléments constitutifs d'une UE si ces éléments sont dotés de crédits spécifiques. Une UE validée par compensation est liée à un parcours et n'est pas transférable à un autre parcours.

Outre le diplôme, les institutions délivrent un supplément du diplôme qui valide le parcours suivi. Le supplément au diplôme fournit des informations détaillées sur les UE, les  stages suivis et les savoir-faire acquis par l'étudiant dans son cursus

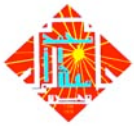 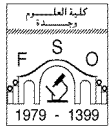 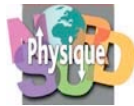 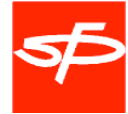

# 3 - La licence

Pour une meilleure intégration dans le marché du travail, il a été institué six cent licences dont les deux tiers concernent la licence appliquée alors le reste sera dédié à la licence fondamentale.

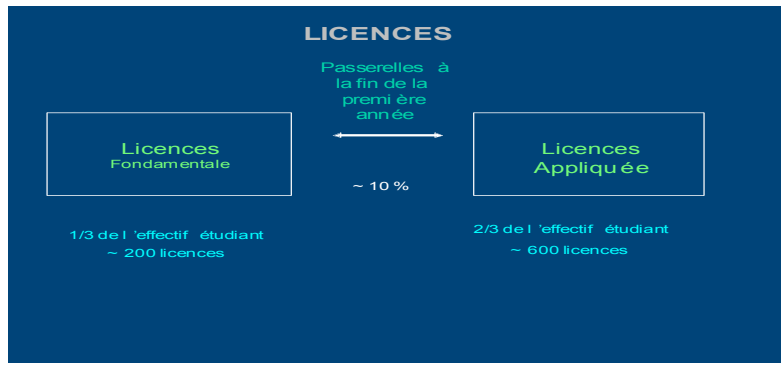

**Licence Fondamentale**

La formation académique est adaptée à l'évolution du progrès scientifique, et habilitée à rejoindre le marché de l'emploi soit directement, soit après une formation complémentaire. Elle permet la préparation du mastère et du doctorat et utilise les outils pédagogiques innovants et les nouvelles technologies de la communication.

**Licence Appliquée**

La formation est définie en collaboration étroite avec le milieu professionnel. L'habilitation des étudiants à rejoindre le marché de l'emploi est conçue. Elle est ciblée et spécialisée tout en étant propice à un changement de parcours. Elle est adaptée aux besoins du marché économique et à la rapidité de l'évolution des métiers et aux nouvelles technologies tout en permettant aux diplômés méritants de préparer le mastère Professionnel

Les unités d'enseignement sont organisées selon le tableau ci-contre :

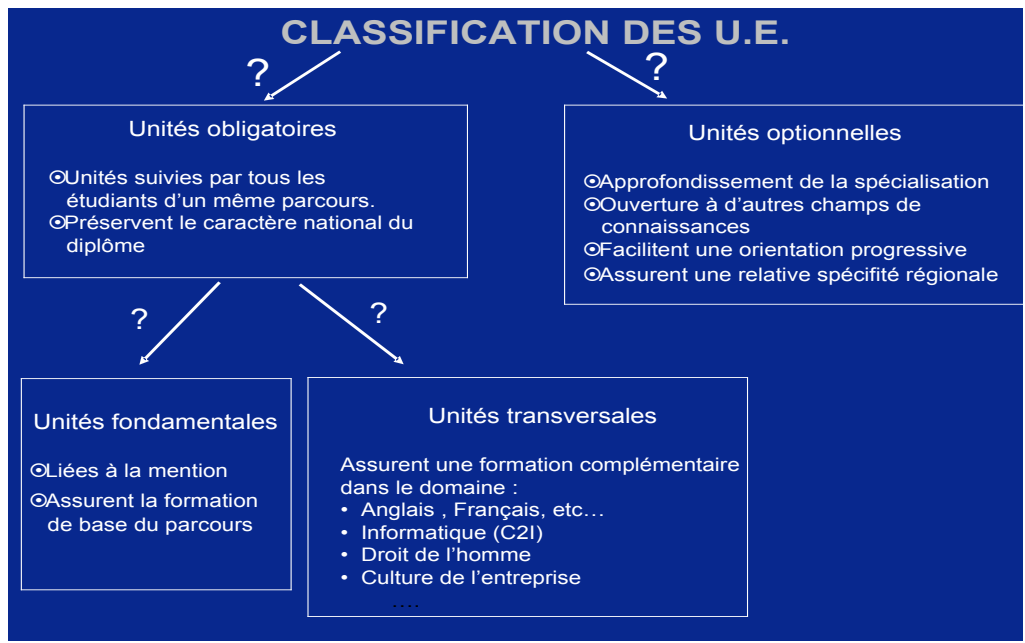

## 4 Conclusion

Parmi les avantages du L. M. D, on retient ls points suivants :
- Nombreux parcours individuels possibles
- Renforcement du rôle fondamental des équipes de formation
- Professionnalisation : facilitée par les modules projets, des stages, sans que les parcours soient obligatoirement identifiés "professionnels"

Les points qui restent à éclaircir
-L'intégration des Ecoles d'ingénieurs, Médecine dans ce système de formation.
-L'intégration des formations conduisant aux préparations des concours…..

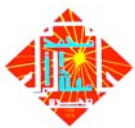 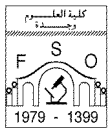 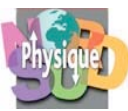 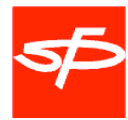

# Expérience du LMD
# à l'Université Mentouri de Constantine
## *Témoignage*

### Jamal MIMOUNI

Université Mentouri, Département de Physique,
Laboratoire de Physique Mathématique et de Physique Théorique (LPMPS), Constantine, Algérie

### Résumé

L'enseignement du type LMD en Algérie a débuté il y a trois ans et a impliqué durant chacune de ces années un nombre croissant d'universités. Malgré que son application aie été décidée en haut lieu sans concertation véritable avec les enseignants, son caractère semble t-il irréversible et sa généralisation ont créés une situation que l'Université Algérienne doit gérer au mieux.
Plutôt que de discuter de l'état de son application, nous apporterons surtout ici un témoignage sur le vécu quotidien à travers notre modeste expérience dans l'enseignement dans le cadre de cette filière.

## 1 – Raisons d'être et objectifs du LMD

Le LMD a été lancé le 25 Mai 1998 à l'occasion du 800e anniversaire de la Sorbonne, et fut adopté en 1999 à Bologne par 29 ministres de l'Education européens. Son but avoué était d'harmoniser l'architecture du système européen d'enseignement supérieur.

Ses objectifs sont généralement formulés ainsi :
- Concurrencer les universités américaines (attraction d'étudiants étrangers…)
- Favoriser la mobilité des étudiants européens, reconnaissance européenne des diplômes
- Rendre l'Europe des Université compréhensible et les cursus lisibles

## 2 - Les Prérequis du LMD

Basé sur 3 grades et trois périodes d'étude (3+2+3) avec :

- Offre de formation avec parcours type
- Nécessités par rapport au système classique:

   - *d'un travail pédagogique plus collectif avec équipe pédagogique*
   - *d'un accompagnement plus actif des étudiants par des enseignants tuteurs*
   - *des moyens humains et matériels considérables*
   - *un développement considérable de la recherche scientifique (accueil prolongé d'étudiants dans des labos en Master…)*

Tous ces prérequis sont absents ou font problème dans le contexte Algérien

## 3 - L'Introduction du LMD en Algérie

Il s'est fait par décision administrative lors de la rentrée universitaire 2004/2005 dans des conditions difficiles :
- 42% de réussite au Bac cette année-là ce qui était presque le double de la moyenne des années précédentes,
- Au milieu d'une crise multidimensionnelle que vivait l'Université algérienne
- Pas de consultations avec enseignants, étudiants: La mise en place du système LMD fut en fait adopté au Conseil des Ministres du 30 Avril 2002
- 10 universités sur 58 l'ont adoptés en 2004
- Généralisation assez large en 2005 (coexistence avec l'ancien système, basculement total dans un certain nombre d'Universités en 2006). Certaines filières notamment dans les sciences sociales et humaines n'ont pas adoptées le LMD. Elles devraient le faire impérativement à la rentrée 2007 prochaine.

Notons que le système du LMD représente la deuxième réforme majeure qu'aura connue l'Université Algérienne depuis l'indépendance.
La précédente, la RES (Réforme de l'Enseignement Supérieur), a eu lieu en 1971. Elle obéissait aux deux impératifs historiques majeurs suivants :
- Le démantèlement de la structure de fonctionnement héritée de l'Université coloniale
- La démocratisation de l'enseignement supérieur avec son ouverture à toutes les couches de la population.
L'université Algérienne a connue cependant des crises aigues durant les années 80' et 90' qui l'ont sérieusement déstabilisée notamment due :
- au flux massif d'étudiants
- à l'arrêt des investissements massifs (universités et secteur économique) du à la chute des cours du pétrole

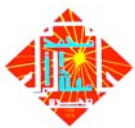 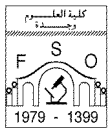 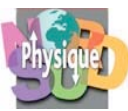 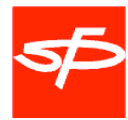

en 1986
- L'exode de nombreux enseignants, et non retour d'autres envoyés en formation doctorale à l'étranger.
- La bureaucratisation de l'Université avec comme à sa base :
• L'arrêt de l'élection des chefs de départements, doyens…
• La réduction drastique des prérogatives pédagogiques des enseignants.
Ajouté à tout cela la diminution dramatique du niveau de vie des enseignants par rapport à celui des années précédentes et aussi celui de leurs collègues des pays voisins.

## 4 - Le LMD à l'Université Mentouri à Constantine

L'Université de Constantine est l'une des dix Universités algériennes qui est entrée dans le système LMD dés la première année de son implémentation. Le système classique des études scientifiques dit du SETI (Sciences Exactes, Technologie et Informatique), a été rapidement circonscrit au fil des trois dernières années devant l'extension du LMD, puis éliminé cette année 2006-2007 au profit exclusif du nouveau système. La période de coexistence a donc été courte.

## 5- Enseignement de certaines matières en LMD

Je relaterais rapidement l'expérience de l'enseignement de deux modules typiquement LMD qui s'inscrivent dans le cadre les modules de découverte, notamment celui de : *Physique et Applications* et celui d'*Histoire des Sciences*.

Le module de **Physique et Applications** est un module difficile car il est sans TD ou TP, et est en une structure prodiguant une formation de qualité à un nombre sans cesse croissant d'étudiants. donc considéré par les étudiants comme un module secondaire.
Une synopsis du module, programme, travaux proposés, est donné à :

http://mentouri.110mb.com/pa.htm

- Comment motiver les étudiants pour la physique? La recette magique qui fut appliquée : Une bonne dose d'astronomie et d'astrophysique !!
- Comment enseigner à des effectifs pléthoriques?
- Comment maintenir un certain niveau d'intérêt alors que le taux de participation dégringole rapidement surtout vers la fin du semestre ?

Le module *d'Histoire des Sciences* est aussi un module basé strictement sur des cours magistraux, et qui intervient au deuxième semestre. Il est affublé du qualificatif douteux de module « littéraire », ce qui est proche de l'ultime insulte parmi les étudiants des filières scientifiques.

http://mentouri.110mb.com/hs.htm

Son enseignement implique là aussi une approche innovante pour intéresser les étudiants et garder un degré de motivation pour un module dont l'inexistence d'autres modules de découverte en parallèle le rende de fait obligatoire.
De plus amples détails sur le LMD des filières d'ingéniorat (Examens, corrigés, programmes, cours…) se trouvent à :http://mentouri.110mb.com/

## 6 - En Guise de Conclusion

Si l'introduction du LMD en Algérie s'est fait au pas de charge et avec un succès relatif du point de vue de son implémentation au niveau national, ce nouveau système fait face à d'immenses défis structurels et pédagogiques : manque criard de structures de support, et inadéquation des enseignements avec les nouvelles exigences du LMD notamment d'individualisation de la formation et d'un taux formateurs- étudiants adéquat.

Il va falloir beaucoup de dévouement ainsi que le déploiement de moyens exceptionnels pour le transformer

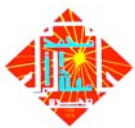 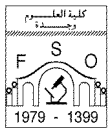 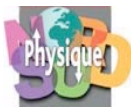 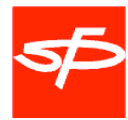

# La mise en place du LMD en France

Luce ABOUAF, professeur émérite
LPMAA, Université Pierre et Marie Curie Paris 6, Paris, France

*Ce travail a été effectué au Ministère de l'Education Nationale, de l'Enseignement Supérieur et de la Recherche, dans le cadre d'une charge de mission à la DGES*

A la suite des Conférences ministérielles de la Sorbonne (1998) et de Bologne (1999), a été créé en France, par décret du 30 août 1999, le grade de Master, conféré de plein droit aux titulaires d'un diplôme de master, d'un DEA, d'un DESS ou d'un diplôme d'ingénieur. L'arrêté du 8 avril 2002 précise l'architecture des nouvelles formations fondées sur les grades de Licence, Master et Doctorat, et le 25 avril 2002 un arrêté crée le Diplôme National de Master (DNM). En parallèle la mission de la Licence Professionnelle, qui avait été créée en 1997, et qui ne fait pas partie du LMD, est précisée, accompagnée d'une liste de dénominations nationales. En 2003, a aussi été créé un Master professionnel pour étudiants étrangers, spécifique aux Ecoles d'Ingénieur.

## 1. Le calendrier

Quelques formations de Licence et de Master ont vu le jour en France, à titre expérimental, dès septembre 2002, mais le grand mouvement de passage au LMD s'est effectué à partir de septembre 2003, avec la mise en place successive des contrats quadriennaux des Etablissements, certains établissements ayant anticipé sur la date du renouvellement de leur contrat, afin de se trouver en phase avec des Etablissements d'académies voisines. Les trois quarts des Etablissements étaient passés au LMD en septembre 2005 et la quasi-totalité en septembre 2006.

La mise en place a été très rapide et l'adhésion des Etablissements vraiment remarquable. Une révision des offres de formation est cependant apparue nécessaire, en particulier afin d'optimiser l'adéquation entre la richesse des offres et les flux d'étudiants.

Les Ecoles d'ingénieur sont passées au master, en général en cohabitation avec des universités, les études d'ingénieur restant en 2+3. Les études médicales sont, à ce jour, inchangées.

## 2. Les procédures

Les habilitations sont décidées par le Directeur Général de l'Enseignement Supérieur (DGES). Les diplômes ont fait l'objet d'une expertise et d'une habilitation nationales. L'appui d'une recherche de qualité, reconnue dans les contrats quadriennaux des Etablissements, a été exigé pour les masters recherche, ainsi que l'implication des secteurs industriels pour les masters professionnels. Les cohabilitations ont fait l'objet d'une grande attention.

Les diplômes se déclinent en :
• Domaines et Mentions pour les Licences (excepté en langues et en sport où des intitulés spécifiques conduisent à des métiers)
• Domaines, Mentions et Spécialités si l'Etablissement le souhaite, ce qui est en général le cas, pour les Masters.

Les intitulés étant libres et proposés par les établissements, une très grande inflation du nombre des diplômes (plus de 1000 en L et de 6000 en M), qui nuit à la lisibilité de l'ensemble des offres de formation, a été observée

Cette nouvelle structuration des formations de l'enseignement supérieur a été effectuée à moyens constants, financiers comme humains. Il en est résulté une très importante surcharge de travail pour tous les personnels, mais surtout pour les enseignants-chercheurs.

## 3. Les cursus

L'un des grands changements, à côté de l'existence des Unités d'Enseignement (UE), capitalisables à vie, réparties, en principe, en semestres de 30 crédits ECTS (European Credit

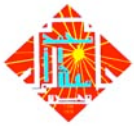
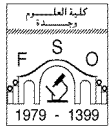
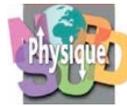
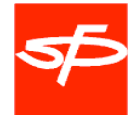

Transfert System), concerne la structure des cursus. Ils sont bâtis sur 3+2+3, alors qu'auparavant, il y avait 1er cycle, 2ème cycle et 3ème cycle (2+2+3). On parle maintenant de cycle L, cycle M et Doctorat. Il n'y a pas de sélection entre les cycles L et M.

Il faut noter que la mission des Ecoles Doctorales, recadrée par un arrêté d'août 2006, a changé : non concernées par les M2, elles ont en charge le Doctorat et la préparation du doctorant à

Les cursus : UE capitalisables, acquises à vie, reparties en semestres de 30 credits (ECTS)

| | Avant | | Actuellement | |
|---|---|---|---|---|
| | 1er cycle | Bac+1 | | |
| | | Bac+2 | **Cycle Licence** | |
| | 2ème cycle | Bac+3 | **180 credits** | |
| | | Bac+4 | **Cycle Master** | |
| Ecoles Doctorales | 3ème cycle | Bac+5 | **120 credits** | |
| | | Doctorat | **Doctorat** | Ecoles Doctorales |

une insertion professionnelle.

## 4. Les conséquences sur les effectifs étudiants

Cette nouvelle approche des cursus universitaires peut susciter un changement de comportement des étudiants. C'est pourquoi nous avons effectué diverses statistiques sur les effectifs d'étudiants de 2001-2002 à 2005-2006. Les tendances nouvelles amorcées en 2003-2004, et confirmées par la suite peuvent être considérées comme un « effet LMD ». Il serait toutefois nécessaire de les consolider par les chiffres de 2006-2007, non encore disponibles à ce jour.

### a) Base des statistiques

Les études qui suivent sont construites à partir des données de SISE (Système d'information pour le Suivi de l'Etudiant), collectées par la Direction de l'Evaluation, de la Prospective et de la Performance (DEPP).

Elles concernent :
• les 83 Universités (81+2 CUFR), ainsi que les 3 Universités Techniques,
• les formations de Bac+1 à Bac+5, LMD ou destinées à être intégrées au LMD (hors DUT, Licence pro, DU, études d'ingénieur et de médecine)

Ces formations sont répertoriées dans une nomenclature disciplinaire, disciplines et secteurs SISE, qui, dans les Sciences, n'est plus adaptée aux formations actuelles et pose un problème de suivi détaillé intelligent des effectifs.

Les statistiques peuvent être présentées sous des angles divers, les paramètres possibles étant les années postbac, les champs disciplinaires et les Etablissements, académies, régions. Dans les données qui suivent, c'est la France dans son ensemble qui est prise en compte.

### b) Statistiques sur les effectifs

L'évolution globale, par année postbac, toutes disciplines confondues, fait apparaître depuis 2003-2004, une décroissance des L1 et une augmentation des M2. Noter que les effectifs L3 sont supérieurs à ceux de L2 à cause de l'entrée en Bac+3 , d'étudiants provenant des DUT, des CPGE et parfois même de BTS.

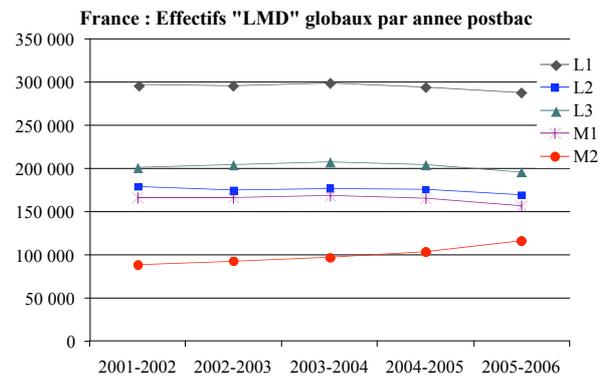

Une étude par champ disciplinaire montre que les effectifs en Sciences et Technologies décroissent dès 2001-2002 mais plus nettement depuis 2003-2004, alors que les effectifs en Sciences Humaines et Sociales croissent continûment.

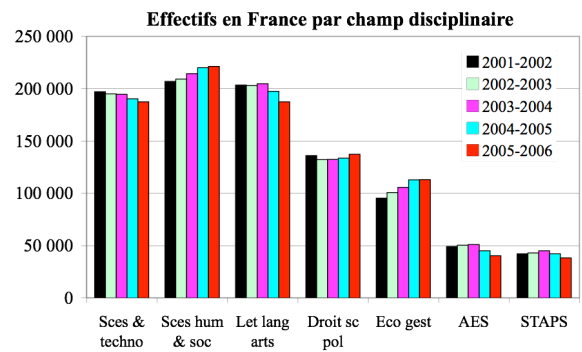

Il faut noter que la décroissance des effectifs depuis 2003-2004 en Administration Economique et Sociale (AES) et en Sciences et Techniques des Activités Physiques et Sportives

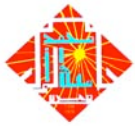
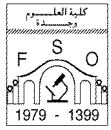
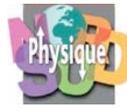
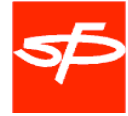

(STAPS) est un résultat de la politique d'habilitation du Ministère, ces 2 disciplines présentent peu de débouchés pour les étudiants.

Ce résumé étant destiné à des physiciens, nous indiquons le détail des Sciences et Technologies par année postbac : une

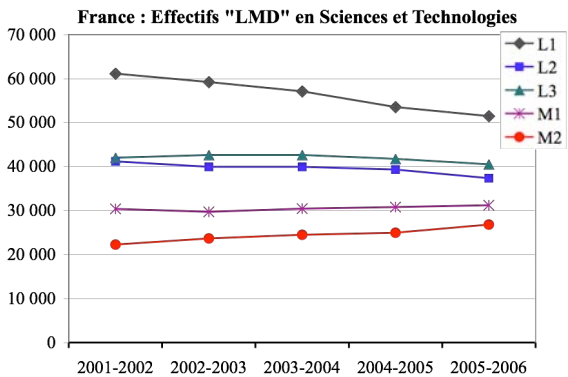

décroissance régulière et importante des effectifs de L1 est observée.

Cependant, la décroissance des effectifs Bac+1 n'est pas aussi prononcée en Sciences, si l'on prend en compte les DUT.

### c) Evolution des rapports d'effectifs.

Il n'est pas possible d'effectuer des suivis de cohortes *stricto sensu*, mais il est intéressant, pour un ensemble ouvert tel qu'un établissement, ou une nation, avec des entrées-sorties d'étudiants à tous les niveaux, d'analyser l'évolution dans le temps du rapport des effectifs Bac+x de l'année n sur les effectifs Bac+(x-1) de l'année n-1. Ce rapport pourrait représenter une sorte de taux global de poursuite d'études, même s'il ne s'agit pas exactement des mêmes étudiants.

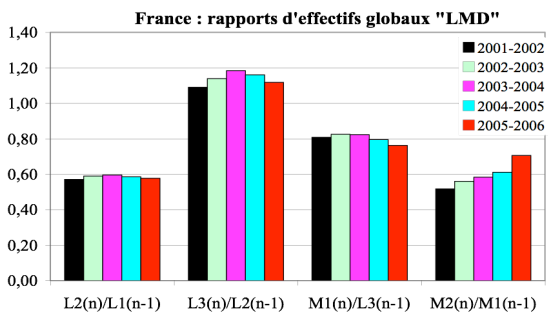

- La valeur très faible (<60%) de L2(n)/L1(n-1) représente l' « évaporation » bien connue des étudiants, au cours ou à la fin de la 1$^{ère}$ année à l'université.
- Le fait que le rapport L3(n)/L2(n-1) soit supérieur à 1 correspond, comme déjà mentionné, à l'entrée à l'université des étudiants de DUT, de CPGE, et même de DEUST et de BTS. La décroissance après 2003 peut correspondre au développement des Licences professionnelles.
- La décroissance du rapport M1(n)/L3(n-1) s'explique probablement par une auto-orientation des étudiants qui hésitent à se lancer dans le cycle M. Avant, ils s'arrêtaient après la maîtrise.
- L'augmentation nette du rapport M2(n)/M1(n-1) est très certainement due à la mise en place du cycle M qui couple Bac+4 et Bac+5. Cette tendance devrait être vérifiée par les statistiques de 2006-2007.

### d) Effet « LMD »

Finalement, on peut conclure à de premiers effets du LMD :
- auto-orientation des étudiants après le L3,
- poursuite d'études en M2.

Il convient aussi de noter que la mise en place du LMD n'a pas nui au développement des formations professionnelles :
- le pourcentage des effectifs de (Lpro)/(L type LMD) est passé de 8,4 à 18,2 de 2002-2003 à 2005-2006,
- le pourcentage des effectifs de Bac+5 (pro) /Bac+5 (rech) est passé de 1,5 à 1,9 de 2002-2003 à 2005-2006.

## 5. Conclusion

La mise en place rapide du LMD en France nécessite des réajustements qui vont s'effectuer sur plusieurs années :
- harmonisation des intitulés des diplômes afin d'augmenter la lisibilité des offres de formation, et regroupements dans des cohabilitations renforcées,
- organisation de sorties dans la vie active à Bac+3, plus larges que celles des Licences professionnelles avec la création de "Licences des métiers", pour ceux qui ne continuent pas en M,
- valorisation du Doctorat auprès des industriels afin de faciliter le recrutement des docteurs par rapport à celui des ingénieurs.

La réussite du LMD, concernant les échanges d'étudiants au sein de l'Europe, dont les règles sont précisées par l'arrêté du 6 janvier 2005 (cotutelle internationale de thèse) et par le décret du 11 mai 2005 (délivrance de diplômes en partenariat international), sera probablement surtout sensible pour les Masters et les Doctorats. Elle ne pourra être évaluée qu'avec des statistiques bien ciblées.

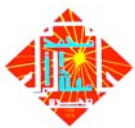 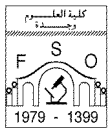 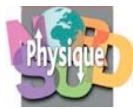 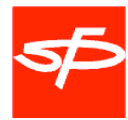

# Synthèse sur la réforme LMD présentée lors du Congrès

## Daniel BIDEAU, Professeur à l'Université de Rennes

Le congrès d'Oujda nous a permis de faire connaissance avec les collègues maghrebins et de mieux appréhender les problèmes auxquels ils sont confrontés.

Au niveau du LMD, il apparaît que l'accent a été mis sur les licences puisque c'est là que l'effort doit être mis au départ. J'ai remarqué une certaine tendance à la dispersion, mais l'élan est donné.

Les masters et Doctorats au Maghreb ont été peu discutés, même si le bilan fait par Luce Abouaf sur la mise en place en France peut servir de base à une réflexion,

Si on veut aller plus loin en recherche et au niveau des Masters et des Ecoles Doctorales, il me semble que se pose un problème de masse critique dans la plupart des Universités prises séparément. Il est donc indispensable d'aller vers la constitution de réseaux, aussi bien pour l'enseignement universitaire que pour la recherche.

Comme l'a dit Michèle Leduc dans sa conclusion, deux thèmes ont émergé, l'un autour de « Sésame », l'autre sur les matériaux « low tech » (matériaux à base de terre, granulaires…).

## Des réseaux thématiques

La constitution d'un réseau thématique sur de tels thèmes (et bien sûr sur d'autres) sur une échelle qui pourrait être celle d'un pays maghrebin ou même être intermaghrebin serait une bonne chose. Ce réseau pourrait être la base d'un master "délocalisé" voire d'une école doctorale également délocalisée.

Cela permettrait de résoudre assez élégamment le problème de la masse critique scientifique, inhérent à la fois à un master et à une école doctorale, voire à un réseau de recherche. Ceci nécessite avant tout un état des lieux relativement précis des forces en présence.

Le rôle de la SFP et des organisations locales pourrait être d'aider à la mise en place de ces réseaux, à l'image par exemple de nos GdR en France et d'apporter aussi une aide à la formation.

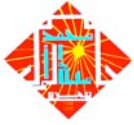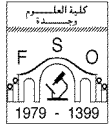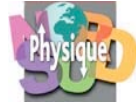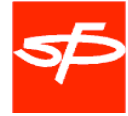

# Le rôle des Sociétés Savantes

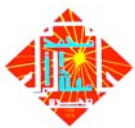 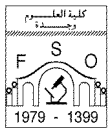 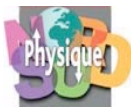 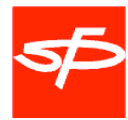

# Rôle des Sociétés Savantes dans la coopération :
# Exemple de la Société Tunisienne de Physique

## Mourad TELMINI , Président
Société Tunisienne de Physique, Tunisie

**Résumé**

Dans cet article, nous décrivons brièvement la Société Tunisienne de Physique (STP) et son rôle dans la promotion des sciences physiques au niveau de l'enseignement et la recherche, ainsi que dans la coopération internationale.

## 1 Présentation de la STP

La Société Tunisienne de Physique (STP) est une association scientifique à but non lucratif dont l'objectif est de créer un cadre scientifique regroupant tous les physiciens tunisiens (toutes disciplines confondues) de toutes les institutions universitaires, scolaires, industrielles et agricoles, dans le but de rehausser les Sciences Physiques et de développer dans tous les domaines et principalement dans l'enseignement, la recherche scientifique et ses applications. Elle a été fondée en 1981 par un groupe de pionniers qui se sont inspirés de l'exemple de la Société Française de Physique.

L'organigramme de la STP se compose d'un Comité Directeur et de trois sections régionales ; la section de Bizerte (Nord), la section du Centre (Monastir) et la section de Sfax pour le Sud. Le Comité Directeur ainsi que les comités des sections sont élus par les adhérents lors d'assemblées générales électives qui se tiennent tous les deux ans. Le mandat actuel est le 11$^{ème}$ et il s'achève bientôt.

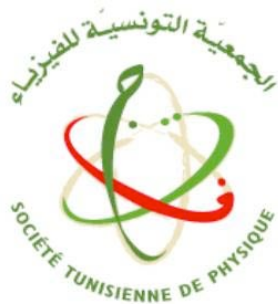

FIG. 1: Logo de la Société Tunisienne de Physique.

## 2 Activités en Recherche

L'un des principaux rôles de la STP est l'organisation de rencontres entre les physiciens tunisiens. Ainsi, depuis sa création en 1981, la STP a organisé :

- 8 éditions du Colloque National de Recherche en Physique. Le prochain colloque est prévu pour Décembre 2008.
- 5 éditions des Journées Tunisiennes des Écoulements et des Transferts Thermiques. La prochaine édition est prévue pour mars 2008.
- 2 éditions des Journées Maghrébines des Sciences des Matériaux.
- 2 éditions du Colloque de Recherche sur les Matériaux Diélectriques.
- Plusieurs autres Colloques spécialisés ou Journées Scientifiques, en collaboration avec les Laboratoires de recherche tunisiens.

Toutes ces manifestations ont été autant d'occasions de tisser des liens de coopération entre les physiciens tunisiens et leurs collègues d'autres pays.

## 3 Activités en Enseignement

La STP est également très impliquée dans l'enseignement de la Physique. Sur le Plan de l'Enseignement :

- Des Journées de Réflexion sur les Programmes de Physique
- Des Concours Nationaux de Physique (Olympiades), soit en moyenne un tous les deux ans. Le dernier en date a eu lieu en avril 2007.
- Des cycles de cours de recyclage à l'intention des enseignants du Secondaire sur des thèmes élaborés par les services du Ministère de l'Éducation Nationale.
- la Fête annuelle de la Physique, lors de laquelle des prix sont attribués aux lauréats des examens de Physique aux baccalauréat, 1e, 2e et 3e cycles de l'Enseignement Supérieur. La prochaine fête est programmée pour juillet 2007.

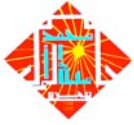 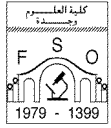 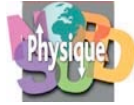 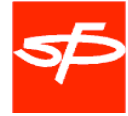

# 4    Coopération Internationale

L'implication de la Société Tunisienne de Phsyique dans la coopération internationale est une constante depuis sa création en 1981. En effet, les membres fondateurs de la STP ont bénéficié du soutien de la Société Française de Phsyique, puisque la majorité de ces pionniers avaient débuté leur carrière de physiciens en France avant de rentrer au pays et ont été pour certains d'entre eux en contact étroit avec les responsables de la SFP. L'ouverture à l'international a continué par la signature d'un accord d'association avec le Centre International de Physique Théorique (Abdus-Salam ICTP, Trieste), qui a permis à un bon nombre de physiciens tuniusiens d'effectuer des séjours scientifiques à Trieste. Par ailleurs, la STP a organisé en 1984 le $3^{ème}$ congrès des physiciens et mathématiciens arabes aisnsi que les journées maghrébines des sciences des matériaux.

En 2005, la STP a déployé beaucoup d'efforts pour la célébration de l'Année mondiale de la Physique, qui a été une occasion pour organiser différentes manifestations (conférences, expositions,…) pour susciter l'intérêt des étudiants et des élèves pour la physique. L'année 2005 a été égalemnt exceptionelle pour la STP sur un autre plan. En effet, c'est lors de l'Assemblée Générale de Cape Town (Afrique du Sud) en octobre 2005, que l'International Union of Pure and Applied Physics (IUPAP) a approuvé la demande d'adhésion de la STP comme membre représentant la Tunisie. Cette adhésion est devenue effective dès l'année 2006. A ce titre, la STP a tissé des liens, avec les Sociétés nationales de Physique de plusieurs pays, dont certains ont abouti à des accords d'association. La STP a pu également présenter avec succès la candidature de quatre physiciens tunisiens comme membres associés dans quatre commissions de l'IUPAP (C9 : magnétisme, C10 : matière condensée ; C13 : Physique et développement et C15 : physique atomique, moléculaire et optique).

Comme première action d'envergure, concrétisant son adhésion à IUPAP, la STP a organisé en mars 2007 la Deuxième Conférence Internationale de Spectroscopie (ISC2007). Cette conférence a été sponsorisée par l'IUPAP, via la commission C13. Elle a regroupé environ 170 chercheurs de 15 pays et a permis de rassembler autour de ce thème fédérateur, qu'est la spectroscpie, plusieurs équipes tunisiennes et des scientifiques de haut niveau de différents pays. L'une des caractéristiques principales de cette conférence est l'inter-disciplinarité. En effet, en plus des physiciens, la conférence a vu la participation de chimistes et de biologistes qui utilisent différentes techniques spectroscopiques et qui ont présenté des communications en lien avec le thème de la conférence. Un autre aspect de la conférence, qui va dans le sens de la coopération inter-maghrébine, est le fait que la première édition de la conférence (CIS2003) avait été organisée par nos collègues marocains de l'Université Cadi Ayyad de Marrakech. Lors de la session de clôture, il a été décidé que la prochaine édition (ISC2010) aura lieu à Alger.

C'est dans le même état esprit, la STP a participé au « $1^{er}$ Congrès Nord-Sud sur l'enseignement et la recherche en physique », co-organisé par la SFP et l'Université de Oujda, dont le présent article et une contribution aux actes. La délégation tunisienne a pu débattre avec les collègues européens et maghrébins de tous les aspects de l'enseignement et de la recherche en physique. Par ailleurs, le débat sur l'intérêt et la manière de mettre en place une Société Marocaine de Physique, voire une Société Magjhrébine de Phsyique a été très stimulant. Nous souhaitons vivement que ce débat ait une suite qui va dans le sens de la consolidation des liens entre les communautés de physiciens maghrébins et européens.

Pour plus d'informations sur la Société Tunisienne de Physique, veuillez consulter son site web :
www.stp.org.tn

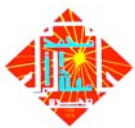 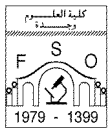 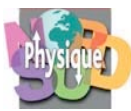 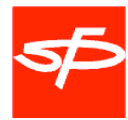

# Physique et Société : la Belgique

Viviane PIERRARD, Dr.
Vice-présidente de la Société Belge de Physique
Institut d'Aéronomie Spatiale de Belgique, Bruxelles, Belgique
Université Catholique de Louvain, Louvain-La-Neuve, Belgique

**Résumé**

Cet article présente brièvement la Société Belge de Physique et ce qu'elle peut apporter au niveau de la recherche et l'enseignement de la Physique ainsi que de la coopération Nord-Sud. Les fonds de coopération attribués par les universités belges aux pays du Sud ainsi que quelques caractéristiques des projets existants sont également présentés.

## 1 Introduction

La Société Belge de Physique (BPS) accorde une importance croissante à la coopération Nord-Sud dans la recherche et l'enseignement de la Physique. Un article a récemment été consacré à ce sujet dans la revue de la BPS [1]. Après une brève description de la Société Belge de Physique, cet article présentera quelques collaborations scientifiques conséquentes que la Belgique a développées avec les pays du Sud. Les Universités en particulier financent de très nombreux projets de coopération en physique et dans d'autres domaines.

## 2 La Société Belge de Physique

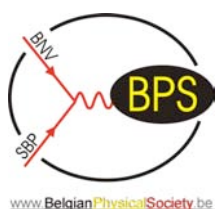

La Société Belge de Physique compte environ 500 membres qui sont principalement des chercheurs des Universités, des Instituts de recherches et des entreprises (www.belgianphysicalsociety.be). Peu d'enseignants du secondaires sont représentés car ils ont leur propre Association des Professeurs de Physique et de Chimie.

Chaque année, la BPS organise un congrès scientifique dans l'une des universités belges. Cette année, le congrès aura lieu le mercredi 30 mai 2007 à l'Université d'Anvers. En 2005, à l'occasion de l'Année Internationale de la Physique, le congrès avait été organisé conjointement avec la Société Française de Physique à Lille, ce qui avait renforcé notre collaboration avec la SFP. En 2006, le congrès a été organisé avec la Société Néerlandaise de Physique (NNV) à Leiden aux Pays-Bas.

Lors des congrès annuels, la BPS décerne des prix pour les jeunes chercheurs : meilleurs mémoires de licence, meilleures présentations orales et posters.

La Figure 1 présente le logo de la Société Belge de Physique en trois langues: SBP (Français), BNV Belgische Natuurkundige Vereniging (Néerlandais) et BPS Belgian Physical Society (Anglais). Les réunions scientifiques se font en général en Anglais. Les membres du bureau, dont les représentants des universités, sont élus chaque année lors de l'Assemblée Générale pour un mandat de 3 ans. Deux revues sont publiées par la BPS : Physicalia Info, bimensuel donnant les dernières nouvelles dans le domaine de la physique, et Physicalia Magazine, trimestriel publiant des articles scientifiques. Une vue d'ensemble des activités de recherches en physique dans les universités belges a récemment été éditée [2].

La Société Belge de Physique est favorable au développement d'une collaboration privilégiée avec les autres sociétés de physique.

## 3 Les Universités Belges

Dix universités belges enseignent la physique. Elles ont toutes un représentant au sein du bureau de la BPS. Les cinq universités francophones sont: l'UCL (Université Catholique de Louvain), l'ULB (Université Libre de Bruxelles), l'ULg (Université de Liège), FUNDP (Facultés Universitaires Notre-Dame de la Paix à Namur) et l'UMH (Université Mons-Hainaut). Les cinq universités néerlandophones sont: la KULeuven (Katholieke Universiteit Leuven), la VUB (Vrije Universiteit van Brussel), l'UA (Universiteit Antwerpen), l'UG (Universiteit Gent) et UH (Universiteit Hasselt). D'autres universités existent également en Belgique. Elles ne forment pas de physiciens, mais peuvent avoir un lien direct avec les matières scientifiques telles que les Facultés Polytechniques de Mons.

## 4 Financements Belges de Coopération

Les scientifiques belges collaborent beaucoup avec les scientifiques des pays du Sud, que ce soit au niveau des recherches, des publications, de l'accueil de doctorants ou de post doctorants ou encore d'échanges au niveau des enseignants. Des financements spécifiques sont attribués par les Universités Belges.

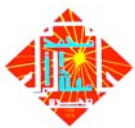
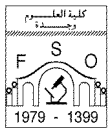
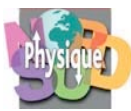
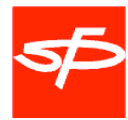

Ces universités belges se sont regroupées par régime linguistique pour mettre en commun leurs ressources et potentialités afin d'augmenter l'efficacité de leur contribution à la coopération internationale et rendre possible des projets qu'aucune institution n'aurait la capacité de réaliser seule. Les neuf universités francophones sont regroupées au sein de la Commission Universitaire au Développement (CUD: www.cud.be). Elle finance des bourses de stages internationaux et de recherches ponctuelles sur des thèmes de coopération pour une période limitée. Elle aide aussi à la consolidation et la formation de groupes de recherche interuniversitaires. L'Université Mohammed 1$^{er}$ à Oujda bénéficie ainsi du financement par la CUD de plus de 200 chercheurs marocains pour un montant de deux millions d'euro. Les universités flamandes proposent également des formations aux ressortissants des pays du Sud et sont regroupées au sein du Vlaamse Interuniversitaire Raad (VlIR : www.vlir.be).

De plus, les universités belges offrent sur fonds propres des bourses de coopération aux ressortissants des pays du Sud pour réaliser des études post-universitaires, généralement au niveau du doctorat. Ainsi, les bourses de développement sont introduites par un professeur belge pour un doctorant étranger qui effectue une thèse en Belgique ou en co-tutelle avec son pays d'origine. Les critères de sélection se basent sur les qualités du candidat, l'avis du promoteur, le sujet et la qualité du projet scientifique, mais aussi sur les perspectives de réinsertion après le doctorat et les possibilités de financements alternatifs. Ces bourses de coopération sont ouvertes seulement à certains pays du sud, qui sont principalement situés en Afrique, quelques pays d'Amérique du Sud et d'Asie.

Indépendamment de leur nationalité, les scientifiques peuvent également bénéficier de financements de recherche belges, qu'ils soient communautaires, fédéraux, régionaux, universitaires, européens ou onusiens.

## 5 Projets en Afrique et Maghreb

Des collaborations privilégiées ont été établies en particulier avec les pays du Maghreb (principalement le Maroc, mais aussi l'Algérie et la Tunisie par exemple), pour des raisons de proximité géographique et aussi parce que ces pays sont francophones ce qui facilite les contacts. Pour des raisons historiques, la Belgique a aussi une coopération privilégiée avec les pays d'Afrique centrale, essentiellement la République Démocratique du Congo, Rwanda, Burundi et aussi le Kénya.

La coopération Nord-Sud concerne tous les domaines de la physique, avec un choix de domaines prioritaires propres à chaque pays et ayant un lien particulier avec l'aide au développement. Beaucoup de projets scientifiques de coopération concernent la géophysique, surtout dans les régions sismiques d'Afrique centrale. L'hydrologie, l'exploitation des énergies renouvelables telles que l'énergie solaire ou encore la physique médicale sont aussi des domaines de prédilection.

Parmi les universités les plus impliquées, on peut citer l'UCL qui consacre 0.77% de son buget à la coopération au développement. Des stages « instrumentation et mesure » ont ainsi permis d'accueillir plus de 200 jeunes chercheurs étrangers en 13 ans. L'UCL aide également à la création de laboratoires et participe à l'échange de professeurs, d'étudiants et de techniciens. Quelques exemples de réalisations concrètes: la création d'une conférence nationale de physique au Maroc en 2006 (Marrakech) et l'installation de lasers ophtalmiques à l'Hôpital de Bukavu ainsi que d'un laboratoire d'optique avec financement de formation d'assistants en optique. Les autres universités participent également activement aux programmes de coopération en physique. Du côté francophone, on peut citer par exemple l'ULg qui a deux projets CUD, un avec l'Université de Bujumbura (Burundi) et l'autre avec l'Université de Kinshasa. L'ULB a une collaboration privilégiée avec le Maroc. Du côté néerlandophone, la KULeuven a développé des collaborations avec l'Afrique du Sud, le Zimbabwe, le Burundi, la RDC et la Tanzanie, en plus de l'accueil de doctorants. L'UA a eu une dizaine de doctorants en physique provenant d'Egypte. La VUB a participé à la création d'un centre de biologie et physique à l'Université de Nairobi au Kénya.

## 6 Conclusion

La Belgique a développé des actions concrètes pour la coopération avec les pays du Sud, notamment dans le domaine de la physique. D'autres domaines de coopération sont par exemple la médecine et l'agronomie. Les universités belges participent de manière conséquente, notamment en finançant des chercheurs directement dans les pays su Sud via la CUD et la VLIR. Les mandats sont généralement à durée limitée, mais avec possibilité de renouvellements après évaluation. La venue de doctorants étrangers est également facilitée. Notons que la mobilité des chercheurs et des étudiants induit une expérience très enrichissante, mais provoque aussi certains problèmes liés à l'expatriation: séparation des familles, chocs culturels, difficultés d'adaptation, d'autres méthodes de travail, difficulté de la langue dans certains cas, limitation des perspectives à long terme...

Les domaines de recherches en physique privilégiés par ces financements sont plutôt appliqués, afin de participer concrètement au développement. D'où le peu de financements de grands instruments. Néanmoins le grand projet international des télescopes de l'ESO installés au Chili a induit une collaboration Nord-Sud fort intéressante.

**Références**
[1] D.K. Callebaut, Physicalia Magazine 28 (2006) 72.
[2] J. Lemonne, Physicalia Magazine 27 (2005) 14

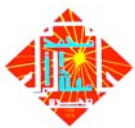 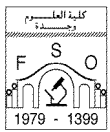 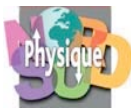 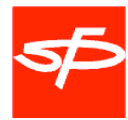

# Vers la crétaion de la Société Marocaine de Physique

## Jahmal DERKAOUI

Faculté des Sciences, Université Mohamed 1$^{er}$ Oujda – Maroc

La physique est souvent sollicitée pour fournir des réponses aux problématiques qui se posent à la société moderne tant dans sa quête d'une vie meilleure et plus compréhensive donc plus respectueuse de la nature que dans ses questionnements métaphysiques et spirituels de toujours. C'est aussi une discipline qui a des connections et des liens très étroits avec la plupart des autres disciplines scientifiques. Elle constitue également une partie importante et indispensable dans le cursus académique et universitaire.

La communauté des physiciens est de ce fait appelée à fournir des avis et des propositions aux décideurs universitaires et institutionnels sur ses divers domaines d'intérêt et/ou d'interaction.

Il est par ailleurs indéniable que les sociétés savantes jouent dans de nombreux pays de part le monde un rôle important. Elles constituent souvent un interlocuteur de valeur pour les pouvoirs publics et de confiance pour la société civile.

La communauté marocaine de physiciens avoisine le millier de membres. De nombreuses sociétés savantes ont vu le jour durant la dernière décennie dans des spécialités de la physique. Il s'agit en particulier des spécialités d'optique et d'optoélectronique, de physique statistique, des polymères, de physique appliquée, de cristallographie, de mécanique, etc. De même, sous l'effet des incitations du ministère en charge de la recherche scientifique, des pôles de compétences liés à la physique ont vu le jour entre 1997 et 2000. Ce sont de regroupements au niveau national des spécialistes d'un domaine particulier autour d'une thématique centrale et d'un projet fédérateur. C'est ainsi que des pôles concernant la physique de la matière condensée, la physique théorique, la physique des hautes énergies et les sciences et techniques de l'espace se sont formés.

L'ensemble de ces sociétés et de ces pôles de compétence ont une activité de recherche importante et reconnue mais n'ont pas beaucoup de poids au niveau décisionnel.

C'est dans ce contexte général que l'idée que créer une société marocaine de physique (SMP) apparaît être très opportune. La SMP pourrait prendre appui sur les sociétés de spécialité existantes et tirer profit de la dynamique qui s'est créée dans les réseaux et les pôles de compétences. Elle devra aussi être ouverte à l'ensemble des physiciens marocains qu'ils appartiennent aux universités, aux instituts de recherches, aux écoles d'ingénieurs ou aux établissements privés d'enseignement supérieurs accrédités. Nous gagnerons beaucoup à associer les physiciens marocains résident à l'étranger à cette entreprise. Ils enrichiront sans aucun doute la SMP par leur expériences professionnelles et assureront l'établissement de liens étroits avec les sociétés savantes de leur pays de résidence.

Elle pourrait ainsi atteindre rapidement une représentativité réelle de la communauté des physiciens. Outre cette représentativité, la SMP pourrait avoir pour missions :

• la réflexion sur les thématiques globales et pluridisciplinaires relatives à l'enseignement de la physique et à la recherche ;

• l'organisation d'activités scientifiques d'envergure (congrès, écoles, conférences, etc..) rendant plus visible l'action de ses membres et des autres sociétés savantes ;

• la réalisation d'actions servant la vulgarisation des sciences dans la société marocaine et pour rendre plus attractive la physique auprès des jeunes (collégiens et lycéens en particulier).

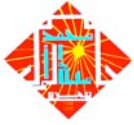 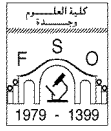 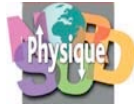 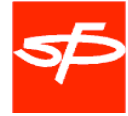

Elle pourrait agir également pour l'établissement de liens confraternels et solides avec les sociétés homologues dans les pays du Maghreb, en Europe et dans le monde. La SMP pourrait aussi réaliser –pour son compte ou pour le compte d'organismes nationaux- des études en relation avec le champ disciplinaire et en conformité avec son statut.

Le débat initié à l'occasion de la préparation et de la tenue du Congrès Nord-Sud sur le Physique devra se poursuivre dans les prochains mois au sein de la communauté des physiciens. Ce débat devra –en tenant compte des recommandations qui ont étés exprimés lors de ce congrès- aboutir à la proposition par un comité préparatoire d'un statut et d'un règlement intérieur qui devront être approuvés lors d'une assemblée constituante durant laquelle devra être élu la premiere équipe dirigeante de la SMP. La constitution de la SMP pourrait avoir lieu dans un délais d'un an après la tenue du CNSP c'est à dire au printemps 2008.

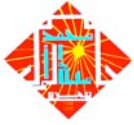 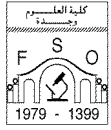 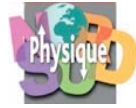 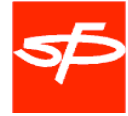

# Physique et Société

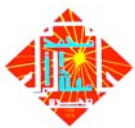 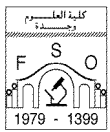 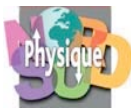 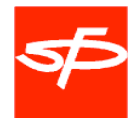

# Hymne à la Physique Fondamentale
## Vicissitudes et Promesses Algériennes

### Jamal MIMOUNI

Université Mentouri, Département de Physique,
Laboratoire de Physique Mathématique et de Physique Théorique (LPMPS), Constantine, Algérie

### Résumé

La physique théorique est une discipline prestigieuse et exigeante à la fois, dont il existe une bonne tradition en Algérie. Nous discuterons tout d'abord de sa centralité en physique. Puis de l'évolution de son statut durant les trois dernières décennies en Algérie, tant du point de vue de la recherche que de celui de l'enseignement. Nous mettrons en exergue le rôle de "pivot" qu'elle joue au niveau des départements de physique. Nous aborderons aussi les facteurs qui ont menés à la crise de vocation que connaît la physique en général. Enfin nous verrons que malgré la faiblesse structurelle de la physique fondamentale en Algérie et certains échecs patents, des développements récents font entrevoir qu'elle a un bel avenir devant elle.

## 1 Hymne à la Physique

**1.1 Comme l'air que nous respirons…**

La physique est en fait cette science fondamentale utilisée par une foule de disciplines. On pourrait dire qu'elle est pour les autres sciences ce que l'air que nous respirons est pour la vie.

- Les Sciences du Globe, de la géophysique à la physique de l'atmosphère, à l'océanographie, à la tectonique des plaques, l'utilisent de manière essentielle.

- Les Sciences biologiques, la biophysique bien sûr, mais aussi la physiologie végétale, la cardiologie… Ainsi par exemple, la cytologie l'utilise pour modeler le mouvement de la cellule grâce au battement des cils, les mouvements à travers la membrane cytoplasmique…

- Idem pour les Sciences de l'Univers, de l'astronomie à la physique stellaire, à la cosmologie. Après tout, l'Univers n'est-il pas ce laboratoire ultime pour étudier la matière aux conditions physiques extrêmes ?

- Pour les sciences de l'ingénieur, leur cas là non plus ne demande pas de longues élaborations.

Ainsi, de même que l'on ne saurait se passer des mathématiques pour analyser et interpréter les données, il est clair qu'il n'est pas possible non plus de se passer de la physique pour faire de la science même si on le voulait.

**1.2 La Physique, tentaculaire mais généreuse**

On a dit de la physique qu'elle était tentaculaire de par les domaines qu'elle s'est appropriée. Ne s'occupe t'elle pas de quasiment tous les champs du monde matériel, du microscopique au macroscopique jusqu'au cosmique.

En fait cet impérialisme est plus une impression qu'une réalité. Les physiciens sont une race de pionniers, qui défrichent constamment de nouveaux territoires aux frontières de l'inconnu, pour laisser ensuite à d'autres disciplines (notamment la technologie) le soin de les faire fructifier. Il en a été ainsi de la mécanique, de l'optique, de la thermodynamique, de l'électromagnétisme, qui ont générés autant de domaines d'ingéniorat. Il en a été de même pour la mécanique quantique qui a donnée naissance à la chimie quantique, à l'électronique ; la physique nucléaire qui a enfantée la technologie nucléaire. La chimie elle-même n'est-elle pas de la physique atomique et moléculaire appliquée ?

La physique est aussi désintéressée, elle ne préoccupe pas des applications immédiates et encore moins du profit à court terme. Les théories et concepts novateurs qu'elle forge ne sont souvent validés que plusieurs décades après leur élaboration, et souvent pour changer la face du monde ou du moins de la technique. Pensons au bout de chemin fulgurant de la formule $E=mc^2$ écrite par Einstein en 1905. La découverte des propriétés quantiques des semi-conducteurs. Le laser découvert durant les années cinquante, découlant des travaux d'Einstein du début du siècle dernier sur l'émission induite. Ce sont ces recherches qui génèrent des connaissances de « rupture », des découvertes susceptibles de révolutionner tous les domaines d'intérêt de l'humanité.

Il est aussi remarquable que nombre de scientifiques de renom, ayant été formés en tant que physiciens et dont la liste serait trop longue pour dresser ici, ce sont déployés dans différents domaines des connaissances allant de la biologie, à la géologie à l'informatique voire la philosophie, et y ont excellés. Deux exemples notables : la découverte de la structure en double hélice de l'ADN ainsi que la conception du Web ont été le fait de physiciens.

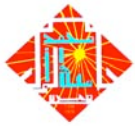 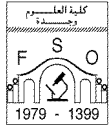 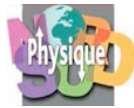 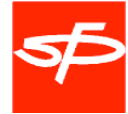

Nous concluons que la physique par sa recherche de lois universelles, par l'utilisation constante des outils les plus efficaces mis au point par la technique et appliqués à des domaines les plus diversifiées, par sa magnanimité et sa manière cavalière à laisser à d'autres le soin de labourer les nouveaux champs de savoir qu'elles à dégagées, à fait preuve au fil des siècles d'une efficacité insolente. Il n'est peut' être pas exagéré de dire que *sa vocation pourrait bien être de former des experts en universalité*.

## 2 La Physique en Crise

Et pourtant la physique traverse une crise aiguë. Il y a une désaffection patente pour la physique de la part de nos lycéens et étudiants pour qui cette discipline est devenue leur bête noire. Ceci se traduit par une grave crise de vocations, La crise de la physique est aussi une crise de l'enseignement de la physique. Une discipline qui s'enseigne comme on enseignerait du latin, à toutes les raisons d'être fuit comme le diable.

## 3 Ode à la Physique Fondamentale

### 3.1- Discipline prestigieuse mais exigeante

La Physique théorique consiste en un mot à mathématiser la description de la Nature

L'aspirant théoricien est confronté à une escalade de demandes … sans fin en vue [1]. Il s'agit de maîtriser :

$MQ \to RQM \to QFT \to$ *Théories de jauge* $\to$ *Gut's* $\to$ *SUSY* $\to$ *Superstrings*

Peut-on être théoricien de nos jours sans être cosmologiste : *GR, inflation, dark matter, dark energy…*
et de là: *GRB (Gamma Ray Bursts), UHECR…*
… ou sans posséder de culture du non linéaire : *Chaos, fractales…*

### 3.2- Les Différentes Convergences Unitaires de la Physique

Au-delà de cette tentation unificatrice sous la forme d'une théorie fondamentale, l'unité de la physique apparaît:
- Par ses lois : Les différentes théories fusent entre elles lorsque l'on passe aux différentes limites (c, h, l $\to$ 0) en une harmonieuse synthèse.
- Le caractère unifié des différentes espèces de particules élémentaires
- La cosmologie contemporaine embrasse l'univers dans sa totalité physique et historique
- L'unité de l'Homme et de la Nature…

### 3.3- Les Limites à la Connaissance en Physique

« I notice that Physics changes
but the world stays the same »

R.Bandler

La pratique de la physique a révélé toutes sortes de limites à la connaissance du monde matériel

**a- Limites subjectives**

- Même si la naïveté de croire que la physique découvre les mécanismes crus et les explications totales et absolues n'est plus de mise, elle fait plus que simplement décrire le monde, elle l'explique… d'une certaine manière non dénudée de subjectivité
- Axiomatiser la physique (Sixième problème d'Hilbert) n'est d'ailleurs plus un objectif atteignable

**b- Limites objectives**

- Imprédictibilité des phénomènes macroscopiques: chaos
- Imprédictibilité microscopique: QM (Relations d'incertitude)
- Théorème de Gödel

### 3.4- Les "Dangers" de la Physique Théorique

- Création d'attentes « déraisonnables » tel que le rêve de TOE
- L'impasse SUSY/String qui perdure :
  • Pas de contact avec l'expérience
  • Le landscape est trop peuplé…
- Mystification à la Hawkins : Frontière non délimitée entre spéculation et théorie consensuelle

### 3.5- Nouvelles postures épistémologiques

- Bâtisseurs de mondes… sur le papier, avec la primauté donnée à l'esthétique, avec comme devise implicite oh combien ambiguë : « La beauté prime sur la vérité ».
- Ouvre la porte à une interprétation à la Wolfram…[2]

## 4 Physique Fondamentale en Algérie

### 4.1- Carte Universitaire Algérienne

Cinq à sept universités géantes dont une hypertrophiée (Bab Ezzouar), une quinzaine de moyennes, et environ 35 centres universitaires.

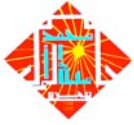 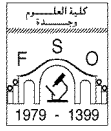 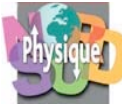 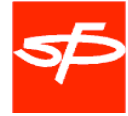

**4.2- A quoi sert un théoricien?**

- La bête de somme des départements due à sa polyvalence…
- Coupure épistémologique grave entre théoriciens et non théoriciens, avec en filigrane des modules « à vie » pour certains enseignants, et une compartementalisation…
    Ex : The Bab Ezzouar's predicament
- La malédiction nucléaire
- Stériliser par les PI (Intégrales de Chemin)

**4.3- On a manqué en Algérie nombre d'occasions: Citons le LEP, le LHC, Sesame…**

**4.4- Un Renouveau de la Physique Théorique?**

- Regain d'intérêt parmi les étudiants et ouverture de sections théorie à Jijel, Oum El-Bouaghi, Tébessa, Msila…
- Le renouveau par le LMD ?!?!

## 5  En Guise de Conclusion

Nous terminerons par ces mots du regretté d'Abdus Salam mettant les choses en contexte :

« …*en dernière analyse, la création, la maîtrise et l'utilisation de la science et de la technologie modernes constituent fondamentalement ce qui distingue le Sud du Nord. C'est de la science et de la technologie que dépend le niveau de vie d'une nation.* »

Il est de notre opinion que la physique peut être le levain 'une telle maîtrise scientifique et technologique.

**Et de plus :**

Φ is Φun
متعة الفيزياء

**Références**

[1] HOW to BECOME a GOOD THEORETICAL PHYSICIST, Gérard 't Hooft
    http://www.phys.uu.nl/~thooft/theorist.html
[2] S.Wolfram, A New Kind of Science, Wolfram Media, 2002

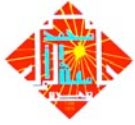 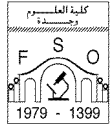 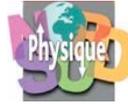 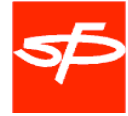

# Réorientation de l'enseignement et de la recherche scientifique pour préparer le décollage économique des pays d'Afrique

## Nahed DOKHANE , maître de conférence
Université de Boumerdès, Algérie

### Résumé

L'enseignement et la recherche, dans nos pays d'Afrique, sont très largement influencés par les tendances et les choix des pays développés du Nord. Rejeter globalement ces choix équivaudrait à se priver du fruit d'une longue et précieuse expérience ; appliquer ces choix sans adaptation aux besoins particuliers de nos pays d'Afrique, qui doivent tôt ou tard œuvrer pour leur décollage économique, équivaudrait à programmer dès le départ son propre échec.

## 1 Introduction

Ne serait-il pas plus raisonnable de parler de développement économique au lieu de décollage économique ? La réponse, c'est l'histoire contemporaine qui nous la donne : cela fait un demi siècle qu'un certain nombre de pays d'Afique essaient et appliquent tant bien que mal diverses stratégies de développement, et quel est le résultat ? L'Afrique est, en ce début du $21^{ème}$ siècle, le continent le plus sous-développé de la planète ! Alors, quand certains scientifiques africains rejettent ces conventionnelles stratégies de développement au profit des audacieuses *stratégies de décollage*, ils ont de bonnes raisons.

Une stratégie de décollage a ses exigences en ce qui concerne l'enseignement et la recherche scientifique. Nous essaierons dans ce qui suit d'en donner un bref apperçu.

## 2 Stratégies de décollage, enseignement et recherche scientifique

### 2.1 En quoi consiste une stratégie de décollage industriel et économique ?

Pour répondre à cette question, nous allons surtout éviter les idées préconçues d'une grande partie des experts qui nous ont trop souvent induit en erreur. Et nous allons simplement interroger l'histoire en suivant le bon sens universel : qu'a donc fait le Japon de Meiji pour bâtir une industrie nationale autonome et performante ? Qu'ont fait tous ces pays émergents qui ont suivi l'exemple du Japan, et ont mis au point une puissante industrie nationale préalablement presque inexistante ?

Tous ces pays émergents, qui ont réussi leur décollage économique, ont commencé par importer la technologie, non pour se contenter de l'utiliser, mais pour l'*imiter* et créer à partir de là leurs propres industries nationales. Cette *politique de l'imitation*, qui est l'un des principaux piliers de la stratégie de décollage, doit être prise très au sérieux par les pays actuellement en voie de développement.

Afin d'assurer la bonne mise en œuvre de cette politique, il faut optimiser la formation scientifique et technologique et favoriser le développement ciblé des compétences, car l'imitation technologique et industrielle est loin d'être une chose aisée.

Le rôle des universitaires, des scientifiques et des chercheurs sera extrêmement important. Et il sera d'autant plus déterminant qu'ils auront la charge d'instaurer, d'assurer et de promouvoir la *veille scientifique* ainsi que la *veille technologique*, tout en orientant une importante partie de l'effort de recherche vers l'*intelligence industrielle*. Celle-ci aura pour tâche le suivi et la consolidation d'*embryons industriels* créés selon les priorités économiques nationales. Pour le choix de ces embryons, la priorité des priorités sera pour l'industrie des *biens d'équipement*, car la production ''autonome'' de machines est le moyen par excellence pour assimiler les diverses technologies, pour favoriser la créativité technologique et pour donner un important ''coup de pouce'' au développement de l'appareil industriel national dans sa globalité. Rappelons à nos lecteurs que les expériences historiques montrent toutes que les tentatives de décollage économique qui ne se sont pas axées sur la *production autonome de biens d'équipements* ont abouti dans la majorité des cas à un échec.

### 2.2 Qu'exige cette stratégie de l'enseignement et de la recherche ?

Ce qui est bien dans une stratégie de décollage c'est qu'elle n'exige pas une révolution dans ces deux domaines, il suffit juste d'une réorientation de ceux-ci. Et cette réorientation est d'autant plus intéressante qu'elle donne enfin une mission claire et un objectif concret à la formation et à la recheche scientifiques, dans des économies qui en ont énormément besoin.

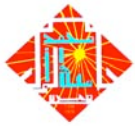 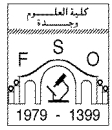 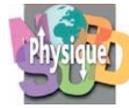 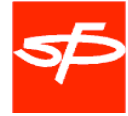

## 3 Contexte International, collaborations

Pourrait-on espérer une collaboration avec les pays du Nord sur une question aussi ''sensible'' ? Il ya quelque temps cette possibilté était, pour notre équipe, tout à fait à exclure, simplement parce que c'était trop demander. Nous nous sommes rendus compte par la suite que nous avions peut être une visions trop étroite de la question. Et cette vision s'est bien élargie à la suite de discussions avec quelques scientifiques européens au 1er Congrès Nord-Sud. Nous avons été agréablement surpris par la réaction de certains vis-à-vis de l'idée de stratégie de décollage incluant politique d'imitation, intelligence industrielle, etc. François Rabelais avait donc raison « les grandes âmes sont capables des plus grandes vertues ». Mais précisons qu'il n'y a pas que la vertue qui doit entrer en jeu, il y a aussi les interêts communs des deux continents, qui doivent eux aussi guider une possible collaboration en ce qui concerne cette idée de stratégies de décollage. Nous sommes de plus en plus convaincus que l'adoption et la réussite de ces stratégies peuvent être tout autant bénéfiques pour les pays d'Europe que pour les pays d'Afrique, car nos intérêts sur plusieurs points sont communs. Le flux migratoire, qui prive l'Afrique d'une partie inestimable de ses cerveaux et de ses jeunes forces actives, n'est-il pas un facteur de plus en plus déstabilisant pour les pays d'Europe ? L'Afrique prospère et florissante ne serait-elle pas un marché économique potentiel pour les industriels européens de demain ? L'Europe développée et l'Afrique émergente ne pourront-ils pas constituer un jour une alliance économique intercontinentale capable de contrer la rude concurrence économique asiatique et américaine ?

Beaucoup de choses sont possibles avec une Afrique émergente et prospère. Peu de choses sont possibles avec une Afrique sous-développée et miséreuse.

Certains craignent, pourtant, qu'une Afrique prospère ne fournisse plus de puissance à la dictature, à l'extrémisme et à la barbarie. Ceci serait sans compter sur le rôle modérateur, régulateur et élévateur du progrès et du développement. Pour appuyer cela je donnerai un exemple qui m'est particulièrement cher : dans les périodes florissantes du Maghreb d'avant le 18ème siècle, dans certaines mosquées à Alger, il y avait une entrée spéciale pour notre communauté juive algérienne ; ainsi, une partie de la mosquée bandée de musulmans, était réservée à nos compatriotes juifs minoritaires pour y célébrer leur culte…Cette exemple lumineux de tolérance et de respect des croyances d'autrui sera plutôt perdu dans les périodes de décadence du Maghreb qui ont suivi.

## 4  Pertinence dans le contexte Maghrébin, Africain et propositions de collaborations

Qu'est ce qui est plus important pour l'Afrique que la question de son *développement* ? (je parle bien sûr d'un développement véritable de l'ensemble des nations africaines, et non de miettes de développement éparpillées dans une misère générale). Ce développement est possible, il suffit que nous ayons le courage et l'audace de l'entamer. Une question pertinente que beaucoup se posent : ce développement par décollage sera-t-il trop lourd financièrement ? Il ne sera certainement pas plus lourd que le poids de nos dépenses pour les stratégies inefficaces et pour les guerres fratricides entre nos pays.

Nous souhaitons vivement une collaboration avec d'autres scientifiques africains sur cette question. Beaucoup de travail est encore à faire, beaucoup de points sont encore à traiter, et toute contribution sera la bienvenue.

## 5    Conclusion

« Une âme qui s'élève élève le monde » , que dire alors d'un continent s'il s'élève… Le sous-dévelopement de l'Afrique n'est absolument pas un atout pour le reste de l'humanité, bien au contraire ce sera toujours une source possible de graves instabilités pour le monde. Les pays du Nord doivent, à cause de l'influence très souvent bénéfique qu'ils exercent sur les pays du Sud, soutenir l'idée du décollage économique et la réorientation de l'enseignement et de la recherche pour servir ce décollage (comme le Japon l'a fait, à partir des années 1960, pour les ex- pays sous-développés d'Asie qui l'entouraient, et ce fut probablement plus par intérêt que par noblesse ─ même si nous ne pouvons négliger la part de noblesse dans tout ces actes qui élèvent les nations…).

### Remerciements

Nous remercions tous ces honorables scientifiques qui ont, si généreusement, dépensé temps et efforts afin que ce 1er Congrès Nord-Sud puisse avoir lieu, et qu'il se soit déroulé dans les meilleures conditions possibles. Nous remercions également tous ceux qui, par leurs divers apports, ont contribué à renforcer les liens humains et scientifiques entre les pays d'Europe et les pays d'Afrique.

### Références

[1]   La Corée du Sud : une sortie du sous-développement, M. Lanzarotti, PUF,1992
[2]   Le Tiers-Monde: les stratégies de développement à l'épreuve des faits, A. Zantman, Hatier, 1991
[3]   Japon: troisième grand, R. Guillain, Seuil, 1981
[4]   Les réussites du Sud-Est asiatique dans le commerce mondial, B. Balassa et J. Williamson, Economica, 1989
[5]   La renaissance de l'Asie, F. Godement, Odile Jacob, 1995

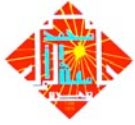 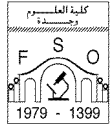 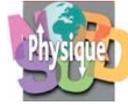 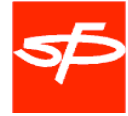

# L'approche du genre et les problèmes spécifiques aux jeunes femmes en recherche et en enseignement

**Monia CHEIKH, Maitre Assistante**
Université Tunis ElManar,Tunis,Tunisie

**Résumé**

L'inégalité dans l'education est une atteinte majeure aux droits des femmes et des filles et un obstacle important au developpement social et économique.Une égalité réelle des sexes dand l'éducation suppose d'offrir aux garçons et aux filles les mêmes possibilités d'accés à l'école,des programmes d'enseignement dépourvus de clichés ,des services d'orientation qui ne soient pas déformés par les préjugés du genre.Elle doit s'accompagner d'une égalité de rendement(durée de scolarisation,réussite,qualification scolaire...) et au-delà d'un accés égal au marché de l'emploi,à la promotion professionnelle,sociale et politique. Les sociétés doivent opérer des changements dans un large éventail de politiques économiques et sociales afin d'assurer l'éducation pour tous,non seulement par équité mais parceque leur progrés économique et social en dépend.

## 1 Introduction

Les spécialistes des sciences sociales et ceux du developpement utilisent deux termes distincts pour marquer,entre hommes et femmes ,les différences déterminées biologiquement et celles construites socialement :il s'agit des mots sexe et genre.Le sexe marque les carctéristiques biologiques permanentes et immuables des hommes et des femmes communes à toutes les sociétés et à toutes les cultures.Alors que le genre se réfère aux différences assignées socialement aux hommes et aux femmes .Les disparités du genre varient selon les cultures et selon les périodes en fonction de l'évolution de la société.Leurs implications dans toutes les sphères de la vie sont multiples : répartition du travail domestique et extra-domestique ,niveau d'éducation ,promotion professionnelle,insertion dans les instances du pouvoir….

Cela fait plusieurs années que les gouvernements et les organisations de développement accordent une priorité importante à la problématique du genre lorsqu'ils arrêtent conjointement leurs stratégies et concoivent leurs politiques.L'intégration de l'approche genre dans leurs plans d'action tend à la réalisation de la justice sociale,l'éfficacité économique et le développement durable.

## 2 Genre et Education

### 2.1 Rapport mondial sur l'éducation[1]

Lors du Forum mondial sur l'éducation à Dakar,en avril 2000 on constate que les deux tiers des 860 millions d'adultes analphabètes étaient des femmes et 57% des 104 millions d'enfants non scolarisés étaient des filles Les trois-quarts de ces enfants vivent dans des pays d'Afrique subsaharienne et d'Asie du Sud et de l'Ouest.

L'écart entre les taux de scolarisation est défavorable aux filles..La réalisation de l'égalité entre les sexes à tous les niveaux de l'éducation d'ici 2015 constitue un défi redoutable pour beaucoup de pays.

### 2.2 Bloquage des filles

Les principaux obstacles à la parité entre les sexes au primaire et au secondaire sont :la pauvreté,le travail des enfants, l'insécurité, le mariage précoce des filles (à 15 ans : 38%en Inde,à8ans en Afrique occidentale…). Dans le supérieur :les préjugés et les réticences familiales freinent l'accés des filles aux filières scientifiques et techniques et la progression de leurs carrières professionnelleLes organisations internationales mettent l'accent sur l'interdépendance entre la réforme éducative et le changement social et économique.L'égalité des sexes dans l'éducation peut contribuer à poser les fondements d'une égalité plus globale dans la société.

## 3 L'exprience Tunisienne

La Tunisie a hérité d'une tradition réformiste favorable aux femmes Des penseurs progressistes et des leaders nationalistes ont été à la tête du courant avant -gardiste et ont apporté chacun sa contribution intellectuelle, politique ou législative à l'édifice légal et socioculturel dans lequel évoluent les femmes aujourd'hui. Le principe d'égalité entre hommes et femmes devant la loi et sur le plan de la citoyenneté est

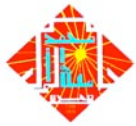 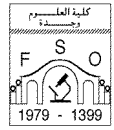 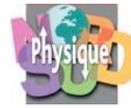 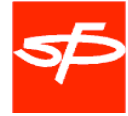

expressément affirmé dans les textes constitutionnels et législatifs tunisiens.

## 3.1 Femme Tunisienne et Education

Les réformes sociales mises en oeuvre, les politiques menées dans le domaine de l'éducation obligatoire et gratuite pour les garçons et les filles ont permis :

- le relèvement du taux d'alphabétisation à 77.1%

- La progression permanante du taux de scolarisation des enfants (99% en 2002).

Les statistiques[2] montrent une augmentation continue du nombre des filles dans les cycles : primaire(47,7% en 2002-2203),secondaire (53% en 2003-2004) et supérieur (57% en 2004-2005). La parité des sexes est donc atteinte, mais de réelles différences existent encore quand à la nature des études embrassés. Les préjugés, les réticences familiales, les habitudes sociales donnent la part belle aux garçons dans les domaines scientifiques et techniques.

## 3.2 Femme Tunisienne et emploi

La législation tunisienne consacre l'égalité entre l'homme et la femme dans tous les domaines du travail.La femme tunisienne a accédé aux divers secteurs du marché de l'emploi,malgré cela les filles sont incitées à opter pour des carrières dites féminines ,l'enseignement est l'une d'entre elles. Dans le corps enseignant du supérieur le nombre d'enseignantes[2] a atteint 40% en 2004-2005.

## 3.3 Carrières scientifiques et académiques des femmes

### Constat

A l'université tunisienne on note de larges différences du taux de féminisation selon les grades ,les disciplines et les institutions.Les femmes occupent en majorité les grades les moins élévés (assistants et maîtres asssistants)

| Position | Professeur | M.C | Ass.Ass | Ass | Agrégé |
|---|---|---|---|---|---|
| Femme/total | 6/70 | 2/76 | 7/334 | 2/240 | 3/43 |

Une fois recrutées à l'université la carrière scientifique et de recherche des femmes enregistre un ralentissement notable ,stagnation dans certains cas ,interruption puis redémarrage tardif dans d'autres.Comme la promotion académique se fait par les travaux de recherche ,le bloquage des carrières des femmes dans l'enseignement supérieur indique leur participation insuffisante à la recherche et donc une déperdition du potentiel scientifique de l'université.

Les femmes sont largement sous représentées dans les instances académiques et syndicales, et dans les organes d'encadrement et de direction de l'université tunisienne. Un faible pourcentage de postes de responsabilité de gestion ou d'administration leur sont accordé.Elles sont absentes dans les instances décisionnelles de la politique scientifique.

### Facteurs de Freinage :

- **Responsabilités Familiales** : En Tunisie comme ailleurs les femmes assument la majeure partie du travail domestique,les principales responsabilités de l'éducation des enfants ,et des soins aux parents. Elles sont obligées de concilier leur rôle de femme au foyer avec celui d'agent économique productif et de citoyenne .L'évolution de leur carrière professionnelle et leur présence dans les instances du pouvoir s'en ressentent. Les inégalités entre hommes et femmes dans la vie publique trouvent leur premier ancrage dans la vie familiale,ou la distribution inégalitaire du travail et des responsabilités circonscrit fortement leur participation dans le domaine publique .

- **Préjugés Sociaux et Culturels** : A cela s'ajoutent les images stéréotypées et les à priori sociaux et culturels qui constituent pour les femmes un obstacle à leur entrée dans les fonctions liées au pouvoir politique et économique. Donc malgré la proportion importante de femmes formées et diplômées, la société reste privée de leur participation active au développement.Dans l'enseignement supérieur en plus des deux facteurs précédents s'ajoute le manque de valorisation de l'effort de recherche scientifique.Les femmes ne sont pas encouragées à continuer cet effort après l'obtention du doctorat ce qui limitent l'évolution de leurs carrières et les cantonnent dand le rôle d'enseignantes uniquement.

## 4 Carrières scientifique et Académique des femmes dans le monde

En Europe la sous représentativitée des femmes dans les disiplines scientifiques et techniques et les inégalités de carrières dans les métiers correspondants est analysée :
70% des chercheurs sont des hommes,10% des femmes dans l'ingénerie et la technologie,89% responsables masculins dans les niveaux élévés de la hiérarchie scientifique. En France ,on note au CNRS de larges différences du taux de féminisation selon les corps et les grades. Il existe aussi Une forte mobilisation des associations et des pouvoirs publics pour promouvoir la place des femmes en science.

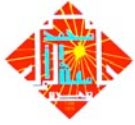
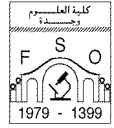
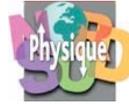
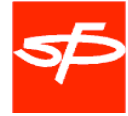

# 5 Conclusion

Les pays industrialisés et les pays en transition ont atteint la parité entre les sexes dans l'éducation depuis 1990. Dans les trois régions où les inégalités entre les sexes sont les plus fortes – Afrique subsaharienne, États arabes et Asie du Sud et de l'Ouest – les disparités ont régressé de manière substantielle. Pour beaucoup de pays l'objectif de la parité dans l'éducation d'ici 2015 ne peut être atteint qu'au prix de profondes réformes économiques et sociales et d'un engagement financier de la communauté internationale. La parité numérique est une étape nécessaire mais pas suffisante sur la voie de l'égalité. Celle ci renvoie à l'égalité des chances dans le marché du travail, l'accès aux ressources,et au pouvoir politique et économique. En Tunisie la parité dans l'éducation est réalisée. Cependant plusieurs chiffres posent des questions qui méritent réponse :

• Les Filles réussissent mieux que les garçons (quelles filles ? et quels garçons ?,dans quelles filières ?)

• Parité en faveur des filles à l'université (sous performance des garçons ?,préférence masculine pour la formation technique et professionnelle ?, métiers différents ?...)

• La féminisation du corps enseignant (carrière compatible avec les responsabilités familiales ? insuffisance des rémunérations pour les hommes ?..)

On constate aussi que cette légère supériorité scolaire des filles n'est pas convertie en une plus grande égalité dans les autres sphères de la vie .Il est nécessaire de produire des données sexuées , afin d'identifier les facteurs affectant la promotion sociale économique et politique des femmes.

## Remerciements

Merci au comité d'organisation et à tous les participants qui ont enrichi par leur présence ce congrés.

## Références

[1] Rapport mondial de suivi sur l'éducation pour tous publié en 2003 par l'Organisation des Nations Unies pour l'éducation,la science et la culture.

[2] www.tunisie.com

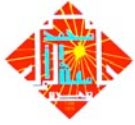 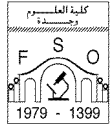 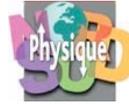 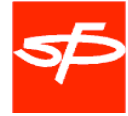

# Forum des journalistes scientifiques

**Bernard MAITTE, Professeur des Universités**
Responsable du master « Journaliste scientifique », Centre d'Histoire des Sciences et d'Epistémologie,
Université des Sciences et Technologies de Lille,
Villeneuve d'Ascq – France

**Résumé**

Les sciences et les nouvelles technologies contribuent à modifier considérablement nos sociétés. Pourtant, la place que leur réserve la presse est faible. Les enjeux du développement des nouvelles technologies peu souvent abordés, le débat démocratique sur celles-ci quasi inexistant. Pour remédier à ces faiblesses, une formation au journalisme scientifique a été mise en place grâce à la coopération de l'Université de Lille 1 et de l'Ecole Supérieure de journalisme de Lille. Depuis 2004, une formation analogue existe à l'Université Saad Dahleb de Blida. Elle vient d'être étendue en 2007 à l'ensemble du Maghreb. Ce sont ces expériences que relate l'intervention.

## 1- Les présupposés

Nouvelles façons de produire, nouvelles manières de concevoir et de fabriquer, nouvelles sources d'énergies mises à disposition, nouvelles formes de cultures induites par les biotechnologies, nouvelles façons de communiquer, nouvelles possibilités offertes à la médecine, nouvelles formes de naissances ….force est de constater que les découvertes scientifiques et les innovations technologiques marquent nos sociétés, contribuent à modifier la vie quotidienne et les habitudes individuelles, à transformer l'économie, la vie sociale, la culture, à interroger l'éthique …

Cette prégnance du scientifique sur la société s'exerce à la fois dans les pays industrialisés, qui sont passés, en une cinquantaine d'années d'une économie nationale à une autre, mondialisée, et dans les pays du Sud où l'on a cru – au mieux – que l'injection et l'utilisation de technologies transférées allaient permettre l'accès du développement. Dans les deux cas cependant, au Nord comme au Sud, les mutations actuelles sont vécues dans un contexte de crise. Malgré les avancées considérables, les dernières décennies sont en train de façonner l'avenir des sociétés et de la condition humaine sans que cet avenir ou cette condition puisse être pensé, souhaité, décidé.

C'est que nous voyons mal, en effet, dans quels buts, pour quelle fin, les progrès scientifiques entraînent ces afflux technologiques, qui induisent les transformations économiques et sociales. Nous vivons l'époque charnière du passage rapide d'une culture tournée vers les hommes et leur devenir à une culture tournée vers les objets et les moyens. Bien que porteuses d'importantes mutations culturelles, économiques, sociales, les sciences – dont la pratique nécessite un langage hautement symbolisé et une spécialisation poussée – restent, même dans les pays industrialisés, un domaine extérieur ou incompris par la grande majorité de la population. Elles sont certes montrées, données à voir, mais non appréhendées ou rendues intelligibles. Pour le plus grand nombre l'Ecole elle-même se révèle incapable à mettre en éveil l'esprit sur les questions scientifiques et techniques, à suivre leur développement, les rythmes de leurs applications, à cerner leurs implications.

Qu'en est-il alors de la presse ? La part la plus importante du faible espace occupé par la science l'est pour des informations ponctuelles et disparates qui ne font pas sens, accroissent la distance existant entre science et public. Les journaux spécialisés s'adressent à des lectorats motivés, ciblés, peu nombreux, qui viennent chercher les informations les réponses à des questions qu'ils se posent. Bien peu de débats ont lieu sur l'utilisation de cette science qui modifie actuellement les frontières qui avaient été tracées depuis des millénaires entre le vivant et le technique, le local et le global, le naturel et l'artificiel. Sans trop caricaturer, on peut proposer que la science ne fait pas partie de la culture de « l'honnête homme » du XXIe siècle, ce qui interdit l'existence de débats démocratiques sur l'introduction des nouvelles technologies dans la société. C'est à ce besoin que veut répondre la formation au journalisme scientifique existant à Lille depuis maintenant une quinzaine d'années.

## 2- La formation en journalisme scientifique à Lille

D'abord conçue comme un Diplôme

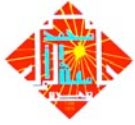 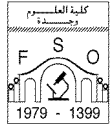 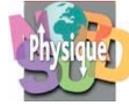 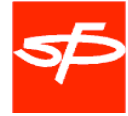

d'Enseignement supérieur Spécialisé (DESS – Bac+5), la formation est maintenant un Master 2 professionnalisé (Bac+5) et donne lieu à la délivrance simultanée d'un certificat d'aptitude professionnelle par l'ESJ. Les étudiants sont sélectionnés par un concours national auquel peuvent s'inscrire tous ceux qui ont, au moins, un diplôme de sciences exactes ou naturelles de niveau bac+4 (maîtrise, master 1, ingénieur …). Le concours écrit comporte surtout, puisque le niveau scientifique initial a été contrôlé par l'obtention d'un titre universitaire, des épreuves de français, de rédaction, de comptes-rendus , sur la connaissance de l'actualité générale des six derniers mois. Les étudiants admissibles après ce concours, passent un oral où ils ont deux épreuves : l'une de motivation, l'autre sur un sujet « Science et société », proposé depuis l'écrit, tiré au sort parmi les six proposés.

## 2–1 Présentation

La filière « Journaliste et Scientifique a pour mission de former, en un an, des scientifiques aux différents métiers du journalisme de presse écrite, que ce soit en presse générale ou spécialisée, en rédaction ou en réalisation. Les étudiants devront posséder des réflexes d'adaptation et des connaissances complémentaires, leur permettant de passer, sans dommage, d'un secteur à un autre. Cette diversité de destination, dans un métier en devenir, amènent à opter pour la polyvalence, aussi bien sur le plan des méthodes et techniques d'information que sur celui des connaissances générales.

## 2–2 Objectifs

Former des journalistes scientifiques ayant une double compétence :
- celle de journaliste capable de recueillir et traiter l'information, connaissant l'économie et le droit de la presse, soucieux de l'éthique professionnelle, sachant s'intégrer dans une équipe.
- Celle de « critique de sciences » soucieux de mettre en perspective les sciences qui se créent, sensibles aux conséquences économiques, sociales, culturelles, politiques, éthiques des applications qui modifient notre quotidien.

L'objectif est de former des professionnels sachant appréhender les modes de production et de diffusion des connaissances scientifiques et technologiques. Il s'agit de mettre en perspective historique, épistémologique et culturelle la science qui se crée dans les laboratoires, qui s'applique dans les entreprises industrielles ; de mettre en débat et en critique les problématiques posées par l'introduction des innovations technologiques dans la société. Il s'agit également de connaître et de maîtriser les techniques d'écriture et de mise en forme des informations à caractère scientifique, connaître le monde des médias, ses contraintes et ses modes de fonctionnement.

## 2–3 Spécificités de la formation

Les enseignements sont organisés conjointement par l'Université des sciences et technologies de Lille et l'ESJ autour de deux axes :
– Epistémologie des sciences ; les technologies dans la société (USTL)
– Enseignements professionnels : généraux et techniques (ESJ)

## 2–4 Contenu de la formation

Les grands axes qui structurent la formation sont les suivants :
- Enseignement généraux : ESJ, USTL) Les grandes questions contemporaines ; étude de l'actualité ; anglais scientifique
- Culture et actualité scientifique et technique (USTL) : Les sciences et les techniques : perspectives historiques et épistémologiques. Histoire de la diffusion des sciences et des techniques vers un large public. Comment présenter la science ?
- Les grandes problématiques scientifiques et techniques (USTL) : Les applications des sciences dans la société, leurs implications, leurs crtitiques, quels intérêts sont en jeu ? Comment se posent débats et confrontations ? Quels sont les succès et les menaces ? Critiques d'ouvrages épistémologiques. Enquêtes sur les représentations des sciences dans les publics.
- Enseignements spécialisés (ESJ) : connaissances des médias, droit de la presse, éthique professionnelle.
- Enseignements techniques et professionnels (ESJ) : sources et collecte de l'information locale ; l'enquête sur l'information scientifique ; les genres rédactionnels et les partis de l'article ; l'expression écrite de l'information ; micro-édition ; multi-média ; initiation à la radio et à la télévision.
- Stages : dans un laboratoire ; dans une structure de communication d'un grand organisme, dans un quotidien, dans une revue, stage de fin d'études de deux mois au minimum dans une entreprise de presse.
- Projets : réalisation d'enquêtes ; critiques de l'information scientifique publiée ; observations de laboratoire, de la presse ; réalisation d'un magazine (projet de groupe en vraie grandeur).

Depuis une quinzaine d'années, ce sont environ 200 journalistes scientifiques qui ont été formés à Lille, dont 90% sont effectivement en poste dans la presse.

## 3 La formation au journalisme scientifique mise en place à l'Université de Blida pour le Maghreb

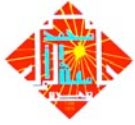 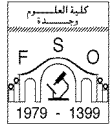 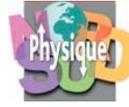 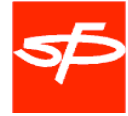

En février 2004, à l'occasion du colloque sur l'histoire des sciences organisé à l'Université Saad Dahleb de Blida, des contacts étaient noués entre cette Université et Lille 1, Madame Mimoune, Rectrice de l'Université de Blida venait à Lille et découvrait notre Université, l'ESJ et la formation J.S. Elle voyait tout de suite l'intérêt de développer une telle formation en Algérie et signait en mai 2004 une déclaration d'intention avec le Président de Lille 1 et le Directeur de l'ESJ.

En 2004-05, cinq séminaires de deux jours étaient organisés à l'Université de Blida. Ils s'adressaient aux enseignants de sciences et de communication des Universités algériennes, aux journalistes, aux étudiants potentiellement intéressés. Ces journées connurent un grand succès par l'affluence (plus de 130 personnes à chaque session), les discussions et les contacts qu'elles permirent. Partant d'abord d'exposés de principe, les cessions se firent de plus en plus précises dans les thèmes, plus techniques quant aux possibilités de mise en place. Elles permirent d'adopter une charpente générale de la future formation, de préciser les rôles, d'associer des enseignants algériens, d'obtenir la labellisation de la formation par les autorités algériennes d'une part, par le Haut-Conseil franco-algérien de l'autre, pour qui l'initiative devint un projet pilote particulièrement soutenu.

## 3–1 Les principes de la collaboration

La formation mise en place à Blida fut au départ à la fois une post-graduation spécialisée (PGS – en un an) et une PG (deux ans, avec mémoire). Elle repose sur quatre partenaires : l'Université Saad Dahleb de Blida, responsable du projet, et la faculté de communication d'Alger pour l'Algérie ; l'Université de Lille 1 et l'ESJ pour la France. La charpente générale de la formation lilloise a été reprise et adaptée : cours, stages, exercices ; les partenaires français prennent en charge les aspects « science et société » et « .écriture journalistique », ainsi que l'encadrement de la réalisation du magazine de fin d'année (qualité : hebdomadaire ou mensuel français). Parallèlement, une formation de formateurs a été mise en place : venue à Lille pendant un an dans la formation J.S d'enseignants de Blida et formation en Algérie : la vocation des partenaires français étant à terme de ne plus intervenir dans la formation de façon directe.

## 3–2 Succès et difficultés

La formation en est à sa deuxième année de fonctionnement. Elle vit de manière satisfaisante. Un grand nombre de candidats se présente au concours. Les étudiants sont motivés, apprenant beaucoup, la presse est particulièrement intéressée, suit la formation, accueille les stagiaires, offre des débouchés.

L'Université de Blida dégage des moyens financiers (missions), a acquis la documentation multimédia et les ouvrages nécessaires, s'est équipée en informatique et en logiciels spécialisés.

Le Haut-Conseil et l'Ambassade de France s'impliquent fortement (missions, aides diverses). Un colloque de fondation maghrébin s'est tenu à Blida les 5-6 mai 2006,avec des représentants marocains et tunisiens. Il est envisagé d'étendre la formation au Maghreb pour les étudiants (lieu Blida) et de faire appel aux compétences en formation journalistique existant au Maroc et en Tunisie. Un centre de formation du journalisme est en projet.

Dès 07-08, l'Université de Blida passe au LMD. La formation journalisme scientifique passera, comme à Lille, en M2 pro. Le M1 n'existant pas dans cette filière, l'année 07-08 sera utilisée pour faire passer les mémoires des étudiants de 06-07 et de se consacrer à la nécessaire formation de formateurs.

## 4 Pertinence dans le contexte Maghrébin et Africain

Dans l'état, une telle formation ne peut actuellement se passer d'interventions très fréquentes des enseignants lillois : 4 jours tous les mois et demi, rythme qui ne peut être poursuivi très longtemps, ni dupliqué ailleurs. La formation de Blida doit rester unique au Maghreb, la montée en puissance progressive et contrôlée. Un magazine mensuel maghrébin en langue française pourrait être édité prochainement grâce à l'AUF. Les pays Sub-saharien pourraient être concernés par la formation.

## 5 Conclusion

La formation « journalisme scientifique » mise en place à l'Université Saad Dahleb de Blida fonctionne de façon tout à fait satisfaisante. Elle va s'étendre au Maghreb et à l'Afrique Sub-saharienne. Elle peut conduire à la création prochaine d'un mensuel en langue française, elle forme des journalistes qui prennent place dans les équipes de rédaction des journaux existants. Très différente dans ses buts de la « communication scientifique » ou de « l'information scientifique spécialisée », elle vise à former des « critiques de science », aptes à mettre en débats les enjeux de l'introduction des nouvelles technologies dans la société. A ce titre, elle contribue à initier un débat démocratique sur les sciences et leurs applications. De nombreuses collaborations sont souhaitées par les initiateurs.

### Remerciements

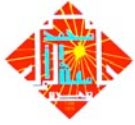 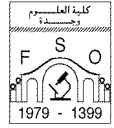 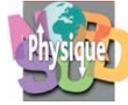 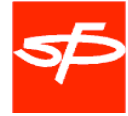

# La physique : un art de l'homme moderne

## Mohammed DIOURI, Professeur

Laboratoire de physique de l'atmosphère, Université Mohammed 1$^{er}$ Oujda, Maroc.
diouri@fso.ump.ma

**Résumé**

Depuis son apparition, l'homme moderne a commencé la diffusion de son savoir, cet art a connu une croissance continûment vertigineuse et une avancée scientifique sans précédent. Parmi les arts, les grandes découvertes de la physique ne se classent pas seulement au niveau des concepts ou des modèles de représentation confinés par des équations mathématiques, elles se classent aussi par la manière de leur exploitation. La beauté d'une équation ou d'un modèle de représentation trouve sa consécration et sa portée dans sa parfaite utilisation : De plus en plus l'idée scientifique constitue dans sa finalité un produit de consommation, cette notion réductrice de l'idée scientifique est en contradiction avec le progrès observé au niveau de l'institution sociale et la notion de **saveur** d'un produit est plus appliquée dans les pays du Nord. Au moment où, à l'aide des techniques de la physique contemporaine, le transport de l'information se fait à la vitesse de la lumière et sa diffusion est facilitée à travers le monde entier, la diffusion du « savoir scientifique » elle, rencontre encore des barrières institutionnelles et culturelles et sa période pour atteindre le niveau de culture générale reste encore très longue dans les pays du Sud.

Depuis son apparition, l'homme moderne a commencé la diffusion de son savoir, le physicien tel un artiste, utilise la pensée, l'intuition, l'imagination pour composer les lois de la physique sous une forme naturelle simple. Cet art humain a connu une croissance vertigineuse et une avancée scientifique sans précédent. Le réacteur de fusion nucléaire ITER[1], l'accélérateur de particules LHC[2], le satellite Envisat[3] ou le télescope européen VLT[4] sont des chef-d'oeuvres d'art ou encore des outils nés de l'imaginaire des physiciens artistes qui permettent entre autre l'amélioration des lois de la physique et leur compréhension.

Tel un art la physique peut être contemplée à travers différents tableaux en relation avec la société ou la pensée vues sous un angle d'esthétique propre. La relation physique société peut être approchée historiquement, un balayage des moments forts de la physique montre l'importance des constantes physiques. Depuis le premier grand pas de Galilée lié à la constante G à nos jours, on peut s'arrêter aux plus remarquables, à savoir la vitesse de la lumière de Römer c, la constante de Boltzmann k et le pas géant lié à la constante de Planck h rattachée à la physique contemporaine.

Les observations du ciel ont été à l'origine des grandes idées nouvelles de Galilée qui ont connues des oppositions manifestes (Et pourtant elle tourne !)[5] et jusqu'à maintenant ce sont encore et aussi les observations du ciel, cette fois-ci avec des moyens technologiques de grandes précisions, qui continuent à apporter les éclaircissements nous permettant une meilleure approche aussi bien de l'infiniment grand que de l'infiniment petit.

Depuis le 19$^{ème}$ siècle l'imaginaire humain n'a pas cessé de grandir et de développer les moyens de son développement, de quelques dizaines de chercheurs au début du vingtième siècle à plusieurs millions au début de ce troisième millénaire avec des moyens d'information de plus en plus importants et rapides et des implications plus larges qui lient la physique aux différents autres arts, en particulier on peut citer l'évolution que connaît actuellement l'arbre de la vie en passant de celle de Haeckel[6] à celle basée sur la connaissance du genom.

La physique continue d'observer le monde et de tenter de le comprendre et de l'expliquer avec une quête incessante d'intelligibilité et de création, cependant les grandes découvertes de la physique ne se classent pas seulement au niveau des concepts ou des modèles de représentation confinés par de belles équations mathématiques qui régissent les phénomènes physiques, elles se classent aussi dans la manière de leur exploitation.

La beauté d'une équation ou d'un modèle de représentation trouve sa consécration et sa portée dans sa parfaite utilisation : De plus en plus l'idée scientifique constitue dans sa finalité un produit de consommation, cette notion réductrice de l'idée scientifique est en contradiction avec le progrès observé au niveau de l'institution sociale et la notion de saveur d'un produit plus que jamais nécessaire est plus et surtout appliquée dans les pays du nord.

Au moment où à l'aide des nouvelles techniques de la physique contemporaine, on développe le monde des environnements intelligents qui incorporent la radioidentification, au moment où le transport de l'information se fait à la vitesse de la lumière et sa diffusion est facilitée à travers le monde entier, la diffusion du « savoir scientifique » elle, rencontre encore des barrières institutionnelles et culturelles et sa période pour atteindre le niveau de culture générale reste encore nettement plus longue dans les pays du Sud.

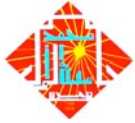 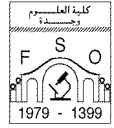 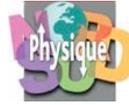 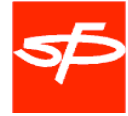

### Recommandation

Il est urgent de penser à la réalisation avec une coopération européenne d'un observatoire de l'atmosphère et du ciel sur les hauteurs des Béniznassen qui représentent beaucoup d'intérêts scientifiques.

[1] LHC sera le plus grand accélérateur collisionneur de particules au monde, un des outils scientifiques les plus grands jamais imaginé et construit par l'homme. Le collisionneur dont la circonférence est de 26,7 km est enterré à une profondeur variant entre 50 et 170 m construit dans 2 pays Suisse et la France. Il va utiliser environ 10 000 aimants supraconducteurs afin de courber la trajectoire des particules. Il sera prêt en 2007.

[2] ITER réacteur de fusion thermonucléaire dont le père est le Joint European Torus de Culham. sa construction débutera en 2008 à Cadarache et les premiers essais sont prévus en 2016.

[3] La mission ENVISAT (ENVIronnement SATellite) a été imaginée par l'Agence Spatiale Européenne (ESA). Satellite dédié à l'étude des ressources terrestres est lancé par Ariane 5 le 1$^{er}$ mars 2002.

[4] VLT (*Very Large Telescope*) est un ensemble de 4 télescopes géants construits au nord du Chili. Ce projet est développé par l'"*European Southern Observatory*" voir image ci-après.

[5] Citation de Galilée : Après des études en Médecine, Mathématiques et Philosophie, il obtient un poste à l'Université de Pise où il étudie alors les mouvements et la chute des corps. Il devient ensuite professeur de mathématiques et se consacre à l'astronomie. Il défend les idées de Copernic et de l'héliocentrisme (c'est le Soleil qui est au centre de l'Univers, et non la Terre), ceci lui vaut d'être attaqué par l'Eglise. Son dialogue concernant les deux principaux systèmes du monde, de Ptolémée et de Copernic, le fait condamner pour hérésie par l'Inquisition. Galilée apparaît comme le fondateur de la physique moderne puisque pour lui les lois physiques doivent être établies sur des expériences. Il inventa notamment la lunette astronomique et le thermomètre...

[6] Haeckel : 1870 présente la généalogie classifiée des espèces vivantes. Première arbre de vie menant des racines avec le règne des Monera (les procaryotes) jusqu'à l'homme au sommet.

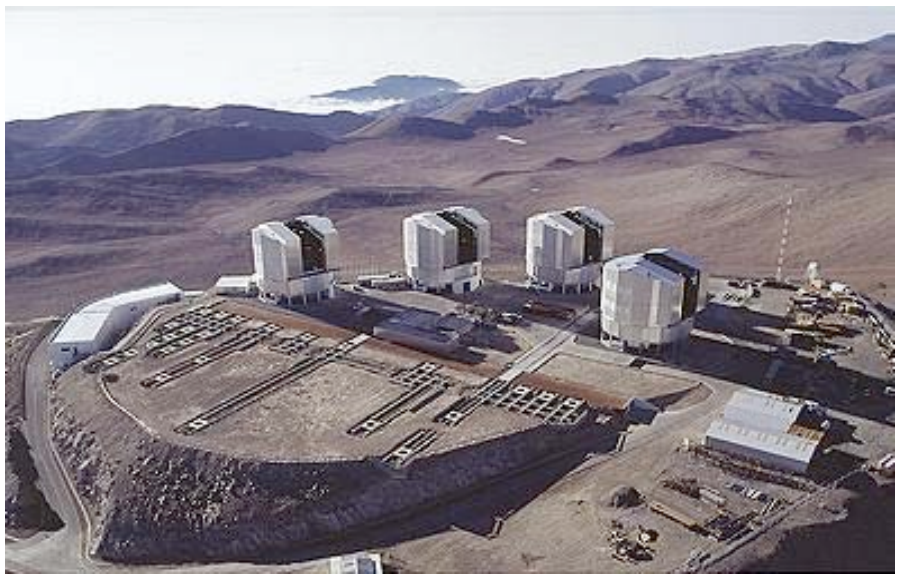

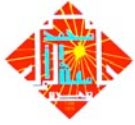 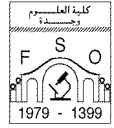 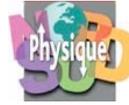 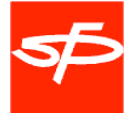

# Politique Scientifique Européenne

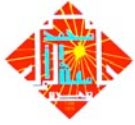 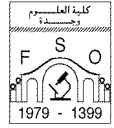 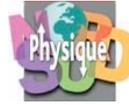 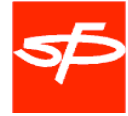

# La coopération scientifique et technique entre le Maroc et l'Union européenne

### Jean-Michel CHASSERIAUX, UE Bruxelles

L'engagement de l'Union européenne (UE) dans la coopération scientifique et technique (S&T) avec le Maroc s'est s'inscrit dès le milieu des années 80 dans le cadre des programmes de recherche STD puis INCO orientés vers les pays en développement. Par ailleurs et bien que la déclaration de Barcelone en 1995 qui visait à établir un partenariat euro-méditerranéen solide ne mentionne pas la recherche, un certain nombre d'activés proches ont pu cependant être conduite dans ce cadre notamment en matière de technologies de l'information et de formation. Aujourd'hui, la coopération S&T entre l'UE et le Maroc s'inscrit dans le prolongement de ces deux initiatives, c'est-à-dire au sein du 7$^{ème}$ Programme Cadre de Recherche et Développement (PCRD) pour l'essentiel et au sein de la « politique de voisinage » qui a profondément renouvelé la politique méditerranéenne de l'UE.

## 1- La Coopération Internationale dans le 7$^{ème}$ PCRD

Si les objectifs visés demeurent assez traditionnels :

- Soutenir la compétitivité européenne à travers des partenariats stratégiques avec les pays tiers dans des domaines scientifiques donnés,

- Résoudre les problèmes spécifiques auxquels les pays tiers font face sur base de l'intérêt mutuel et le bénéfice mutuel,

- Utiliser la coopération en S&T pour renforcer les relations extérieures et les autres politiques de la Communauté,

Une nouvelle approche est suivie pour les atteindre. Le programme INCO consacré exclusivement à la coopération S&T internationale est supprimé. A l'exception du programme consacré à la sécurité, l'ensemble du 7$^{ème}$ PCRD est désormais ouvert aux pays tiers (c'est-à-dire non-membres de l'UE), chaque programme ayant son propre volet international dans les limites qui lui sont assignées par son objet :

- Les projets de recherche en collaboration internationale sont soutenus par les thématiques du programme Coopération ;

- Les activités internationales de formation et mobilité sont gérées par le programme Personnes ;

- Les activités horizontales de soutien à la politique de coopération internationale sont développées dans le Programme Capacités ;

- Dans le programme "Idées" qui soutient l'ERC, les chercheurs étrangers peuvent à titre individuel participer aux projets que présentent leurs collègues européens et s'ils sont établis en Europe soumettre eux-mêmes des propositions et bénéficier d'un financement.

Les pays tiers sont divisés en trois catégories: les pays industrialisés, les pays associés et candidats, les pays partenaires de coopération internationale (ICPC) qui regroupent la quasi-totalité des pays à revenu faible ou intermédiaire dont les Pays méditerranéens.

Ils peuvent participer de deux façons différentes au *programme "Coopération"* qui devrait être le principal vecteur de la coopération :

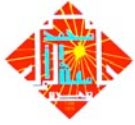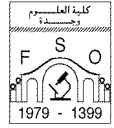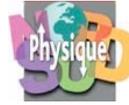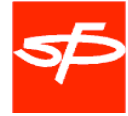

• **Dans le cadre d'actions spécifiques de coopération internationale (SICA) ciblées vers des pays identifiés dans les appels à propositions et dont les thèmes devraient à l'avenir être choisis en concertation avec les pays concernés. Dans ce cas, les propositions doivent réunir au moins 4 participants de pays différents dont deux de pays à revenu faible ou intermédiaire (dits pays ICPC) et deux Etats membres de l'UE ou associés au $7^{ème}$ PCRD.**

• **Dans le cadre des autres projets (hors sécurité). Les conditions de participation sont alors celles de ces projets à avoir au minimum, trois participants de trois pays membres de l'Union ou associés au $7^{ème}$ PCRD différents. Au-delà de ce minimum, tout pays tiers peut participer. Les ICPC sont financés, les pays industrialisés ne pouvant bénéficier d'un financement que si leur participation est jugée indispensable à la réalisation du projet.**

Les critères d'évaluation des propositions portent essentiellement sur la qualité scientifique et technique du projet, son impact sur les objectifs visés par le programme, la qualité de sa mise en œuvre. En ce qui concerne les SICA, s'y ajoutera naturellement le principe de l'intérêt mutuel.

A tire d'exemple de SICA, le programme "Coopération" pourra ainsi financer :

• **Dans le cadre du thème "santé" des actions sur les maladies de la pauvreté, les maladies orphelines, les politiques de santé, etc.**

• **Au sein du thème "alimentation et agriculture" des recherches sur la génomique pour amélioration des récoltes, l'épidémiologie, le contrôle et le développement de vaccins pour les maladies animales, la malnutrition et l'obésité, la biomasse, etc.**

• **Dans les technologies de l'information et de la communication des travaux sur la prévention des catastrophes naturelles, la réduction de la fracture numérique, etc.**

*Le programme "Personnes"* vise deux objectifs en matière internationale :

• **Renforcer la recherche européenne en attirant des chercheurs étrangers ;**

• **Bâtir des liens durables à travers la mobilité des chercheurs.**

Il comprend notamment d'une part des bourses dites "sortantes" pour les chercheurs européens qui veulent aller à l'étranger avec un retour obligatoire, et d'autre part, des bourses "entrantes" pour les chercheurs de pays tiers qui veulent venir en Europe. Des dispositions sont prévues pour limiter la fuite des cerveaux et faciliter le retour de ces chercheurs dans leur pays d'origine. Comme dans le $6^{ème}$ PCRD des chercheurs des pays tiers pourront également être accueillis en nombre limité au sein des réseaux de formation. En outre des moyens spécifiques seront mobilisés pour les pays du voisinage et les pays ayant un accord de coopération scientifique et technique avec l'UE. Enfin un effort est prévu pour soutenir les diasporas, celles des chercheurs européens à l'étranger comme celles des chercheurs étrangers en Europe.

*Le programme "Capacités"* comporte un volet "Activités de coopération internationale" doté de 185 millions d'Euros et destiné à financer des mesures horizontales de soutien à la coopération internationale. Il s'agit non seulement d'assurer la coordination des actions internationales conduites dans les différents thèmes du $7^{ème}$ PCRD mais aussi de définir une approche cohérente et de développer des synergies avec les autres politiques communautaires (relations extérieures,

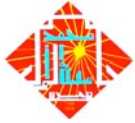 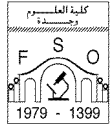 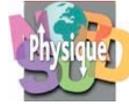 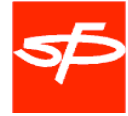

développement, environnement, commerce, etc) et enfin de renforcer la coordination des politiques nationales des Etats membres. Les activités prévues à cet effet peuvent être regroupées en trois catégories :

*1. La Coordination des coopérations bi-régionales* au travers d'un instrument nouveau : les INCO-NET. Il s'agit de plates-formes permettant d'établir un dialogue politique avec les décideurs et les partenaires d'une région/pays afin de :

• **Définir les orientations de la coopération et, en particulier, d'identifier des thèmes prioritaires dans le respect du principe de l'intérêt mutuel ;**

• **Mettre en œuvre des activités spécifiques destinées à promouvoir, organiser et structurer la participation des pays concernés au 7$^{ème}$ PCRD ;**

• **Suivre les performances et les effets de la coopération internationale dans le 7$^{ème}$ PCRD.**

*2. Le développement des partenariats bilatéraux dans le contexte des accords de coopération conclu entre l'UE et un certain nombre de pays tiers.*
L'objectif est ici de :

• **Renforcer les échanges d'information sur les programmes et les financements destinés à promouvoir la coopération entre l'Europe et les pays concernés, notamment en mettant en place des plates-formes d'information;**

• **Faciliter la formation des partenariats de recherche ;**

• **Mettre en valeur l'intérêt et le bénéfice mutuel qui peut être retirés de la coopération scientifique et technique ;**
• **Partager les bonnes pratiques à travers l'organisation de forums, de groupes de travail sur l'état de l'art et les perspectives de la coopération ;**

• **Établir des liens entre les collaborations existantes et les programmes de l'UE.**

*3. La coordination des politiques et activités nationales* par le canal des ERA-NETs.
Ce dispositif qui s'est mis en place au cours du 6$^{ème}$ PCRD vise à augmenter la coordination des programmes de coopération internationale scientifique et technique gérés au niveau national ou régional dans les pays membres. Les activités mises en œuvre dans ce cadre permettront de :

• **Définir une stratégie de coopération scientifique internationale au sein des membres de l'UE ainsi que des objectifs communs avec les pays tiers ;**

• **Encourager la mise en place de programmes conjoints ou coordonnés ;**

• **Renforcer l'efficacité et l'impact des coopérations bilatérales en cours entre les pays membres et les pays tiers.**

## 2. La politique de voisinage et les autres politiques

L'élargissement de l'UE à dix nouveaux États membres d'Europe de l'Est en 2004, puis à la Roumanie et à la Bulgarie, en 2007 a rendu nécessaire la redéfinition de ses relations avec ses voisins les plus proches. Elle doit notamment répondre aux attentes qu'elle suscite et qui ont été exprimées en particulier par le Maroc qui avait conclu des 2000 un accord d'association avec l'UE.

La politique européenne de voisinage (PEV) définie en mai 2004 s'adresse aux pays du Sud de la Méditerranée, aux pays de l'Europe orientale (Biélorussie, Ukraine, Moldavie), aux trois républiques du Caucase du Sud (Géorgie, Arménie, Azerbaïdjan).

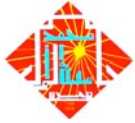 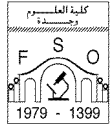 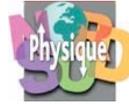 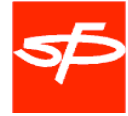

Elle bénéficie, à partir de 2007, d'un nouvel instrument: "l'instrument européen de voisinage et de partenariat" (IEPV) doté de 11 181 M€ pour la période 2007-2013, en augmentation de 32% par rapport à la période précédente 2000-2006. Sur le plan technique, l'attribution des crédits s'inscrit dans le cadre d'un certain nombre de documents étroitement reliés :

1. Le plan d'action voisinage qui s'inscrit dans le cadre des accords d'association et qui définit à court et moyen terme (3 à 5 ans) un agenda commun sur les réformes politiques et économiques et propose des orientations pour la programmation de l'assistance financière et technique. Le plan d'action entre l'UE et le Maroc a été adopté en 2005.

2. Le document de stratégie pays 2007-2013 qui établit un diagnostic de la situation du pays concerné.

3. Le programme indicatif national qui porte sur la moitié de la période de programmation i.e. 2007-2010 et qui définit projet par projet les grandes lignes de l'action communautaire. Le programme indicatif du Maroc est ainsi doté de 654 MEuros dont 110 pour l'éducation et l'alphabétisation.

4. Les plans de financement annuels qui permettent de lancer des mesures «concrètes» telles que des appels d'offres ou des appels à propositions.

L'approche retenue initialement était proche de celle mise en œuvre pour la gestion des fonds structurels et relativement peu adaptée à la coopération S&T. Mais la Commission a envisagé récemment, dans une communication en date du 4 décembre 2006 diverses mesures susceptibles d'y remédier. Elle propose notamment d'encourager les échanges de personnes dans des domaines tels que l'éducation, la culture, la recherche, de développer la dimension thématique dans des secteurs comme les transports, l'énergie l'environnement et de consolider l'approche régionale qui avait été encouragée par le processus de Barcelone puis relative délaissée au profit du cadre bilatéral.

Enfin, en matière d'éducation, le Maroc est éligible aux programmes Erasmus Mundus et Tempus qui connaissent actuellement un très fort développement lié à la prise de conscience de la nécessité de développer la dimension internationale de la stratégie de Lisbonne qui vise à faire de l'Europe l'économie de la connaissance la plus compétitive au monde et pour cela à renforcer le "triangle de la connaissance" : formation, recherche, innovation.

## 3. Conclusion

Malgré les évolutions intéressantes notées dans la politique de voisinage où la recherche apparaît désormais explicitement, l'outil privilégié de la coopération S&T entre l'UE et le Maroc demeurera le $7^{ème}$ PCRD. Il est encore trop tôt pour savoir si la phase d'apprentissage passée, les nouvelles dispositions qu'il comporte et dont beaucoup restent encore à affiner permettront de compenser et au-delà la disparition du programme INCO. L'affirmation récente de l'importance accordée à la coopération internationale dans le PCRD, l'apparition de la recherche dans la politique de voisinage permettent d'être raisonnablement optimistes.

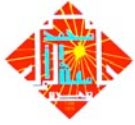 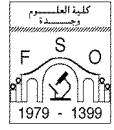 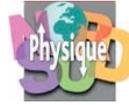 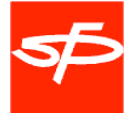

# Les Ateliers

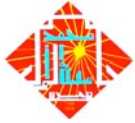 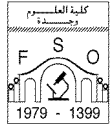 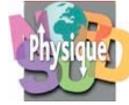 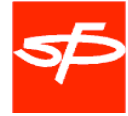

# Atelier n°1
## La Démarche expérimentale en physique

### Modérateur : Étienne GUYON

L'insuffisance des travaux pratiques ou d'une démarche d'expérimentation active dans de nombreuses universités du Sud nous a conduit à examiner les divers moyens de créer celle ci dans les établissements d' enseignement supérieur tout en reconnaissant une grande hétérogénéité suivant les établissements.

L'intérêt des très nombreux jeunes marocains qui se sont arrêtés devant les expériences de mosaïque de la physique a souligné tout à la fois leur curiosité pour des approches expérimentales mais aussi leur ignorance de cette pratique dans de nombreux cas.

La démarche peut se conduire à divers niveaux comme cela a été présenté dans l'atelier 1 :
- Mise en place de « TP traditionnels » avec un accompagnement matériel et technique
- Démarche de projet piloté
- Projets originaux en fin du cycle L débouchant sur une activité de recherche expérimentale (Couder à Paris bulletin SFP 40 1981)

## 1- Expériences de pratiques et d'outils pour les TP de DEUG

### Laurence CHERIGIER et Francis SECCIA
Université de Provence

L'expérience conduite sur 7 ans par l'Université de Provence pour mettre en place un service de TP dans des Universités de Madagascar est présenté et souligne les conditions dans les quelles une telle collaboration peut être efficace (matériel de rechange ; entretien ; formation de techniciens et ingénieurs de laboratoire).

La mise en place d'un service commun de TP sur l'ensemble de l'Université de Provence introduit une plus grande efficacité et économie dans sa gestion et peut servir de référence pour des services de TP dans les pays du sud. Nous avons souligné la nécessité d'un service technique de maintenance.

Notre expérience antérieure avec Madagascar a montré la nécessité de prévoir, en plus de l'armoire de TP, une formation et un accompagnement des techniciens de l'université destinataire. Un réel échange doit exister, sous forme de stage d'au moins deux semaines en France pendant une période d'enseignement des TP. Des visites sur place dont la périodicité est à définir pour assurer un suivi et un soutien technique sont aussi nécessaires.

Il nous faut trouver la personne ressource sur place et en faire le contact privilégié pour la partie technique du système. Ainsi la collaboration sera efficace et permettra les échanges de documents techniques et les astuces pour régler certains problèmes. En effet les travaux pratiques de l'armoire fonctionnent dans nos universités et sont l'émanation d'un travail réalisé dans nos ateliers.

Il est nécessaire que l'université destinataire soit équipée d'un atelier minimum comprenant :
- Un équipement de base pour effectuer les mesures utiles dans la détection d'un défaut ou d'une panne.
- Un équipement de soudage et dessoudage.
- Un pont de mesures pour pouvoir réaliser les prototypes.
- Une perceuse et mini tour avec l'équipement.
- L'outillage de l'électronicien.
- Une documentation technique.
- Un fond de roulement de composants et de matériaux de rechange pour effectuer les dépannages de premier niveau.

Des stages d'été sur la mise en œuvre et la maintenance des enseignements expérimentaux pourraient venir compléter ce processus en accueillant dans nos universités les enseignants et techniciens. Cela permettrait de présenter les nouveaux travaux pratiques et de leurs assurer des séances de manipulation en situation d'élèves.

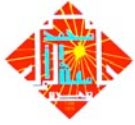 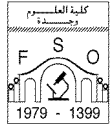 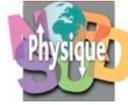 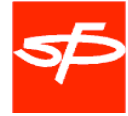

## Atelier n°1
## 2- De l'expérience de laboratoire et des projets de travaux pratiques jusqu'aux modules d'exposition

Patrice Jenffer - Université Paris XI - Orsay

## But : initier les étudiants à la pratique expérimentale

Pour cela nous avons choisi la mécanique des fluides.
**Pourquoi ce choix ?**

- la mécanique des fluides est un domaine neuf pour les étudiants
- les Manipulations peuvent être simples ("de coin de table"),
- les Observations sont souvent spectaculaires
- ce sont aussi les expériences de la nature et de la vie de tous les jours
- cela sert aussi à démystifier l'approche théorique directe de l'hydrodynamique
- autre critère important ces projets expérimentaux associent souvent le Palais de la Découverte, le département de physique enseignement d'Orsay ainsi que les laboratoires de recherche CNRS (PMMH et FAST) et également le Centre•Sciences , CCSTI de la région Centre et l'Agora des sciences, Marseille.

## Déroulement de l'enseignement

Attention ce ne sont pas des TP classiques. Cet enseignement se déroule dans le cadre de l'enseignement en Licence 2 S4 de la faculté des sciences d'Orsay (http://www.lmd.u_psud.fr/licence/sts/Modules/OPT1PH1.html)

- **Cours et ou conférences d'introduction (10h)**
- **Visites de Laboratoires de Recherches du CNRS**
- **Projets expérimentaux (40h) : choix du projet et une étude de faisabilité, (recherche bibliographique)**
   Exemples de projets proposés : cuve à vagues, soufflerie, ascension de bulles, chute de feuilles, stabilité et écoulement de mousses, films de savon, rides de sable, gouttes et jets d'eau, avalanches de sable, formation de méandres, tornade et tourbillon, etc.
- **Réalisation du montage expérimental**
- **Mesures, traitement des données et interprétation.**

- **Fonctionnement** : Une demi journée par semaine pendant un semestre.

- **Encadrement** : 2 enseignants, 1 ingénieur et 1 technicien.

- **Evaluation :** contrôle de connaissances écrit sur le cours d'introduction,
présentation orale de l'étude de faisabilité, conduite du projet, rapport écrit et soutenance orale finale.

- **Conclusion** : les étudiants découvrent un nouveau domaine de la physique et s'initient à la pratique expérimentale.

## Du projet (étudiant) au prototype (musée)

Un exemple *« Clepsydre et sablier »*. vise à comparer en direct le vidage d'une horloge à eau et d'un sablier.
**Il faut ajuster les expériences de recherches ou des projets pour en faire une exposition.**

**Les Éléments Communs sont :** l'expérience de coin de table, l'originalité, le recours analogique, le moindre coût.

**Les Différences sont : la** robustesse, la fiabilité, l'autonomie (éviter les interventions), l'entretien minimum, le spectacle (le design),l'interrogation (le so what), l'effet de surprise (maintenir une attention quelques minutes!).

### Références
- Expérimentation en hydrodynamique : une expérience pédagogique.
Bulletin de l'Union des Physiciens 786 1259 (1996)
C. ALLAIN, C. BÉTRENCOURT, P. BINETRUY, M. CLOITRE, J. C. DEROCHE, E. GUYON, P. JENFFER
- De l'expérience de laboratoire et des projets de travaux pratiques jusqu'aux modules d'expositions interactives
Bulletin de l'Union des Physiciens 101 455 (2007)
É. GUYON, P. JENFFER

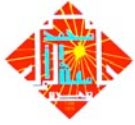 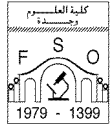 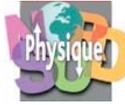 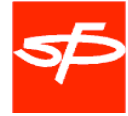

**Atelier n°1**
### 3- Une armoire à TP de physique pour la licence
**Présentation à Oujda en avant première
par l'équipe du CIRUISEF**

**Moktar RAY** - Université Paris XII

## *Constitution d'un ensemble minimal d'expérimentation physique en Licence*

Depuis deux ans, des physiciens sous l'égide de la **CIRUISEF** (Conférence Internationale des Responsables des Universités et Institutions à dominante Scientifique et technique d'Expression Française) l'un des réseaux l'**AUF** (Agence Universitaire de Francophonie) travaillent pour mettre sur pied un ensemble de travaux pratiques (TP) à destination des collègues des universités du sud qui éprouvent souvent de très grosses difficultés à bâtir un enseignement expérimental, en particulier en Physique.

Après une opération "Mallette TP de Chimie", la CIRUISEF propose de faciliter des enseignements expérimentaux de Physique au cours des deux premières années de la licence : échanges de savoir-faire, constitution d'un ensemble de kits comprenant le matériel de base et des textes initiaux de travaux pratiques adaptables aux conditions locales.

Dans le cadre de ce colloque, nous avons présenté, en avant première, le projet de l'armoire qui est complètement validé à ce jour.

Après avoir souligné l'importance de l'expérimentation dès la première année universitaire, nous avons décrit brièvement le cahier des charges concernant la constitution de l'armoire : Les différents TP proposés doivent, en effet, couvrir les principaux thèmes abordés en licence. Compte tenu des effectifs d'étudiants élevés en premières années universitaires, le matériel doit être robuste et d'une maintenance relativement facile. Les manipulations proposées doivent également être évolutives et adaptables à l'environnement local et d'un coût abordable pour qu'elles puissent être dupliquées en plus grand nombre.

Le projet résulte des premiers travaux collaboratifs entre plusieurs universités européennes et africaines. L'armoire est constituée à ce jour d'une vingtaine de manipulations, regroupées en « tiroirs » selon les thèmes (électricité, mécanique, optique, thermique, sciences de la matière…) dont une quinzaine sont opérationnelles et complètement validées.

En parallèle, nous avons présenté sous forme de posters l'ensemble des manipulations.

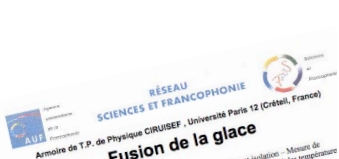 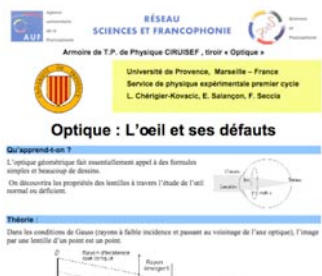 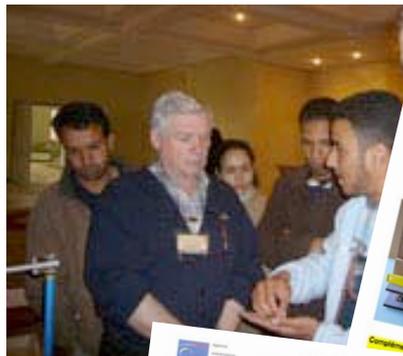 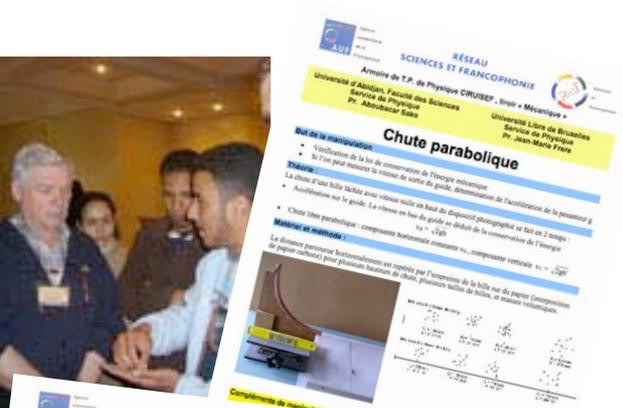 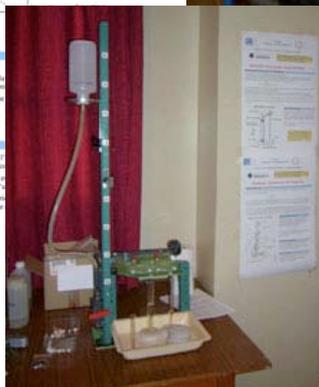

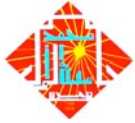 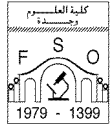 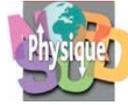 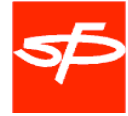

**Atelier n°2**
# Démarches expérimentales innovantes :
# « Secondes : Options Sciences »

Férial TERKI

Groupe « enseignement scientifique » IREM
Université Montpellier II Groupe d'Etude des Semiconducteurs, CC074,
Place E. Bataillon 34095 Montpellier Cedex 05, France

**Résumé**

Dans le cadre de la promotion des sciences, l'académie de Montpellier a mis en place, à titre expérimental depuis la rentrée 2004 un enseignement scientifique de détermination en Seconde appelé « Options Sciences ». L'objectif premier visé étant de sensibiliser à la pratique de la science des élèves qui à priori ne se destinaient pas à une filière scientifique espérant ainsi augmenter le nombre d'étudiants « scientifiques » dans le supérieur. Une équipe d'enseignement scientifique a été consistituée à l'IREM de Montpellier. Elle se compose d'enseignants du secondaire et d'universitaires. Le rôle de ces options sciences est de proposer un travail interdisciplinaire entre physique, mathématique et sciences de la vie et de la terre.

## 1 Contexte

Depuis 2000, plusieurs rapports[1] et comptes rendus d'études à l'echelle nationale mais aussi internationale témoignent d'une baisse significative des effectifs des étudiants dans les domaines scientifiques. Les prévisions quant au renouvellement des cadres supérieurs dans les années à venir sont extrêmement préoccupantes.

Les constats pré-bac font état d'une absence de curiosité, de plaisir de comprendre et de recherche personnelle chez les élèves. Par ailleurs, la réussite en sciences des élèves n'est pas souvent suivie par une poursuite d'études scientifiques : la réputation de difficulté en est le principal verrou. D'autre part, en classe de seconde, les enseignements de détermination relèvent largement des langues vivantes ou anciennes et des arts. Aucune option de détermination ne relève du domaine des sciences de la vie et de la terre ou des mathématiques. Certaines options sont associées à des projets d'orientation vers les filières technologiques ou générales, donc sur des métiers scientifiques et technologiques: MPI (Mesure Physique et Information), ISI (Initiation aux Sciences de l'Ingénieur), STL (Sciences et Techniques de Laboratoire, moins fréquente dans les établissements). Ces options sont mono-disciplinaires et aucune option n'est associée à la culture scientifique au sens large. Cependant, le désir de réussir avec les sciences peut prendre appui, chez les élèves, sur la perception d'une science sans vrai cloisonnement entre discipline. C'est dans cet esprit qu'a été mis en place le projet expérimental des secondes options sciences dans l'académie de Montpellier.

## 2 Mise en place institutionnelle

S'appuyant sur la demande de collègues du secondaire (J. P. Richeton et *al*)[2] qui ont expérimenté cette option de détermination à Strasbourg 1997 et profitant du plan académique de promotion de l'enseignement scientifique en lycée : un *dispositif expérimental pilote* dans l'académie de Montpellier a été crée à la rentrée 2004 . la mise en place a été précédée par les étapes suivantes :
- 2002-2003 mission de l'académie pour la promotion de l'enseignement scientifique en lycée (mission IA-IPR B. Dirand)
- 2003 : Plan académique de formation pour la promotion des sciences : comité de pilotage secondaire-supérieur (Président du CEVU : J. P Merle an partenariat avec IA-IPR B. Dirand)
- 2004 : ouverture de classes de secondes option sciences dans 9 établissements pilotes dans l'académie.
- Depuis 2005 généralisation de ce dispositif dans quelques lycées au niveau national : académie d'Aix-Marseille, Caen, Lille, Nice…
- En 2006 : 35 établissements de l'académie de Montpellier soit près de 1000 élèves.

## 3 Objectifs de l'option science

Parmi les différentes missions de l'option sciences[2-4] nous pouvons retenir les points suivants :
- Inciter au choix de la série scientifique, au delà des clivages disciplinaires,
- Réduire le "saut" méthodologique et conceptuel entre

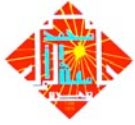
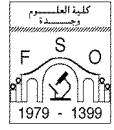
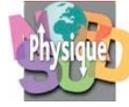
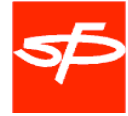

2nde et 1ère S.
- Permettre aux élèves de s'orienter de manière positive vers les sciences ou, s'ils ne se dirigent pas vers des études scientifiques, d'avoir une image positive de la science et une véritable "culture scientifique" comme futurs citoyens
- Donner plus de sens à l'enseignement « des sciences »,
- Montrer comment les sciences sont impliquées tant dans la culture (élaboration de la pensée et du discours scientifique) que dans l'approche interdisciplinaire de situations concrètes ou de problèmes de société.
- Mettre les élèves en situation de recherche, d'expérimentation, de réussite et d'acquisition d'autonomie.
- Approfondir la démarche scientifique.
- Permettre aux élèves une appropriation des concepts et techniques introduits en cours sans alourdir le contenu des trois matières
- Occasion d'affirmer un projet et de convaincre des familles pas toujours décidées à engager leurs enfants vers des études réputées difficiles.

## 4 Modalités générales

La mise en place de l'option science s'est faite à moyen constant en terme de DHG (Dotation Horaire Globlale)[2,4].

Les élèves de seconde peuvent désormais choisir l'option sciences comme option de détermination de 3 heures (comme SES (Sciences Économiques et Sociales) MPI, ISI, STL, LV3 , …). La coexistence des trois disciplines est indispensable pour croiser les regards sur un même thème. Comme dans la rénovation du collège les mathématiques sont sur le même plan que les deux autres matières. La différence essentielle entre un atelier et une option est que l'option entre dans l'horaire officiel des élèves. Par contre pour un atelier, les horaires se situent en dehors des heures de cours (heures supplémentaires).

Une des caractéristique de l'Option Science est qu'Il n'y pas de programme spécifique mais un *cadrage national* concernant les objectifs, les méthodes, les compétences à développer… L'enseignement s'effectue par demi-classe. Cette option ne constitue pas un critère pour le passage en 1ère S. L'évaluation des élèves privilégie les progrès dans la maîtrise des méthodes et l'acquisition des compétences propres aux disciplines scientifiques.
Le programme est essentiellement axé sur :
- le questionnement, l'initiative, l'expérimentation, la culture scientifique, l'histoire des sciences...
- la lecture et la production de textes scientifiques: composantes essentielles de la démarche mathématique comme des sciences expérimentales pour promouvoir une image des sciences dynamique et motivante

## 5 Méthode

Dans cet enseignement, l'interdisciplinarité [2]va permettre la mise en œuvre de plusieurs problématiques croisées autour d'un même objet central. En mettant l'objet d'étude en commun dans les différentes disciplines, on créé une mobilisation autour d'un objet. La transversalité de l'approche et le croisement disciplinaire est la clef de la réussite de cette option. Cet enseignement des sciences permettra en effet d'approcher chaque objet d'étude selon plusieurs angles d'étude, permettant ainsi un repositionnement de chaque objet d'étude. Finalement chaque objet apparaît dans sa complexité. Cela met en relief le fait que chaque science est une construction de l'esprit. Pour un même objet d'étude, l'approche multi-disciplinaire permet de poser plusieurs questions et de proposer différents outils (mathématiques, techniques physiques, procédé biologiste), tout en gardant un unique objectif.

L'élève comprend ainsi que même si le questionnement est dépendant de la matière, la démarche scientifique est la même. On lève ici le cloisonnement artificiel entre discipline. L'élève analyse plus facilement que certains outils (mathématiques) peuvent être communs aux trois disciplines et comment la rigueur du développement mathématique peut être moduler (approximation) dans les sciences appliquées. On se doit d'aiguiser son sens critique vis à vis d'un résultat obtenu. Il peut de lui même faire des allers retours entre discipline selon le problème posé. L'élève « voit » ainsi qu'un problème donné vu par différentes disciplines peut être résolu grâce à une démarche scientifique commune qui repose sur des méthodes identiques entre disciplines même si les outils nécessaires peuvent être différents. Ensuite le replacement des résultats dans le contexte scientifique général, sera encore une fois le résultat d'un savoir faire acquis par l'expérience professionnelle, et souvent caractéristique de la pensée de chacun.

La discussion scientifique requiert une bonne connaissance du contexte. C'est à ce niveau que l'acquisition des connaissances propres à chaque discipline prend une importance majeure. Il faut donc effectuer un retour sur la discipline, qui se fera naturellement dans la mesure où il n'y a pas un « prof » d'option sciences, mais des profs de maths, physique et de SVT. Dans cette expérience, le milieu universitaire se doit de s'intéresser aux initiatives issues de l'enseignement secondaire. L'implication directe et institutionnelle des établissements universitaires et de recherche permet la création d'un lieu d'échanges et de mutualisation des expériences en apportant un éclairage sur les thèmes de recherche abordés.

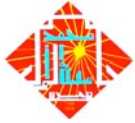
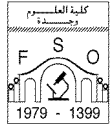
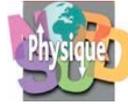
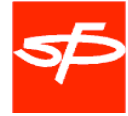

# 6 Conclusion

Les réactions très favorables[3] des élèves qui s'y pressent montrent qu'il répond manifestement à une attente. Les élèves y ont appris l'autonomie et le plaisir de chercher. L'augmentation du nombre d'établissement intéressés (facteur 4 en 2 ans). Cette option facilite grandement le passage en 1ère S tant redouté. Pour quelques élèves qui s'orientent dans une autre section : ils disent avoir acquit un regard plus positif sur les sciences.

Plus de filles choisissent cette option par rapport aux autres options scientifiques. Par ailleurs, Il a été constaté que ces classes options sciences participent activement à la semaine de la fête de la science ou à d'autres concours par projets tels que les olympiades de physique mais aussi les exposciences régionales et nationales. Citons par exemple, la selection d'un groupe élèves d'une option science de Bagnols sur Cèze pour faire partie de la délégation Française à l'expo-sciences internationale qui aura lieu du 8 au 13 Juillet 2007 à Durban en Afrique du Sud http://esi2007.milset.org/.

Les perspectives à plus long terme serait une généralisation de cette iniave à l'echelle nationale et pourquoi pas internationale.

## Références

[1] Rapports, Ourisson, Porchet

[2] Articles Repères APMEP 429 (2000), APMEP 65 (2006) p91

[3] Grid Network, Etude de cas Option sciences en seconde Lycée Mas de Tesse Montpellier.

[4] Collectif-Action-sciences : http://www.sfc.fr/SocietesSavantes/Option-sciences.pdf

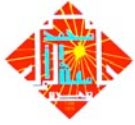 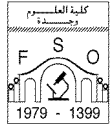 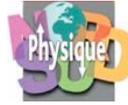 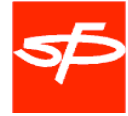

## Atelier n°2
# La formation britannique des physiciens et des ingénieurs

### Peter MELVILLE, international director
Institute of Physics, London, UK

### Résumé

Au Royaume-Uni beaucoup de physiciens travaillent en industrie et le système de formation des ingénieurs leurs permet de devenir également ingénieurs.

## 1 Introduction

Quoiqu'il existe en France une grande différence entre les ingénieurs et les physiciens, peut-être à cause du système des grandes écoles, au Royaume-Uni beaucoup de physiciens travaillent en industrie et plusieurs deviennent ingénieurs, parfois même des ingénieurs célèbres. Par exemple la moitié des présidents de l'Institution of Electrical Engineers, maintennant l'Institution of Engineering and Technology, étaient à l'origine des physiciens. L'industrie reste toujours l'employeur le plus important de physiciens, dont beaucoup travaillent comme ingénieurs électriciens, ingénieurs mécaniciens, ingénieurs du génie chimique, ingénieurs des constructions civiles, etc.

## 2 La formation des ingénieurs

Le système de formation des ingénieurs aide les physiciens. Pour devenir Chartered Engineer (CEng) (ingénieur professionnel) on a besoin non seulement d'un diplôme d'ingénieur, en ce cas une MEng (maîtrise d'ingénieurie) mais aussi de quatre ou plus d'années de formation professionnelle et d'expérience d'un niveau ingénieur professionnel. Peut-être il faut clarifier que la MEng n'est pas une maîtrise de Bologne (3+2) mais une formation universitaire de quatre ans.

Les physiciens peuvent obtenir le CEng s'ils démontrent qu'ils ont acquis toutes les compétences necessaires pour une MEng et aussi, naturellement, la formation professionnelle et l'expérience en tant qu'ingénieur professionnel. En practique ce n'est pas très difficile. Si l'on travaille en industrie, comme beaucoup de physiciens, on acquiert très rapidement la connaissance spécialisée dans les technologies. On travaille seul et en groupe sur les projets et l'on peut ainsi acquérir non seulement des compétences MEng mais aussi la formation professionelle et, bien entendu, l'expérience. Comment peut-on démontrer cela ? Tout simplement– par les raports qu'on a rédigé au cours du travail technique. C'est le système développé par l'Institute of Physics et adopté par l'Engineering Council (UK) comme le système standard pour démontrer équivalence d'une MEng. Par exemple c'est le cas des ingénieurs qui n'ont qu'une BEng (Bachelor of Engineering – 3 ans).

## 3 Physicien et ingénieur

Il y a des avantages d'être à la fois – physicien et ingénieur [1]. A partir d'une formation de base en physique on acquiert la connaissance de plusieurs technologies différentes et, peut-être de façon plus cruciale on a l'aptitude à resoudre des problèmes nouveaux et des principes de base. De l'ingénieurie on a l'aptitude à transferer les resultats dans la practique. On observe souvent en industrie que quand un problème nouveau apparaît on requiert l'aide de physiciens.

## 4 Conclusion

C'est evident que ce serait très difficile de transposer le système britannique dans les autre pays. Cependant c'est le principe qui est plus important que la practique. Au Royaume-Uni il y a beaucoup de physiciens qui traivaillent comme ingénieurs. Ces derniers préfèrent l'identité seulement de physiciens et ne veulent pas devenir ingénieurs, bien qu'il y en ait d'autres qui sont fiers de l'indentité double de physicien et ingénieur. Ce qui est le plus important de préciser c'est la possibilité offerte aux physiciens de travailler dans l'industrie aux côtés d'ingénieurs et d'être même souvent considérés comme ingénieurs.

### Références

[1] P. Melville, Eur. J. Eng. Ed. 28 (2003) 13

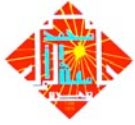 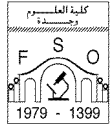 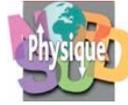 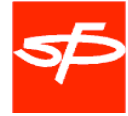

# Atelier n°3
## de l'enseignement et la recherche
## à la communication grand public

### Modérateur : Michel Darche – Centre•Sciences

S'appuyant sur la présentation de l'exposition "Mosaïque de la physique" sur le lieu du congrès, cet atelier a présenté différentes activités de communication de la physique vers les publics jeunes – lycéens et étudiants -, leurs enseignants et leurs parents réalisées à l'occasion de la présentation d'expositions de physique au Maghreb depuis 2oo5.

## "La caravane de la physique" en Algérie
**Jamal Mimouni** (Université Mentouri - Constantine)

Amorcée en 2oo5 avec l'année mondiale, "la caravane de la physique" a connu un très vif succès avec la présentation de l'exposition Mosaïque dans 14 villes du pays de janvier à décembre 2oo6 (http://yp05.110mb.com/mosaique.htm).

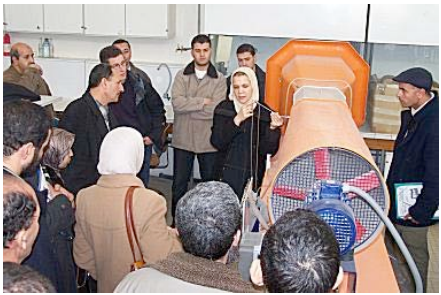 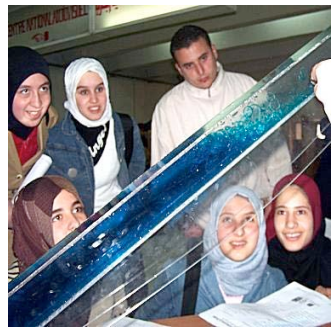 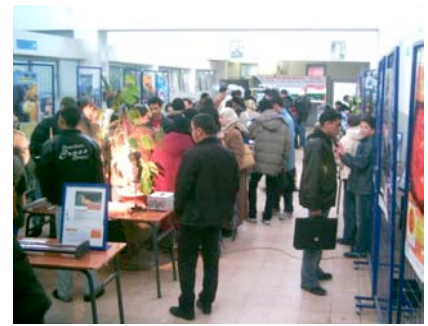

L'opération s'est faite à l'initiative du Comité d'Organisation de l'Année Mondiale de la physique de l'Université Mentouri avec le concours du CCF de Constantine et le soutien de l'Ambassade de France à Alger (SCAC). Présentée dans les universités ou les centres-villes, elle accueillait lycéens, étudiants et tout public.

Elle a permis aussi aux physiciens de montrer ici leurs labos, là des manips extraites des labos. Des prolongements sont envisagés avec la réalisation de "manips d'expo" telles qu'on peut les voir dans les expos de Centre•Sciences.

La caravane reprend sa route cette année avec l'exposition "**Quand les sciences parlent arabe**" (cf www.SciencesArabExpo.org et http://yp05.110mb.com/sa.htm) qui va aussi traverser l'Algérie d'est en ouest. Elle sera suivie en 2oo8 de l'exposition "**Jeux de grains**".

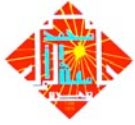 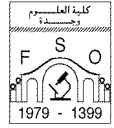 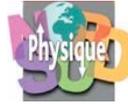 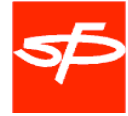

## La formation de médiateurs à l'Université
**Fouad Lahlou** (Université Ibn Tofail – Kenitra)

Les ateliers "Arts & Sciences » conduits par les scientifiques (mais pas seulement) de l'université débouche chaque année sur une semaine de la science animée par les étudiants ainsi que sur la formation validante à la communication scientifique.

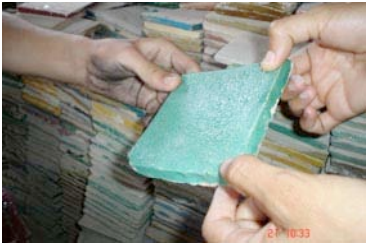 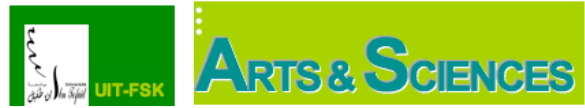 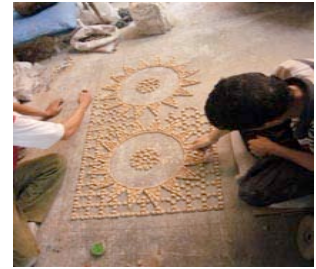

Travaux de recherche sur le thème y alternent avec la réalisation et l'animation d'expositions interactives (Jeux de grains en 2oo5, Sciences arabes en 2oo6) ou de posters (Science au Sud en 2oo6) ainsi que les journées "Les sciences & les jeunes" initiées par l'Académie Hassan II des Sciences et Techniques.

Ils réalisent actuellement une exposition interactive ayant pour titre "**Quand les Sciences rencontrent les Arts**" avec le soutien du PCST (programme d'un FSP Afrique) et de Centre·Sciences.

## Des expositions à l'association "Avenir pour la Science et le savoir"
**El Hassan Tahri** (Université Mohamed 1er – Oujda)

Comment l'organisation et la gestion d'expositions interactives nous a conduits à créer pour la région (l'Oriental) une association "Avenir pour la Science et le Savoir" composée d'universitaires, de scientifiques (ingénieurs, médecins…) et d'étudiants pour avoir une structure souple et autonome visant à développer des outils de communication grand public (par exemple autour de l'astronomie) mais aussi du matériel de TP (en physique, biologie et même mathématiques) pour permettre aux enseignants des lycées démunis de pouvoir développer une approche expérimentale.

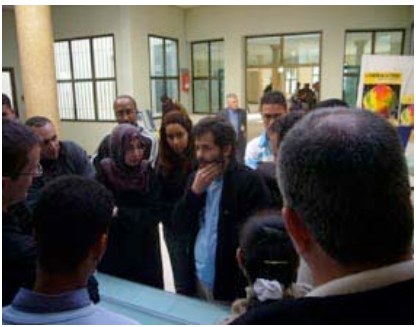 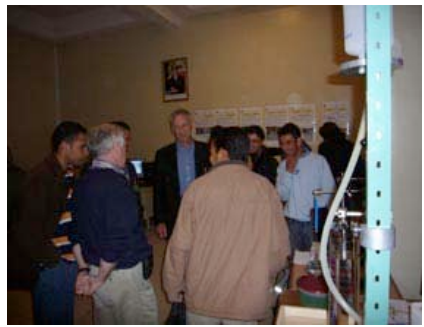 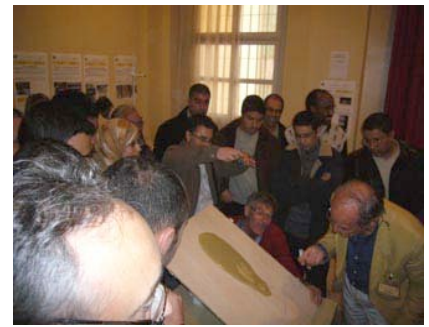

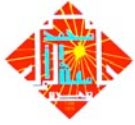 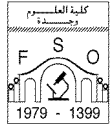 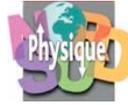 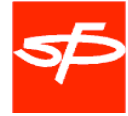

Travaux de recherche sur le thème y alternent avec la réalisation et l'animation d'expositions interactives (**Jeux de grains** en 2oo5, **Sciences arabes** en 2oo6) ou de posters (**Science au Sud** en 2oo6) ainsi que les journées "Les sciences & les jeunes" initiées par l'Académie Hassan II des Sciences et Techniques.

## Des TP de labo à la communication grand public
Patrice Jenffer (Université Paris XI – Orsay)

Patrice Jenffer (Université Paris XI) a montré comment les physiciens pouvaient passer des manips de "recherche" et des projets expérimentaux d'étudiants de licence 2 à des manips d'exposition ou de cours. Ces manips sont présentées parfois par ces mêmes étudiants au grand public l'occasion de la semaine "fête de la science ", aux journées "portes ouvertes de l'université, au forum "faites de la science". Elles font également l'objet d'une présentation dans un des stands itinérants du Palais de la Découverte.

Pour le public l'expérience sera "in vivo" ou en "direct" après avoir été conçue réalisée par les étudiants ou les chercheurs.

## Comment se conçoivent ces expositions ?
Etienne Guyon (Espci)

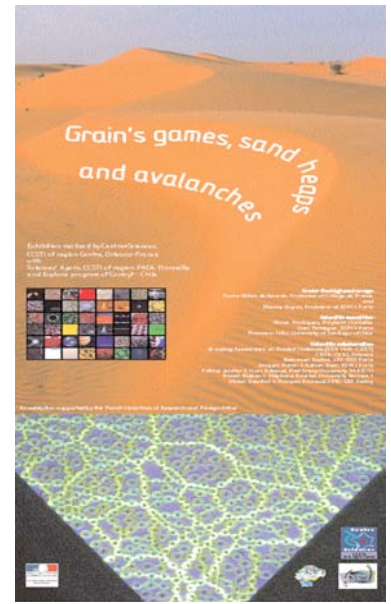

Etienne Guyon a rappelé comment on pouvait passer des expériences de labo (exemple de l'armoire de TP présentée au congrès) a une exposition comme "Jeux de grains" et "Mosaïque" en fédérant une vingtaine de chercheurs de différentes universités (une douzaine) en France et au Sud et en permettant ainsi à des professionnels (ici Centre•Sciences) de choisir et traduire les expériences proposées en manips d'expo.
C'est, pour les chercheurs, une valorisation forte de leur travail sans qu'ils aient à réaliser la manip totalement tout en apportant leurs conseils de praticiens.

## La circulation des expositions de physique depuis 3 ans
Michel Darche (Centre•Sciences)

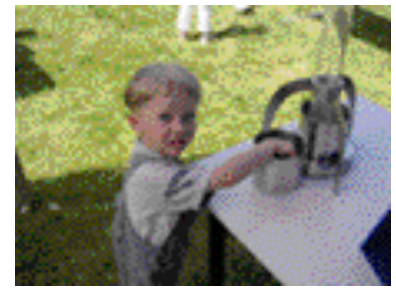

Europe (12 villes dans 4 pays), Moyen-Orient (10 villes de 5 pays), Afrique (Ouest, Est et Australe et bien sûr Maghreb), Asie du Sud-Est ont ou vont bientôt accueillir l'une de ces expositions. Seule l'Amérique latine n'est pas encore sur la liste, bien que nous ayons co-réalisé l'expo "Jeux de grains" avec les Chiliens et qu'ils en ont une version chilienne.

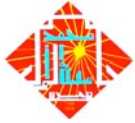 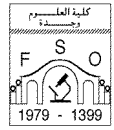 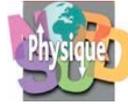 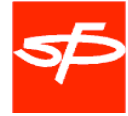

Ces expériences d'expo permettent à chaque visiteur de se questionner, formuler des hypothèses, tester, observer et… essayer de comprendre pourquoi ! Pourquoi l'expérience se déroule-t-elle ainsi ? Pourquoi n'est-elle pas conforme à ce que je pensais ? Qu'apporte le déroulement de cette expérience à ma compréhension des sciences sous-jacentes ?...

**Quels moyens ?**

Ces expositions circulent à l'initiative des universités ou des musées et des CCF et Instituts français ou Alliances françaises. Nous nous efforçons de planifier des circulations sur une durée de 6 à 12 mois. Les coûts intègrent le transport, la venue d'un formateur au 1er montage, la maintenance et sur place, la communication et l'animation, parfois la prise en charge du transport des scolaires.

**Quels prolongements ?**

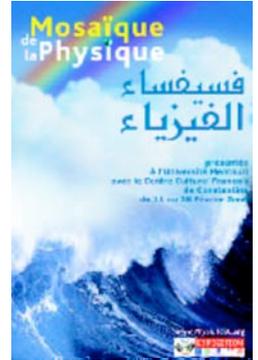

Ces présentations peuvent trouver naturellement deux prolongements :
- dans la création d'expositions voire d'un centre de sciences régional ou national,
- dans la formation initiale ou continue des enseignants à la démarche
expérimentale avec un matériel simple et peu coûteux en y rajoutant cependant des appareils de mesure et d'observations plus fines que dans une expo.

Contact : mldarche@free.fr

## Les expositions de physique en

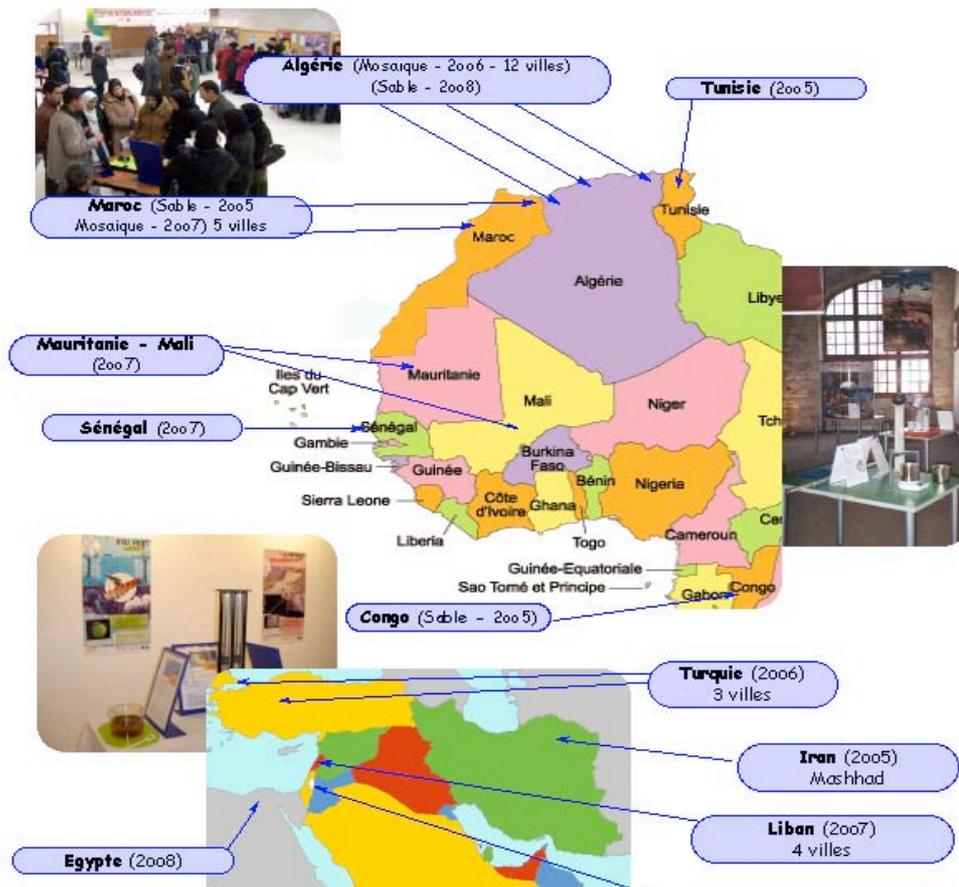

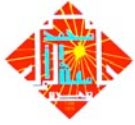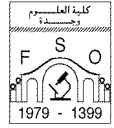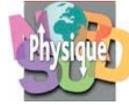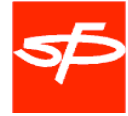

# Bilan et Perspectives

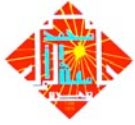 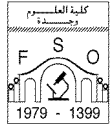 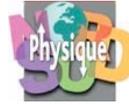 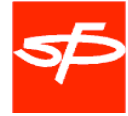

# Le congrès Nord-Sud à Oujda, synthèse et perspectives

## Michèle LEDUC
### Présidente de la Société Française de Physique

Le congrès d'Oujda a été conçu selon une formule originale, regroupant les problèmes de l'enseignement supérieur, de la recherche et de la diffusion des connaissances en physique. Il a réuni plus de 120 personnes qui ont suivi toutes les sessions. En dehors de la quarantaine de participants de la délégation française, l'assistance provenait de diverses universités du Maroc, de Tunisie et dans une moindre proportion d'Algérie. On notait ausssi la présence d'un collègue de Mauritanie et d'un autre d'Afrique du Sud. De très nombreux jeunes marocains et marocaines sont venus assister aux démonstrations de physique accompagnant l'exposition « Mozaïque de la physique », qui a rencontré un très vif succès avec les expériences sur les milieux granulaires et les écoulements, assorties de panneaux explicatifs. Toutefois les étudiants ont été peu nombreux à assister aux conférences. Les posters présentés par les jeunes chercheurs de l'université d'Oujda ont permis de fructueuses discussions. La visite de l'Ecole Centrale de Physique d'Oujda a montré le dynamisme de la formation d'ingénieurs de haut niveau au Maroc.

De très précieuses informations sur la situation de l'enseignement supérieur et de la recherche au Maroc et en Tunisie ont été fournies par de hauts responsables de ces pays. Au Maroc la création de réseaux nationaux s'accompagne d'un effort financier modéré mais en nette croissance (0.8% du PIB). Le démarrage est encore plus net en Tunisie (1.2% du PIB), où l'on enregistre une progression spectaculaire du nombre d'étudiants dans les sciences physiques (350 000 dont 58% de femmes). Ces pays sont maintenant engagés dans ce qu'ils nomment « la bataille de la qualité ».

Les présentations du LMD ont montré une bonne progression des systèmes qui se mettent en place, avec les mêmes difficultés qu'en France, dans les pays du Maghreb. Les filières d'ingénieurs y semblent bien intégrées et les masters sont évalués comme de bon niveau. Les écoles doctorales restent à créer. La grande similarité avec les filières LMD en France founit une grande opportunité pour créer des passerelles pour les diplômes. Il y a là des pistes de coopération future, avec la possibilité de création de masters régionaux éventuellement trans-méditerranéens. Au niveau de la thèse, la question du placement des jeunes docteurs est une préoccupation générale, mais moindre en Tunisie qu'en France, compte tenu du très grand besoin de recrutement dans l'enseignement supérieur. Le dispositif des cotutelles de thèse est maintenant très opérationnel et offre une grande flexibilité. Il sera à exploiter au maximum en conjonction avec la constitution de vrais réseaux thématiques de recherche.

Le contenu scientifique du congrès est apparu très éclectique dans les thèmes abordés, mais riche et de haut niveau. Les sujets ont concerné notamment les matériaux (matériaux de grande diffusion, matériaux pour l'électronique, nanomatériaux, matériaux pour les énergies renouvelables), l'informatique avec les grilles de calcul et l'exploitation des grandes masses de données (en astrophysique et en hautes énergies), l'optique et ses applications à la physique atomique et à la métrologie, la physique théorique et ses limites. La participation des pays du Sud à l'utilisation des grands instruments a été discutée avec l'exemple de la coopération du Maroc au détecteur ATLAS du CERN. Les grandes questions de société comme la physique pour la médecine et les problèmes de l'énergie ont également été abordées. Un débat s'est finalement ouvert sur la pertinence de la recherche fondamentale pour les pays du Sud, actuellement confrontés à de grands besoins pour leur développement économique. Il a certes manqué au congrès une vision globale des points forts de la recherche dans les différents pays et un bilan des coopérations des pays du Maghreb avec la France, avec les autres pays européens et aussi entre eux. Il apparaît néanmoins que la France est un interlocuteur privilégié et que les potentialités à exploiter sont grandes.

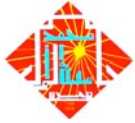 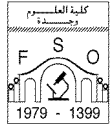 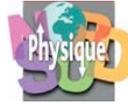 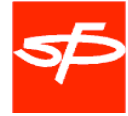

De l'ensemble du congrès il ressort plusieurs perspectives d'action :

1-  La diffusion de la culture scientifique à travers les expositions itinérantes et le matériel didactique pour l'enseignement de la physique (affiches, mallettes de travaux pratiques, armoire de la physique, etc.).

2-  La création d'une école doctorale Nord-Sud du type de l'Ecole de Physique des Houches, située au Maghreb dans un lieu fixe bien choisi et destiné à la formation des enseignants du supérieur au niveau de la recherche actuelle (voir mon texte dans les actes du colloque).

3-  L'utilisation des grands équipements structurants la recherche, existants ou à créer. L'utilité pour les physiciens du Maghreb de participer à la construction et à l'utilisation du synchroton SESAME en Jordanie a été débattue. Les installations en France (ILL, ESRF, SOLEIL) sont à exploiter et devraient favoriser la formation des chercheurs du Maghreb. Des projets structurants pourraient voir le jour en Afrique (par exemple une horloge à atomes froids).

4-  L'exploitation des grandes masses de données distribuées, comme celles des grandes missions spatiales en astrophysique ou encore celles de la physique des particules. Ceci ne nécessite pas la construction d'équipements lourds mais suppose la participation scientifique à haut niveau des chercheurs du Maghreb aux groupes de recherche existants en Europe.

5-  La mise à jour de notre connaissance des réseaux de recherche en coopération, en commançant par les PHC (programmes Hubert Curien) et les échanges via les universités. En effet le congrès n'a pas permis de se faire une idée complète de l'existant. Ceci permettrait de regrouper des ensembles de coopérations bilatérales par grands domaines.

6-  L'identification de nouveaux thèmes de recherche hors des sentiers battus offrant des coopérations équilibrées sur des sujets pertinents et originaux. Un exemple est fourni par les matériaux de grande diffusion (les granulaires, les bétons, les terres crues, les sables…).

Le congrès s'est conclu par la décision de constituer un comité de suivi Nord-Sud rassemblant des représentants de la France et des trois pays du Maghreb, en associant des observateurs de l'Afrique du Sud et de la Mauritanie. La SFP et la STP en Tunisie sont partis prenantes de ce comité et sont convenues de signer ensemble une convention de coopération. On attend que l'Algérie délègue des représentants au comité de suivi. Pour ce qui concerne le Maroc, le congrès d'Oujda aura été l'occasion de rassembler diverses associations de physique marocaine, qui ont pu discuter l'intérêt de se constituer en Société Marocaine de Physique, fortement encouragées en cela par l'exemple positif de la STP.

La première réunion du comité de suivi aura lieu en Juillet 2007 au congrès général de la SFP à Grenoble. Son objectif est de mettre en oeuvre sans tarder les différentes perspectives d'action évoquées plus haut. En outre elle permettra de commencer à préparer une demande de Fonds de Solidarité Prioritaire sur le thème « physique et développement », centrée pour plusieurs années sur le Maghreb et l'Afrique sub-saharienne. Cette forme d'action, inspirée de celle mise en place avec succès pour les mathématiques par le Ministère des Affaires Etrangères (le CIMPA - www.cimpa-icpam.org), sera destinée à combler le fossé Nord-Sud pour tout ce qui concerne la diffusion des connaissances, la formation, la recherche et ses applications pour le développement durable. Une extension de ce projet aux autres pays du pourtour de la Méditerranée avec le soutien de l'Union Européenne a également été évoquée au congrès.

Les participants dans leur ensemble se sont déclarés très satisfaits de ce premier congrès Nord-Sud, qui a été une occasion remarquable de contacts très chaleureux entre physiciens des différents pays. Ils remercient vivement les organisateurs et souhaitent qu'un congrès suivant soit organisé, par exemple dans deux ans, à l'initiative des pays du Maghreb et avec l'aide de la SFP pour fédérer les initiatives.

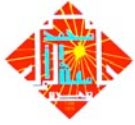 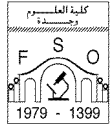 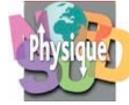 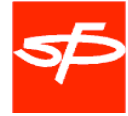

# Pour une école de Physique Nord-Sud,
# un modèle : l'école prédoctorale aux Houches

**Michèle LEDUC**
Présidente de la Société Française de Physique
Directrice de recherche au CNRS
Laboratoire Kastler Brossel à l'Ecole Normale Supérieure, Paris
leduc@lkb.ens.fr

Le but principal de cette conférence Nord-Sud de physique à Oujda était de consolider et de développer les liens scientifiques entre la France et les pays du Maghreb, dans une perspective plus large associant les pays du Nord de la Méditerranée et ceux de l'Afrique tout entière. Il est clair, pour des raisons qui tiennent à l'histoire, que le Maroc, l'Algérie et la Tunisie sont des partenaires privilégiés pour ces échanges scientifiques dans la mesure où nos universités en France accueillent toujours un nombre très important d'étudiants maghrébins. Beaucoup des cadres du système académique de ces pays se sont formés en France et ont gardé de forts liens avec leurs laboratoires d'origine. La langue française est un vecteur de communication qui favorise considérablement le rapprochement. Le remarquable accroissement des moyens mis à la disposition de la recherche et de l'eneignement supérieur en Tunisie et maintenant au Maroc crée depuis peu des conditions particulièrement favorables pour l'accélération et la coordination des coopérations avec la France.

L'année mondiale de la physique en 2005 a favorisé les contacts entre physiciens français et maghrébins. La Société Française de Physique a été invitée au congrès de la Société Tunisienne de Physique, l'exposition itinérante « Mosaïque de la physique » a circulé au Maghreb et le congrès final sur le dévelopement durable à Durban a facilité les rencontres avec les collègues physiciens des pays africains. Des discussions de Durban est ressortie le constat du grand besoin de formation complémentaire des étudiants et surtout des enseignants chercheurs africains à leur discipline, de façon à les porter au niveau de la recherche contemporaine et leur permettre de développer avec succès des thèmes scientifiques pertinents. Les enseignants africains en effet, après de premières études souvent menées en France, souffrent au retour du poids de très lourdes tâches de cours et de l'éloignement des centres majeurs de la création scientifique. C'est ainsi qu'est née l'idée de créer une école de physique Nord-Sud sur le territoire africain, en suivant l'exemple des mathématiques et des écoles d'été du CIMPA ou celui du centre de formation de haut niveau établi au Cap - l'African Institute of Mathematical Science (AIMS) - avec l'appui, au Sud, des universités de Stellenbosch, Western Cape et Cape Town, et au Nord, des universités de Cambridge, Oxford et Paris Sud-Orsay. C'est donc à la suite des réflexions qui ont suivi le congrès de Durban et avec le vif encouragement du Ministère des Affaires Etrangères à Paris que s'est fait jour le projet d'école proposé à Oujda par la Société Française.

Il nous a semblé qu'un modèle prestigieux pouvait être avancé, celui de l'école de physique des Houches, qui a servi de cadre à la formation de presque tous les grands physiciens de ce monde depuis plus d'un demi siècle. Tout particulièrement certains enseignements spécialement destinés aux jeunes chercheurs (les cours de l'école prédoctorale) pourraient fournir des exemples fructueux. L'école des Houches est une école nationale française mais à vocation internationale. Un de ses grands charmes est le cadre isolé et grandiose dans laquelle elle est située, au-dessus du village des Houches et face à la chaîne du Mont-Blanc dans les Alpes françaises. De fait de son éloignement de la vie urbaine, le lieu est très propice aux échanges scientifiques intenses et au travail personnel en profondeur. Une des caractéristiques permanentes de l'Ecole des Houches est la vie communautaire qu'y mènent les élèves, enseignants et participants de tous âges et de tous pays. L'hébergement en chalets est confortable mais sans luxe. Les repas sont pris en commun sur place. La montagne permet des activités de détente quelques heures par jour. Les programmes des cours et séminaires sont destinés à la formation initiale et permanente des physiciens. Différentes activités s'y déroulent : les traditionnelles écoles d'été de très haut niveau qui durent un mois entier, les « workshops » du Centre de Physique, et enfin les écoles dites prédoctorales de deux semaines destinées aux chercheurs débutants.

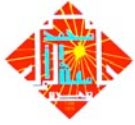 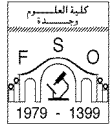 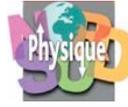 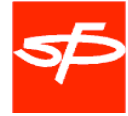

Le modèle de ces écoles prédoctorales est celui qui paraît le mieux adapté pour un développement dans les pays du Sud. Elles ont pour but d'aider les jeunes chercheurs à acquérir une culture qui leur permettra de mieux situer leur travail de thèse dans un contexte plus large et d'évoluer plus facilement dans leur carrière scientifique future, en échappant à une spécialisation trop précoce. Dans les pays du Sud leur public pourrait être étendu aux enseignants du supérieur désireux d'augmenter leur culture générale. Les sujets traités couvrent de larges pans de la physique tels que : physique mésoscopique, introduction à la biophysique, les bases de la physique statistique, les lasers et l'optique atomique, la modélisation atmosphérique, etc. Chaque professeur invité fait un cours de 4 à 5 heures, donnant les bases essentiellement théoriques du sujet abordé.

L'école Nord-Sud proposée pourrait ainsi se situer au Maghreb, dans un lieu fixe hors université que pourraient proposer les sociétés de physique concernées. L'emplacement choisi devrait se situer dans un cadre paisible et agréable, propre à la concentration. Les cours de niveau « master-plus » dureraient deux semaines, avec la présence obligatoire des professeurs pour une semaine au moins et celle des participants pour la totalité de l'école. L'auditoire proviendrait de pays du Sud sans exclure des participants de pays du Nord. La langue de travail serait le français, avec certaines exceptions pour l'anglais. Les professeurs seraient donc francophones, mais quelques anglophones pourraient aussi être invités. Une date convenable serait l'été, peu avant la rentrée universitaire. Une puis deux écoles par an pourraient être ainsi organisées. Les thèmes, les programmes et les professeurs invités seraient choisis par un comité scientifique Nord-Sud, chargé également de la sélection des participants conjointement avec les organisateurs de la session. Quelques premières idées de thématiques résultant des travaux du congrès de Oujda ont été avancées : les matériaux de grande diffusion, l'énergie et l'environnement, les applications de l'optique, les nanosciences, la physique pour la médecine.

Si une proposition est faite par un pays du Maghreb au comité de suivi de l'après-Oujda, et si les soutiens nécessaires sont recueillis, la mise sur pied de l'école de physique Nord-Sud ici proposée bénéficiera de l'appui de la Société Française de Physique, à laquelle pourrait s'associer la Société Française de Chimie. Le parrainage des grandes institutions du même type sera sollicité, en particulier celui de l'ICPT à Trieste et évidemment celui de l'école de physique des Houches. Celle-ci a d'ailleurs fait savoir qu'elle est susceptible de fournir une caution et une aide, à condition d'être consultée sur les programmes scientifiques de l'école du Sud.

La discussion à Oujda a fait ressortir que la création de l'école de physique Nord-Sud ici présentée est de nature plutôt théorique et ne couvre pas tous les besoins des pays du Sud. En effet il est clair que les chercheurs de ces pays manquent aussi gravement de pratique de la physique expérimentale : d'autres initiatives de formation proches des « manips » pourraient être développées en parallèle, par exemple dans les grands centres universitaires français et tout particulièrement auprès des grands instruments. Le projet de l'école de physique Nord-Sud adapté de l'école des Houches n'est qu'un élément, à notre avis fondamental, d'un ensemble plus vaste de développement de la coopération entre la France et l'Afrique francophone dans le domaine des sciences de la matière. Il a d'ailleurs vocation à être étendu à d'autres pays européens du pourtour de la Méditerranée, ce qui augmenterait son audience et assurerait des moyens supplémentaires.

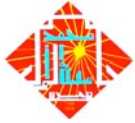 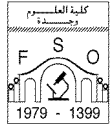 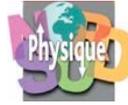 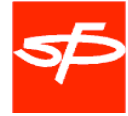

# Bilan et perspectives de la conférence d'Oujda

**Roger MAYNARD,** Vice-président de la SFP
et **Jean-Claude TOPIN**, Conseiller au MAE

L'idée d'organiser une conférence sur la recherche en physique et son enseignement, est venue très naturellement lors de l'année mondiale de la physique qui a eu un grand retentissement aussi bien en France que dans les pays du Maghreb. L'Université d'Oujda, à travers l'heureuse initiative du doyen de la faculté des sciences, a été la première à proposer d'accueillir cette rencontre. Que les collègues d'Oujda soient ici remerciés de leur accueil et de l'organisation remarquable de cette conférence, qui a également attiré de nombreux étudiants grâce à l'exposition « Petite mosaïque de la physique » présentée avec l'appui de l'Institut Français de l'Oriental.

Cette rencontre a permis de mieux se connaître et de commencer à dresser un panorama de la recherche et des domaines d'intérêt susceptibles de rapprocher diverses équipes, plus particulièrement au Maroc, en Tunisie, en Algérie, en Mauritanie et en Afrique du Sud. La mise en place des filières de l'enseignement supérieur organisées en Licence-Maîtrise-Doctorat, selon des exigences communes à l'ensemble de nos pays, va permettre de renforcer nos coopérations et les échanges d'étudiants. Elle marque une étape importante dans la compréhension de nos sociétés, appelées à être plus solidaires dans le développement des connaissances.

Le contexte mondial est favorable au développement de la recherche dans les pays émergents, comme l'a souligné la conférence sur la physique et le développement, qui a eu lieu à Durban en Afrique du Sud en 2005. Grâce aux interventions des Directeurs de la Recherche du Maroc, de la Tunisie et de l'Afrique du Sud, nous avons ainsi appris que le budget de la recherche, avec une part d'environ 1% du PIB, était en augmentation sensible dans ces trois pays.

Les axes de la recherche et de la coopération scientifique que nous avons identifiés au cours de ces journées se situent dans un continuum allant de la recherche fondamentale et à la recherche appliquée. Citons :
• les grands équipements comme la participation à Atlas au CERN de Genève, aux expériences utilisant le rayonnement synchrotron à Soleil en France, et au Sésame, en Jordanie
• les réseaux de recherche sur les matériaux pour l'électronique et les nanotechnologies ainsi que les grilles de calcul
• des développements théoriques variés (physique atomique et moléculaire, matière condensée, magnétisme…)
Des recherches originales, particulièrement adaptées aux contexte africain, ont été mises en évidence : comme les propriétés physico-chimiques des matériaux de construction tels que la terre crue ou encore la dynamique des dunes.

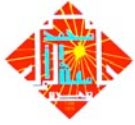 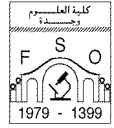 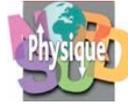 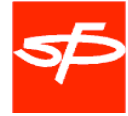

Une coopération forte pourrait s'établir sur la formation doctorale, non seulement des étudiants en thèse, mais aussi des jeunes maîtres-assistants qui ont besoin de consolider leurs connaissances. Un projet d'Ecole Doctorale au Maghreb du type « Les Houches », a été évoqué avec enthousiasme. Cette école Nord-Sud, à travers ses programmes et son organisation, aurait vocation à s'inscrire dans l'espace universitaire et scientifique euro-méditerranéen.

Par ailleurs, il convient d'encourager la constitution de sociétés de physique, telles en Tunisie et en France, pour assurer une représentation des communautés scientifiques et universitaires dans ce domaine. Une société marocaine de physique pourrait ainsi prochainement voir le jour.

La diffusion de la culture scientifique reste également un point sensible sur lequel nos pays doivent pouvoir œuvrer conjointement.

Enfin, la mise en place d'un comité de suivi de cette rencontre d'Oujda paraît à tout le moins nécessaire si l'on souhaite, à l'échelle méditerranéenne, voire pan-africaine, faire émerger des équipes et des projets et trouver les moyens d'un travail coopératif en réseau sur les applications de la physique au développement durable. Le lancement d'un programme mobilisateur, appuyé par le Fonds de Solidarité Prioritaire, pourrait apporter une meilleure visibilité internationale à nos efforts communs pour promouvoir les sciences de la matière et affirmer leur rôle déterminant dans le développement.

*\* Laboratoire de Physique et Modélisation des Milieux Condensés*
*CNRS/Université Joseph Fourier UMR 5493*
*BP 166*
*38042 - GRENOBLE cedex 09*
*Tel: +33 4 76 88 10 19*
*Fax: + 33 4 76 88 79 83*
e-mail: roger.maynard@grenoble.cnrs.fr
http://lpm2c.polycnrs-gre.fr/

*\*\* Ministère des Affaires Étrangères et Européennes*
*DGCID/DCSU - Recherche*
*244, bd Saint Germain*
*75303 PARIS 07 SP*
*Tel: +33 1 43 17 84 10*
*Fax: +33 1 43 17 89 37*
e-mail: jean-claude.topin@diplomatie.gouv.fr

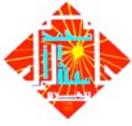 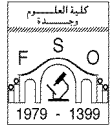 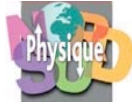 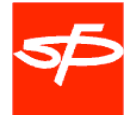

# Annexe

# Contribution Posters

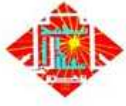 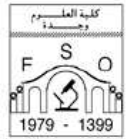 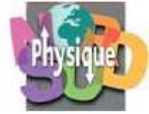 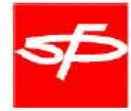

# POSTERS

## Introduction de FATIHA MAAROUFI

Une image panoramique construite de plusieurs communications affichées durant toute la période du déroulement du congrès a mis en évidence l'enseignement des sciences, en particulier de la physique, et la recherche en physique, chimie et mathématiques au Maroc, en Algérie et en Tunisie.

L'enseignement des sciences physiques, l'enseignement du concept temps en physique, l'enseignement de la physique et la difficulté des étudiants, la femme en physique, l'application des méthodes physiques en biomédicale, le phénomène d'agrégation des colloïdes, l'expérience Atlas, la cosmologie à cinq dimensions, les symétries quantiques en graphes, la simulation d'une machine asynchrone, la méthode de simulation RMC, la modélisation numérique des écoulements et du transfert de chaleur, la spectroscopie moléculaire haute résolution, les noyaux déficients en neutrons, les vibrations mécaniques, les plasmas froids, la technique LIBS appliquée à l'eau, un montage expérimental de diffusion de la lumière, la relativité spatiale, l'extrapolation du champ d'une antenne, les capteurs solaires, le transfert de chaleur, l'étude du confort thermique dans l'habitat…

Sujets variés mettant en valeur les sciences physiques que les docteurs et les chercheurs maghrébins ont exposé, expliqué et réexpliqué avec acharnement et amour à la physique.

Bravo !!

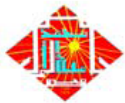 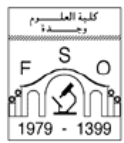 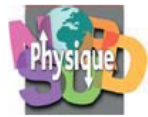 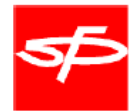

# APPORT DES METHODES PHYSIQUES A L'ETUDES DE QUELQUES BIOMATERIAUX A USAGES MEDICAUX

E. ABBAOUI, M. Elgadi, A. Essaddek, L. L. Elansari, E. Mejdoubi

Laboratoire de chimie du solide minérale et analytique, Univiversité Mohamed Ier, Oujda, Maroc

**Résumé**

Les ciments phosphocalciques sont largement utilisés en chirurgie orthopédique et odontologique. Ces ciments sont généralement préparés à partir d'un mélange de deux phosphates de calcium, ou plus, les uns à caractères acides et les autres à caractères basiques. L'évolution de ce ciments, après prise et durcissement, est due à des mécanismes physicochimiques très complexes. La compréhension de ces mécanismes nous facilitera l'optimisation et l'utilisation appropriée de ces biomatériaux. L'apport des méthodes d'analyses physiques tel que la diffraction des rayons X, la microscopie électronique à balayage, la spectroscopie d'absorption infrarouge, etc..,sont nécessaires pour suivre l'évolution des ciments dans chaque étape.

## 1 Introduction

Afin de pallier les inconvénients que présentent les autogreffes et les allogreffes dans le domaine orthopédique et odontologique, différents biomatériaux de substitution osseux ont été proposés, dont les ciments phosphocalciques [1,2]. Ces derniers présentent outre l'avantage d'être totalement biocompatible, et maniable pour combler entièrement les cavités osseuses, la possibilité de faire prise et durcir *in vivo* en évoluant vers l'hydroxyapatite [3], le matériau le plus proche par sa structure cristalline de la phase minérale de l'os, ou mélange d'hydroxyapatite et de phosphate tricalcique de type β[4]. La prise et le durcissement de ces ciments phosphocalciques sont deux phénomènes physico-chimiques très complexes. La compréhension de leurs mécanismes est essentielle aux activités biologiques des ciments. Elle conditionne leurs solubilités, et donc leurs interactions avec le milieu biologique[5].
Les méthodes d'analyses physiques tel que : la spectroscopie d'absorption infrarouge, la diffraction des rayons X, la microscopie électronique à balayage, etc.. Sont d'une grande importance dans notre travail, elles aident à mieux suivre l'évolution physico-chimique du ciment après gâchage, et donc identifier les différentes phases intermédiaires et finales. Ainsi, grâce aux résultats obtenus, nous pouvons proposer des mécanismes chimiques qui pourraient être responsable des réactions acido-basiques entre les différents constituants du ciment phosphocalciques.

## 2 Méthode d'analyse

Le ciment est un mélange de deux parties: une partie solide formée de l'hydroxyde de calcium ($Ca(OH)_2$), du phosphate tricalcique de type α ($\alpha\text{-}Ca_3(PO_4)_2$) et du phosphate tétracalcique ($Ca_4(PO_4)_2O$) et une partie liquide constituée d'acide orthophosphorique ($H_3PO_4$) (85% en poids), d'eau et de glycérophosphate. Le mélange des deux parties liquide et solide fait prise et durcit
Les échantillons prélevés aux cours de l'évolution du ciment sont analysés par spectroscopie d'absorption infrarouge à l'aide d'un diffractomètre PERKIN-ELMER de fréquence 4000-600$cm^{-1}$. Environ 1 milligramme de l'échantillon à analyser est homogénéisé et broyé avec 200 milligrammes de KBr sec, dans un mortier en agate. Le mélange est ensuite comprimé sous vide en lui appliquant une pression de 10 tones/$cm^{-2}$ pour former une pastille translucide. Les diagrammes de diffraction des rayons X ont été effectués à l'aide d'un diffractomètre DATA MP-SIEMENS D501. La radiation utilisée est $k_\alpha$ du cuivre ($\lambda_{Cu}$=1.5405Å) sélectionnée à l'aide d'un monochromateur à lame de graphite. L'échantillon à analyser se présente sous forme de poudre fine formée de cristaux que l'on a obtenue après broyage dans un mortier en agate. Cette poudre est déposée sur une lame de verre et étalée uniformément en une couche mince, pour obtenir une épaisseur inférieure à 0,1 mm.
L'imagerie est assurée par microscopie Electronique à balayage de type JEOL, en mode S.E.I (Secondary Electron image). Le principe d'observation consiste à coller l'échantillon sur une grille porte-objet, qui sera ensuite métallisé avec du carbone puis de l'or. La métallisation est effectuée sous vide secondaire par pulvérisation cathodique du carbone puis par évaporation de l'or. Des couches de très faibles épaisseurs sont disposées sur l'échantillon ce qui le rend conducteur.

## 3 Résultats et discussion

Les figures 1 et 2 qui présentent successivement la diffraction des rayons X et le spectre d'absorption infrarouge du mélange après prise et durcissement montrent la présence du phosphate dicalcique dihydraté ($Ca(HPO_4)_2,2H_2O$) (spectre 1A, 2A) dans le ciment. Ce phosphate dihydraté, responsable de la prise du ciment, réagira ensuite avec le phosphate tétracalcique en excès, pour donner l'hydroxyapatite (spectre 2A, 2B).

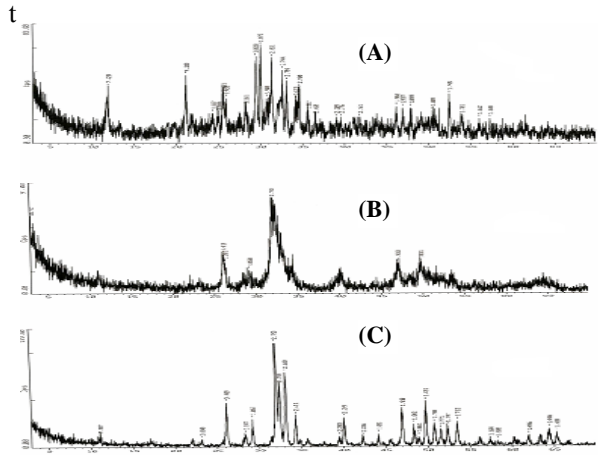

**Fig.1**: Diffractogrammes des rayons X du ciment à base du système $Ca(OH)_2/\alpha\text{-}Ca_3(PO_4)_2/Ca_4(PO_4)_2O$
Spectre (A) : prélèvement juste après gâchage.
Spectre (B) : prélèvement après durcissement du ciment.
Spectre (C) : prélèvement après durcissement et calcination du ciment à 900 °C.

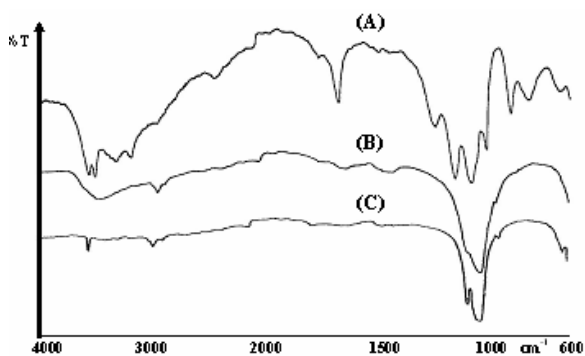

**Fig. 2** : Evolution du ciment en fonction du temps : (A) deux minutes après gâchage, (B) après durcissement et (C) après calcination

La figure (3) présente l'analyse par microscopie électronique à balayage du ciment phosphocalcique calciné à 900 °C pendant 2 heures. Elle montre des cristaux d'hydroxyapatite sous forme de fleures, bien ordonnés et possèdent la même taille (1-2 μm). La distribution des grains d'hydroxyapatite apparaît bien homogène et la surface semble relativement poreuse

## 4 Conclusion

Dans ce travail, nous avons pu mettre en évidence les mécanismeisss physico-chimiques responsables de la prise et du durcissement du ciment. En effet, la prise est due à la formation in situ des cristaux du phosphate dicalcique dihydraté, tandis que le durcissement est dû à la formation de l'hydroxyapatite.

## Références

[1] J.R. Woodard et al., Biomaterials, 28 (2007) 45;
[2] E. Abbaoui et al., J. Phys. IV, 123 (2004) 225;
[3] S.B. Baker, J. Weinzweig, R.E. Kirschner, S.P. Bartlett. Plast Reconstr Surg 109 (2002) 109;
[4] C. Damien, R. Parson, J. Appl. Biomater. 2 (1991) 187.

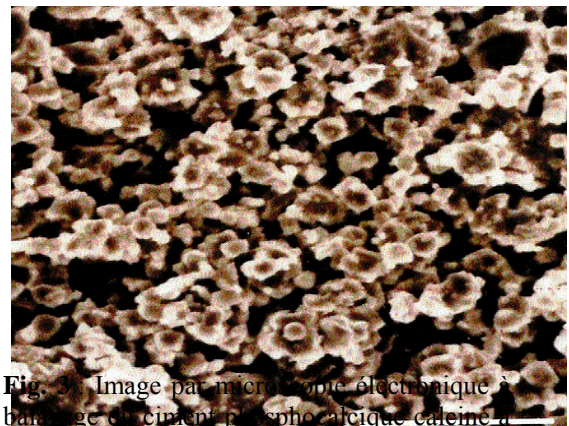

**Fig. 3** : Image par microscopie électronique à balayage du ciment phosphocalcique calciné à 900 °C pendant 2 heures (scale 5μm)

# SCREENED POTENTIAL CONSTRAINT IN A REVERSE MONTE CARLO (RMC) SIMULATION

**Abbes Oukacha**[2][*], **kotbi Mohamed**[1],

[1] Department of physics, A.B. Belkaïd University, BP.119,Tlemcen , Algeria
[2] Department of physics, T. Moulay University center Saïda, Saida , Algeria

**Abstract**

A recent simulation method called Reverse Monte Carlo (RMC) applicable without interaction potential is used to study the aqueous electrolyte system LiCl-6H$_2$O. Artifacts are appeared in some pair functions of radial distribution particularly a small pick near the first coordination of $g_{OO}(r)$ and also near the $g_{OCl}(r)$ one. One try to remedy for that artifact with introducing screened coulomb potential for the ensemble of the atomic species with equivalent charges fraction for the water ones. An improvement in the first coordination of this function is noticed suggesting a useful test of an interaction potential model for classical methods as Monte Carlo (MC) and Molecular Dynamic (MD).

**Key words**: Potential constraint, RMC simulation, LiCl-6H$_2$O

## 1 Introduction

By a structural modelisation, the aqueous electrolyte of type LiCl-6H2O is studied. This system presents the property of forming a glass in passing via the metastable state-supermelt when the temperature decreases. The method of simulation called Reverse Monte Carlo or RMC (3) presents the interest being applicable without specifying interactions (interatomic and/or intermolecular).

## 2 Implementation of the RMC Method

Details of the RMC method have been given before (1,2). The aim is to produce three dimensional structural models of a system that are consistent with the avaible diffraction data, within the errors of that data. To do this we use a modification of the standard Metropolis Monte Carlo (MMC) method (4) where, instead of minimising the system energy, we minimiose the difference between the calculated and experimental spectra.

## 3 Simulation Details

Some experimental results obtained by the neutrons scattering experiment and the technique of isotopic substitution could be used (5, 6). In order to improve the results obtained in RMC (7), one could introduce an additional constraint as a potential of interaction with a weight parameter ω chosen in our case to be unit. We studied the structure of LiCl-6H2O using the RMC with four functions taken from the experimental partial distribution functions (PDF).

## 4 Results and discussions

The calculated partial functions and the experimental ones show a net concordance and so there is no mismatch between RMC without or with potential constraint. So there is no conflict between the method used in our study and the constraint of potential introduced..

## 5 Conclusion

The RMC method allows exploring a certain number of structural features of the system based on experimental data limited to four partial distribution functions (PDF).The results one obtains could include some artifacts. To remedy for this, we could make a propose choice of potential. One must take into account the mismatch between the interaction potential between charges and the method of RMC simulation based on experimental data. So following the atomic or molecular species in presence, introducing a potential like additional constraint in the RMC simulation, one could obtain some better results. Consequently, we could suggest that the choice of the interaction model as a function of atomic or molecular properties forming the system could bring a meaningful improvement to the results .

## References:

[1] R.L. Mc Greevy and M.A. Howe, Phys. Chem. Liq., 24, 1 (1991).
[2] R.L. Mc Greevy and Pusztai, Mol. Simulation, 1, 359 (1988).
[3] R.L. Mc Greevy, M.A. Howe and J.D. Wicks, RMCA Version 3, "A General Purpose Reverse Monte Carlo Code", (1993).
[4] M.N.Rosenbluth, A.H teller , E.Teller, J.Phys.Chem.,21,1087 (1953).

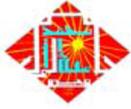 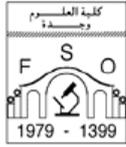 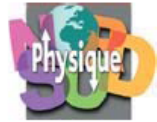 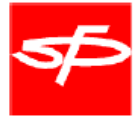

[5] J.F.Jal, K. Soper, P. Carmona, J. Dupuy, J. Phys. Cond. Matter (3) 551-567, (1991).
[6] B. Prével, J.F. Jal, J. Dupuy-Philon, A.K. Soper, 1995, J. Chem. Phys., 103, (1986)
[7] M.Kotbi , H.Xu , Molecular Physics, Vol. 98, No 2, 373-384, (1998)

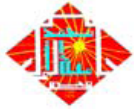 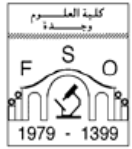 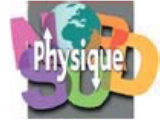 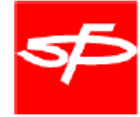

# Pour une intégration réussie des capteurs solaires thermiques

## Samir AMRAQUI[1], Ahmed MEZRHAB[1], Cherifa ABID[2]

[1] Faculté des sciences, Département de physique, Laboratoire de Mécanique & Energétique, Oujda, Maroc
[2] Ecole polytechnique Universitaire de Marseille, IUSTI U.M.R. Marseille cedex 13, France

### Résumé

Le besoin pressant de réduire la consommation en énergie fossile dans les bâtiments fait de l'intégration des technologies solaires dans l'enveloppe du bâtiment un thème d'une grande actualité. L'intégration doit être conçu dans un lien étroit avec le bâtiment.

D'un autre côté le capteur solaire thermique aussi pourrait/devrait être conçu à la base comme élément de construction et développé en tant qu'élément architectural. Cependant, son utilisation reste le plus souvent confinée à une installation en toiture du fait de son manque d'esthétisme du point de vue architectural.

L'objectif de ce travail est intéressante à deux titres: pour l'effort d'intégration architecturale des capteurs solaires d'une part, et l'utilisation optimisée et diversifiée de la chaleur produite par les capteurs solaires d'autre part (production d'eau chaude sanitaire, chauffage de la maison et de la piscine découverte).

## 1 Pourquoi utiliser l'énergie solaire?

- Préserver les ressources énergétiques,
- Faire des économies,
- Valoriser une énergie locale et inépuisable,
- Réduire les émissions de gaz à effet de serre.

## 2 Quelle est la manière la plus simple d'utiliser l'énergie solaire ?

Un chauffe-eau solaire individuel (**C.E.S.I**) de 4m² de capteurs solaires thermiques suffit à apporter les 2/3 de l'énergie, en moyenne annuelle, au chauffage de l'eau sanitaire d'une famille de quatre personnes.

## 3 Y-a-t-il d'autres applications domestiques de l'énergie solaire ?

Un système solaire combiné (S.S.C), équipé d'une dizaine de m² de capteurs solaires thermiques, peut fournir la moitié de l'énergie nécessaire au chauffage et à l'eau chaude sanitaire d'une maison neuve de 130 m².
Enfin, un système photovoltaïque, raccordé au réseau public et équipé d'une vingtaine de m² de capteurs solaires thermiques photovoltaïque, suffit à compenser les consommations électriques d'une famille de 4 personnes.

## 4 Comment intégrer les capteurs solaires dans le bâtiment?

L'intégration des capteurs solaires dans le bâtiment doit être étudiée précisément, tant pour assurer une bonne efficacité énergétique des équipements que pour obtenir une qualité esthétique satisfaisante tout en respectant les règlements d'urbanisme.

## 5 Où installer les capteurs solaires?

Pour obtenir les économies attendues, les capteurs doivent être correctement exposés au rayonnement solaire.
**1.** En toiture, terrasse ou en pente,
**2.** En façade,
**3.** Au sol…

## 6 Les caractéristiques et spécificités techniques

### 6-1 Comment fonctionne un C.E.S.I ?

Il assure un ensemble de fonctions :
- Captage de l'énergie solaire
- Transfert de la chaleur des capteurs
- Stockage de l'énergie
- Distribution de l'eau chaude

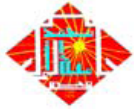
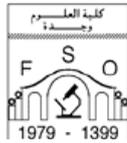
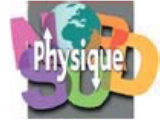
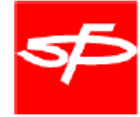

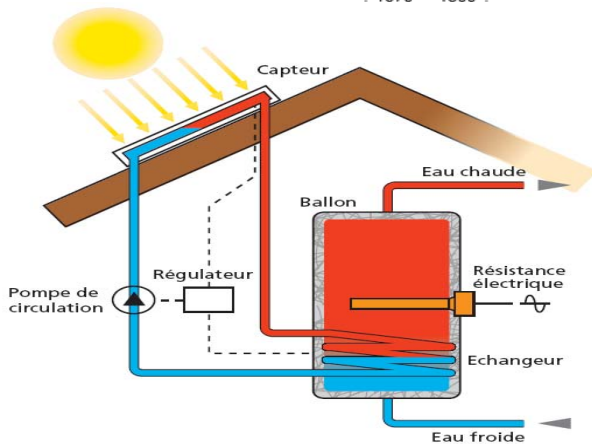

**6-2 Comment obtenir un bon compromis entre performance énergétique et intégration architecturale ?**

| Type d'implantation | Inclinaison du capteur | Plein sud | Sud est ou sud ouest | Est ou ouest |
|---|---|---|---|---|
| En toiture terrasse et au sol | 45° | 73% | 69% | 61% |
| En toiture en pente et en auvent | 15° | 68% | 67% | 63% |
| | 20° | 69% | | |
| | 25° | 70% | 68% | |
| En façade et en garde corps | 90° | 56% | 53% | 42% |

Pour mémoire:
15°=27%
20°=36%
22°=40%
25°=47%

(Autonomie pour différentes orientations et inclinaisons pour une consommation quotidienne de 200 litres d'eau chaude à 45°C, avec 4m² de capteurs et un ballon solaire de 300 litres à appoint intégré)

**Inclinaison préférentielle des capteurs:**
- ❖ 45° pour les chauffe-eau solaires individuels (C.E.S.I)
- ❖ 60° pour les systèmes solaires combinés de chauffage (S.S.C)

**6-3 Comment positionner les capteurs solaires ?**

Une orientation pleine sud et une inclinaison par rapport à l'horizontale proche de 45° sont idéales.

**6-4 Quels sont les différents types de capteurs solaires utilisables pour les C.E.S.I et S.S.C ?**

**1.** Capteurs plans ou capteurs coffre indépendants de la structure du bâtiment,
**2.** Capteurs plans à intégrer en toiture ou façade de batimant,
**3.** Capteurs à tubes sous vide.

**6-5 L'intégration architecturale des capteurs solaires**

Il est donc à placer sur un bâtiment comme une souche de cheminée, une fenêtre ou une porte… et se doit d'être considéré comme un élément de composition architecturale.

# 7 Conclusion

Le capteur solaire est un élément de composition architecturale à condition de tenir compte de :
- ❖ La forme
- ❖ La position
- ❖ La proposition
- ❖ L'association

L'intégration architecturale des capteurs solaires est plus facile avec des architectures contemporaines qu'avec l'architecture néo-provençale dont la toiture en tuiles rondes reste une composante essentielle. Différentes technologies, dimensions et formes de capteurs existent et permettent divers intégration en association à une autre fonction : pergola, garde-corps, brise-soleil…

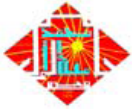 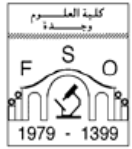 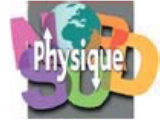 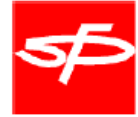

# La spectroscopie moléculaire haute résolution et la télédétection atmosphérique

## Mohammed Badaoui, Professeur

Institut Agronomique et Vétérinaire Hassan II
B. P. 6240, Madinat Al Irfane, 10101 Rabat les Instituts, Maroc

**Résumé**

Pour appréhender le phénomène du réchauffement planétaire, il est nécessaire d'effectuer des mesures précises de concentrations de constituants chimiques de l'atmosphère. Un des outils utilisés est la spectroscopie moléculaire haute résolution par transformée de Fourier. Dans ce domaine, la télédétection spatiale par satellites est particulièrement performante à condition de disposer de paramètres spectroscopiques précis [1]. Il est donc indispensable d'entreprendre au laboratoire une étude spectroscopique à haute résolution en position, en intensité, en élargissement et en déplacement par la pression de raies des spectres de rotation et de rotation-vibration des molécules. Lors de la détermination de ces paramètres, nous tenons compte des effets physiques fondamentaux, et en modélisant « correctement » si nécessaire la fonction d'appareil du spectromètre à transformée de Fourier [2]. Nous participons ainsi à l'enrichissement ou à la mise à jour des banques de données moléculaires d'intérêts atmosphériques terrestre ou planétaires: GEISA [3], HITRAN [4]…etc.

## 1 Méthodologie

La transmission de la lumière à travers une succession de N milieux absorbants m (cas de l'atmosphère), homogènes, dont les grandeurs thermodynamiques sont stationnaires, chacun de longueur $\ell_m$ et de température $T_m$, et contenant $n_m$ gaz j de pressions partielles $P_{mj}$, est donnée par :

$$T_{th}(\sigma) = \exp\left[-\sum_{m=1}^{N}\sum_{j=1}^{n_m}\ell_m \sum_i k_{ij}(\sigma - \sigma^0_{mji}, \gamma^L_{mji}, P_{mj})\right] \quad (1)$$

où $\sigma^0_{mji}$ est la position absolue de la raie i du constituant j absorbant dans le milieu m, et $\gamma^L_{mji}$ est sa largeur collisionnelle à mi-hauteur. $k_{ji}(\sigma - \sigma^0_{mji}, \gamma^L_{mji}, P_{mj})$ est le profil de la raie i du constituant j dans le milieu m. Son intensité est $S_{mji} = \int_{raie} k_{ij}(\sigma - \sigma^0_{mji}, \gamma^L_{mji}, P_{mj})\,d\sigma$. La relation (1) fait partie de l'équation de transfert radiatif de l'atmosphère qu'il faut inverser pour extraire ses caractéristiques chimiques et physiques selon le principe: Atmosphère = $f^{-1}$(raies, spectre mesuré par satellite (MIPAS, Envisat…)). Il faut donc connaître au préalable les paramètres de raies par un travail de laboratoire. En fait, la transmission calculée est le produit de convolution de la transmission théorique par la fonction d'appareil [2]:

$$\frac{T_{cal}(\sigma)}{T_b(\sigma)} \approx T_{th}(\sigma) \otimes f_{app}(\sigma) = \int_{\Delta\sigma, raie} T_{th}(\sigma') f_{app}(\sigma' - \sigma)\,d\sigma'$$

$f_{app}(\sigma)$ est modélisée selon le domaine spectral. $T_b(\sigma)$ est l'enveloppe (ou réponse instrumentale globale) au niveau de la raie que nous modélisons par un polynôme en $\sigma$. Nous calculons les paramètres par les techniques de moindre carrées en ajustant la transmission calculée à la transmission observée (figure) [5]. Nous élaborons ensuite des modèles qui permettent de calculer les paramètres d'autres raies dans différentes conditions expérimentales.

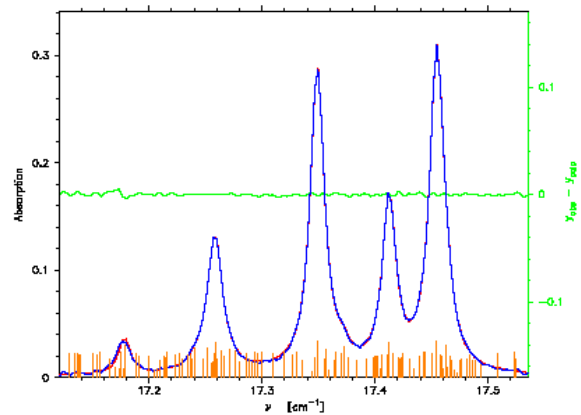

Les courbes bleu et rouge sont respectivement le spectre observé et le spectre calculé de rotation pure de l'ozone. Les barres orange indiquent les raies de l'ozone dont les paramètres pris de HITRAN sont des valeurs initiales.

## 2 Pertinence dans le contexte Marocain et proposition de collaborations

Les problèmes de la pollution atmosphérique concernent toutes les nations. Nous cherchons la collaboration nationale et internationale dans ce domaine et que les étudiants doctorants adhèrent aux efforts de calculs associés: spectroscopie de laboratoire, dépouillements de données satellitaires, développement de codes…etc.

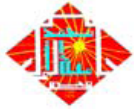 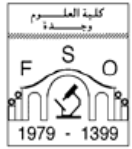 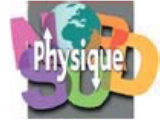 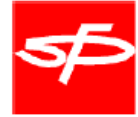

Au Maroc, la télédétection spatiale par satellite (spot) est un outil incontournable de développement de l'agriculture, de l'océanographie, de l'aménagement territorial,…et de la sécurité. Le traitement des images satellitaire nécessite des corrections atmosphériques. Nous pensons participer dans le développement de ses codes en introduisant des paramètres relatifs à notre réalité géographique (effets de tempêtes de sables…etc.).

# Références

[1] Badaoui, Doctorat de l'Université de Paris VI, 1993, Mesures de paramètres de raies d'absorption dans l'infrarouge à partir de spectres obtenus par transformée de Fourier.
[2] M-Y Allout, J-Y. Mandin, V. Dana, N. Piqué and G. Guelachvili, FTS Generalised apparatus function, J. Quant. Spectrosc. Radiat. Transfer. Vol. 60, N 6, pp. 979-987, 1998.
[3] http://ara.lmd.polytechnique.fr/htdocs-public/products/GEISA/HTML-GEISA/
[4] http://www.cfa.harvard.edu/hitran//
[5] F. Schreier and B. Schimpf, Least squares fitting of molecular line parameters from high resolution Fourier transform spectra, DLR (Germany), internal report, 1995.

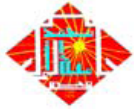 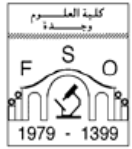 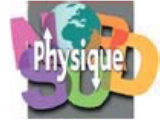 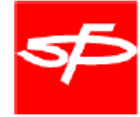

# Une stratégie d'enseignement des sciences physiques

## Djanette BLIZAK, chargée de cours

Université UMBB, Faculté des sciences, Boumerdès.

**Résumé**

En ce début du 3$^e$ millénaire, les sociétés contemporaines se développent selon des rythmes accélérés subissant des mutations voir des révolutions provoquées en partie par les Technologies d'Information et de Communication (TIC).
Le pari consiste à réduire la place qu'occupent les méthodes d'enseignement dogmatique au détriment de méthodes actives centrées sur l'apprenant et qui tiennent compte de ses préacquis, de ses conceptions, de ses besoins et de son rythme d'apprentissage.
Cette nouvelle mission qui incombe aux enseignants chercheurs de l'université, nous a interpellé pour étudier, expérimenter et proposer un modèle pédagogique sous forme d'une stratégie d'enseignement.
Dans ce travail nous exposons notre modèle d'enseignement que nous avons appelé « stratégie d'enseignement de motivation-historique ». Bien qu'elle se base sur l'histoire des sciences du savoir enseigné pour motiver les étudiants et les rendre responsable de leur propre apprentissage, elle encourage aussi l'apprentissage coopératif.

# 1 Introduction

Avec les TIC (Technologie d'Information et de Communication) appliquées à l'enseignement, les contenus des matières, eux-mêmes, ne sont plus considérés comme une fin en soi, mais comme une ressource que l'apprenant doit s'approprier pour développer des compétences. Le rôle de l'enseignant n'est d'ailleurs plus de transmettre ces contenus-matières comme tels, mais de concevoir et de gérer des séquences d'apprentissage dans lesquelles les apprenants sont confrontés à des situations nouvelles et motivantes qui les amènent à interagir pour chercher et traiter l'information nécessaire.
Dans cet article nous parlerons de l'intérêt d'une nouvelle stratégie d'enseignement des sciences physiques, et plus spécialement de l'optique géométrique, pour les étudiants en première année universitaire en Algérie. Par la suite, nous exposant notre modèle de stratégie d'enseignement avec le cadre théorique qui nous a inspiré et guidé.

# 2 Problématique

Déjà en 1938 Gaston Bachelard [1] écrivait :
« *J'ai souvent été frappé du fait que les professeurs des sciences […] ne comprennent pas que leurs élèves ne comprennent pas les sciences […]. Les professeurs de sciences imaginent que l'esprit commence comme une leçon […]. Et qu'on peut faire comprendre une démonstration en la répétant* ».
On peut sans doute dire que la question de savoir pourquoi ils ne comprennent pas fut la question centrale des débuts de la didactique, pratiquement jusqu'à nos jours.

## 2.1 Question de recherche

**-Quelle stratégie d'enseignement des sciences physiques peut remplacer une stratégie d'enseignement traditionnelle dans le but d'un meilleur apprentissage ?**

Pour répondre à cette question nous essayons de mettre au point un nouveau modèle de stratégie d'enseignement: que nous appelons stratégie d'enseignement de motivation historique. Le choix de ce nom est lié au fait que notre stratégie se base principalement sur l'histoire des sciences pour motiver les étudiants et les amener à prendre en main leur propre apprentissage.
L'histoire des concepts scientifiques est un facteur de motivation très important.
« *Quand l'histoire des sciences ne survivrait qu'à faire surgir, stimuler ou satisfaire une élémentaire curiosité intellectuelle, elle apporterait déjà beaucoup à leur enseignement.* » (Saroui. Eric, 2001)[2]
*N*ous pensons que la présentation de l'histoire des sciences en utilisant la TIC comme le CD-ROM, attire l'attention et la curiosité des étudiants. Cela permettra aux étudiants de prendre facilement la place d'un chercheur pour mieux acquérir des nouvelles connaissances scientifiques

## 2.2 Stratégie d'enseignement

Avant de donner le scénario de notre modèle de stratégie d'enseignement nous avons préféré d'abord, définir le concept « stratégie d'enseignement ».
« *Une stratégie est un art de coordonner des actions et de manœuvrer pour atteindre un but*» (petit la rousse 1969)[3].

# 3 Scénario de la stratégie d'enseignement

Nous l'avons appelée stratégie d'enseignement de motivation-historique, pour la raison que nous allons utiliser l'histoire des concepts scientifiques comme un facteur de motivation. Notre stratégie comporte trois étapes: l'étape de motivation, l'étape de la recherche et l'étape de l'exposé. Le rôle de l'enseignant dans cette

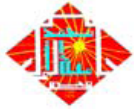
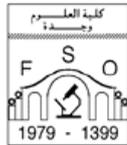
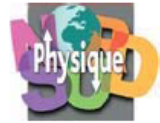
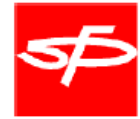

stratégie c'est de gérer et de contrôler le déroulement des trois étapes.

### 3.1 l'étape de motivation

Il s'agit d'élaborer une séquence d'enseignement dans une classe artificiel, où l'enseignant fait passer un CD-ROM sur l'histoire du concept scientifique enseigné et invite les étudiants à se concentrer sur le contenu. Durant cette période, l'enseignant n'intervient pas. A la fin cette phase, l'enseignant ouvre un débat sur la contenue du CD-ROM.

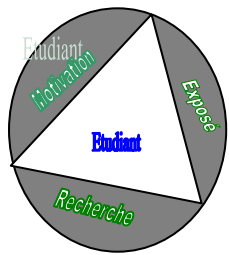

FIG. 1: Les trois étapes de la stratégie

### 3.2 l'étape de la recherche

L'enseignant propose aux étudiants des sujets de recherche sur des concepts scientifiques qui présentent un certain obstacle à leurs apprentissages. Il leur demande de se mettre par paire et de préparer un document sur le sujet choisie. Durant la période de la préparation de ce document, qui ne dépasse pas deux semaines, l'enseignant joue le rôle d'un tuteur.

### 3.3 l'étape de l'exposé

Les étudiants donnent une copie de leur travail à l'enseignant et une autre à un petit groupe d'étudiants qui jouera le rôle du jury. Ils exposent, après, devant le jury présidé par l'enseignant et à la présence des autres étudiants. A la fin de cette séance, les étudiants doivent répondre à touts les questionsdu jury.

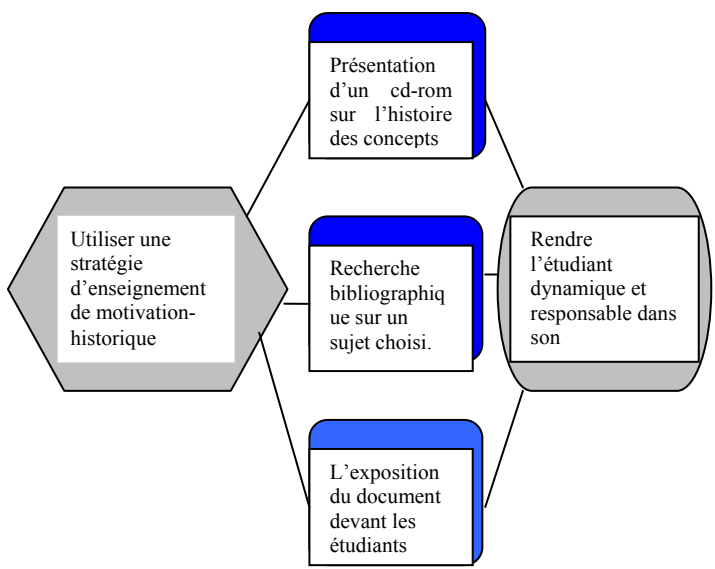

FIG. 2: La stratégie d'enseignement de motivation-historique

## 4 L'expérimentation retenue

### 4.1 Nécessité d'une étude de cas en continu

Nous faisons le choix d'étudier le cheminement qu'un étudiant peut suivre au cours de son apprentissage. Il s'agit de suivre de près les évolutions d'un étudiant.

### 4.2 Chois d'une étude comparative

Pour montrer l'efficacité de notre stratégie, nous comparons les acquis des étudiants, que nous avons choisi dans notre expérience, avec ceux des étudiants qui suivent des séquences d'enseignement traditionnelles dans des classes réelles.

### 4.3 . Nécessité de construire d'une séquence d'enseignement

La construction d'une séquence d'enseignement nous apparue comme indispensable dans le cadre de notre étude. Elle doit être en accord avec les hypothèses d'apprentissage que nous avons adoptées sur le rôle du facteur motivation.

## 6 Conclusion

Les universités sont confrontées à une situation sans précédent résultant de la diffusion des technologies de l'information et de la communication qui s'imposent à l'enseignement universitaire.
Puisque les universités apparaissent, aujourd'hui, confrontées non seulement à une évolution des médias mais aussi et surtout à une émergence d'une autre culture et d'un rapport au savoir transformé, la nouvelle reforme LMD s'impose, et les stratégies d'enseignement doivent être changées.
Nous avons essayé d'élaborer une stratégie d'enseignement dans l'espoir de répondre aux nouveaux défis que pose l'enseignement universitaire, particulièrement en ce qui a trait à la motivation des étudiants et à la construction de compétences visant l'autonomie d'apprentissage. Notre stratégie d'enseignement se base sur un facteur affectif, très important, souvent oublié par les chercheurs en didactique des sciences, et montrer de doigts par les chercheurs en psychologie cognitive qu'on nome motivation.

Lors d'un travail de recherche future nous présenterons nos résultats sur l'application de notre stratégie sur l'enseignement des concepts d'optique géométrique aux

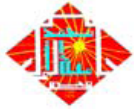 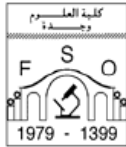 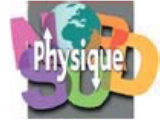 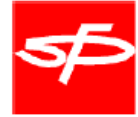

étudiants de première année universitaire en Algérie.

# Références

[1] G.J Posner. & al, Science. Education, Vol. 66, (1982) 211-227.
[2] E. Sarou « L'histoire des sciences dans l'enseignement » BUP, Vol.95 (2004).
[3] petit la rousse (1969).

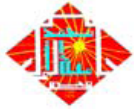 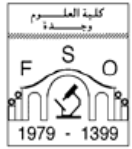 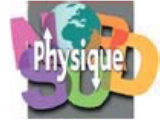 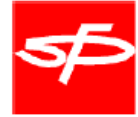

# Etude des vibrations mécaniques émises dans l'environnement.

**J.BOUSTELA[1*], M.BENELMOSTAFA[1], Y.BENELMOSTAFA[2]**

[1]UFR Mécanique des Fluides et Énergétique, Laboratoire de Physique Théorique et de Particules et Modélisation. Faculté des Sciences, Oujda.
[2]Ecole des Ingénieurs en Génie des Systèmes Industriels, Casablanca

## 1 Introduction

Les vibrations mécaniques émises dans l'environnement par les installations classées pour la protection de l'environnement peuvent constituer un problème pour la protection des populations riveraines. L'évaluation des effets des vibrations mécaniques transmises dans l'environnement par ces installations est actuellement faite avec des moyens et des méthodes qui varient considérablement. L'aspect vibratoire est un aspect primordial dans l'étude des systèmes mécaniques. En effet, dans un grand nombre d'applications industrielles le système est soumis à des excitations qui le font vibrer. Cette étude est assez complexe dans le cas général. Cette complexité peut avoir pour origine le système lui même ou/et l'excitation.

## 2 Mouvement de vibration d'un parallélépipède

Un parallélépipède en mouvement vibratoire, relié au sol par des liaisons viscoélastiques appliquées aux 4 sommets A, B, C, D d'un rectangle horizontal, de largeur 2a et de longueur 2b. On admet qu'il y a symétrie par rapport au plan longitudinal $xG_0z$ passant par le centre $G_0$ du parallélépipède.

Soit $K_1$, $C_1$ la raideur et le coefficient d'amortissement aux sommets A et B dans les directions $G_0x$, $G_0y$ et $G_0z$, $K_2$, $C_2$ la raideur et le coefficient d'amortissement aux sommets C et D dans les directions $G_0x$, $G_0y$ et $G_0z$.

La masse du parallélépipède est m. On donne les moments d'inertie $j_1$, $j_2$ et $j_3$ par rapport aux axes $G_0x$, $G_0y$ et $G_0z$ que l'on suppose axes principaux d'inertie. $G_0x$ est dit axe d'avance et de roulis, $G_0y$ est dit axe de ballant et de tangage, $G_0z$ est dit axe de pompage et de lacet. $G_0$ est à une distance h au dessus du plan horizontal ABCD [1].

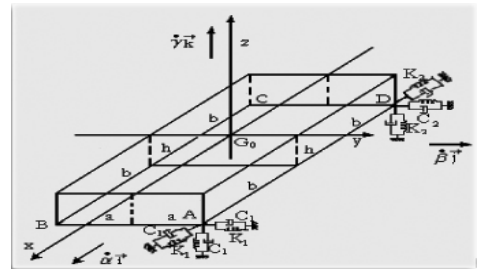

**Figure1** : Parallélépipède en liaisons viscoélastiques.

## 3 Equations de mouvements

Pour trouver les équations de mouvements, on applique le théorème de Lagrange [2] soit :

$$\frac{d}{dt}\left(\frac{\partial L}{\partial \dot{q}^i}\right) - \frac{\partial L}{\partial q^i} + \frac{\partial R}{\partial \dot{q}^i} = Q_i^A \quad (1)$$

Tel que : $L = T - V_S$ avec :

L : Lagrangien ;  R : La fonction de
T : L'énergie cinétique ;  dissipation de Rayleigh
$V_S$: L'énergie potentielle ;  $Q_i^A$ : Force généralisée

On a :
$$2T = \begin{pmatrix}\dot{x}\\\dot{y}\\\dot{z}\\\dot{\alpha}\\\dot{\beta}\\\dot{\gamma}\end{pmatrix}^T \begin{pmatrix} m & 0 & 0 & 0 & 0 & 0\\ 0 & m & 0 & 0 & 0 & 0\\ 0 & 0 & m & 0 & 0 & 0\\ 0 & 0 & 0 & j_1 & 0 & 0\\ 0 & 0 & 0 & 0 & j_2 & 0\\ 0 & 0 & 0 & 0 & 0 & j_3\end{pmatrix}\begin{pmatrix}\dot{x}\\\dot{y}\\\dot{z}\\\dot{\alpha}\\\dot{\beta}\\\dot{\gamma}\end{pmatrix},$$

$$2V_s = \begin{pmatrix}x\\y\\z\\\alpha\\\beta\\\gamma\end{pmatrix}^T \begin{pmatrix} 2K & 0 & 0 & 0 & -4Kh & 0\\ 0 & 2K & 0 & 4Kh & 0 & -4K'b\\ 0 & 0 & 2K & 0 & 4K'h & 0\\ 0 & 4Kh & 0 & 2K(a^2+h^2) & 0 & 4K'bh\\ -4Kh & 0 & 4K'h & 0 & 2K(b^2+h^2) & 0\\ 0 & -4K'b & 0 & 4K'bh & 0 & 2K(a^2+b^2)\end{pmatrix}\begin{pmatrix}x\\y\\z\\\alpha\\\beta\\\gamma\end{pmatrix}$$

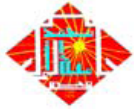
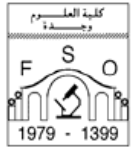
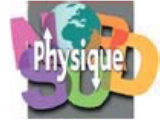
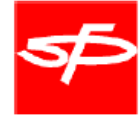

$$2R = \begin{pmatrix} \dot{x} \\ \dot{y} \\ \dot{z} \\ \dot{\alpha} \\ \dot{\beta} \\ \dot{\gamma} \end{pmatrix}^T \begin{pmatrix} 2C & 0 & 0 & 0 & -4Ch & 0 \\ 0 & 2C & 0 & 4Ch & 0 & -4Cb \\ 0 & 0 & 2C & 0 & 4Ch & 0 \\ 0 & 4Ch & 0 & 2C(a^2+h^2) & 0 & 4Cbh \\ -4Ch & 0 & 4Ch & 0 & 2C(b^2+h^2) & 0 \\ 0 & -4Cb & 0 & 4Cbh & 0 & 2C(a^2+b^2) \end{pmatrix} \begin{pmatrix} \dot{x} \\ \dot{y} \\ \dot{z} \\ \dot{\alpha} \\ \dot{\beta} \\ \dot{\gamma} \end{pmatrix}$$

On introduit T, V et R dans l'équation (1) on trouve une équation au second ordre :

$$[M]\{\ddot{q}(t)\} + [C]\{\dot{q}(t)\} + [K]\{q(t)\} = \{F(t)\}$$

## 4 Résolution des équations

La résolution de la plupart des équations différentielles requiert l'utilisation des méthodes numériques. Chacune de ces méthodes peut être appliquée à la résolution de la majorité des équations différentielles. Dans notre étude, on va utiliser les méthodes de Runge Kutta car elles présentent plusieurs avantages, (facilité de programmation, stabilité de la solution, modification simple du pas et la connaissance de y(0) suffit pour intégrer l'équation différentielle). Les inconvénients de cette méthode se résument au temps de calcul lent et à la difficulté de l'estimation de l'erreur locale [3], [4].

La méthode de Runge Kutta consiste à résoudre un système de premier ordre:

$$y' = f(y(t), t)$$

On réduit le système (2) en un système du premier ordre en posant :

$$y_1 = x(t), y_2 = \dot{x}(t),\ y_3 = y(t), y_4 = \dot{y}(t),$$
$$y_5 = z(t), y_6 = \dot{z}(t)$$
$$y_7 = \alpha(t), y_8 = \dot{\alpha}(t),\ y_9 = \beta(t), y_{10} = \dot{\beta}(t),$$
$$y_{11} = \gamma(t), y_{12} = \dot{\gamma}(t)$$

Dans la plupart des cas, la méthode de Runge Kutta utilisé est celle d'ordre 4 :

$$k_1 = hf(t_i, y_i)$$
$$k_2 = hf(t_i + \tfrac{1}{2}h, y_i + \tfrac{1}{2}k_1)$$
$$k_3 = hf(t_i + \tfrac{1}{2}h, y_i + \tfrac{1}{2}k_2)$$
$$k_4 = hf(t_i + h, y_i + k_3)$$
$$y_{i+1} = y_i + \tfrac{1}{6}(k_1 + 2k_2 + 2k_3 + k_4)$$

## 5 Résultats

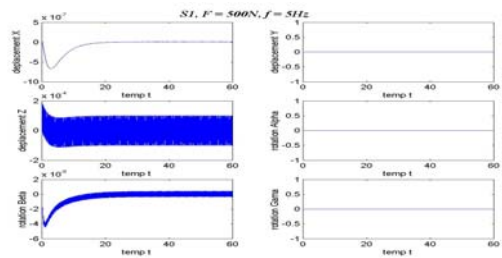

**Figure2** : Déplacement du parallélépipédique pour S1, F=500N, f=5Hz

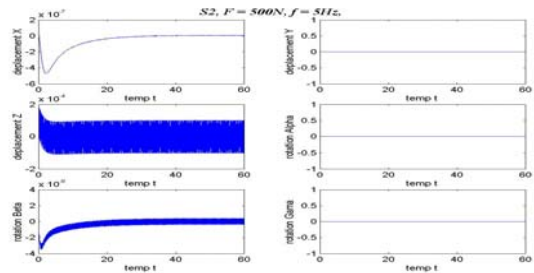

**Figure3** : Déplacement du parallélépipédique pour S2, F=500N, f=5Hz .

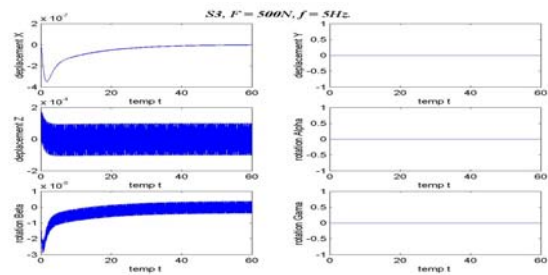

**Figure4** : Déplacement du parallélépipédique pour S3, F=500N, f=5Hz .

## 6 Conclusion

On rappelle que le but de ce travail est de minimiser les vibrations transmises par des installations mécaniques à leur voisinage.
On a simulé ces installations via un parallélépipède, et on a étudié l'influence du coefficient d'amortissement et de la raideur sur le mouvement vibratoire du système. On remarque, à travers les différentes courbes, que les vibrations diminuent au fur et à mesure que le coefficient d'amortissement augmente et la raideur décroît. Dans un prochain travail, nous allons appliquer cette étude à une éolienne installée au voisinage des habitations.

## Références

[1]. S.Laroz, Résistance des matériaux et structure Tome

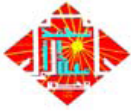 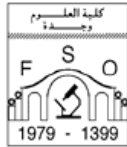 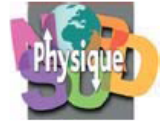 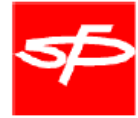

3, Eyrolles, 1979.
[2]. J. Boustela, M. Benelmostafa, Y. Benelmostafa, Vibration d'une éolienne, 5$^{\text{éme}}$ Rencontre Nationale des Jeunes Chercheur en physique, Casa 2006.
[4]. J.H.E. Conwright and O. Prio, The dynamics of Runge-Kutta methods, Queen Mary and Westfield college, University of London, 1992.
[5]. R. Vaillancourt, Numerical Methods with Matlab, Department of mathematics and statistics, University of Ottawa, ON, Canada, KIN 6N5.

# ETUDE DE L'EFFET CASIMIR ENTRE NANO-PARTICULES DANS UN MELANGE DE POLYMERES

## N.Chafi[1], M.Benelmostafa[1], M.Benhamou[2], M.El Yaznasni[2], H.Ridouane[2], E-K.Hachem[2]

[1] Équipe de la Physique de Polymères et Matières Molles, LPTPM, Département de Physique Faculté des Sciences, Université Mohamed Premier Oujda, Maroc

[2] Laboratoire de Physique des Polymères et Phénomènes Critiques, Faculté des Sciences Ben M'sik, Université Hassan II-Mohammedia P.B.7955. Sidi Othman, Casablanca, Maroc

**Résumé**

Le phénomène d'agrégation des colloïdes immergés dans les solutions de polymères est dû aux forces de Casimir critiques, qui tirent leur origine des fortes fluctuations du paramètre d'ordre autour de la température critique de démixion. Ce travail s'organise comme suit:

❖ Force de Casimir entre les colloïdes immergés dans une solution ternaire de polymères.

❖ Étude cinétique de l'agrégation des colloïdes immergés dans une solution binaire de polymères.

# 1 Force de Casimir entre colloïdes dans les mélanges ternaires

## 1.1 Introduction

Nous considérons des particules colloïdales sphériques en suspension dans un mélange ternaire de polymère A et B de nature chimique dufférents, en dolution dans un bon solvant. Nous supposons que au voisinage immédiat de la température critique de démixion $T_k$ du mélange ternaire, les colloïdes manifestent une affinité attractive vis-à-vis de l'un des deux polymères. Comme conséquence, les particules s'agrégent dans la phase non préférée. Le but de cette partie est précisément la détermination de la force induite responsable de l'agrégation des colloïdes.

## 1.2 Force de Casimir entre colloïdes dans les mélanges binaires

Le sytème étudié, dans ce cas, est un assemblage de n colloïdes immergés dans un mélange binaire de polymères A et B (voir figure1). La force responsable de l'agrégation des colloïodes dans ce système s'écrit [1] :

$$\frac{F(r)}{k_b T_c} = -\frac{64\pi^2}{27} N \frac{d_0^2}{r^3} \qquad (r > d_0) \quad (1)$$

$T_c$ est la température critique de démixion du mélange binaire, $N$ est le degré de polymérisation, $d_0$ est le diamètre des colloïdes, $r$ est la distance inter-particule.

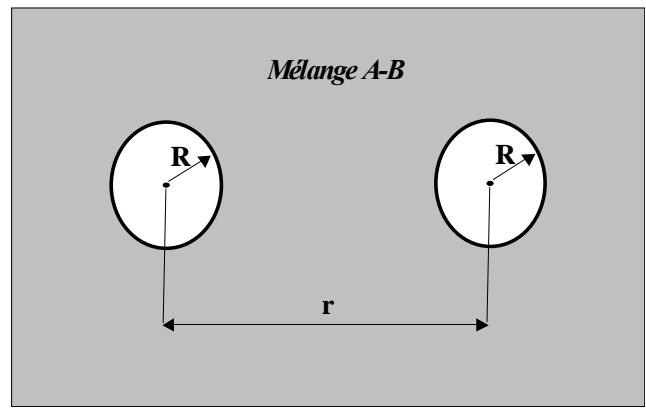

FIG. 1: Deux particules colloïdales sphériques en susppension dans un mélange binaire de polymères.

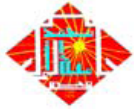 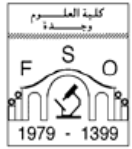 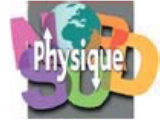 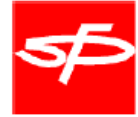

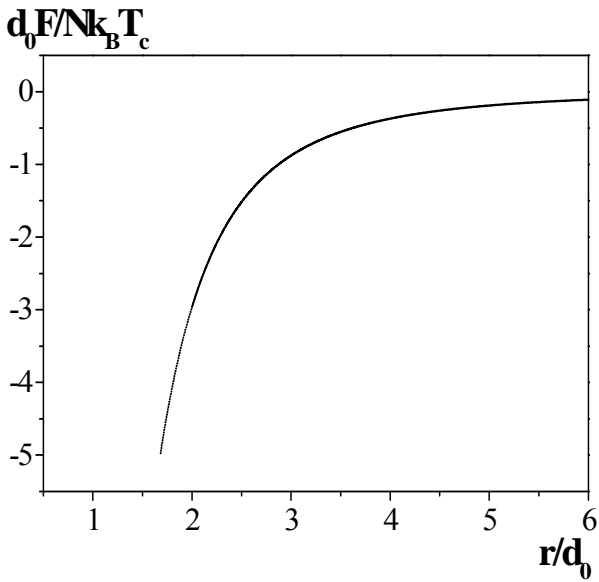

FIG 2 : Variation de la force induite en fonction de la distance inter-particule *r*

### 1.3 Force de Casimir entre colloïdes dans les mélanges ternaires

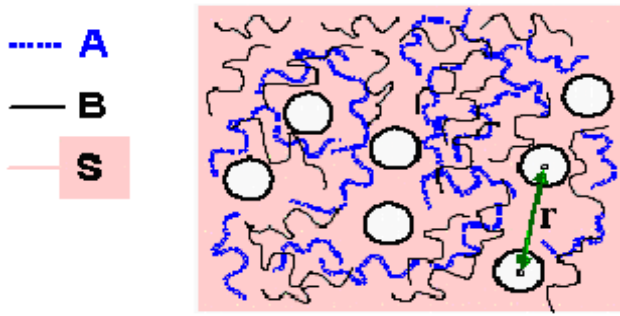

FIG. 3: Deux particules colloïdales sphériques en susspension dans un mélange terinaire de polymères

Le système étudié, dans ce cas, est un assemblage de n colloïdes immergés dans un mélange terinaire de polymères, formé de deux polymères A et B, en solution dans un bon solvant (voir FIG 3). La force induite entre les colloïdes qui est réspnsable de l'agrégation des coloïdes, a été calculée selon deux régimes différents :

**i-** Régime à faible distance (D<<R), la force induite suit la loi :

$$\frac{RF}{k_B T_K} = -\pi \Delta \left(\frac{R}{D}\right)^2 \quad (2)$$

**ii-** Régime à forte distance (r>>$d_0$), la force iuduite suit la loi :

$$\frac{RF}{K_B T_K} = -B \left(\frac{R}{r}\right)^{2.04} \quad (3)$$

Avec :

$$\Delta = 0.326 \qquad B = 8.04$$

### 1.4 Conclusion

Le solvant induit un changement de l'expression de la force. Le gonflement des chaînes modifie la dépendance de la force avec la distance, à travers l'apparition d'un terme en $r^{-2}$. En effet qu'en présence d'un bon solvant, les fluctuations de composition, près du point critique, sont assez fortes.

## 2 Etude cinétique de l'agrégation des colloïdes dans un mélange binaire de polymères

### 2.1 Introduction

Dans cette partie, nous nous proposons d'étudier le problème de l'agrégation des colloïdes dans un mélange de deux polymères, d'un point de vue cinétique. Nous nous intéressons à la relaxation du système colloïdale au cours du temps en partant d'un état d'équilibre à une température initial $T_i$ vers une température finale $T_f$ proche de la température critique $T_c$. Dans l'intervalle $T_i \leq T \leq T_f \approx T_c$. Le paramètre d'ordre varie en fonction du temps d'une valeur d'équilibre initiale à la valeur finale. Le but est de voir la manière dont le paramètre d'ordre relaxe.

### 2.2 Résultats

L'écart du paramètre d'ordre à sa avleur d'équilibre décroît avec le temps selon la loi :

$$\delta \varphi = E \, e^{-t/\tau} \quad (4)$$

$$\tau \propto \left| t - t^* \right|^{-1} \quad (5)$$

τ est le temps de relaxation.

### 2.3 Conclusion

l'écart du paramètre d'ordre à sa valeur d'équilibre décroît exponentiellement avec le temps. Le temps de relaxation τ diverge à la température critique t*.

## Références

[1] H. Ridouane, E.K. Hachem, M. Benhamou, J. Chem. Phys. 118, (2003) 10780

[2] H.Ridouane, E.K.Hachem, M.Benhamou, J.Cond.Matter Phys, 7 (2004) 63

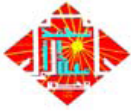
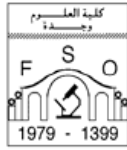
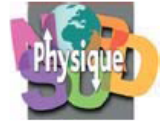
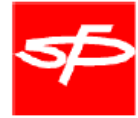

[3] T.W. Burkhardt, E. Eisenriegler, Phys. Rev. Lett. 74, 3189 (1995).

[4] E. Eisenriegler, U. Ritschel, Phys. Rev. B 51, 13717 (1995)

[5] F. Schlesener, A. Hanke, and S. Dietrich, J. Stat. Phys., 110 (2003) 981.

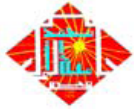 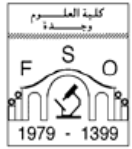 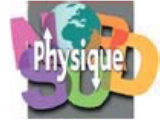 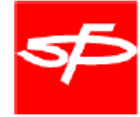

# Les plasmas froids hors équilibres

**Hassan CHATEI**, Professeur

Laboratoire de Physique Théorique et des Particules et Modélisation
Faculté des Sciences, Université Mohamed I, Oujda- Maroc

**Résumé**

Les plasmas sont présents dans de nombreuses installations industrielles et se sont placés en première ligne de la recherche ces dernières années. Dans plusieurs secteurs tels que la microélectronique ou le traitement de surface, leurs applications sont d'ores et déjà implantées. Des technologies en émergence, comme les nanotechnologies, en seront largement utilisatrices. Nous présentons ici brièvement le plasma froid et par un exemple nous montrons la possibilité d'utilisation du plasma froid pour élaborer des nano-structures de carbone.

# 1 Introduction

Le plasma, quelquefois appelé quatrième état de la matière, est un milieu partiellement ou complètement ionisé. Il est constitué d'un nombre très voisin d'électrons, d'ions coexistant éventuellement avec des atomes et des molécules neutres ou radicaux le plus souvent dans un état excité. Historiquement le terme « plasma » a été utilisé en physique pour la première fois par le physicien américain Irving Langmuir en 1928 lorsqu'il observait le comportement du gaz ionisé dans des tubes à décharge. Ce quatrième état de la matière, que l'on retrouve dans les étoiles et le milieu interstellaire, constitue 99 % de notre univers. Sur Terre, on ne le rencontre pas à l'état naturel, mais on le produit artificiellement par ionisation d'un gaz. Il existe deux types de plasmas. Le plasmas chaud où la pression du gaz est relativement élevée et la température peut atteindre plusieurs milliers de degré, et le plasma froid qui apparaît sous des pressions réduites et dont la température ne dépasse pas 400°.

# 2 Qu'est ce qu'un plasma froid?

## 2.1 Un état hors équilibre thermodynamique

Un plasma froid est un gaz ionisé, en état de non équilibre thermodynamique, dont seuls les électrons sont portés à haute température ($10^4$-$10^5$K), les autres particules lourdes restant à température ambiante. La violation de l'équilibre thermodynamique est une particularité très importante de l'état plasma qui le distingue des autres états de la matière (gaz ou liquide) et qui permet d'expliquer sa réactivité et donc ses différentes applications dans l'industrie.

## 2.2 Comment générer un plasma froid ?

Dans une enceinte confinée, en général sous vide partiel, dans laquelle on injecte un gaz plasmagène, on peut créer un plasma en fournissant de l'énergie à ce gaz par l'action d'une décharge électrique. La décharge peut être obtenue de diverses manières: Soit par un système avec électrodes (décharge sous champ électrique), soit par un système sans électrodes (décharge sous champ électromagnétique variable de hautes fréquences ou micro-ondes).

# 3 Utilisation du plasma froid

L'application la plus vulgarisée de l'utilisation des plasmas froids est notre éclairage au néon. Au niveau industriel, les applications liées au plasma froids sont très diverses. Si la microélectronique a été la première, les secteurs utilisant des procédés plasmas froids se diversifient de plus en plus. L'agroalimentaire, le monde médical, le textile, l'environnement, etc…s'ouvrent progressivement à ces procédés et bientôt le monde des nanotechnologie. La figure1 montre un exemple de nanostructure ( les nano-pointes de carbone) dont les domaines d'application sont envisageable en nanoélectronique. Ces nano-pointes de carbone sont élaborées à partir d'un plasma froid micro-onde dans un mélange de gaz $CH_4$-$CO_2$, en utilisant des particules catalytiques de fer.

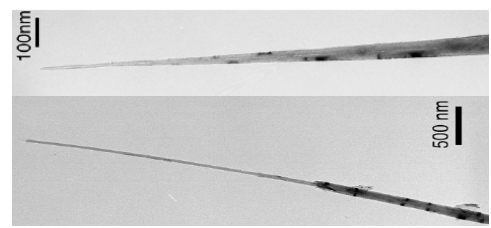

FIG.1: Dépôt de nano-pointes de carbone par plasma [1]

# 4 Conclusion

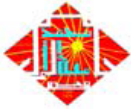 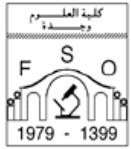 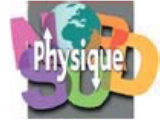 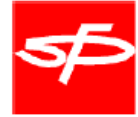

L'intérêt grandissant pour la méthode plasma fait de cette technique l'avenir en matière d'élaboration de nanomatériaux pour les applications en nanotechnologie.

## Références

[1] L. Le Brizoual, M. Belmahi, H. Chatei, M.B. Assouar et J. Bougdira, Diamond & Related Materials 16 (2007) 1244–1249.

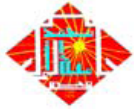 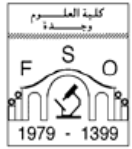 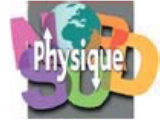 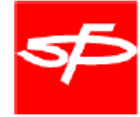

# Le Bouchon du calorimètre électromagnétique d'ATLAS et sa réponse aux muons

**B. Dekhissi**, **H. Dekhissi**, **J. Derkaoui**, **A. Elkharrim**, **F. Maaroufi**

Laboratoire de Physique théorique, Physique des Particules et Modélisation, Département de Physique, Faculté des Sciences, Université Mohamed Premier, 60000 Oujda, Maroc

**Résumé**

Ce document consiste à étudier quelques caractéristiques du bouchon du calorimètre électromagnétique. Dans la première partie, on décrira le comportement des muons dans le calorimètre. Dans la seconde partie, on présentera l'analyse des données de tests en faisceau des muons du module ECC0.

## 1 Introduction

L'expérience ATLAS est l'une des principales expériences auprès du futur LHC. ATLAS s'intéressera à l'étude de nombreux canaux en physique de particules, et notamment la recherche du boson de Higgs [1,2] (par exemple le canal H→γγ). Pour réaliser cet objectif, un calorimètre électromagnétique performant a été construit. Ce calorimètre est composé d'un tonneau et de deux bouchons. Chaque bouchon est constitué de huit modules. Des modules de série du bouchon du calorimètre électromagnétique (ECC0, ECC1 et ECC5) ont été testés au CERN sous faisceaux d'électrons, de positrons et de muons afin d'étudier la réponse de ce bouchon. Le présent travail vise à analyser les données du test en faisceau des muons.

## 2 Le comportement des muons Dans le calorimètre

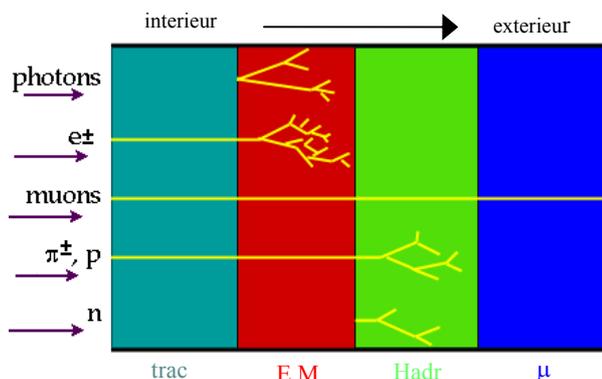

FIG. 1: Représentation schématique du comportement des particules dans le détecteur ATLAS.

Les muons agissent très différemment aux électrons. Ils ne produisent pas de gerbes de particules dans le calorimètre (Fig 1). On peut dire que ces particules se comportent comme des MIP (Minimum Ionising Particles), ce qui est très intéressant car les muons touchent peu de cellules du calorimètre. On obtient anisi, des informations plus précises, cela n'est pas réalisé dans le cas des électrons car ces derniers se répartissent dans plusieurs cellules (gerbes électromagnétique).

## 3 Analyse des données des muons de test en faisceau

Pour étudier certaines caractéristiques du module ECC0, un balayage en η allant de 1.7 jusqu'à 2.3 ($\eta_{cell}$ de 11 à 36) a été effectué avec des muons de 120 GeV. Dix milles événemets ont été enregistrés pour chaque cellule, sauf cinq points $\eta_{cell}$ = 12, 17, 27 et 32 qui ne comptent que 40 000 événements.

### 3.1 Energie dans les cellules du module ECC0

Toute l'étude est traitée dans le deuxième compartiment du module de bouchon du calorimètre électromagnétique qui collecte la majeure partie de l'énergie des muons. Dans ce compartiment (S2) deux cellules en φ et une cellule en η sont presque toujours touchées (Fig 2). Dans cette étude on a utilisé certaines corrections qui ont été appliquées dans le tonneau du calorimètre électromagnétique [3].

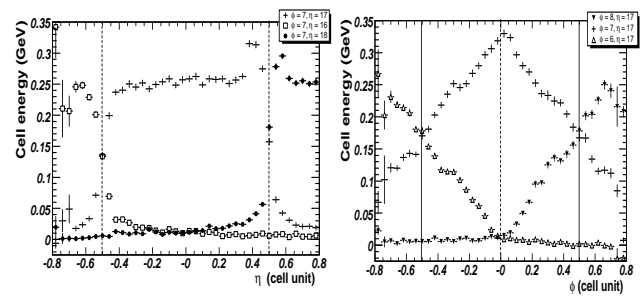

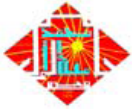 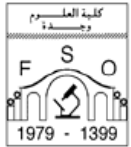 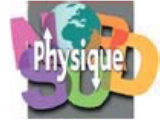 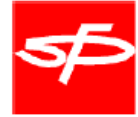

FIG.2 : Energie déposée par les muons dans des cellules centrées sur (η,φ)=(17,7) en fonction de η (à gauche) et de φ (à droite).

La figure 2 présente l'évolution de l'énergie déposée par les muons dans les cellules de S2 selon η et φ.

### 3.2 Modulation en η

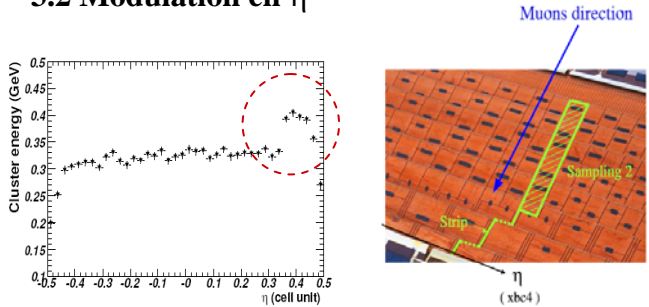

FIG.3 : Modulation en η (à gauche) et représentation shématique de la bande sur l'éléctrode (à droite).

La figure 3 montre la réponse du calorimètre des muons de 120 GeV en fonction de η pour un amas centré sur la cellule (17,7)

La bande (Strips) est due au fait que le compartiment S2 se prolonge par une bande empiétant sur le compartiment S3 (Fig.3 à droite)

### 3.3 Modulation en φ

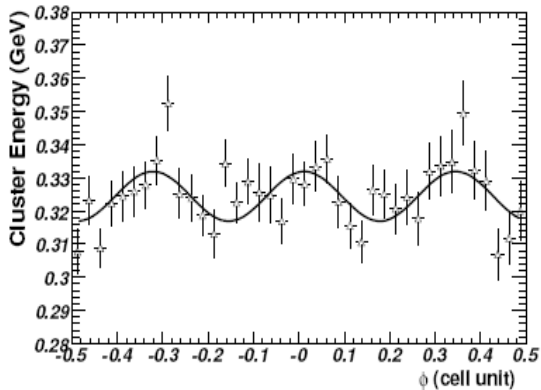

FIG.4 : Modulation en phi

Une réponse périodique selon φ des muons des données du test en faisceau de 120 GeV dans la cellule (17,7) (Fig4).

### 3.4 Le signal de muons et le bruit

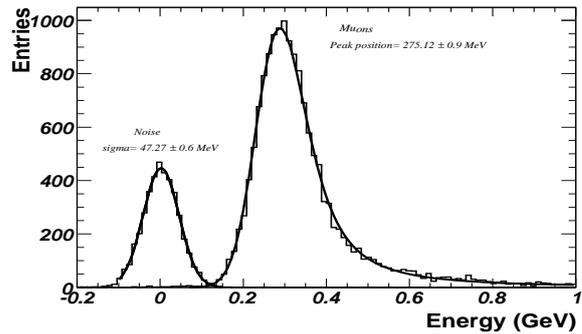

Fig5 : Le spectre des muons et du bruit dans une cellule du ECC0.

La figure 5 montre la distribution de l'énergie déposée dans une cellule du ECC0 par des muons de 120 GeV. Cette distribution est ajustée par une fonction de landau convoluée avec gaussienne. On constate que les muons déposent peu d'énergie dans le calorimètre.

### 3.3 La rèponse du module ECC0 en eta

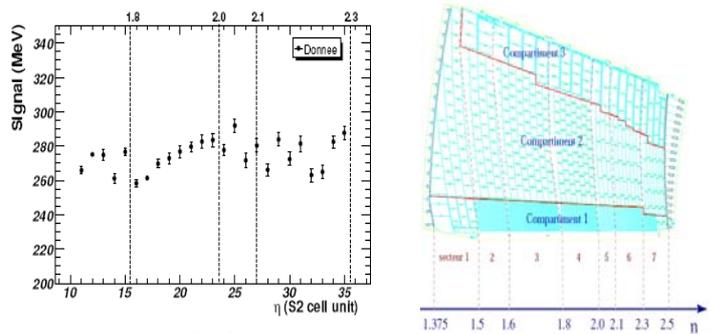

FIG.6 : L'énergie déposée par les muons en fonction de η (à gauche) (Les lignes verticalesen pontillé séparent les secteurs haute tension), représentation shématique qui montre la variation des pistes sur l'électrode (à droite)

La figure 6 montre que le signal augmente linéairement avec η dans chaque secteur de haute tension. Mais dans les deux secteurs [2.0,2.1] et [2.1,2.3], le signal augmente linéairement pour chacune des 2 positions successives. Ceci peut s'expliquer par la variation de longueur de piste entre une cellule à l'autre (Fig 6 à droite). Plus la piste est long, plus l'énergie qui y récupérée est grande.

## 4  Conclusion

L'analyse des muons issue des donnée prises du test en faisceau a montré:

- Le comportement des muons est différent de celui des électrons (Les muons se comportent comme des MIP).
- L'énergie déposée par les muons est faible devant leur énergie incidente.

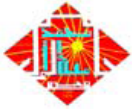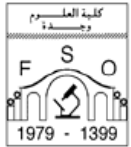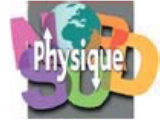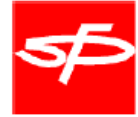

- Deux cellules en φ et une cellule en η sont suffisantes pour reconstruire l'énergie des muons (pour les électrons 3X3 où 5X5). Ceci nous permet d'isoler la réponse de chaque cellule, de distinguer les signaux et de fournir des informations plus précises par rapport aux électrons.

## Références

[1] ATLAS Collaboration, ATLAS Calorimeter performance, CERN/LHCC/96-40.
[2] ATLAS Collaboration, ATLAS Liquid Argon Calorimeter Technical Design Report, CERN/LHCC/96-41.
[3] A. Camard, F. Hubaut, Stady of the Em Barrel Module 0 with muons, ATL-LARG-2001-017.

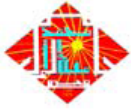 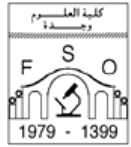 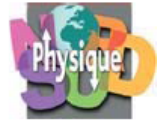 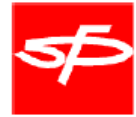

# Modélisation d'une décharge plasma microonde pulsée en vue de la synthèse de films de diamant

M. El Bojaddaini[1], H. Chatei[1], M. El Hammouti[2], H. Robert[3], J. Bougdira[3]

[1]Laboratoire de Physique Théorique et des Particules, Faculté des Sciences, Université Mohamed I, Oujda, Maroc
[2]Faculté pluridisciplinaire de Nador, Université Mohamed I, Nador, Maroc
[3]Laboratoire de Physique des Milieux Ionisés et Applications, Université H. Poincaré, Nancy, France.

**Résumé**

Les procédés assistés par les plasmas froids sont devenus indispensables dans les processus industriels et présentent des intérêts importants dans des domaines innovants et de hautes technologies telle que la microélectronique.

# 1 Introduction

La modélisation d'un plasma de décharge est relativement complexe à cause des nombreux phénomènes mis en jeu et de leur fort couplage.
Dans ce travail on met l'accent sur les différents approches existantes pour modéliser un plasma, puis on cite quelques difficultés liées à cette modélisation, et finalement, on présente quelques résultats concernant la densité électronique, obtenus par le modèle fluide dans un réacteur dédié aux dépôts de diamant par plasma micro-onde fonctionnant en régime pulsé.

# 2 Pourquoi modéliser les plasmas froids?

## 2.1 Intérêts scientifiques

Les plasmas froids utilisés pour les procédés industriels sont très différents (sources d'excitation et géométries très diverses) et comportent plusieurs phénomènes physiques et chimiques fondamentaux mal connus.

## 2.2 Intérêts commerciaux

Les plasmas froids sont à la base de nombreuses technologies fortement innovantes et prometteuses dans plusieurs secteurs stratégiques (traitement de surfaces, production de nanomatériaux…). Ils jouent un rôle majeur en micro-électronique, ils sont responsables des processus de gravure et de dépôt.

# 3 Différentes approches existantes

## 3.1 Approche microscopique
Ce modèle utilise des méthodes particulaires et se base sur la résolution du système équation de Boltzmann-équation de Poisson dans l'espace des phases. Le problème de cette approche est qu'elle nécessite un temps de calcul très important.

## 3.2 Approche fluide

Le plasma est décrit à l'aide des grandeurs macroscopiques; l'équation de Boltzmann est remplacée par les moments. Le problème de cette approche est dû notamment au terme source de l'équation de continuité.

## 3.3 Approche hybride fluide-particulaire

Les électrons froids sont décrits d'une manière fluide et ceux rapides de manière particulaire. Cette approche a l'avantage de donner un terme source réaliste et un coût en temps de calcul bien moins élevé que pour un modèle entièrement particulaire.

# 4 Résultats du modèle fluide

Dans un réacteur micro-onde destiné au dépôt de diamant [1], nous avons trouvé que la distribution de la densité électronique est symétrique par rapport à l'axe du tube où l'intensité du champ est maximale. La densité électronique diminue en fonction de la pression totale pour déférents instants de la décharge et pour une puissance micro-onde fixe et croit lorsqu'on augmente la puissance micro-onde.

# 5 Conclusion
La conception de nouveaux réacteurs et procédés nécessite une bonne compréhension du plasma à travers le choix d'un modèle adéquat permettant de décrire le plasma, et par la construction de codes numériques du fait de la complexité des phénomènes mis en jeu.

# Références

[1] H. Chatei et al, Diamond relat. Mater., 6 (1997) 107.

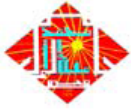 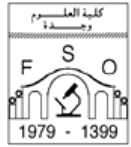 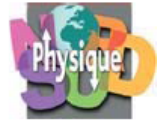 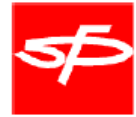

# SIMULATION D'UNE MACHINE ASYNCHRONE DANS LE LOGICILE ORCAD/PSPICE

**M.L. Elhafyani** [(1)], S. Zouggar [(1)], M. Benkaddour [(2)]

[(1)] Université Mohammed 1er, Ecole Supérieure de Technologie LEEP BP. 473. 60000 Oujda Maroc
[(2)] Université Mohammed 1er, Faculté des Sciences, Département de physique Oujda Maroc.

**Résumé**

Dans cette communication, nous montrons comment créer dans l'environnement Orcad-Pspice une librairie de modèle d'une machine asynchrone. A partir de son schéma équivalent et ces équations différentielles, nous établissons et implantons dans Orcad-Pspice le schéma électrique de la machine asynchrone auto-excitée en régime transitoire. Ceci permet de modéliser le fonctionnement de la machine asynchrone. Aussi, nous étudions par simulation l'effet de la variation de la vitesse et de la capacité sur la tension de sortie dans les deux cas de fonctionnement à vide et en charge.

## 1 Introduction

La modélisation et la simulation d'une génératrice asynchrone auto-excitée en régime dynamique dans le référentiel αβγ (Toutes les pertes magnétiques et mécaniques sont supposées négligeables, La force magnétomotrice est sinusoïdale le long du stator) nécessitent l'utilisation des simulateurs de circuits performants [1,4]. Nous avons choisi le simulateur Orcad-Pspice, parce qu'il est performant pour implanter et simuler les circuits électriques les plus complexes et parce que sa bibliothèque est riche en composants électriques et électroniques et il permet aussi de mélanger, sans aucun problème, des composants numériques et analogiques.

Mais ce simulateur ne permet pas de simuler directement les grandeurs mécaniques telles que les machines tournantes. Pour surmonter ce problème, nous avons appliquée une analogie entre les grandeurs mécaniques et électriques. Généralement, pour simuler le comportement d'une machine asynchrone en régime transitoire, on doit tout d'abord, implanter le schema equivalent de la machine asynchrone (Fig1) afin de la appeler directement dans l'environnement Orcad-Pspice.

Dans ce résumé, nous présentons le symbole de la machine asynchrone dans l'environnement Orcad-Pspice dans le pargraphe.2. En fin les différents résultats de symbolisation et de simulations seront présentés.

## 2 Symbolisation De La Machine Asynchrone

### 2.1 Schema Equivalent

A partir des équations caractéristique de la machine asynchrone son modèle peux se traduire sous forme du schéma équivalent suivent :

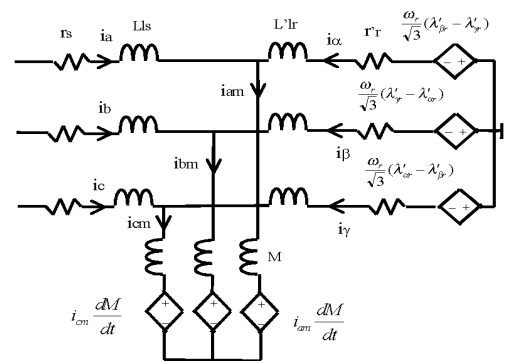

FIG. 1: SCHEMA EQUIVALENT DE LA MACHINE DANS LE MODELE αβγ [3].

$r_s$ et $r_r$ : résistances statorique et rotorique.

$L_{ls}$ et $L'_{lr}$ inductances statorique et rotorique.

### 2.2 Symbole De La Machine Asynchrone Sur Orcad/Pspice

Il n'existe pas en électrotechnique de logiciel qui permet de simuler directement les grandeurs mécaniques. La simulation, nous l'avons fait sur le logiciel Orcad-Pspice en utilisant les analogies suivantes :

| Grandeur mécanique | Grandeur électrique |
|---|---|
| Vitesse | Tension |
| Couple | Courant |
| Moment d'inertie | Capacité d'un condensateur |
| Frottement visqueux | Inverse d'une résistance |

Nous avons implanté le modèle de la génératrice asynchrone (Fig 2) dans l'environnement Orcad-Pspice. Ce schéma peut être symbolisé sous forme de bloc donné par la figure 3.

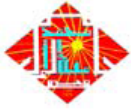 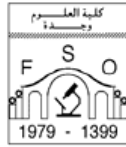 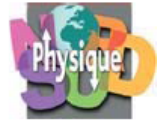 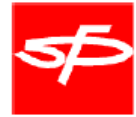

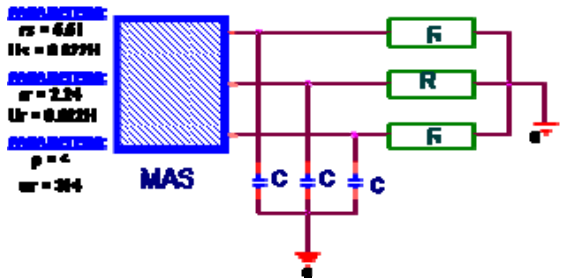

FIG 2 : SYMBOLE D'UNE MACHINE ASYNCHRONE DANS L'ENVIRONNEMENT ORCAD

## 3 Resultats De Simulation

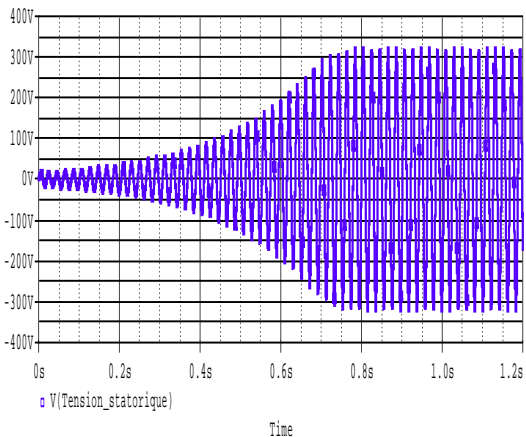

FIG 3 : FORME D'ONDE DE TENSION DE SORTIES Vs POUR $\omega_R$= 314RAD/S ET C= 68µF.

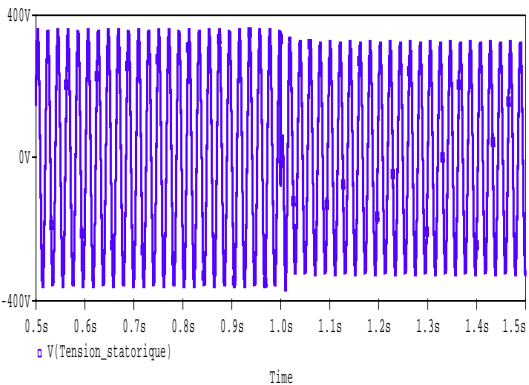

FIG 4 : EVOLUTION DE LA TENSION Vs LORSQUE LA CAPACITE VARIE DE −14%.

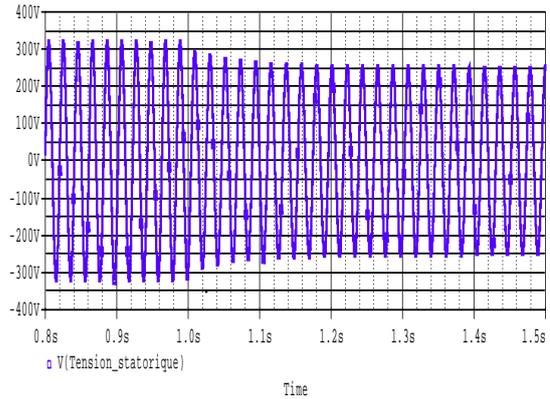

FIG 5: TENSION STATORIQUE LORSQUE LA VITESSE DIMINUE

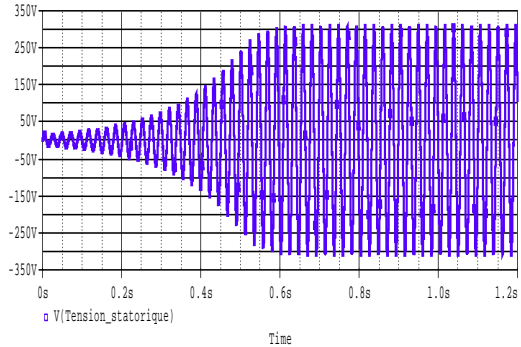

FIG6: EVOLUTION DE LA TENSION STATORIQUE EN FONCTION DU TEMPS POUR $\omega_R$= 314RAD/S, C= 107µF, RL= 115Ω.

## 4 Conclusion

Dans ce travail nous avons réalisé dans l'environnement Orcad-Pspice un modèle d'une génératrice asynchrone auto-excitée. A partir de ce modèle nous avons analysé le comportement de la machine en régime transitoire dans les deux cas de fonctionnement à vide et en charge et pour différentes valeurs des paramètres de la génératrice. En perspective et grâce à la technique de symbolisation de la génératrice asynchrone dans les librairies du logiciel Orcad/Pspice, nous poursuivons l'étude de la réalisation des convertisseurs d'énergie (redresseur, onduleur, hacheur) afin de réguler la tension de sortie de la génératrice asynchrone.

## Références

[1] R. Szczesny et al, "A New equivalentcircuit approach to simulation of converter induction machine associations," Proceedings of the European Conference on Power Electronics and Applications 4/356-4/361, Italy, 1991.

[2] E. G. Marra et al, « Self-excited induction generator controlled by a VS-PWM bidirectional converter for

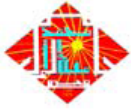 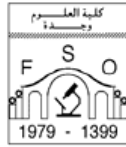 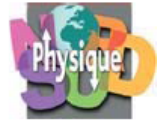 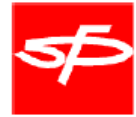

rural applications » IEEE Trans on IA, vol. 35, 877-883, 1999.

[3] E. G. Marra et al, "Spice-Assisted Simulation of the αβγ Model of Cage-RotorInduction Machines including Saturation" IND2000_FDI05, 497-502

[4] A.M. Stankovic et al, "Dynamic Phasors in Modeling and Analysis ofUnbalanced Polyphase AC Machines" IEEE TRANS, ON ENERGY CONVERSION, VOL. 17, NO. 1, MARCH 2002.

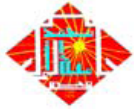 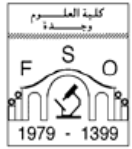 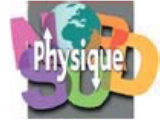 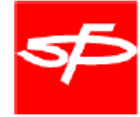

# Le DCS de la haute tension du détecteur TilCal de l'expérience ATLAS

A. Elkharrim*, B. Dekhissi, H.Dekhissi, J.Derkaoui, F. Maaroufi, Y. Tayalati

Laboratoire de Physique Théoriques, Physique des Particules et Modélisation,
Université Mohamed 1$^{er}$, Oujda, Maroc

**Résumé**

ATLAS est une des quatre expériences proposées auprès du CERN qui sera opérationnelle en 2008 révélant ainsi aux physiciens, d'autres dimensions inaccessibles qu'à l'échelle énergétique des TeV. Dans le présent papier, on décrira brièvement le travail réalisé au CERN en collaboration avec les responsables du DCS (Detector Control System) destiné à contrôler le système de la haute tension alimentant les PMTs du TileCal, le sous détecteur hadronique d'ATLAS. Ce travail est de nature informatique, il consiste à un développement d'une librairie en langage C++.

## 1 Introduction

### 1.1 DCS, SCADA, PVSS, ETM

Le rôle du DCS est d'assurer un fonctionnement cohérent et sain du détecteur ATLAS. Le DCS d'ATLAS est un système distribué, constitué d'un SCADA (Supervisory Control And Data Acquisition) et d'une interface. Le SCADA permet:
- L'acquisition des données de l'équipement électronique
- Le traitement, la présentation et l'archivage des données
- Prise en charge des commandes, des messages et des alarmes.

En 1999, le CERN a adopté le SCADA commerciale, PVSS, pour bâtir le DCS de ses quatre grandes expériences auprès du LHC. PVSS est un software développé par la société autrichienne ETM.

### 1.2 TileCal

Le TileCal est un des sous détecteur d'ATLAS destiné aux hadrons. Il est fait de 4 roues (Fig.1), dont chacune est faite de 64 modules contenant chacun un tiroir. Un tiroir (Fig.2) héberge 48 Photomultiplicateurs (PMT). Chaque PMT est alimenté par une haute tension qui doit être maintenue, contrôlée et surveillée. D'où la nécessité d'un DCS. La haute tension est distribuée dans le tiroir par le biais d'une carte *HVOPTO*, contrôlée par une carte *HVMICRO*. Cette dernière possède un Microprocesseur capable de procéder les commandes qui lui sont envoyées via CANBUS, par le system de supervision.

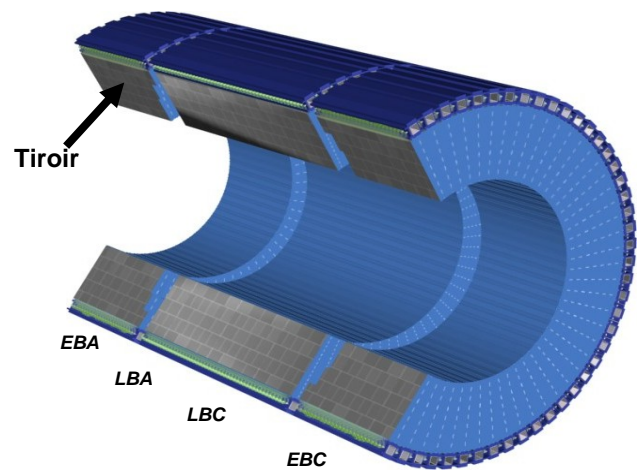

Fig.1 : Vue du TileCal d'ATLAS

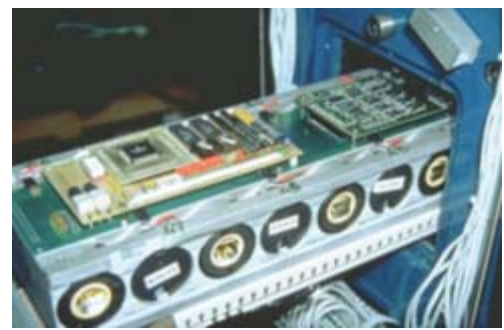

Fig.2 : Une partie du tiroir à l'entrée d'un module du TileCal. On peut voir les cartes HV_MICRO et HV_OPTO et 7 PMTs.

## 2 Le DCS de la haute tension

HvAc*t* est un programme en C++ (réalisé par le groupe ATLAS à Clermont Ferrand) qui permet de lire et d'envoyer des commandes aux tiroirs (HVMICRO) via des CANBUS. HvActManager et HvActCommands sont des interfaces pour PVSS (API), i.e. des managers de PVSS. Ils permettent au PVSS de communiquer avec HvAct via deux mémoires partagées de windows. La figure 3 montre la structure de cette communication entre ces différents programmes. On a remarqué que l'usage de ces mémoires, efficaces soient-elles, provoque des lenteurs indésirables. En plus, ces mémoires rendent ces processus

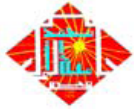 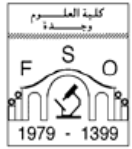 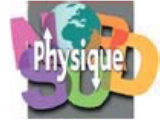 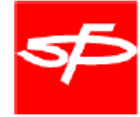

dépendants, ceci a l'effet négatif suivant: si HvAct crache, les managers crachent.

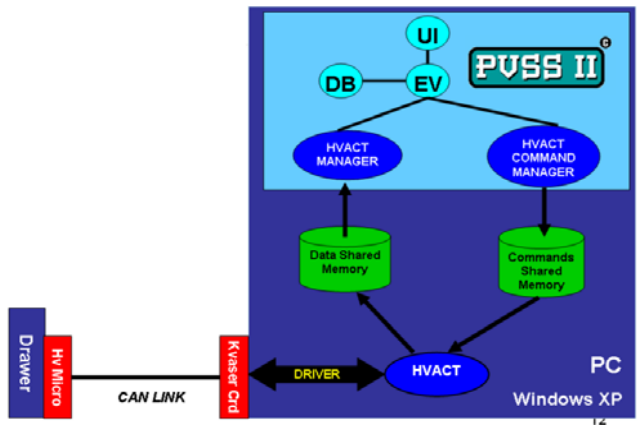

Fig.3: Communication des différentes Composantes de système de haute tension

## 3 DIM (Distributed Information Managment)

DIM est un système de communication déstiné aux environnements distribués ou mixtes. Il fourni une base pour une communication transparente entre les processus. Il est basé sur le paradigme *client/serveur*. Le concept principal de cette approche est celui du *service*. Le *serveur* publie le *service* en l'enregistrant dans le *Domaine Name Server*. Le *client* s'enregistre ensuite au service en demandant au DNS quel est le serveur qui le publie pour le contacter ensuite directement. La figure 4, montre la structure de cette communication entre processus au sein de la structure DIM.

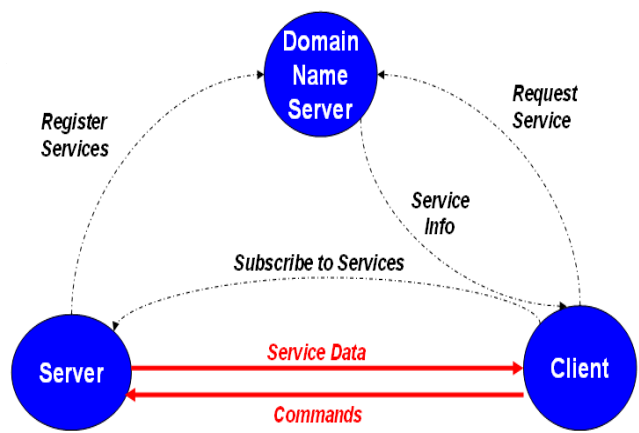

Fig.4 : Les différentes composantes de DIM

## 4 Le DCS de la haute tension et DIM

On était chargé d'écrire un serveur DIM pour le system de la haute tension, puis de configurer un client *PVSS-DIM* (interface entre DIM et PVSS) pour accéder à l'information publiée par ce serveur. DIM va assurer une conversation plus transparente et plus indépendante entre les processus en question. Ainsi, ils peuvent dès lors être distribués, i.e. être dans des stations différentes (dotées même de système différents) connectées par un réseaux Ethernet et indépendants (le crache d'un processus n'affecte pas l'autre grâce au DNS qui figure dans l'architecture de DIM).

A cet égard, nous avons écrit une librairie en langage C++, qui prend en charge les commandes du client (PVSS-DIM) et publie les services (les informations venant des tiroirs) dans le DNS.

La figure 5, montre la nouvelle structure DIM appliquée au DCS de la haute tension du TileCal d'ATLAS.

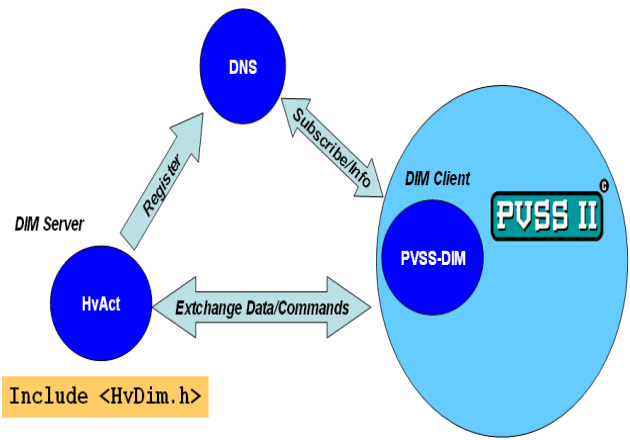

Fig.5 : Les différentes composantes de DIM

## 5 Conclusion

*HvAct* est maintenant un *serveur* DIM qui peut lire les tiroirs et *publier* l'information lue comme étant des *services* dans le *DNS* du DIM. Il peut aussi procéder les *commandes* envoyées par le *client*, PVSS-DIM. On a testé avec succès (lecture/écriture) certains services. Il nous reste de couvrir tous les autres services et finaliser l'ensemble des composantes pour assurer les bonnes performances exigées par les responsables du DCS.

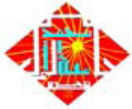 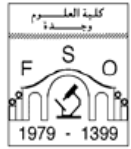 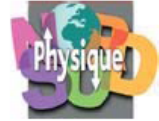 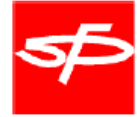

# L'univers en expansion accéléré en 3-brane Brans-Dicke théorie

**Ahmed Errahmani**, **Taoufik Ouali**

Laboratoire de Physique Théoriques, Physique des Particules et Modélisation, Faculté des sciences, Oujda

**Résumé**

Dans le cadre de la théorie de Brans-Dicke nous étudions la cosmologie à 5 dimensions en négligeant le terme mixte et le terme quadratique dans l'équation d'Einstein modifiée.
Les équations des champs ainsi obtenues sont mises sous forme d'un système dynamique, les résultats obtenus sont comparés avec les données observationnelles.

## 1 Introduction

Dans ce papier, l'univers est un fluide cosmologique homogène et isotrope, La gravitation gouverne la cosmologie et la description du contenu énergétique se fait par des quantités moyennes. La gravitation est décrite par la relativité générale, On postule que les lois physiques sont partout les mêmes en tout point de l'Univers.

La Métrique de Friedman-Lemaitre-RobertsonWalker est donnée par:

$$ds^2 = dt^2 - a^2(t)\left(\frac{dr^2}{1-kr^2} + r^2(d\theta^2 + \sin^2\theta d\phi^2)\right)$$

Le fluide cosmologique est barotropique, son équation d'état est: $p=(\gamma-1)\rho$.
$\gamma=1$ domination de la matière $p=0$
$\gamma=0$ domination de l'énergie du vide $p=-\rho$
$\gamma=4/3$ domination de la radiation $p=\rho/3$
Le tenseur energie-impulsion associé est:

$$T^\mu_\nu = diag(-\rho(t), p(t), p(t), p(t))$$

La conservation de l'energie $(T^{\mu\nu}_{;\nu}=0)$ donne:

$$\dot\rho + 3\frac{\dot a}{a}(p+\rho) = 0$$

## 2 Paramètres cosmologiques

Pour une notation harmonieuse, on défini la densité critique $\rho_{critique}$, qui sera utilisée comme unité cosmologique de densité d'énergie par :

$$\rho_0 = \frac{3(H_0)^2}{8\pi G} = \rho_{critique}$$

Pour toute forme d'énergie r, on posera $\Omega = \rho/\rho_{critique}$.
Par exemple, $\Omega_{matière} = \rho_{matière}/\rho_{critique}$
On discute deux cas :
● $\rho^2 \ll \rho$ les équations du champ deviennent celles de la théorie BD à 4-dim.
● $\rho^2 \gg \rho$ les équations du champ sont celles de la théorie BD à 4-dim à condition de remblacer le $\gamma$ par $\gamma/2$.

Donc il suffit d'étudié le premier cas avec le changement de variable:

$$H = \dot a/a, \quad F = \dot\phi/\phi, \quad H' = dH/da \text{ et } F' = d\phi/da$$

On trouve le système dynamique suivant :

$$\begin{cases} aH(2\omega+3)H' = -3(\gamma\omega+2)H^2 - \omega\left(\frac{\omega(2-\gamma)+1}{2}\right)F^2 - \omega(3\gamma-4)HF \\ \qquad -\frac{k}{a^2}(\omega(3\gamma-2)+3) + \Lambda_4(\gamma\omega+1) \\ aH(2\omega+3)F' = -3(3\gamma-4)H^2 - \left(4\omega+3-\frac{3\omega\gamma}{2}\right)F^2 - 3(3\gamma-1+2\omega)HF \\ \qquad \frac{3k}{a^2}(3\gamma-4) + \Lambda_4(3\gamma-2) \end{cases}$$

### 2.1) Solution d'équilibre

Les observations ont montrer que l'univers semble plat $k \approx 0$ de plus $(k/a^2)$ décroît dramatiquement avec l'expansion. Donc on prend l'équilibre du système pour $(k/a^2 \approx 0)$, les points d'équilibre du système sont :

$$(H_\infty, F_\infty) = \sqrt{\frac{2\Lambda_4}{(2\omega+3)(3\omega+4)}}(\omega+1, 1)$$

### 2.2) Linéarisation du système dynamique

Soient $H=H_\infty+h(a)$ et $F=F_\infty+f(a)$ avec $h(a)$ et $f(a)$ des fonctions de perturbation qu'on va les déterminé ultérieurement. On injecte ces expressions dans le système et on élimine les termes en $h^2, f^2$ et $hf$ on obtient

$$\begin{pmatrix} \dot h \\ \dot f \end{pmatrix} = -\frac{H_\infty}{\omega+1}\begin{pmatrix} 3\gamma\omega+4 & (\gamma-1)\omega \\ 9(\gamma-1) & 3\gamma+3\omega+1 \end{pmatrix}\begin{pmatrix} h \\ f \end{pmatrix} - \frac{k}{a^2}\begin{pmatrix} \frac{(3\gamma\omega-2\omega+3)}{(2\omega+3)} \\ \frac{3(3\gamma-4)}{(2\omega+3)} \end{pmatrix}$$

La solution de ce système est :

$$\begin{cases} H = H_\infty - \frac{k}{a_0^2 H_\infty}\frac{(\omega+1)(\omega+3)}{(\omega+2)(2\omega+3)}\left(\frac{a_0}{a}\right)^2 + H_0 C_1 \left(\frac{a_0}{a}\right)^{3+\frac{1}{\omega+1}} \\ \qquad + H_0 C_2 \left(\frac{a_0}{a}\right)^{3\gamma+\frac{1}{\omega+1}} \\ F = F_\infty + \frac{3k}{a_0^2 H_\infty}\frac{\omega+1}{(\omega+2)(2\omega+3)}\left(\frac{a_0}{a}\right)^2 + H_0 C'_1 \left(\frac{a_0}{a}\right)^{3+\frac{1}{\omega+1}} \\ \qquad + H_0 C'_2 \left(\frac{a_0}{a}\right)^{3\gamma+\frac{1}{\omega+1}} \end{cases}$$

### 2.3) Discutions

Dans la limite $\omega \gg 1$ on trouve:

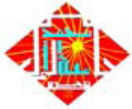 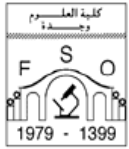 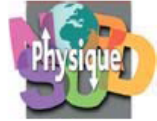 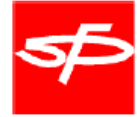

$$\frac{H}{H_0} = \frac{H_\infty}{H_0} - \frac{k}{2a_0^2 H_0 H_\infty}(\frac{a_0}{a})^2 + C_1(\frac{a_0}{a})^{3+\frac{1}{\omega}} + C_2(\frac{a_0}{a})^{3+\frac{1}{\omega}}$$

Avec

$$\frac{H_\infty}{H_0} = \sqrt{\frac{\Lambda_4}{3H_0^2}}$$

Pour chaque valeur de γ on fait $(H/H_0)^2$ et on compare le

$$\Omega_\Lambda = \frac{\Lambda}{3H_0^2} \quad \Omega_M = \frac{8\pi G \rho_M}{3H_0^2}$$

$$\Omega_k = \frac{-k}{a_0^2 H_0^2} \quad \Omega_R = \frac{8\pi G \rho_R}{3H_0^2}$$

Donc :

$$H^2(a) = H_0^2(\Omega_\Lambda + \Omega_k \frac{a_0^2}{a^2} + \Omega_M \frac{a_0^3}{a^3} + \Omega_R \frac{a_0^4}{a^4})$$

Λ est la canstante cosmologique, a est le facteur d'echelle
$H_0$ est le paramètre de Hubble

$$H_0 = (\frac{\dot{a}}{a})_0$$

Aujourd'hui,($a=a_0$) on néglige $\Omega_R$

$$\Omega_\Lambda + \Omega_k + \Omega_M = 1$$

Le paramètre de décélération est :

$$q_0 = -(\frac{a \cdot \ddot{a}}{\dot{a}})_0 = \Omega_k + \frac{1}{2}\Omega_M - \Omega_\Lambda$$

Des mesures récentes ont trouvé:
- $H_0$=65 km s$^{-1}$ Mpc$^{-1}$
- $q_0$=-0.58 l'univers est en expansion accélérée
- La densité critique $\rho_c$=10$^{-47}$Gev$^4$
- $\Omega_\Lambda$≈0.75, $\Omega_M$≈0.25, $\Omega_k$≈0

## 3 Brane Brans-Dicke théorie

La théorie BD représente une généralisation de la Relativité générale, elle est basée sur un champ scalaire Φ. L'action en BD est :

$$S_{BD} = \int dx^4 \sqrt{-g}\left(\phi[R - 2\Lambda] - \frac{\omega}{\phi}\partial_\mu \phi \partial^\mu \phi - 16\pi L_m\right)$$

**R**: Scalaire de Ricci
**L$_m$**: Densité de Lagrangien de tous les champs de matière
**Φ**: joue un rôle analogue à **G$^{-1}$**
**ω** :est une constante sans dimension

A 5 dimensions on généralise la théorie BD Dans un espace-temps de Friedman- Robertson-Walker homogène et isotrope on considérant la même action. Le pricipe variationnel δS=0 donne les équations du champ

$$-\frac{1}{a^3}\frac{d(\dot\phi a^3)}{dt} = \frac{8\pi}{3+2\omega}((3\gamma-4)\rho + \frac{k_5^4\phi}{48\pi}(3\gamma-2)\rho^2 - \frac{\Lambda_4}{4\pi}$$

$$(\frac{\dot a}{a} + \frac{1}{2}\frac{\dot\phi}{\phi})^2 + \frac{k}{a^2} = \frac{3+2\omega}{12}\left(\frac{\dot\phi}{\phi}\right)^2 + \frac{8\pi\rho}{3\phi} + \frac{k_5^4}{36}\rho^2 + \frac{\Lambda_4}{3}$$

$$\dot\rho + 3\frac{\dot a}{a}(p+\rho) = 0 \quad\quad p = (\gamma-1)\rho$$

résultat avec l'équation suivante :

$$(\frac{H}{H_0})^2 = \Omega_\Lambda + \Omega_R(\frac{a_0}{a})^4 + \Omega_M(\frac{a_0}{a})^3 + \Omega_k(\frac{a_0}{a})^2$$

On touve l'expression des canstantes d'integration comme dans le tableau

| γ | -1/3 | 0 | 1/3 | 1/2 | 2/3 | 1 | 4/3 | 2 |
|---|---|---|---|---|---|---|---|---|
| $C_1$ | $\frac{\Omega_M}{2\sqrt{\Omega_\Lambda}}$ | $\frac{\Omega_M}{2\sqrt{\Omega_\Lambda}}$ | $\frac{\Omega_M}{2\sqrt{\Omega_\Lambda}}$ | $\frac{\Omega_M - C_1^1}{2\sqrt{\Omega_\Lambda}}$ | $\frac{\Omega_M}{2\sqrt{\Omega_\Lambda}}$ | $\frac{\Omega_M}{2\sqrt{\Omega_\Lambda}} - C_1$ | $\frac{\Omega_M}{2\sqrt{\Omega_\Lambda}}$ | $\frac{\Omega_M}{2\sqrt{\Omega_\Lambda}}$ |
| $C_2$ | 0 | 0 | 0 | $\Omega_M - 2C_1\sqrt{\Omega_\Lambda}$ | 0 | $\frac{\Omega_M}{2\sqrt{\Omega_\Lambda}} - C_1$ | 0 | ∀ |

Le paramètre de décélération est

$$q_0 = -(\frac{a \cdot \ddot a}{\dot a})_0 = \Omega_R + \frac{1}{2}\Omega_M - \Omega_\Lambda$$

Dans la suite on néglige $\Omega_R$ et on caractérise l'univers par (matière/énergie).
- la ligne $\Omega_\Lambda$=1-$\Omega_M$ correspond à un univers en expansion uniforme ($q_0$=0)

$$C_1 = \frac{\Omega_M}{2\sqrt{\Omega_\Lambda}} = \sqrt{\Omega_\Lambda}$$

$C_1 < \sqrt{\Omega_\Lambda}$ ⟶ ($q_0$<0) accélération

$C_1 > \sqrt{\Omega_\Lambda}$ ⟶ ($q_0$>0) décélération

- la ligne $\Omega_\Lambda$=$\Omega_M$/2 correspond à un univers plat (k=0)

$C_1 < \frac{1-\Omega_\Lambda}{2\sqrt{\Omega_\Lambda}}$ ⟶ univers ouvert ($\Omega_k$<0)

$C_1 > \frac{1-\Omega_\Lambda}{2\sqrt{\Omega_\Lambda}}$ ⟶ univers fermé ($\Omega_k$>0)

## 6 Conclusion

Avec les résultats observationnels

$\Omega_\Lambda \simeq 0.75, \quad \Omega_M \simeq 0.25, \quad \Omega_K \simeq 0$

- $C_1 = \frac{\Omega_M}{2\sqrt{\Omega_\Lambda}} \simeq 0.15$ donc: $C_1 < \sqrt{\Omega_\Lambda} \simeq 0.86$

Ce qui est en accord avec un univers en expansion accéléré

- $C_1 = \frac{\Omega_M}{2\sqrt{\Omega_\Lambda}} \simeq \frac{1-\Omega_\Lambda}{2\sqrt{\Omega_\Lambda}}$ accord avec un univers plat

On conclut donc que La théorie est en bon accord avec les observations expérimentales

## Références

[1] C.Brans and R.H.Dicke,Phys.Rev.124(1961) 925.
[2] S. Weinberg, Gravitation and Cosmology (Wiley) 1972.
[3] S. J. Kolitch, Annal. Phys. 246 (1996) 121-132.
[4] T. Shiromizu et al, Phys. Rev. D62 024012(2000).
[5] Physics Letters B 641 (2006) 357–361.

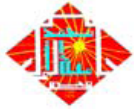 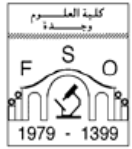 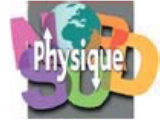 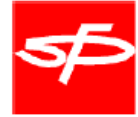

# Etude de la qualité de la qualité de l'eau par la technique LIBS (Laser Induced Breakdown Spectroscopy)

**F. FETHI[a]**, Z. Kamar[ab]. M. DAHMANI[b], M. AZIZI[a], Z. BENLAKHDAR[c] et G. TAIEB[d]

[a]LDOM et [b]LPTPM, Faculté des Sciences, Université Mohamed I[er], Oujda (Maroc)
[c]Département de Physique, Faculté des Sciences,
Campus Universitaire Tunis, 1060 (Tunisie)
[d]LPPM, Bat 210, Université de Paris Sud, 91405, Orsay cedex (France)

## 1 Introduction

Nous avons utilisé la technique LIBS (Laser Induced Breakdown Spectroscopy) ou bien LIPS (laser Induced Plasma Spectroscopy) pour analyser l'eau en détectant les sels minéraux comme le magnésium (Mg), le calcium (Ca), le potassium (K) et le sodium (Na). Cette technique qui est basée sur l'ablation laser, permet d'analyser directement l'échantillon en évitant les phases longues et coûteuses pour le préparer.

Dans la première phase de ce travail [1], nous avons tracé les courbes d'étalonnages des sels mentionnés ci-dessus dans l'eau, afin de les utiliser pour analyser n'importe quel échantillon liquide. Nous avons analysé deux échantillons inconnus: l'eau de l'océan atlantique et celle d'une nappe friatique (Moulay Yaakoub, au centre du Maroc).

## 2 Description du projet de recherche

La LIBS est une méthode de diagnostic optique (figure 1). Elle est facile à mettre en œuvre, efficace, rapide, non destructive et non polluante. Elle peut être utilisée en temps réel, in situ et On line pour analyser n'importe quel échantillon (liquide, solide ou gazeux). Elle peut être employée aussi pour détecter les métaux lourds comme le Plomb (Pb), le mercure (Hg), le cadmium (Cd) et d'autres éléments qui sont toxiques.

Son principe consiste à analyser l'échantillon ablaté directement à partir du rayonnement lumineux qu'elle émet en utilisant les méthodes spectroscopiques. Elle permet de remonter à la composition chimique de l'échantillon

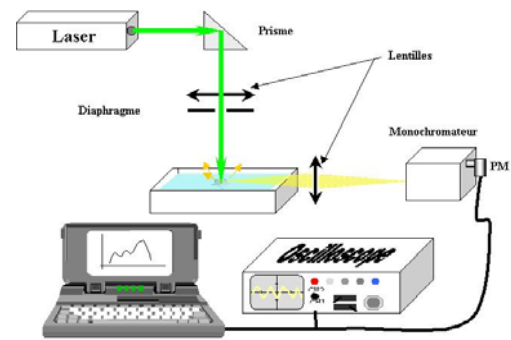

FIG. 1: Dispositif expérimental de la technique LIBS.

## 3 Contexte International, collaborations

Ce travail a été réalisé au laboratoire de Photophysique Moléculaire (à Orsay) grace au projet d'une acion intégrée CNRS (France) et CNRST (Maroc) (N° SP105/99) et une collaboraion avec la Tunisie.

## 4 Résultats

Pour tracer les courbes d'étalonages, nous avons étudié la variation de l'intensité des raies Ca (I), Ca(II), Mg(I), Mg(II), Na(I) et K(I) en fonction de la concentration de $CaCl_2$, $MgCl_2$, NaCl et KCl respectivement dissous dans l'eau pure. Le signal a été optimisé de tel sorte d'assurer la reproductibilité du signal et des résultats. C'est pourquoi, les expériences sont répétées plusieurs fois pour toutes les concentrations et à toutes les longueurs d'ondes.

Le tableau ci-dessous résume les différentes concentrations des sels dans l'océan atlantique en comparant avec celles obtenues par P. THOMAS [2] par une méthode d'analyse chimique classique.

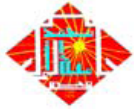 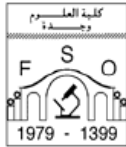 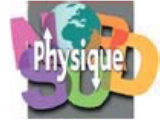 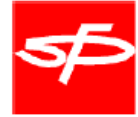

|  | Ca (g/l) | Mg (g/l) | K (g/l) | Ca (g/l) |
|---|---|---|---|---|
| Présent travail | 0,3 | 1,2 | 0,6 | 9 |
| P. THOMAS [2] | 0,4 | 1,2 | 0,4 | 10 |

Pour la nappe de Moulay Yaakoub, on a obtenu les concentrations suivantes : Ca (0,28g/l), Mg (0,1g/l), K (0,12 g/l) et Na (6,89g/l)

## 5 Conclusion

La création du plasma par la technique LIBS en surface de l'eau, nous a permis de tracer les courbes d'étalonnages et de déterminer les concentrations des éléments inorganiques Ca, Mg, K et Na dans deux échantillons différents : l'océan d'atlantique et la nappe friatique de Moulay Yaakoub (Maroc)

Cette technique qui est puissante et sensible peut être généralisée et appliquée dans d'autres secteurs, afin de déterminer la composition chimique de n'importe quel échantillon.

## Références

[1] F. FETHI, Thèse d'état sotenue le 16 février 2002 à la Faculté des Sciences d'Oujda (Maroc)
[2] P. THOMAS, http://www.ens-lyon.fr/Planet-Terre/Forum/Climats/Ocean/sel.htm

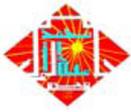 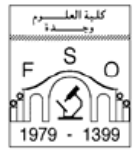 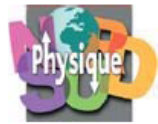 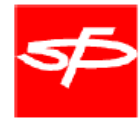

# Quantum Symmetries of Higher Coxeter Graphs:
# The algebraic and conformal aspects

**Dahmane HAMMAOUI** Professeur agrégé de l'enseignement secondaire, E. TAHRI

Laboratoire de Physique Théorique, Physique des Particules et Modélisation, Faculté des Sciences, Oujda

## 1 Introduction and General Context

Along the last fifteen years or so, investigations performed in several fields of research belonging to theoretical physics or mathematics suggest the use of fundamental objects such that the ADE Dynkin diagrams or their higher patterns.
Such mathematical or physical fields are: statistical mechanics, string theory, quantum gravity, conformal field theory, theory of bimodules, Von Neumann algebras, sector theory, modular tensor categories, weak Hopf algebras…
 - In an algebraic viewpoint, higher Coxeter graphs of the SU(N) type are related to the classification of affine Lie algebras and describe their representation theory.
 - In a conformal viewpoint higher Coxeter graphs of the SU(N) type are related to the classification of modular invariants of conformal models and desscribe the different aspects of Q CFT.
 - Quantum geometry on graphs (introduced by A. Ocneanu) describes quantum symmetries of these graphs together with the corresponding Ocneanu graphs wich index the defect lines of the CFT and brings out new structures of weak Hopf algebras called algebras of double triangles.

## 2 Fusion algebras and $A_k$ graphs

Fusion algebras describe fusion of primary fields in CFT and their structure constants are nimreps (non-negative integer matrix representations) denoted $(N_i)_j^k$ and are related to the OPE of fields. $A_k$ graphs are the Weyl alcoves of SU(N) type truncated at a level k (see Fig1). Their vertices i,j,k… (or lambda, mu, nu …) represent (irreps) irreducible representations of quantum groups $SU(N)_k$ at a root of unity. The $A_k$ graphs encode the tensor product of irreps inhereted from fusion of fields.
Examples: $A_4$ graph of SU(2), $A_3$ (SU(3)) and $A_2$ (SU(4)).

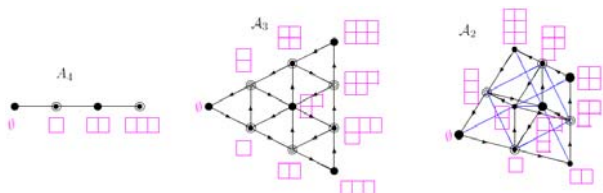

## 3 Graph algebras and module graphs

The other graphs G are considered as module over the fusion algebras A(G). This action of A-module is encoded in nim-reps $(F_i)_a^b$ which provide solutions to the Cardy equation in BCFT and where a,b,c… denote vertices of the G graphs and represent the boundary conditions of the theory. $(F_i)_a^b$ gives also the number of horizontal essential paths on the graph G from a to b associated to the Young frame i.
The following figure shows two examples: the $E_6$ graph of SU(2) type and the $D_3$ graph of SU(3) type.

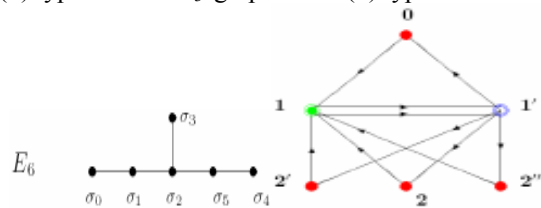

## 4 Quantum symmetries

In twisted chiral CFT, vertices x,y,z… of an Ocneanu graph $\Gamma(G)$ associated to a higher Coxeter graph G encode defect lines of the theory. These vertices are the generators of an algebra structure Oc(G) called the algebra of quantum symmetries. A representation of Oc(G) is carryed by nimreps $(O_x)_y^z$. In almost cases Oc(G) is realized as a square tensor product of the graph algebra G, or as semi-direct product of G and the discret symmetry group $Z_N$. On one hand, Oc(G) is an A-A bimodule whoose action is encoded in nimreps $W_{xy}$ called toric matrices and enable to determine the twisted generalized partition functions of the corresponding CFT. On the other, G is an Oc(G)-module s.t its action is encoded in nim-reps $(S_x)_a^b$ which give the number of vertical paths on the graph G from a to b associated to the vertex x.
 Example: The Ocneanu graph $Oc(E_6)$

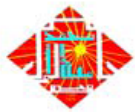 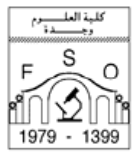 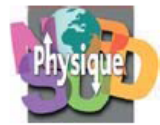 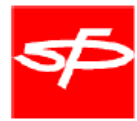

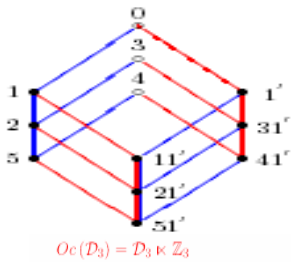

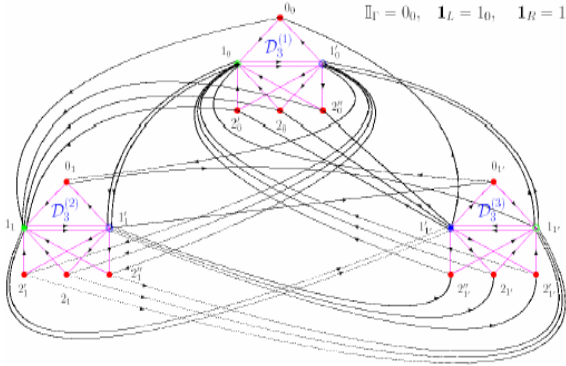

## 5 Nim-reps, graphs and CFT

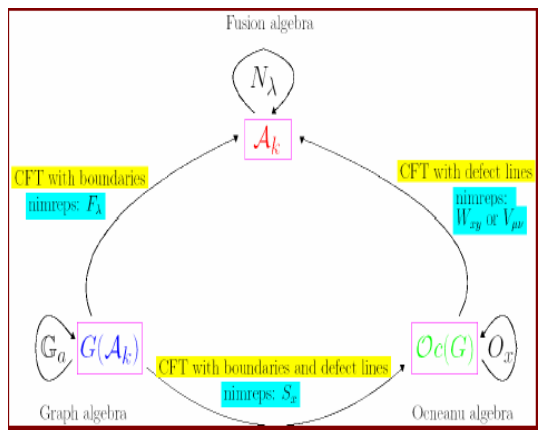

## 6 Notion of Ocneanu quantum groupoïds

Following A. Ocneanu, to each graph G one can associates a weak Hopf algebra (WHA) structure called the algebra of double triangles denoted BG. The algebra B (respectively its dual B') of horizontal double triangles (of vertical double triangles) is endowed with a product and a coproduct operations. Axioms of WHA are ensured by a system of or generalized 6j-symbols called Ocneanu cells. Representation theory of the bi-algebra B is described by the fusion algebra A(G) and the representation theory of its dual B' is encoded by the algebra of quantum symmetries Oc(G). The WHA BG has two block-diagonal decompositions: the first blocks are indexed by the young frames i,j,k… (of $A_k$) and the second one are indexed by vertices x,y,z… of $\Gamma(G)$.

## 7 Conclusion and outlooks

Mathematical aspects of generalized Coxeter-Dynkin graphs on lie in the formulation of branching rules between representation theories of quantum groups and the corresponding representations of their sub-groups, and in the structure of WHA that we can associate to each graph. Physical aspects of these families of graphs are shown in the description of solvable critical two-dimensionnel models in statistical mechanics. Other features of higher Coxeter graphs and their patterns of quantum symmetries arise from the description of CFT in many environnements such as, conformal invariance, boundaries, defect lines,… In string theory and quantum gravity, it seem that many algebraic tools such as generalized quantum 6j symbols can be provided by the data of Ocneanu quantum groupoïds associated to SU(N) graphs.

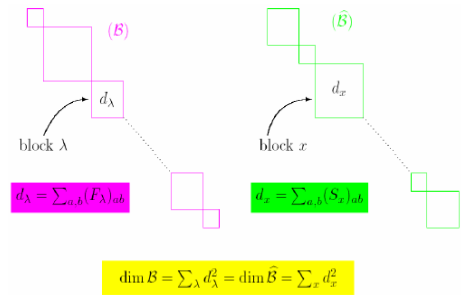

## References

[1] R. Coquereaux, "Racah-Wigner quantum 6j symbols, Ocneanu cells for AN diagrams and quantum roupoïds," J. Geom. Phys. (2006).

[2] D. Hammaoui, E. H. Tahri et al, "Comments about quantum symmetries of SU(3) graphs," J. Geom. Phys. 57 (2006) 269.

[3] R. Coquereaux et al, "On quantum symmetries of ADE graphs," Adv. Theor. Math. Phys. 8 (2004) 1.

[4] P. Di Francesco et al, "SU(N) lattice integrable models associated with graphs,"Nucl. Phys. B 338 (1990) 602.

[5] D. Hammaoui, "Géométrie quantique d'Ocneanu des graphes de Di Francesco-Zuber associés aux modèles conformes de types su(3)'', Thèse de Doctorat National, 2007.

[6] D. Hammaoui, "SU(3) graphs, quantum symmetries, quantum groupoïds and CFT," Meeting on coherent states on generalized Heisenberg algebras: Algebraic and geometrical aspects, Rabat, June 12th-14th, (2006).

[7] D. Hammaoui et al, "Higher Coxeter graphs associated to affine su(3) modular invariants," J.

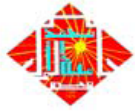 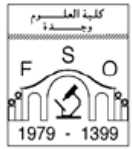 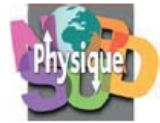 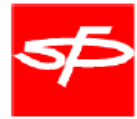

Phys. A: Math. Gen. 38 (2005) 8259.

[8] A. Ocneanu: "The classification of subgroups of quantum SU(N)", Summer School, Jan. 2000, Argentina, AMS Contemp. 294.

[9] A. Ocneanu: "Paths on Coxeter diagrams: From platonic solids and singularities to minimal models and subfactors", Operator theory. Fields Institute Monographs Providence (1999) 243.

[10] V. B. Petkova et al, "The many faces of Ocneanu cells," Nucl. Phys. B 603 (2001) 449.

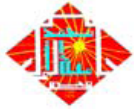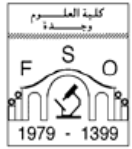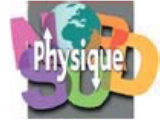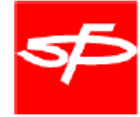

# Modélisation numérique des écoulements à surface libre bidimensionnels par la méthode des volumes finis.

**M. Hammouch, M. Boulerhcha, N. Salhi, I. El Mahi**

U.F.R .Doctorat « Mécanique et Énergétique », Université Mohamed Premier Oujda, Maroc

**Résumé**

La présence de plus en plus fréquente des inondations dans les rivières incite à développer des outils de simulation numérique permettant de déterminer très précisément les zones inondables.
La nature tridimensionnelle de l'écoulement n'est pas importante dans toutes les situations, notamment, quand le rapport largeur profondeur est très grand. Dans de tels cas, l'écoulement est régi par les équations de Saint Venant qui découlent des lois de bilan pour un fluide incompressible, Ces équations sont résolues par le schéma de Roe basé sur la méthode des volumes finis appliquée à un maillage structuré et non structuré et constitué de triangles. Les applications les plus courantes du modèle concernent la repture de barrage, les crues, l'hydraulique des fleuves, les écoulements marins….

## 1 Introduction

Les équations de Saint Venant représentent généralement les écoulements à surface libre en eau peu profonde, et permet de modéliser les problèmes liés à l'environnement tels que la prévention et le contrôle des inondations. Ces équations découlent des lois de bilan pour un fluide incompressible, soumis à la gravité. Elles sont obtenues en intégrant les équations de Navier-Stokes selon la verticale et en supposant différentes hypothèses fondamentales dont celle de la pression hydrostatique.

Nous proposons la résolution des équations de Saint Venant par un schéma volumes finis bidimensionnel (schéma de Roe) qui est basé sur le solveur approché de Riemann, dont la précision est améliorée grâce à la technique MUSCL.

## 2 Discrétisation de la partie convective

### 2.1 Solveur de Roe [1], [3]

Afin de construire un schéma décentré basé sur un solveur de Riemann approché de type Roe, nous considérons la partie hyperbolique du système de loi de conservation qui représente le système de Saint-Venant :

$$\frac{\partial W}{\partial t} + \frac{\partial F}{\partial x} + \frac{\partial G}{\partial y} = 0 \quad (1)$$

Où :
$$W = \begin{Bmatrix} h \\ hu \\ hv \end{Bmatrix}, \quad F = \begin{Bmatrix} hu \\ hu^2 + 0.5gh^2 \\ huv \end{Bmatrix}, \quad G = \begin{Bmatrix} hv \\ huv \\ hv^2 + 0.5gh^2 \end{Bmatrix}$$

W représente le vecteur des variable conservatives, h est la hauteur d'eau, u et v sont respectivement les vitesses d'écoulement dans les directions x et y et g est la constante de la gravité.

Nous intégrons les équations (1) sur une cellule $T_i$ du domaine en espace de frontière $\partial T_i$.
En appliquant le théorème de Green, on déduit :

$$A_i \frac{\partial W_i}{\partial t} + \int_{\partial T_i} (Fn_x + Gn_y) d\Gamma = 0 \quad (2)$$

$\vec{n}(n_x, n_y)$ est le vecteur unitaire normal dirigé vers l'extérieur, $G_i(x_i, y_i)$ et $G_j(x_j, y_j)$ sont respectivement les barycentres des triangles $T_i$ et $T_j$. $W_i$ et $A_i$ sont respectivement la restriction de W et l'aire de $T_i$.

$$\partial T_i = \bigcup_{j \in k(i)} \Gamma_{ij}$$

$\Gamma_{ij}$ est l'interface entre le volume $T_i$ et $T_j$, k(i) est l'ensemble des triangles qui ont une arrête commune avec le triangle $T_i$. La forme (2) devient :

$$A_i \frac{\partial W_i}{\partial t} + \sum_{j \in k(i)} \int_{\Gamma_{ij}} (Fn_x + Gn_y) d\Gamma = 0, \quad E(W,n) = Fn_x + Gn_y$$

Le flux convectif est donné par :

$$\int_{\Gamma_{ij}} E(w, \vec{n}) d\sigma = \Phi(W_i, W_j, \vec{n}_{ij}) mes(\Gamma_{ij}) \quad (3)$$

$\Phi$ est le flux numérique, $W_i$ et $W_j$ sont respectivement les valeurs de W sur les cellules $T_i$ et $T_j$.
La moyenne de Roe doit vérifier la relation de conservation : $E(W_j) - E(W_i) = A(\tilde{W})(W_j - W_i)$

Où :

$$\tilde{W}(W_i, W_j) = \begin{cases} \tilde{h} = \frac{1}{2}(h_i + h_j) \\ \tilde{u} = \frac{u_i\sqrt{h_i} + u_j\sqrt{h_j}}{\sqrt{h_i} + \sqrt{h_j}} \\ \tilde{v} = \frac{v_i\sqrt{h_i} + v_j\sqrt{h_j}}{\sqrt{h_i} + \sqrt{h_j}} \end{cases}$$

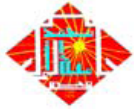 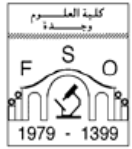 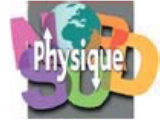 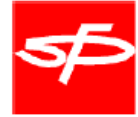

$$A(W, \vec{n}) = \begin{bmatrix} 0 & n_x & n_y \\ (gh - u^2)n_x - uvn_y & 2un_x - vn_y & un_y \\ (gh - v^2)n_y - uvn_x & vn_x & 2vn_y + un_x \end{bmatrix}$$

$$\Phi(W_i, W_j, \vec{n}_{ij}) = \frac{1}{2}\left[E(W_i, \vec{n}_{ij}) + E(W_j, \vec{n}_{ij})\right] - \frac{1}{2}\left|A(\tilde{W}, \vec{n}_{ij})\right|(W_j - W_i)$$

$\Phi(W_i, W_j, \vec{n}_{ij})$ représente la fonction du flux numérique bidimensionnel.

La matrice jacobienne $A(\tilde{W})$ doit avoir des valeurs propres réelles.

## 2.2 Extension à l'ordre deux en espace (MUSCL). [2], [3]

Nous appliquons une démarche qui se décompose en deux étapes : étape de reconstruction des états gauche et droite, séparés par l'arête $\Gamma_{ij}$, et une étape de limitation des gradients afin d'éviter les oscillations.

$$\begin{cases} W_{ij}^l = W_i + \frac{1}{2}\vec{\nabla}W_i \overrightarrow{G_iG_j} \\ W_{ij}^r = W_j - \frac{1}{2}\vec{\nabla}W_j \overrightarrow{G_iG_j} \end{cases}$$

Le problème de cette étape est l'évaluation des gradients de la solution sur chaque volume de contrôle. Pour contourner cette difficulté, on utilise le limiteur MinMod pour déterminer $\vec{\nabla}W_i$ et $\vec{\nabla}W_j$ :

$$\frac{\partial W_i^{lim}}{\partial x} = \frac{1}{2}\left[\min_{j \in k(i)} \text{sgn}(\frac{\partial W_j}{\partial x}) + \max_{j \in k(i)} \text{sgn}(\frac{\partial W_j}{\partial x})\right] \min_{j \in k(i)} \left|\frac{\partial W_j}{\partial x}\right|$$

De même pour $\frac{\partial W_i^{lim}}{\partial y}$. Cette opération peut être étendue sur l'ensemble k(i) des triangles ayant un sommet commun avec $T_i$.

## 2.3 Discrétisation temporelle [1]

Pour obtenir un schéma de second ordre, le terme temporel est intégré selon un schéma explicite de type Runge Kutta. En combinant les équations (2) et (3) on déduit:

$$\frac{\partial W_i}{\partial t} = -\frac{1}{A_i} \sum_{j \in k(i)} (\Phi(W_i, W_j, \vec{n}_{ij}) \text{mes}(\Gamma_{ij}))$$

On peut l'écrire sous la forme: $\frac{\partial W}{\partial t} = H(W)$, on aura alors:

$$\begin{cases} \overline{W}^{n+1} = W^n + \frac{\Delta t}{2}H(W^n) \\ W^{n+1} = W^n + \Delta t H(\overline{W}^{n+1}) \end{cases}$$

## 3 Application

**Ressaut hydraulique oblique**

Il s'agit d'un courant circulant dans un canal de largeur 30m. A l'entrée: $u_1 = 8.57\,m/s$, $h_1 = 1\,m$, $v_1 = 0\,m/s$, Q=257.1m³/s et un nombre de Froude $F_{r1} = 2.74$, ce courant est dévié par une paroi latérale d'angle $\theta = 8.95°$, théoriquement cela donne naissance à une onde d'angle $\beta = 30°$ et un courant aval parallèle à la paroi, de hauteur $h_2 = 1.5$m et une vitesse $v_2 = 7.76\,m/s$. Numériquement nous résolvons les équations de Saint Venant sans terme source sur un domaine maillé en éléments triangulaires contenant 2149 nœuds et 4098 éléments (fig.1), le pas de temps est $\Delta t = 3.80.10^{-3}$ et correspond à un nombre Cfl=0.27.

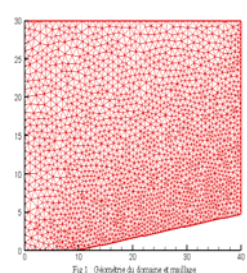
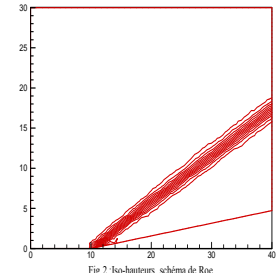

Fig 1 : Géométrie du domaine et maillage     Fig 2 : Iso-hauteurs, schéma de Roe

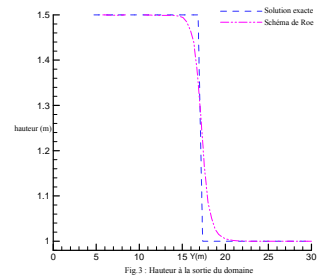

Fig. 3 : Hauteur à la sortie du domaine

La figure 2 représente les iso-hauteurs calculés par la méthode des volumes finis, l'angle du choc est de l'ordre 30.2°, la capacité à capturer le choc est satisfaisante. Nous présentons également la hauteur à la sortie du domaine (fig.3), nous constatons que la méthode des volumes finis est légèrement diffusive.

## 4 Conclusion

On a étudié les écoulements à surface libre régis par les équations de Saint Venant en variables conservatives, connues pour leurs caractères hyperboliques. La résolution numérique est effectuée par la méthode des volumes finis couplée au schéma de Roe. Elle montre des performances très satisfaisantes, en particulier, les chocs droits sont capturés de façon précise. Le schéma de Roe montre toutefois un léger surplus de diffusion.

## References

[1]. E. Audusse. Modélisation hyperbolique et analyse numérique pour les écoulements en eaux peu profondes. Thèse, Septembre 2004.
[2] M. Boulerhcha et al, Résolution des équations de Saint Venant par un schéma éléments finis et un

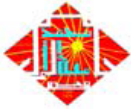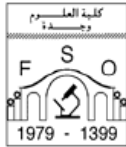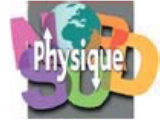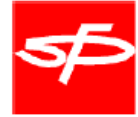

schéma volume finis. Revue Européenne des éléments finis, 14(8), 999–1013, 2005.

[3] I. Elmahi, Schéma Volumes finis pour la simulation Numérique de Problèmes à fronts Raides en maillages Non structurés Adaptatifs. Ph.D.thesis, Rouen, 1999.

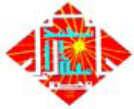 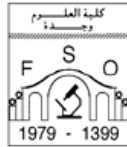 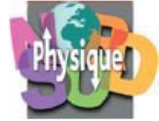 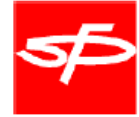

# Structure nucléaire et états de bas spin dans le noyau $^{125}$La par la méthode du couplage Quasiparticule-Phonon plus Rotor

**J. INCHAOUH**[1], O.Jdair[1], K. JAMMARI[2], H.Chakir[1], A.Morsad[1], A. ZAAFA[1]

[1] Université Hassan II-Mohammedia, Faculté des sciences Ben M'Sik, Casablanca, Maroc
[2] Université Hassan II, Faculté des sciences Ain chock L.P.T.N BP. 5366, Casablanca, Maroc

## 1 Introduction

Les noyaux déficients en neutrons de la région de transition autour de la masse A~130, ont fait l'objet de nombreux travaux tant expérimentaux que théoriques. La spectroscopie γ et e⁻ de ces noyaux a été étudiée au cours d'un programme de recherches initié à SARA (Système Accélérateur Rhône-Alpes) pour déterminer les états intrinsèques, à bas spin, peuplés par désintégration bêta [1-2]. Les techniques utilisées (jet d'Hélium, IGISOL [3]) ont permis d'apporter de nouvelles informations sur les spectres d'excitation à basse énergie de ces noyaux.

La bande découplée basée sur le niveau $I^\pi=11/2^-$ (attribué à l'orbitale de Nilsson π 1/2⁻ [550]) est généralement bien reproduite en couplant une quasiparticule à un cœur rigide triaxial (PTRM [4,5]). Par ailleurs, les modèles utilisés pour décrire les noyaux impairs $^{125}$La et $^{127}$La, en particulier les calculs self-consistents du "Total Routhian Surface" (TRS) ont fait apparaître une mollesse importante vis à vis du paramètre d'asymétrie γ tant pour reproduire les niveaux de parité positive que ceux de parité négative [5]. En conséquence, il nous est apparu important de prendre en considération de façon explicite les degrés de liberté de vibration.

Dans ce travail, la structure du noyau $^{125}$La à basse énergie d'excitation est décrite en utilisant une approche théorique basée sur le couplage quasiparticule-phonon plus rotor [6]. Dans cette méthode de calcul, les états intrinsèques résultent de la prise en compte simultanée des quatre types d'interactions: champ moyen de Nilsson, force d'appariement, force quadripolaire-quadripolaire et la force de recul. Le spectre des états excités est obtenu en incluant aussi la force de coriolis. Cette méthode nous a permis de déduire la structure des cinq bandes à parité positive et de deux bandes à parité négative pour le noyau $^{125}$La.

## 2 Formalisme théorique

L'hamiltonien total est défini par l'expression suivante :
$$H = H_{int} + H_I + H_C \quad (1)$$
Où $H_{int} = H_{sp} + H_P + H_Q + H_J$
$H_I = A_R(I^2 - I_3^2)$, $H_C = -A_R(I_+J_- + I_-J_+)$, $H_J = A_R(J^2 - J_3^2)$
Avec $I_\pm = I_1 \pm iI_2$, $J_\pm = J_1 \pm iJ_2$ et $A_R = \hbar^2/2\bar{\Im}$

Il est séparé en trois termes, l'hamiltonien intrinsèque $H_{int}$, l'hamiltonien rotationnel $H_I$ et celui de coriolis $H_C$ qui couple les mouvements rotationnel et intrinsèque. $A_R$ est le paramètre d'inertie nucléaire. L'hamiltonien intrinsèque quant à lui, contient quatre parties: $H_{sp}$, décrivant le champ moyen déformé de Nilsson, le terme $H_P$ représente l'interaction d'appariement monopolaire (approximation BCS) qui décrit la superfluidité de la matière nucléaire, la force quadripolaire-quadripolaire $H_Q$ rend compte de la vibration du noyau [7] et le dernier terme $H_J$ représente la force de recul résultant de l'hypothèse d'un mouvement de rotation à symétrie axiale. Avec l'inclusion de la force de coriolis $H_C$, la matrice de l'hamiltonien H total doit être construite et diagonalisée dans l'espace des fonctions rotationnelles symétriques [8].

$$|IMK\rangle = \left(\frac{2I+1}{16\pi^2}\right)^{\frac{1}{2}} \{D^I_{MK}|K\rangle + (-)^{I+K} D^I_{M-K}|\overline{K}\rangle\} \quad (2)$$

Où $|\overline{K}\rangle$ est l'état symétrique par renversement du temps de l'état intrinsèque $|K\rangle$ obtenu suite à la résolution du problème séculaire

$$H_{int}|K\rangle = (H_{sp} + H_P + H_Q + H_J)|K\rangle = E^{int}_{K\rho}|K\rangle \quad (3)$$

L'état intrinsèque s'écrit donc comme une combinaison linéaire des états à 1-quasiparticule et des états du couplage quasiparticule-phonon.

$$|K\rangle = (\sum_\nu C_\nu \delta_{\Omega_\nu,K} \alpha^+_\nu + \sum_{\nu\gamma} D_{\nu\gamma}\delta_{K=\Omega_\nu+\gamma}\alpha^+_\nu B^+_\gamma)|BCS\rangle \quad (4)$$

Où $B^+_\gamma$ (avec γ=±2) est l'opérateur de création d'un phonon γ dans l'approximation Tamm-Dancoff.

## 3 Résultats et discussions

Les calculs effectués pour le noyau $^{125}$La, à l'aide de la méthode du couplage quasiparticule-phonon plus rotor (QPRM) montrent l'existence à basse énergie d'excitation de cinq bandes à parité positive et deux bandes à parité négative. Les paramètres choisis sont: la déformation quadripolaire $\varepsilon_2 = 0.250$ [9], le gap d'appariement de 1077 KeV qui est déterminé par la relation phénoménologique ($\Delta_p = \Delta_n = 12/A^{1/2}$), le paramètre d'inertie déterminé en fonction de l'énergie du premier état excité $E(2^+)$ ($A_R = E(2^+)/6$) et l'énergie de l'état de phonon qui est ajustée de façon à obtenir l'énergie expérimentale de la tête de bande γ du cœur pair-pair

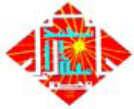 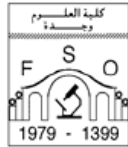 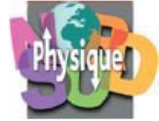 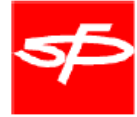

$^{124}$Ba $(E(2_\gamma^+) = 887.8 keV)$. En additionnant la force quadripolaire-quadripolaire et le terme de recul à l'interaction d'appariement, un nouvel arrangement des états intrinsèques est obtenu pour le noyau $^{125}$La sur la figure 1 on constate que la force quadripolaire agit spécifiquement sur les états à parité positive, tandis que la force de recul agit fortement sur les états à parité négative. Le calcul théorique de quasiparticule-phonon plus rotor montre que l'état intrinsèque à parité positive 1/2$^+$[420] de la bande 3 (fig 2) représente l'état fondamental du noyau $^{125}$La. Sa configuration est dominée par l'orbital $g_{7/2}$ (99,5% $g_{7/2}$ +0.3% $d_{5/2}$), tandis que le premier état excité 5/2$^+$ de cette même bande présente un fort mélange de coriolis (47% $d_{5/2}$ + 42% $g_{7/2}$). L'énergie d'excitation (calculée par rapport au 3/2$^+$) de l'état excité 7/2$^+$ de la bande 2 est de l'ordre de 145 keV. Cette énergie paraît supérieure à celle de la transition de 297KeV observée par spectroscopie sur faisceau [5, 11] (bande 2) de la figure 3. D'autre part, ce même niveau 7/2$^+$ calculé à une énergie de 176 keV présente un fort mélange de coriolis [45%[422]3/2$^+$+30%[420]1/2$^+$]. Compte tenu de cette configuration, sa désexcitation vers l'état 5/2$^+$ [69%[420]1/2$^+$+14%[422]3/2$^+$] de la bande 3 devrait être favorisée. D'après nos calculs, cette transition devrait avoir une énergie de l'ordre de 76KeV alors que le schéma de niveau expérimental du $^{125}$La indique la présence d'une transition de 168KeV de l'état 7/2$^+$ vers l'état 5/2$^+$. En plus, les états excités 1/2$^+$, 3/2$^+$, 9/2$^+$, 7/2$^+$ et 11/2$^+$ rassemblés dans la bande 3 (fig 2) ne sont pas identifiés expérimentalement. Ils pourraient correspondre à des éventuels niveaux alimentés suite à la désintégration β$^+$/CE ($^{125}$Ce→$^{125}$La [10]) et dont la connexion avec les niveaux observés dans les expériences "sur faisceau" [11] est pour l'instant mal élucidée. Par ailleurs, une bande construite sur un niveau excité 9/2$^+$ est élucidée [11]. D'après nos calculs, cette dernière correspond à la bande 7 bâtie sur l'état intrinsèque [404]9/2$^+$ (issu de l'orbitale $\pi g_{9/2}$) et dont la position en énergie est calculée à 713.6KeV.

La structure des fonctions d'onde des états excités à parité négative (bande 1) est dominée par l'orbitale déformée de Nilsson [550]1/2$^-$ avec un mélange non négligeable de [541]3/2$^-$. Ceci donne lieu en particulier à la bande découplée 11/2$^-$, 15/2$^-$, 19/2$^-$,…Il est à noter que nos calculs fait apparaître des nivaux 7/2$^-$ et 3/2$^-$ non encore observés expérimentalement.

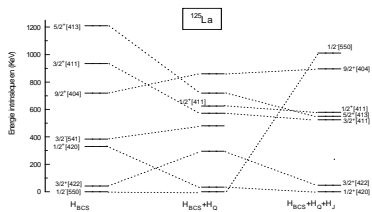

Fig. 1: L'évolution en énergie des états intrinsèques du $^{125}$La suivant l'ajout de la force quadripolaire et de la force de recul.

## 4 Conclusion

La structure à basse énergie du noyau impair $^{125}$La a été étudiée en utilisant la méthode du couplage quasiparticule-phonon plus rotor (QPRM). Nos calculs indiquent la contribution du couplage quasiparticule-phonon joue un rôle important dans la configuration des états intrinsèques. Les états excités à basse énergie sont influencés par le mélange dû à la force de coriolis. Les résultats de nos calculs montrent aussi que la méthode du couplage quasiparticule-phonon est une bonne approche théorique, reproduisant de façon satisfaisante le spectre d'excitation des noyaux appartenant à la région de transition A~130 et mettant en évidence la prépondérance des niveaux issus des orbitales $d_{5/2}$ et $g_{7/2}$ pour les états à parité positive et de l'orbitale $h_{11/2}$ pour les états à parité négative.

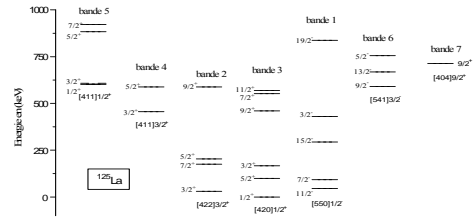

Fig. 2: Schéma des premiers niveaux excités du noyau $^{125}$La obtenu à l'aide de la méthode du couplage quasiparticule-phonon plus rotor.

## Références

[1] J. Inchaouh et al, Contribution to the In. Conf. On Nuclear Shapes and Nuclear Structure Cargèse (1991)22
[2] A. Gizon et al, Phys. A359, (1997) 11 and A. Gizon, J. Genevey et al, Nucl. Phys. A605 (1996) 301
[3] R. Béraud et al, Nucl. Inst. And Meth. In Phys. Rev. A346 (1994) 196
[4] J. Meyer-ter-Vehn, Nucl. Phys A 249 (1975) 111, A249 (1975) 141
[5] K. Starosta et al Phy. Rev. C.53 (1996) 137 and K. Starosta et al, Phy. Rev. C.55 (1997) 2794
[6] J. Inchaouh al, Eur. Phys. J. A 7 (2000) 317
[7] P. Ring and P. Schuck, the Nuclear Many -Body Problem (Springer, New-York, 1970)
[8] A. Bohr et al, Nuclear Structure, Vol. 2 (1975)
[9] P. Möller et al, Atomic data and Nuclear Data tables, Vol.59, N°. 2 (1995) 185
[10] G. Canchel et al, Eur. Phys. J. A5 (1999) 1
[11] D. J. Hartley et al, Phys. Rev. C60, 014308.

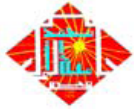 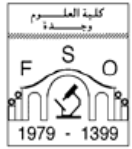 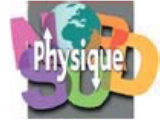 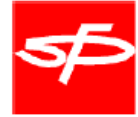

# Montage d'une expérience de diffusion de la lumière

**Zakaria KAMAR**\*°, Fouad FETHI° et Mohammed DAHMANI\*

°Laboratoire de Physique Théorique, de Physique des Particules et Modélisation
\*Laboratoire d'Optique et Dynamique des Matériaux
Département de Physique, Faculté des Sciences,
Université Mohamed I$^{er}$ Oujda, (Maroc)

## 1 Introduction

Ce projet consiste à réaliser un montage expérimental original à l'université Mohamed I$^{er}$. Il est important dans le sens où la technique de la diffusion de la lumière est pluridisciplinaire, car il est au carrefour de toutes les disciplines scientifiques, physique, chimie et biologie.

La diffusion est le phénomène par lequel un faisceau de rayonnement (lumineux, acoustique, neutronique, rayon X, etc.) est dévié dans de multiples directions. Le phénomène de diffusion peut se produire quand une onde rencontre un obstacle, dont la surface n'est pas parfaitement plane et lisse.

## 2 Rappels théoriques

le rayon lumineux d'intensité $I_0$ traverse l'échantillon puis, il sera diffusé avec une intensité $I<I_0$.

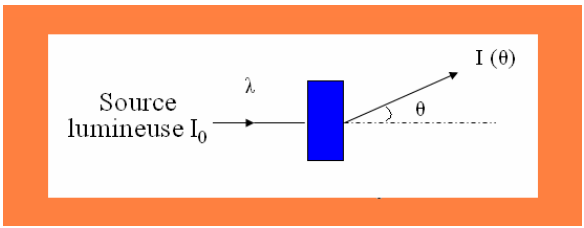

Vecteur de diffusion : $q = \dfrac{4\pi n_0}{\lambda} \sin \dfrac{\theta}{2}$

Équation générale de la diffusion de lumière :

$$\frac{KC}{\Delta R(\theta)} = \frac{1}{M_n}\left[1 + \frac{16\pi^2 n_0^2 R_G^2}{3\lambda_0^2} \sin^2\left(\frac{\theta}{2}\right)\right] + 2A_2 C$$

L'extrapolation de $KC/\Delta R(\theta)$, à $\theta=0$ pour chaque concentration :

$$\left[\frac{KC}{\Delta R(\theta)}\right]_{C=0} = \frac{1}{M_n}\left(1 + \frac{16\pi^2 n_0^2 R_G^2}{3\lambda_0^2} \sin^2\left(\frac{\theta}{2}\right)\right)$$

L'extrapolation de $KC/\Delta R(\theta)$, à $C = 0$ pour chaque angle :

$$\left[\frac{KC}{\Delta R(\theta)}\right]_{\theta=0} = \frac{1}{M_n} + 2A_2 C$$

## 3 Dispositif expérimental

Le montage expérimental utilisé est composé par : un laser, un goniomètre, un photomultiplicateur et un système d'acquisition.

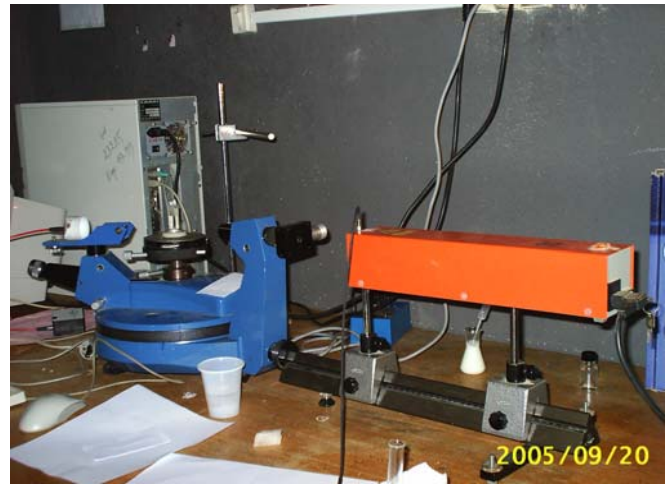

En utilisant le matériels de travaux pratiques d'optique existant au département de physique, nous avons pu faire certains tests préliminaires sur des échantillons très diffusants comme le lait et le toluène.

## 4 Résultats

On effectue plusieurs étapes pour tracer le diagramme de Zimm puis, on calcule les trois les paramètres ($M_w$, $R_g$ et $A_2$) en faisant des extrapolations.

a) On calcule la différence du rapport de Rayleigh entre la solution du polymère et le solvant aux différentes angles.

$$\Delta R(\theta) = R_{solution}(\theta) - R_{solvant}(\theta)$$

b) On calcule la ratio, **KC/ΔR(θ),** pour chaque angle et concentration. **K**(mol.cm$^2$/g$^2$) est une constante optique définie par: 
$$K = \frac{4\pi^2 n_0^2}{\lambda_0^4 N_A}(\frac{dn}{dc})^2$$

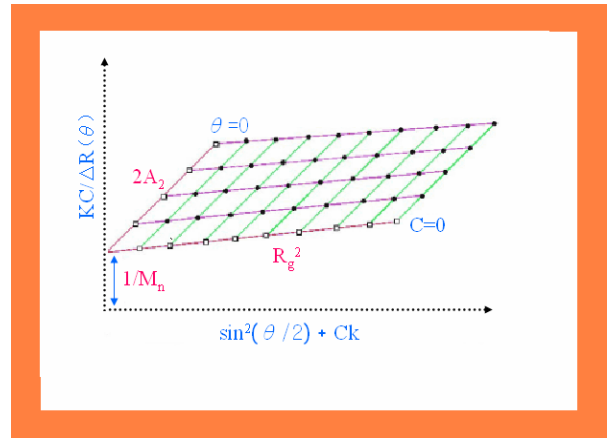

c) On déduit la quantité, **sin2(θ/2) + kC** où la constante arbitraire est choisi pour améliorer la lisibilité du diagramme de Zimm. k(ml/g).

d) On trace le diagramme de Zimm en faisant le graphique de **KC/ΔR(θ)** en fonction de **sin2(θ/2) + kC.**

e) Le calcul la masse molaire **M$_w$**, le 2$^{ème}$ coefficient du viriel **A$_2$** et le rayon de giration moyen **R$_g$** du polymère s'obtiennent en faisant des extrapolations et en utilisant l'équation gérale de la diffusion:

$$\frac{KC}{\Delta R(\theta)} = \frac{1}{M_n}[1+\underbrace{\frac{16\pi^2 n_0^2 R_g^2}{3\lambda_0^2}\sin^2(\frac{\theta}{2})}_{q^2 R_g^2/3}] + 2A_2 C$$

## 5 Conclusion

Grâce à la technique de diffusion de la lumière, on pourra avoir des renseignements précieux sur plusieurs quantités d'intérêt physique ou industriel, tel que, la masse molaire (M$_n$), les dimensions de l'échantillon (R$_g$), et l'interaction polymère-solvant (A$_2$).

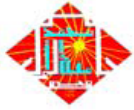 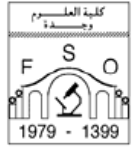 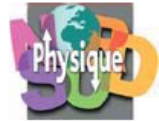 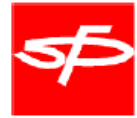

# L'Enseignement du Concept « Temps » en Physique

**MOHAMED KHADRAOUI, M.A.CHARGE DE COURS.**
**CHAMS EDDINE KHIARI, MAITRE DE CONFERENCE**

LABORATOIRE DE DIDACTIQUE DES SCIENCES, E.N.S, KOUBA, ALGERIE

**Résumé**

Une grande problématique se pose dans l'enseignement des concepts en physique, parmi lesquels le concept de temps. La définition même du concept de temps en physique est un problème.

## 1 Introduction

Selon le philosophe EL GHAZALI (5ème siècle), toutes les sciences sont innées chez l'être humain. L'acte de l'enseignement consiste alors à faire passer les connaissances de l'état potentiel à l'état actualisé. Généralement le mot « concept » possède plusieurs définitions parmi lesquelles nous retenons la suivante : « un concept est une abstraction mentale des propriétés communes d'un ensemble d'expériences et de choses ». Cette définition convient bien à la pratique didactique.
Concernant le concept temps, il faut convenir que c'est un concept difficile à enseigner. Les questions immédiates qui se posent sont : le temps a-t-il une existence réelle ?
Avant de répondre à cette question remarquons que l'existence est de deux types :
- Une existence réelle, unique et absolue.
- Une existence relative et éphémère.
 Le temps n'entre dans aucune de ces    catégories.

## 2 Enseignement du concept « temps »

Il existe plusieurs méthodes d'enseignement des concepts en physique :
- Les sciences sont des connaissances que l'apprenant doit acquérir.
- Les sciences sont des méthodes de recherche et d'investigation.
- Méthodes par extrapolation : le savoir est découpé en éléments et on enseigne les règles d'utilisation de ces éléments.
- Méthodes par interpolation : l'apprenant construit une « boite à outils » dans laquelle il trouvera plus tard les éléments dont il aura besoin.

## 3 Conclusion

La méthode la plus naturelle pour enseigner le concept de temps consiste à le faire à travers l'étude du mouvement. Dans cette perspective, le temps est introduit comme étant le facteur d'enchainement d'événements qui se succèdent.

### Remerciements

## Références

المراجع :

1 ــ ابراهيم العاني ، ط1 ،1993م **الزمن في الفكر الإسلامي**،دار المنتخب العربي.

2 ــ د.عبد اللطيف الصديقي، ط1،1995 م ،**الزمان أبعاده وبنيته**، المؤسسة الجامعية للدراسات والنشر والتوزيع.

3 ــ مجموعة من المختصين،ترجمة د.مصباح الحاج عيسى،ط2،1984م، **التقنيات التربوية في تدريس العلوم للمعاهد العليا والجامعات**،مؤسسة الكويت للتقدم العلمية.

4 ــ الغزالي،ط1 ،2003 م ، **مجموعة رسائل الغزالي**،دار الفكر للطباعة والنشر والتوزيع، بيروت،لبنان.

5ــ د.حميد الطرير،**قضية الزمن**،ط1 2004 م، دار وحي القلم ،سوريا.

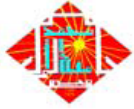 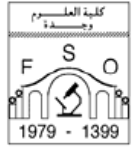 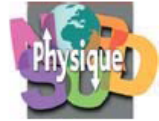 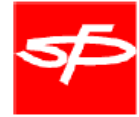

# The reality about length contraction and time dilation in Special Relativity and the ways students interpret them

**R. LADJ, M. OLDACHE, C.E.KHIARI**

Ecole Normale Supérieure de Kouba-Alger
Laboratoire de Didactique des Sciences

**Abstract**

Among the most important subjects, which were the reason for the appearance of the relativity theory is that, the Galilean transformation does not conserve the form of Maxwell's equations. Because of that, a relativity theory with a new transformation called, "Lorentz transformation" appeared in the beginning of the 20th century.
The purpose of this transformation is to make physical laws invariant; when we move from an inertial frame to another. But the results of measuring physical quantities, arising from using the above transformation have faced a great debate and different arguments among physicists and philosophers as well. Through teaching this field of physics, teachers face great difficulties to make students believe in the result of this theory, especially the reality of length contraction and time dilation, i.e., whether the latter are real facts or just manifestations due to the frame chosen. We try to give an easy interpretation which we hope will be useful in teaching.
**Keywords**: Epistemology, education, relativity, time dilation, length contraction.

## I Introduction

As it is very well known in physics, Special Relativity is based on the Lorentz transformation, which leaves the laws of physics invariant in different inertial frames. However; when this transformation is used to measure certain physics quantities such as the time interval between two events taking place in two different places in a certain inertial frame, we measure, from the other frame, a different time interval (*the paradox of time*). Such phenomena arise also with measured lengths (length contraction). We should remind first that the main invariants of the Lorentz transformation are:

1- *All inertial frames are equivalent.*
2- *The universal time is common and identical for all frames.*
3- *The universal space is common and identical for all frames.*
4- *The velocity of light is constant in all frames and is independent of the velocity of the source.*

Let us focus first on the problem of time dilation which is called **the paradox of time.**

### A- Time dilation

This paradox is the hallmark of Special Relativity. We try to discuss this paradox in details.
We assume that we have two clocks, **A** and **B**, exactly similar in every way, which are fixed in two different frames subleased by a signal generator which sends out light pulses at equal intervals of time. These frames are moving one relatively to the other with uniform velocity along a line joining them.

a) If the frames are at rest relatively to one another (i.e.,V=0), then, the two observers, in the different frames, note the same interval of time:

$$\Delta t = \Delta t' = \Delta t_1 = \Delta t_1^{/} = \Delta t_0$$

where : $\Delta t$ and $\Delta t'$ are respectively the time intervals which are measured by the observers in two different frames.

$\Delta t_1$ the time interval in the first inertial frame which is measured by the observer in the second frame. $\Delta t_1^{/}$ the time interval in second inertial frame which is measured by the observer in the first frame.

b) Let the reference frames, where the clocks **A** and **B** are fixed, move one relatively to the other with a uniform velocity (i,e,V ≠ 0).
In this case, the rate of time can be changed due to the Lorentz transformation.
The observer, in the frame with the clock **A,** measures $\Delta t = \Delta t_0$ and the observer in the frame with the clock **B,** measures, $\Delta t^{/} = \Delta t_0$. The time intervals $\Delta t$ and $\Delta t^{/}$ do not depend on the relative velocity of the frames. Hence, $\Delta t = \Delta t_0$ and $\Delta t^{/} = \Delta t_0$ are parameters of the essence.

The observer, in frame **A,** sees that $\Delta t_1^{/}$, in frame **B**, depends on V, and observes that $\Delta t_1^{/} \rangle \Delta t$. The observer in frame **B,** with the clock **B,** sees that $\Delta t_1$, in frame **A**, depends on V and that $\Delta t_1 \rangle \Delta t^{/}$. Hence, $\Delta t_1$ and $\Delta t_1^{/}$ are the parameters of the phenomenon which appear due to the motion. We notice here four parameters arising from the Lorentz transformation: $\Delta t, \Delta t^{/}, \Delta t_1$ and $\Delta t_1^{/}$. The Lorentz transformation gives two inequalities:

$$\Delta t \langle \Delta t_1^{/} \quad ; \quad \Delta t^{/} \langle \Delta t_1$$

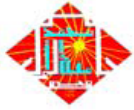 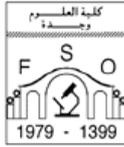 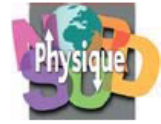 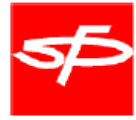

We would like to notice, that some physicists and even Einstein assumed that:

- The time interval $\Delta t_1$ is the real interval in frame A, (means essence in A), while we see it as phenomena observed by B.

- The time interval $\Delta t_1^{/}$ is the real interval in frame B, (means essence in B), while we see it as phenomena observed by A.

We believe here that Einstein some physicists made a mistake in their interpretations when they mistook the phenomenon (apparent Lorentz time transformation) for the essence (real time).

We shall write the following version of the logical connections which preserve the equality of inertial frames and present no epistemological error.

$$\Delta t \langle \Delta t_1^{/} ; \ \Delta t^{/} \langle \Delta t_1 ; \ \Delta t = \Delta t^{/}; \Delta t_1 = \Delta t_1^{/}$$

Now the interpretation of the connections is clear:

- $\Delta t = \Delta t^{/} = \Delta t_0$. The proper times in all inertial frames are common and identical to world time.

- $\Delta t_1 = \Delta t_1^{/}$. The inertial frames are equivalent and the phenomena are reflected from one frame to the other symmetrically.

### B- Length contraction

For the length contraction, we discuss it in the same way we followed when dealing with the clock paradox.

We assume that we have two segments $\Delta L$ and $\Delta L^{/}$ exactly similar and which are fixed in two different frames subleased by observers **A** and **B**.

a) If the frames are at rest relatively to one another. the two observers, in the different frames, measure the same length for both segments, i.e. $\Delta L = \Delta L^{/} = \Delta L_1 = \Delta L_1^{/} = \Delta L_0$ Where: $\Delta L$, and $\Delta L^{/}$ are respectively the lengths which are measured by the observers in two different frames. $\Delta L_1$: the length in the first inertia frame which is measured by the observer in second frame. $\Delta L_1^{/}$: the length in the second inertia frame which is measured by the observer in first frame.

b) Let the reference frames, where the segments are fixed, move one relatively to the other with a uniform velocity. The Lorentz transformation gives $\Delta L \rangle \Delta L_1^{/} ; \ \Delta L^{/} \rangle \Delta L_1 \ ; \Delta L = \Delta L^{/}, \Delta L_1 = \Delta L_1^{/}$ We say the similar idea about length contraction, as the idea which we noticed in "time dilation":

- $\Delta L = \Delta L^{/} = \Delta L_0$. The proper lengths in all inertial frames are common and identical.

- $\Delta L_1 = \Delta L_1^{/}$ The inertial frames are equivalent and the phenomena are reflected from one frame to the other symmetrically.

Einstein and some physicists assumed that:

- $\Delta L_1$ is the real length in frame A, (means essence in A), while we consider it as phenomena observed by B.

- the say the thing for $\Delta L_1^{/}$.

## II Conclusion and suggestions

The Lorentz transformations are the basic equations of Special Relativity. They are purely mathematical. In so far as the theory is considered to have physical implications (physics deals with the essence, not with phenomenon), these implications must be the result of the interpretation of mathematical expressions in physical terms.

We suggest that the phenomena which are discussed above rise because of the following:

- The two frames are linked with a constant velocity (light velocity)

- The results are interpreted as if the observer were in both frames at the same time.

We suggest, finally, that before teaching relativity to students, we should give a brief philosophical introduction with many geometrical examples in order to clarify the differences between the essence and the phenomenon.

## References

[1] G. Burniston Brown What is wrong with Relativity.. Modern physics p.1-14.
    Albert Einstein : Philosopher Scientist . Harper and Brother. 1959 New York.
[2] Einstein, A . Principle of Relativity. Ann Phys. Lpz, 17,891 . 1905 Voronezh University Press, pp 89-105 1987.
[3] Kulingin,V.A., Kuligina, G.A. and Korneva Paradox of relativity Mechanics and Electrodynamics, . deposited with VINITI ,July 24 1990 Moscow (in Russian).
[4] Kulingin,V.A., Kuligina, G.A Causality And physical interactions in: Determinism and Modern Science ,
[5] Fock, V .Theory of Space, Time and Gravitation.., 1959 (London: Pergamon Press ). P.401

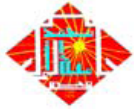 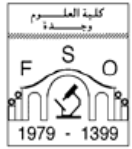 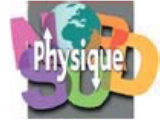 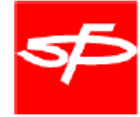

# La recherche et l'enseignement de la Physique
# En féminin à l'université Mohamed Premier

**Fatiha MAAROUFI, Professeur**

Laboratoire de Physique Théorique, Physique des Particules et Modélisation, Faculté des Sciences, Université Mohamed Premier Oujda, Maroc

**Résumé**

Le nombre de femmes physiciennes dans l'enseignement supérieur au Maroc et au maghreb, reste faible par rapport à celui aux pays du nord [1].
Dans cette communication nous donnons pour le première fois une étude comparative du nombre d'enseignantes par rapport à celui des enseignants à l'université d'Oujda.

## 1 Introduction

A l'université Mohamed Premier nous observons une sous représentation des femmes. Parmi 611 enseignants 75 sont du sexe féminin soit un pourcentage égale à 12.5%. Seules six enseignantes pratiquent dans le domaine de la physique soit 1% du nombre total d'enseignants chercheurs et 9% des femmes enseignantes.

## 2 Enseignement féminin en chiffre

### 2.1 Université

| Faculté ou Ecole | Hommes et Femmes | Femmes | Pourcentage |
|---|---|---|---|
| Lettres et Sciences Humaines | 158 | 32 | 20 |
| Sciences Juridiques Economiques et Sociales | 77 | 6 | 8 |
| Sciences | 264 | 28 | 11 |
| Pluridisciplinaire de Nador | 34 | 5 | 15 |
| Ecole Supérieure de Technologie | 31 | 3 | 10 |
| Ecole des Sciences Appliquées | 23 | 1 | 4 |
| Ecole de Commerce et de Gestion | 24 | 0 | 0 |
| Total | 611 | 75 | 12.5 |

TABLEAU. 1: Pourcentage des femmes enseignantes à l'université Mohamed Premier.

### 2.2 Faculté des Sciences

| Département | Hommes et Femmes | Femmes | Pourcentage |
|---|---|---|---|
| Biologie | 60 | 11 | 18 |
| Chimie | 58 | 6 | 10 |
| Géologie | 32 | 4 | 12.5 |
| Mathématiques et Informatique | 50 | 2 | 4 |
| Physique | 64 | 5 | 8 |

TABLEAU. 2: Pourcentage des femmes enseignantes à la Faculté des Sciences.

La comparaison des femmes enseignantes par rapport aux enseignants à l'université montre une représentation inégale des femmes dans toutes les disciplines.
Le tableau 1, montre une forte inclinaison des femmes vers les Lettres et les Sciences Humaine à l'échelle de l'université. Alors qu'à léchelle de la Faculté des Sciences (tableau 2), nous observons une nette orientation des femmes vers la biologie et la chimie. Les motifs et les facteurs qui déterminent le choix de ces formations par les femmes pourraient être expliqués par une bonne correspondance avec 'leurs qualités'; capacité d'apprendre par cœur, minutie, amour du concret et du vivant….

## 3 Femmes en Physique

La physicienne assume les mêmes tâches d'enseignement et de responsabilité que sont collègue cependant, elle est souvent confrontée à des réactions négatives de la part des étudiants et à un blocage moral de la part de ses collègues hommes et femmes. Ces réactions sont fortement liées au système éducatif marocain; la physique est perçue comme une discipline contraire à l'identité attendue de la femme. L'enseignante doit continuellement montrer ses capacités intellectuelles et fournir plus d'effort pour imposer sa personnalité.
Le nombre de femmes diminue encore plus dans le domaine de la recherche. Après l'obtention du doctorat d'état donc une nette amélioration de son salaire, la physicienne abandonne la recherche comme beaucoup d'autres collègues de sexe masculin. 50% des enseignantes physiciennes à l'université Mohamed premier ont un doctorat d'état.
La problématique de la non adhésion de la femme dans les travaux de recherche est complexe [2]. Pour expliquer cet abandon nous citons quelques causes qui restent à notre,

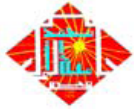 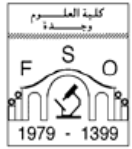 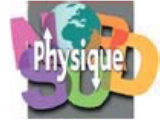 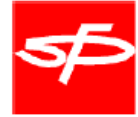

avis non suffisantes:
- les responsabilités familiales: c'est une cause universelle. Au Maroc en plus de sa petite famille, la femme reste le soutien moral de ses parents et souvent aussi de ses frères et sœurs.
- la complexité des travaux de recherche, les femmes et les hommes ne diffèrent pas quant aux difficultés éprouvées face à la recherche à l'université: manque de moyens, manque de soutien moral et matériel, les sujets de recherche restent loin des besoins sociaux, absence d'encadrement socio-économique…
- le manque d'ambition,
- l'absence de l'encouragement de l'entourage (famille, société…)

D'autre part, on remarque une sous représentativité des femmes enseignantes dans les structures de décision de l'université, 33% des physiciennes interviennent dans des postes de décision; une directrice de l'Ecole Supérieur de Technologie et une élue au Conseil de la Faculté des Sciences (Auteur de la présente communication).

## 4  Recommandations

Afin d'augmenter le nombre de femmes enseignantes dans le domaine de la physique, il faut œuvrer à éliminer les facteurs de freinage et les préjugés sociaux et culturels notamment [1,2,3]:
- Donner à la fille les mêmes chances de scolarisation que le garçon.
-
- Améliorer l'information aux études et encourager les filles à étudier la physique: parents, enseignants du secondaire, professeurs universitaires, encadrants de thèses.
- Placer les femmes dans les structures de décision et des instituts de recherche: Ministère de l'Enseignement, Présidences des Universités et Facultés des Sciences et Ecoles, Conseils et Comité de Recherche des établissements…
- Améliorer l'ambiance professionnelles pour les femmes en créant des structures d'encadrement pour les enfants.

## Références

[1]  Conférence sur les femmes en physique, organisée par l'Union Internationale de Physique Pure et Appliquée (2002), http:/www.if.ufrgs.br/iuap.
[2]  Monia Cheikh, communication orale au 1er congrès 'Nord-Sud' sur la recherche et l'enseignement de la Physique, Oujda 9-13 avril 2007.
[3]  A.Gilbert et al, Rapport de recherche et recommandations, 'Promotion des femmes dans les formations supérieures techniques et scientifiques' (2003).

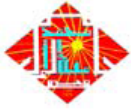 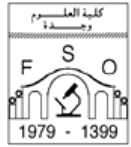 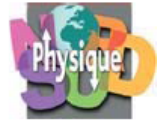 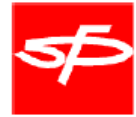

# Extrapolation du champ des antennes utilisant un développement modal

**Zouhir MAHANI**[(1)], Abdeliah HAKIM[(1)], Abdelhak ZIYYAT[(2)]

(1) F.S.T Gueliz, Université Cadi Ayyad Marrakech, Maroc
(2) Faculté des Sciences, Université Mohammed Premier, Oujda, Maroc

**Résumé**

Ce papier présente l'extrapolation du champ d'une antenne, à partir des données sur une surface fermée englobant cette antenne, cette extrapolation repose sur le développement du champ en ondes sphériques.

## 1 Introduction

Un des problèmes les plus courant en électromagnétisme est l'extrapolation du champ électromagnétique d'une structure rayonnante. Selon le principe de Huyguens, la connaissance du Champ électromagnétique au voisinage d'une structure rayonnante permet de définir ce champ n' importe où en espace. Différentes méthodes mathématiques existent et permettent de faire ce calcul, et elles donnent de bons résultats. Cependant lorsqu'on traite les grandes fréquences, on aura des systèmes d'ordres $1/\lambda^3$ ($\lambda$ est la longueure d'onde), dont la résolution est très coûteuse, D'autre part pour évaluer les intégrales il faudrait échantillonner avec un pas d'ordre $\lambda/10$, ce qui les rend sans intérêt.

## 2 Méthode

La méthode proposée ici, repose sur un développement modal du champ. La base du développent est issu de la géométrie sphérique, et permet ainsi la sépparation de la dépendance du champ. L'équation qui régit le rayonnement est celle de Maxwell :

$$\nabla \otimes E = -iwB \qquad \nabla . B = 0$$
$$\nabla \otimes H = -iwD + J \qquad \nabla . D = Q_v \qquad (1)$$

Avec :

$E$ : Le champ électrique  $H$ : Le champ magnétique,
$B$ : La densité du flux magnétique. $D$ : La densité du flux électrique. $J$ : La densité du courant électrique.
$Q_v$ : La densité de charge électrique.

Dans un milieu libre de source, homogène, linéaire et isotrope, l'équation de Maxwell se réduit en équation de Hélmholtz. Tout d'abord on va la résoudre en considérant une décomposition spectrale du champ :

$$\psi(r,\theta,\varphi) = \sum_{m,n} Q_{mn} z_n(kr) \overline{P}_n^m(\cos\theta) e^{im\varphi} \quad (2)$$

## 3 Conclusion

Dans ce travail, une attention toute particulière a été donnée aux aspects de la précision et de la rapidité de cette technique afin de répondre aux exigences industrielles. En effet, une implémentation classique de cette méthode nécessitait plusieurs heures d'exécution. L'algorithme proposé par contre, sollicite un temps d'exécution de moins d'une minute pour les plus grandes antennes testées. Le nombre de points de mesure a été limité au minimum en prenant en compte la bande spatiale nécessairement limitée de l'antenne sous test (échantillonnage à la demi-longueur d'onde). En coordonnée sphérique, cette technique de mesure indirecte a bien montré tout son intérêt notamment en procurant une excellente précision pour des antennes de géométrie variées et de rapports dimensions à longueur d'onde très élevés.

L'équation (2) reste valable presque partout dans l'espace et non uniquement en champ lointain. Cette étude se poursuivra en s'intéressant à l'évaluation du champ au voisinage de l'antenne. Cette caractérisation en zone proche répond actuellement à un besoin économique et sociétale. On cite en particulier la nécessité d'évaluer les périmètres de sécurité des antennes de télécommunication mobile (antennes de station de base et de téléphone mobile) en vue de leur conformité aux recommandations sanitaires, nationales et internationales, fixant les limites d'exposition des personnes aux champs électromagnétique.

## Références

[1] A.Ziyyat, L.Casavola, D.Picard and J.Ch.Bolomey, "Prediction of BTS Antennas Safety Perimeter from NF to NF Transformation: An Experimental *Validation*", , AMTA'2001 Symposium (Antenna Measurement Techniques

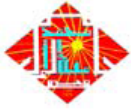 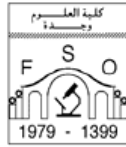 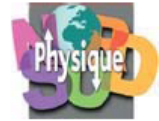 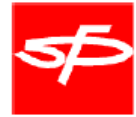

Association), Denver, Colorado, USA, October 21-26, 2001.

[2] J.Ch.Bolomey, O.M.Bucci, L.Casavola, G.D'Elia, M.D. Migliore and A.Ziyyat, "Reduction of truncation error in near-field measurements of antennas of base-station mobile communication systems ", IEEE Trans. Antennas and Propagation, Vol 52, No 2, February 2004

[3] E. Roubine, J.Ch. Bolomey, "Antennes", Masson, vol. 1, p. 35, 1986.

[4] J. Hansen, "Spherical near-field measurements", Peter Peregrinus, 1988

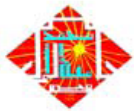 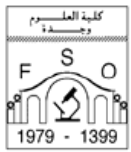 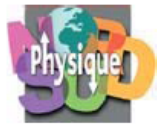 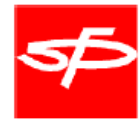

# SYNTHESE ET CARACTERISATION PHYSICO-CHIMIQUE DE NOUVEAUX POLYMERES CONJUGUES

**Rachid Mahy**[(a)], B. Bouammali[(a)], A. Oulmidi[(a)], A. Chellioui[(a)], D. Derouet[(b)]

(a) Laboratoire de Photochimie et Chimie Macromoléculaire, Université Mohamed I, Oujda, Maroc.
(b) LCOM-Chimie des Polymères (UMR du CNRS UCO2M N°6011), Université du Maine, Faculté des Sciences, Avenue Olivier Messiaen, 72 085 Le Mans Cedex 9, France

**Résumé**

Le travail proposé consiste en la synthése et l'étude physico-chimique de nouveaux polymères photoréticulables porteurs en chaîne latérale des groupes cinnamates comme unité photosensible.

## 1 Introduction

Les polymères contenant des groupes photoréactifs et conduisant à la formation des réseaux tridimensionnels (polymères réticulés) sous l'action des radiations lumineuses, sont appelés polymères photoréticulables (schéma).

Ce type de macromolécules suscite l'intérêt industriel dans des domaines très variés tels que l'art graphique, l'électronique, les revêtements, etc.

En art graphique, les résines photoréticulables interviennent principalement dans l'élaboration des couches photosensibles pour les plaques d'impression offset.

Dans le domaine de l'électronique, ils jouent un rôle vital dans la fabrication des circuits imprimés ou intégrés. Pour ce qui est des revêtements, les résines photosensibles peuvent améliorer nettement les propriétés de surface d'un matériau. En effet, divers types de matériaux (bois, plastique, métaux, fibres optiques, papier, textile, etc.) sont actuellement protégés à l'aide de vernis, d'où l'augmentation de la durabilité de ces matériaux en ralentissant les phénomènes de dégradation, d'oxydation, de corrosion, etc.

Parmi les polymères photoréticulables potentiellement utilisables dans la technologie des photorésists, ceux portant des groupes α,β-insaturés tels que les dérivés cinnamiques, les chalcones, les coumarines, les α-furylacryliques, les styrylacryliques, l'α-phénylmaléimide ont été largement explorés [1-3].

La propriété photochimique recherchée pour les applications citées ci-dessus, est la réaction de photodimérisation entre les groupes α,β-insaturés par cycloaddition [2Π +2Π] qui conduit à la formation de structures cyclobutaniques, et donc à la réticulation des polymères supports.

Notre projet de recherche consiste à synthétiser de nouveaux polymères photoréticulables porteurs en chaîne latérale des groupes cinnamates ou dérivés comme unité photosensible.

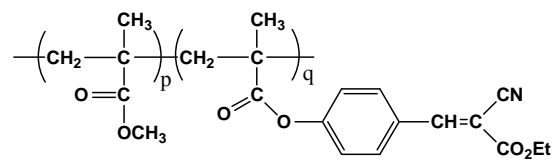

Homopolymère (p=0, q#0) et Copolymers (p#0, q#0)

**Exemple d'application**

Le processus de photolithographie :

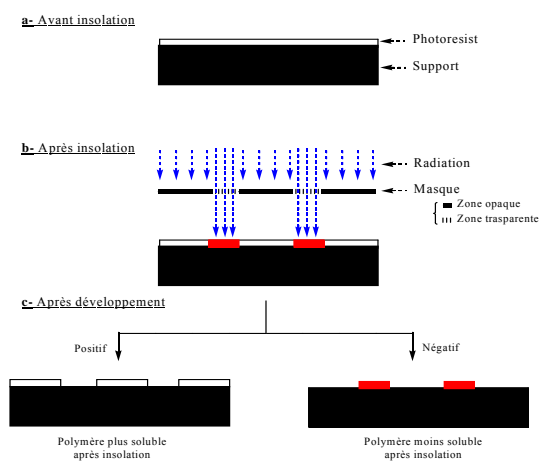

**Fig1 :** Principe de formation de l'image par la techniquede photolithographie

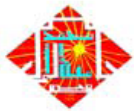 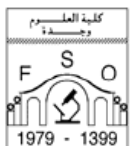 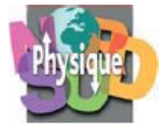 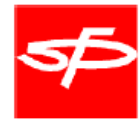

## 2 Partie expérimentale :

Les polymères photoréticulables préparés sont identifiés et analysés par différentes méthodes spectroscopiques : chromatographie exclusion stérique, la résonance magnétique nucléaire et IR, l'homopolymère poly(2-cyano-3-(4-(méthacrylate) phényl) prop-2-ènoate d'éthyle) est pris comme exemple pour illustrer les différentes méthodes d'analyse utilisées:

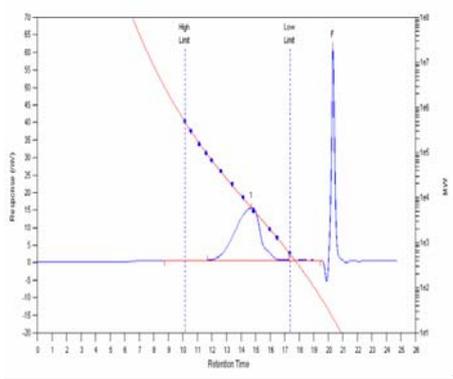

**Chromatographie exclusion stérique**

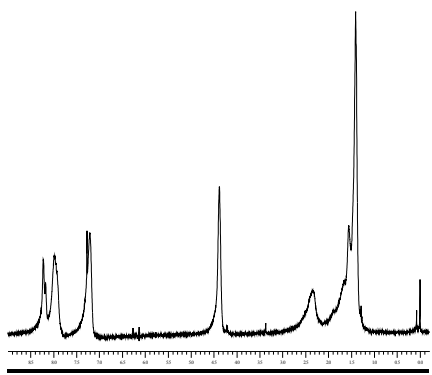

**RMN $^1$H**

**IR**

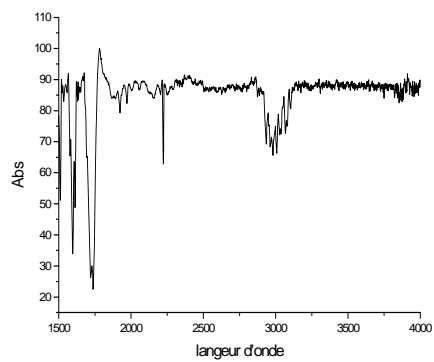

### Etude photochimique :

Les polymères préparés sont soumis à des radiations lumineuses. La progression de la photoréticulation a été suivie uniquement par spectrométrie UV et IR. La RMN n'a pas été utilisée en raison de l'insolubilité des photoréticulats obtenus dans les solvants.

## 3 Conclusion

Divers homo- et copolymères photoréticulables à unités cinnamiques pendantes ont été synthétisés par polymérisation radicalaire à partir de monomères portant à la fois un site polymérisable (le groupe méthacrylique) et un site photoréactif (le groupe cinnamique). La photoréticulation est également confirmée par le fait que le film irradié devient insoluble.

### Remerciements

## Références

[1] (a) M.S. Mizra, N.G. Navale, D.S. Sadafule, C.G. Kumbar, S.P. Panda, *J. Macromol. Sci., Chem. A*. **27**, (1):1(1990).

(b) D. Feng, M. Tsushima, T. Matsumoto et T. Kurosaki, *J. Polym. Sci. A: Polym. Chem*. **36**, 685-693(1998).

[2] (a) M. Obi, S. Morino et K. Ichimura, *Macromol Rapid Commun*. **19**, 643-646(1998).

(b) M. Tsuda et H. Nakanishi, *J. Polym. Sci. A-1*. **7**, 259-264(1969)

[3] K. Ichimura, S. Watanabe et H. Ochi, *J. Polym. Sci : Polym. Lett,* **14**, 207-209(1976

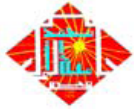 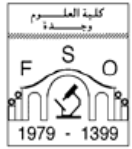 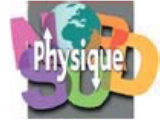 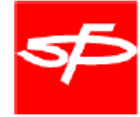

# MESURE DE LA COCENTRATION DU RADON DANS LES HABITATIONS DE LA VILLE D'OUJDA

**A. Moussa**, H. Dekhissi, J. Derkaoui, M. Hamal

Laboratoire de Physique Théorique, de physique des Particules et Modélisation
Faculté des Sciences d'Oujda

**Résumé**

Le radon est un gaz rare radioactif, c'est un produit de filiation de l'uranium et du thorium, il apparaît dans les roches et les sous-sols.

Le radon ou ses produits de désintégration présentent un grand risque dû à la radiation ionisante. Le degré de danger dépend essentiellement de la concentration du radon et de la période de temps d'exposition à cette concentration.

Nous Présentons les premieres mesures de la concentration du radon à Oujda. Les mesures ont été faites en utilisant le détecteur à chambre à ionisation AlphaGuard.

## 1 Introduction

Le radon appartient à la famille des gaz rares (hélium, néon, krypton...) et en possède les propriétés chimiques: inodore, incolore, sans saveur, ne réagissant pas chimiquement avec les autres éléments.

C'est le seul gaz rare à être naturellement radioactif. Il provient de la chaîne de désintégration de l'uranium 238, naturellement présent dans toute l'écorce terrestre. Le radon 222 est le descendant direct du radium 226.

Le radon et les produits de filiation du radon (connus par leur durée de vie courte) sont présents partout dans le sol, dans l'eau et dans l'air, en quantités plus ou moins grandes.

Le radon s'infiltre dans les habitations par les fissures, les jointures, les canalisations... Il provient principalement du sous-sol (roches, failles, eaux....), mais aussi des matériaux de construction qui contiennent, en plus ou moins grandes quantités, du radium.

Étant donné que le radon est un gaz, les variations de pression atmosphérique influent aussi sur son taux d'émission à partir du sol et sur son taux d'accumulation dans l'air des bâtiments.

Le plancher et les murs en béton des sous-sols ralentissent l'infiltration, dans les bâtiments, du radon contenu dans le sol. Cependant, les fissures dans le plancher, les jonctions dalle/mur et la tuyauterie d'évacuation permettent au radon de s'infiltrer dans un bâtiment.

Les teneurs en radon à l'intérieur des bâtiments sont presque toujours plus élevées que celles à l'extérieur. Une fois qu'il a pénétré dans un bâtiment, le radon ne peut pas s'échapper facilement. Les teneurs en radon sont généralement les plus élevées dans les caves et les sous-sols, car ces endroits sont ceux qui se trouvent le plus près de la source et ils sont habituellement mal ventilés.

## 2 L'accumulation du radon

Le radon émane du sol et se dilue rapidement dans l'air extérieur. Sa concentration moyenne est alors plus importante à proximité des gisements uranifères et des secteurs miniers.

Il va par contre s'accumuler dans tous les espaces qui sont peu ou mal aérés (les cavités naturelles mais aussi les habitations). Cette accumulation représente un risque pour la santé.

Les zones à risque correspondent à un sol contenant des roches riches en uranium (roches magmatiques,« granites »).

L'inhalation du radon et surtout de ses descendants radioactifs provoque des lésions au niveau des cellules pulmonaires, ce qui augmente le risque de développer un cancer du poumon. Le risque augmente avec la concentration et la durée d'exposition.

Le ministère de l'Environnement aux États-unis conseille d'agir dès que la concentration dépasse 150 Bq/m3. Sur un plan international, la CIPR (Commission Internationale de Protection Radiologique) recommande un seuil d'intervention entre 200 Bq/m3 et 600 Bq/m3.

Elle considère que passer 80% du temps dans un habitat à 200 Bq/m3, peut conduire à la survenue d'un cas de cancer du poumon supplémentaire pour 5000 personnes exposées.

Beaucoup d'étude systématique ont été faites à travers le monde pour évaluer la contamination par le $^{222}$Rn. Il est important de rappeler que la moitié de la radioactivité environementale est dûe au radon.

## 3 Mesure de la concentration du radon

La mesure de la concentration du radon a été faite en utilisant le detecteur actif AlphaGuard. Il fonctionne par diffusion dans une chambre d'ionisation, il permet de mesurer en même temps la température, la pression et l'humidité; il permet aussi d'effectuer des mesures dans différents milieux: l'air, l'eau, le sol.

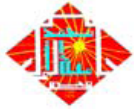 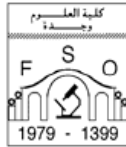 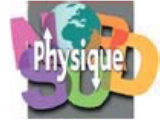 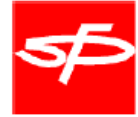

Il nous permet aussi d'éffectuer l'étude de variation de la concentration en fonction des facteurs géologiques, météorologiques, temporels…

## 4  Résultat et discussion

Des mesures préliminaires de la concentration du radon ont été effectuées dans différentes habitations de la ville d'Oujda.

La concentration du radon augmente dans les cavités et diminue graduellement dans les niveaux les plus hauts. Les matériaux de construction peuvent contribuer dans la concentration du radon.

La ventilation joue un rôle important pour diminuer la concentration. La concentration élevée a été mesurée dans des caves avec de faible ventilation et quand les portes et les fenêtres sont gardées fermées le long de la période de mesure.

La figure 1, représente la corrélation entre la concentration du radon et la température.

La figure 2, représente la corrélation entre la concentration et l'humidité, on observe que la concentration augmente quand l'humidité diminue.

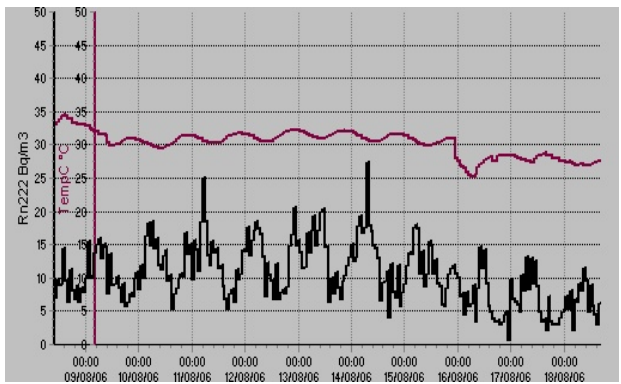

**Figure 1: Corrélation entre la concentration du radon (noir) et la température (rouge) durant la période de mesure (09/08/06 au 18/08/06)**

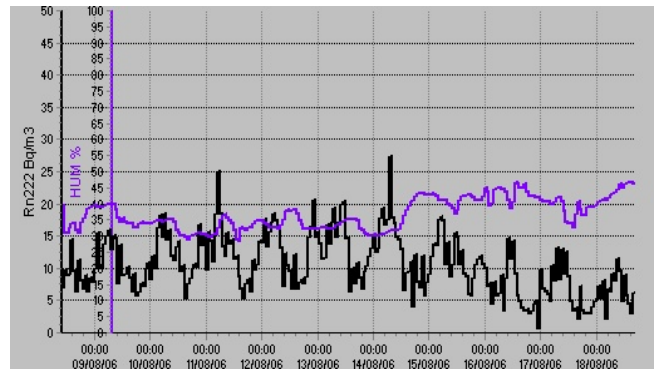

**Figure 2: Corrélation entre la concentration du radon (noir) et l'humidité (violet) durant la période de mesure (09/08/06 au 18/08/06)**

## 4 Conclusion

La concentration à l'intérieur des constructions d'Oujda est faible, elle dépend fortement de la ventilation. Les sous-sols sont les principales sources du radon dans les caves, alors que les matériaux de construction contribuent principalement dans les niveaux supérieurs.

## Références

[1]  UNSCEAR (1988) Effects and risks on ionizing radiation, *united Nations*, *New. York, pp.* **49-134**
[2]  ICRP (1993) Protection against Rn-222 at home and at work, *ICRP publication*, 65
[3]  Giacomelli, G. et al. (1990) Misure di radon nelle abitazioni di bologna, *Il Nuovo Saggiatore*, **4**, pp. 5-8 Beozzo, M. et al (1991), *Acqua Aria*, **4**, pp. 371-378.
[4]  Bassignani, A. et al. (1994) Study of radon concentration in the Gran Sasso underground lab, Review of long term radon studies, *Rad. Meas.*, **25,** pp. 557-562

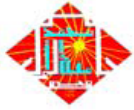 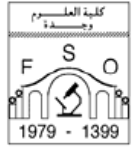 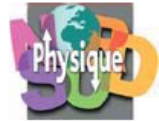 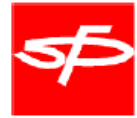

# Etude numérique de la convection forcée à l'aide de la méthode de Boltzmann sur réseau

**Mohammed Amine MOUSSAOUI**, Mohammed JAMI, Ahmed MEZRHAB

Laboratoire de Mécanique & Energétique, Université Mohamed Premier, Oujda

**Résumé**

Une étude numérique de l'écoulement et du transfert de chaleur par convection forcée, dans un canal en présence de trois obstacles chauffés ($T_c$) montés sur sa paroi inférieure et supérieure est réalisée (voir Fig.1). On a choisi comme fluide caloporteur l'air et l'écoulement est supposé laminaire, bidimensionnelle et incompressible. Un couplage entre la méthode de Boltzmann sur réseau et la méthode des différences finies [1] est utilisée comme schéma numérique pour déterminer la vitesse d'écoulement et la température de l'air. Le nombre de Reynolds et l'espacement entre les obstacles ($w$) sont variés pour analyser leurs influences sur l'écoulement et le transfert de chaleur.

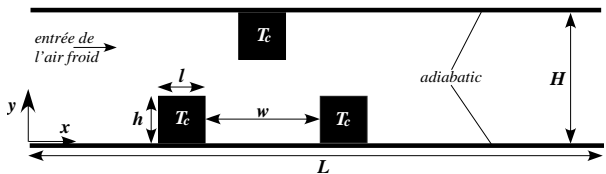

Fig.1 : Configuration étudiée

## 1 Méthode numérique

Nous considérons un modèle bidimensionnel de la méthode LBE avec neuf vitesses discrètes (modèle D2Q9) sur une grille carrée de pas $\delta x$. Dans l'étape d'advection, les particules fluides se déplacent d'un noeud de la grille vers le noeud voisin. Les vitesses discrètes sont données par :

$$e_i = \begin{cases} (0,0), & i=0 \\ (\cos[(i-1)\pi/2], \sin[(i-1)\pi/2])c, & i=1-4 \\ (\cos[(2i-9)\pi/4], \sin[(2i-9)\pi/4])\sqrt{2}c, & i=5-8 \end{cases} \quad (1)$$

Nous supposons fixée la vitesse de la grille $c = \delta x/\delta t$ avec $\delta t$ le pas de temps. Ainsi, en prenant $c=1$, dans les unités de $\delta x = 1$ et $\delta t = 1$, dans tous ce qui suit, toutes les quantités sont données en unités non dimensionnelles.

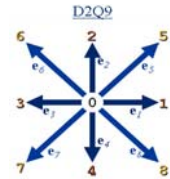

L'équation d'évolution temporelle de l'état du fluide suit l'équation générale :

$$f_i(x+e_i, t+1) = f_i(x,t) + \Omega f_i(x,t), \quad i=0,1,...,8 \quad (2)$$

où $f_i$ est la fonction de distribution d'une particule et $\Omega$ est l'opérateur de collision. La linéarisation de cet opérateur autour de la fonction de distribution à l'équilibre locale $f_i^{eq}$ apporte une simplification importante de la méthode LBE. Avec l'approximation de BGK [2], l'équation (2) s'écrit :

$$f_i(x+e_i,t+1) = f_i(x,t) - \frac{1}{\tau}\big(f_i(x,t) - f_i^{eq}(x,t)\big), \quad i=0,1,...,8 \quad (3)$$

Le développement de Chapman-Enskog est utilisé pour obtenir les équations de NS à partir de la méthode LBE.

A chaque nœud du domaine, on calcule un ensemble de neuf moments venant des neuf fonctions de distribution et qui sont liés par la transformation linéaire [3]

$$m = Mf \quad (4)$$

où la matrice **M** d'ordre 9 est donnée dans [3]. Les collisions modifient les moments comme suit: quelques uns sont conservés (la densité et l'impulsion mais pas l'énergie). Les six moments restant sont calculés à partir d'une simple équation de relaxation linéaire vers les valeurs d'équilibre qui dépendent des quantités conservées. Les nouvelles fonctions de distribution $f^c$ sont calculées à partir des nouveaux moments

$$f^c = M^{-1} m^c \quad (5)$$

La densité et la vitesse sont donnés par:

$$\rho(x,t) = \sum_i f_i(x,t) \quad (6)$$

$$\vec{u}(x,t) = \sum_i f_i(x,t) e_i / \rho(x,t) \quad (7)$$

## 2 Résultats et discussion

*- Effet du nombre de Reynolds :*
La figure 2 montre l'apparition de tourbillons derrière chaque bloc, ces tourbillons augmentent en longueur avec l'augmentation du nombre de Reynolds, ce qui entraîne une augmentation du transfert de chaleur (voir figure 3).

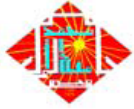
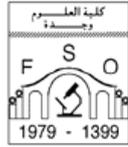
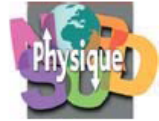
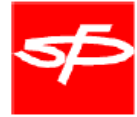

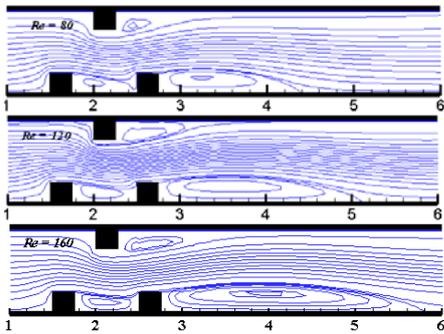
Fig2 : Lignes de courant pour Re = 80, 120 et 160

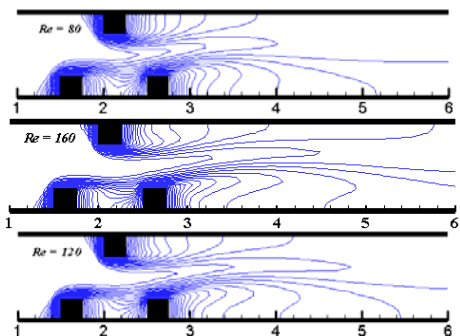
Fig3 : isothermes pour Re = 80, 120 et 160

**-Effet de la distance :**
Sur les figures 4 et 5, La réduction de l'espacement entre les deux blocs inférieurs défavorise leur refroidissement, car un espacement réduit fait obstruction au passage du fluide, minimisant ainsi le contact entre celui ci et les parois des blocs.

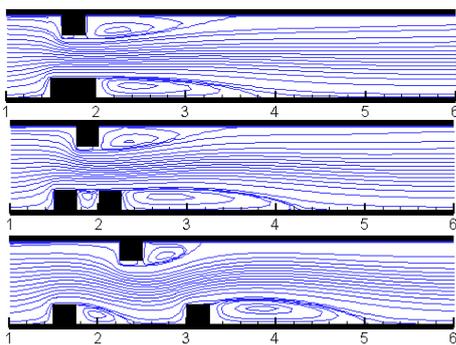
Fig4 : Lignes de courant et isothermes pour w=0, H/4 et H/2

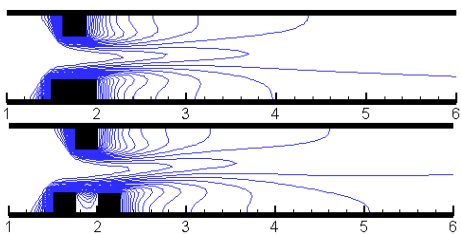
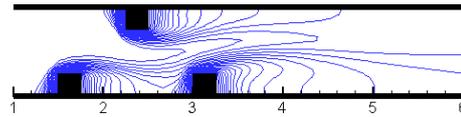
Fig5 : Lignes de courant et isothermes pour w=0, H/4 et H/2

**- Nusselt local :**
La figure 6 illustre la variation du nombre de Nusselt sur les faces d'échange du $1^{er}$ bloc. L'effet du nombre de Reynolds sur le transfert est apparent. On note aussi que la face horizontale des blocs est mieux refroidie que les autres faces.

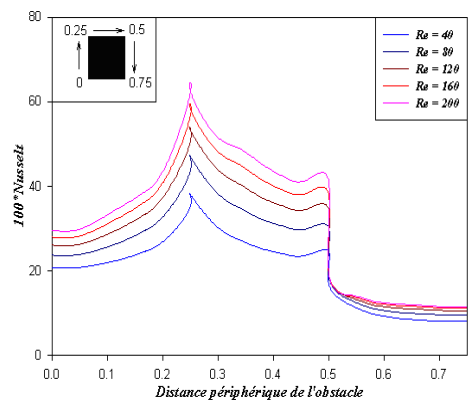
Fig6 : Nombre de Nusselt sur le $1^{er}$ bloc

## 3 Conclusion

En conclusion, le refroidissement des blocs est d'autant meilleur que le nombre de Reynolds est important.
La face horizontale est mieux refroidie que les autres faces

## References

[1] A. Mezrhab, M. Bouzidi, P. Lallemand, Hybrid lattice Boltzmann finite-difference simulation of convective flows, *Computer and Fluids*, 33, (2004), 623-641.
[2] P. Bhatnagar, E.P. Gross, M. Krook, Phys. Rev., vol 94, p511, 1954.
[3] P. Lallemand, L. S. Luo, Theory of the lattice Boltzmann method: dispersion, dissipation, isotropy, galilean invariance and stability, *Phys. Rev.*, E61, (2000), 6546-62.

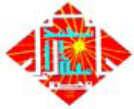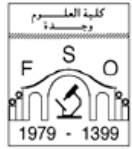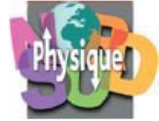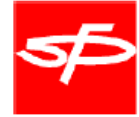

# Enseignement de la physique moderne et difficultés des étudiants

**M. OLDACHE, R. LADJ et C.E.KHIARI**

Ecole Normale Supérieure de Kouba - Alger
Département de Physique - Laboratoire de Didactique des Sciences
16 050 Vieux Kouba - Alger

**Résumé**

La transition de la Physique Classique à la Physique moderne nécessite plus qu'un effort d'abstraction ou une capacité à manier des formalismes mathématiques sophistiqués. Elle requiert une véritable rupture épistémologique et un changement radical de paradigmes. L'histoire des sciences nous apprend que tous les efforts déployés par les physiciens pour "expliquer" les phénomènes se déroulant à l'échelle microscopique dans le cadre de la physique classique ont été vains. Après avoir posé la problématique de l'enseignement de la physique moderne, nous procédons à un survol historique puis nous exposons les difficultés les plus courantes tout en essayant d'en déterminer les causes et, enfin, nous proposons des solutions de remédiation.
Mots clés : didactique - physique moderne - épistémologie - concepts - mécanique quantique - relativité

## 1 APERCU HISTORIQUE ET FONDEMENTS EPISTEMOLOGIQUES

La physique moderne s'appuie sur la philosophie de l'expérience, autrement dit une attitude intellectuelle consistant à ne retenir que les concepts qu'on peut mesurer expérimentalement, au moins en principe. C'est pourquoi Einstein, Bohr, Heisenberg les autres fondateurs de la moderne ont abondamment utilisé les expériences de pensée. Cette attitude amena notamment Einstein à rejeter les notions d'espace et de temps absolus pour les remplacer par un espace et un temps relatifs à un observateur donné.
L'élaboration de la mécanique quantique, quant à elle, commença par l'introduction du concept de quantification pour aboutir à l'équation de Schrödinger. Le principe d'incertitude Heisenberg ruina le déterminisme en physique et ouvrit la voie à l'Ecole de Copenhague.

## 2 BSTACLES DIDACTIQUES

Les principales difficultés rencontrées sont les suivantes :
-Les étudiants tiennent pour "exact" le modèle de Bohr et ce, malgré notre insistance sur les hypothèses sur lesquelles repose le modèle. Ceci est du peut être au fait que le concept "modèle" est mal acquis.
**-**Les étudiants ont du mal à comprendre le concept du corps noir**.**
-Difficulté d'admettre un espace-temps à 4 dimensions, preuve que les étudiants restent "attachés" aux notions d'espace et de temps absolus.
- Difficulté à admettre la dualité onde-corpuscule, les étudiants étant habitués à décrire une « réalité » donnée par un concept unique.

## 3 PROPOSITIONS DE REMEDIATION

L'analyse des problèmes et difficultés relevés ci-dessus montre que ceux-ci sont liés surtout à la compréhension des concepts de base de la mécanique quantique et de la relativité. Pour y remédier, nous proposons :
- De donner l'importance nécessaire à des thèmes tels que les relations d'Ehrenfest, les approximations semi-classiques, les conditions de validité des relations de la physique classique, etc.
- Enseigner le principe d'incertitude de Heisenberg graduellement tout en approfondissant son sens à chaque fois.
- Introduire la notion de fonction d'onde en tenant compte des différentes écoles d'interprétation et en suivant l'évolution historique de ce concept.

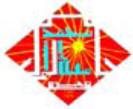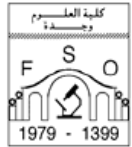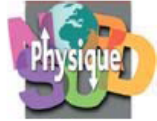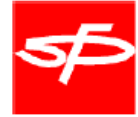

# 4 CONCLUSION

Les difficultés d'enseignement de la physique moderne sont dus au fait que le formalisme mathématique est trop favorisé au détriment de la discussion des idées de base. Nous pensons que l'introduction d'éléments d'épistémologie et d'histoire de la physique moderne pourrait aider à renforcer la compréhension chez les étudiants. Par ailleurs, les problèmes et difficultés soulevés ici et concernant l'enseignement de la physique moderne à l'université sont, pour l'essentiel des témoignages d'enseignants ayant une certaine expérience dans l'enseignement de cette discipline. Le diagnostique que nous avons établi a été étayé partiellement par des tests et des analyses de copies d'examens. Mais il reste un diagnostique partiel (et peut-être partial) qui mérite d'être confirmé ou infirmé par des questionnaires précis et systématiques.

## Références

[1]  A.Einstein, L.Infeld, l'évolution des idées en physique, Trad. M. Solovine, Ed. Champs Flammarion, Paris 1983
[2] M. Planck, Verh. De Deutsh. Physik Gesellscaff, dec. 1900, 237
[3]  N.Bohr, Phil. Mag. t.26(1913)1, 476
[4]  L. de Broglie, Recherches sur la théorie des quanta, thèse insérée aux Annales de Physique, $10^{\text{ème}}$ série, t.III(1925)22
[5]  D'Abro, The rise of the new physics, Dover Publications 1939
[6]  Paul Davies, The new physics, Cambridge University Press 1993
[6]  R.Ladj et M.Oldache, in "Cinquième Congrès de la Physique et ses Applications", Université Hadj Lakhdar, Batna 2002
[7]  L.I. Schiff, Quantum Mechanics, Third Edition, McGraw-Hill Inc. 1968
[8]  C.E.Khiari and M.Oldache, in "Quatrième Congrès de la Physique et ses Applications", Sidi Fredj 2000
[9]  M.Oldache, in "Première Conférence Nationale sur la Didactique de la Physique", Université Hadj Lakhdar, Batna 1999
[10]  M.Oldache, in "Séminaire National sur le Rapport entre la Culture Scientifique et le Savoir à transmettre", Université Essénia, Oran 06-07 novembre 2001
[11] Wichmann, Quantum Physics (Berkeley Physics Course), Vol.4, McGraw-Hill 1967
[12] Feynman, The Feynman Lectures on Physics, California Institute of Technology 1965
[13]  J.M. Lévy-Leblond et F.Balibar, Quantique, InterEditions, Paris 1984

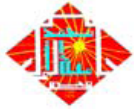 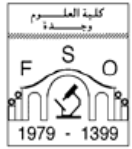 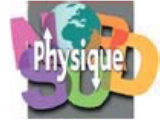 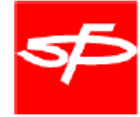

# Etude numérique du couplage convection naturelle - rayonnement dans un dispositif du type mur Trombe

**Mohammed RABHI, Ahmed MEZRHAB, Hassan NAJI**

Faculté des sciences, Laboratoire de Mécanique & Energétique, Oujda, Maroc

## 1 Introduction

Plusieurs études ont été réalisées sur le transfert de chaleur par convection naturelle et par rayonnement thermique dans des cavités rectangulaires partitionnées [1-3].
Nous étudions l'effet de deux ouvertures situées prés de la paroi chaude, l'une en bas et l'autre en haut de la cavité (figure 1) sur le transfert de chaleur au sein d'une cavité différentiellement chauffée en tenant compte des couplages entre la conduction thermique à travers les parois, les échanges par convection naturelle dans et entre les compartiments et les échanges par rayonnement entre les surfaces radiatives constituant l'enceinte.

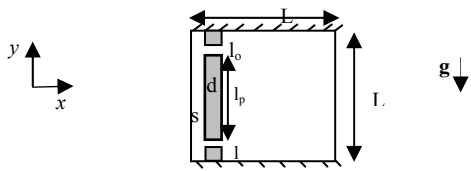

Figure 1. Schéma de la cavité de type mur Trombe

## 2 Formulation mathématique et procédure numérique

Nous supposons sur la figure 1, la géométrie bidimensionnelle, l'écoulement laminaire, les surfaces radiatives grises et isotropes en émission/réflexion et les propriétés physiques de l'air, à part sa densité, constantes et prises à la température moyenne $T_o$. Seules les surfaces solides participent aux échanges par rayonnement thermique. Les parois horizontales de la cavité sont adiabatiques, tandis que les parois verticales gauche et droite sont respectivement chaude et froide.
Les équations adimensionnelles gouvernant le transfert de chaleur et l'écoulement dans la cavité s'écrivent sous la forme :

$$\frac{\partial U}{\partial X} + \frac{\partial V}{\partial Y} = 0 \quad (1)$$

$$U\frac{\partial U}{\partial X} + V\frac{\partial U}{\partial Y} = -\frac{\partial P}{\partial X} + \lambda Pr\left(\frac{\partial^2 U}{\partial X^2} + \frac{\partial^2 U}{\partial Y^2}\right) \quad (2)$$

$$U\frac{\partial V}{\partial X} + V\frac{\partial V}{\partial Y} = -\frac{\partial P}{\partial Y} + \lambda Pr\left(\frac{\partial^2 V}{\partial X^2} + \frac{\partial^2 V}{\partial Y^2}\right) + RaPr\theta \quad (3)$$

$$U\frac{\partial \theta}{\partial X} + V\frac{\partial \theta}{\partial Y} = R_k\left(\frac{\partial^2 \theta}{\partial X^2} + \frac{\partial^2 \theta}{\partial Y^2}\right) \quad (4)$$

avec $\lambda = 1$, $R_k = 1$ dans la région fluide et $\lambda = \infty$, $R_k = k_m/k_a$ dans le solide. $k_a$, $k_m$ sont les conductivités thermiques respectivement de l'air et du mur.
Les conditions aux limites sont : sur les parois de la cavité: $U = V = 0$

$X = 0$, $0 \leq Y \leq 1$ : $\theta = 0.5$
$X = 1$, $0 \leq Y \leq 1$ : $\theta = -0.5$

Le long du mur :

$$R_k \frac{\partial \theta_m}{\partial n} = \frac{\partial \theta_a}{\partial n} - Nr\,Qr$$

$Qr = qr/\sigma T_c^4$, densité du flux radiatif adimensionnel.
$Nr = \sigma T_c^4/(k_a \Delta T/L)$, nombre de rayonnement.

Avec sur les parois adiabatiques

$$0 = \frac{\partial \theta_a}{\partial Y} - Nr\,Qr$$

***n*** : direction normal à la surface radiative considérée
Les équations (1-4) sont discrétisées à l'aide de la méthode des volumes finis. Le couplage pression-vitesse est traité à l'aide de l'algorithme SIMPLER. Les facteurs de forme, avec effet d'obstacles, ont été déterminés par les méthodes des éléments de frontières et de Monte Carlo [4]. Les équations, du système algébrique, obtenues ont été résolues par la méthode des gradients conjugués.
Après étude, nous avons trouvé que le maillage 60×60, irrégulier et fin aux voisinages des parois de la cavité et des faces du mur, permet d'avoir un bon compromis entre la précision des résultats et le temps de calcul. Lorsqu'on prend en compte le rayonnement thermique, nous avons considéré une température moyenne $T_0 = 290K$. Afin de respecter la validité de l'approximation de Boussinesq, la différence de température $\Delta T$ entre les parois chaude et froide est choisie inférieure ou égale à $15K$. Le nombre de Prandtl a été fixé à $Pr = 0.71$. $R_k$ est fixé à 50 dans tous les calculs, alors que le nombre de Rayleigh $Ra$ et la largeur adimensionnelle de l'ouverture $L_0$ ont été variés de $10^3$ à $10^7$ et de 0 à 0.2, respectivement. Les extrémités haute et basse du mur central sont à la même distance des deux parois adiabatiques et sa hauteur est fixée à $0.6L$. En conséquence, l'augmentation des largeurs des deux ouvertures provoque la diminution des largeurs l des murs haut et bas attachés respectivement aux parois adiabatiques.
Notre objectif dans ce travail, est d'étudier les effets du rayonnement thermique, du nombre de Rayleigh et de la largeur des ouvertures sur le nombre de Nusselt moyen,

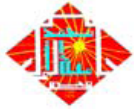 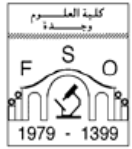 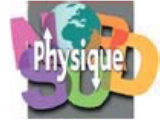 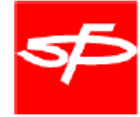

$$Nu = \int_0^1 \left( -\frac{\partial \theta}{\partial X}\bigg|_{X=0,Y} + Nr\, Qr(X=0,Y) \right) dY$$

et le débit d'air adimensionnel à travers les ouvertures:

$$Q = -\int_{l/L}^{0.2} U(X=S,Y)\,dY = \int_{0.8}^{(0.8+L_0)} U(X=S,Y)\,dY$$

La comparaison entre les résultats obtenus en convection naturelle pure et en convection naturelle combinée au rayonnement thermique permet de mettre en évidence l'influence du rayonnement thermique.

## 3 Remarques et discussions

Chaque configuration dépend au minimum de quatre paramètres adimensionnels ($Ra$, $R_k$, $S$ et $D$). La distance entre la paroi chaude et le mur, l'épaisseur du mur ont été fixés respectivement à $S = 0.05$ et $D = 0.15$. Le pas choisi, pour tracer les lignes de courant, est égal à 1.6 ($\Delta\psi = 1.6$) alors que celui choisi pour tracer les isothermes est égal à 0.1 ($\Delta\theta = 0.1$).

### 3-1 Isothermes et lignes de courant en absence des échanges radiatifs ($\varepsilon=0$)

Les isothermes relatifs au cas $Ra=10^6$, $R_k=50$, sont présentés sur la figure 2 pour $\varepsilon=0$ et pour ($0 \leq L_0 \leq 0.2$) et la cavité vide (sans mur) ont été tracées dans cette étude.
Dans chaque cas, le fluide circule dans la direction des aiguilles d'une montre, il est monocellulaire à cause de la position des parois isothermes chaude et froide. Nous remarquons que les isothermes dans le milieu de la cavité sont presque horizontales, donc on a une stratification thermique de la température sous l'effet des forces volumiques, aussi les isothermes sont denses au voisinage de la paroi chaude et au voisinage de la paroi gauche du mur, ainsi que celles de la paroi froide de la face droite du mur c'est-à-dire que les températures sont rapprochées.

### 3-2 Isothermes et lignes de courant en présence des échanges radiatifs ($\varepsilon = 1$)

Nous constatons sur les tracés des isothermes relatifs au cas $Ra=10^6$, $R_k=50$, et $0 \leq L_0 \leq 0.2$ pour $\varepsilon=1$. que les échanges radiatifs rapprochent les températures de la paroi chaude et de la face gauche du mur, ainsi que celles de la paroi froide et de la face droite du mur. L'inclinaison des isothermes près des parois adiabatiques est due à l'importance des flux radiatifs. Le long de la paroi froide, les gradients de température sont nettement moins importants en partie haute qu'en partie basse de la cavité.

Le nombre de Nusselt convectif moyen est plus important quand on tient compte du rayonnement, nous observons aussi l'augmentation du débit à travers les ouvertures. En effet, l'écart entre les températures moyennes de l'air dans les parties gauche et droite de la cavité s'élève sous l'effet du rayonnement.

## 6 Conclusion

La simulation numérique portant sur le couplage entre la convection naturelle et le rayonnement thermique dans une cavité différentiellement chauffée et partitionnée conduit aux conclusions suivantes :
1) Les lignes de courant et les isothermes sont considérablement affectées par la présence du mur.
2) Le rayonnement thermique uniformise les températures dans les deux parties de la cavité et augmente l'écart entre leurs températures moyennes.
3) Le débit d'air à travers les ouvertures est augmenté en présence du rayonnement thermique.
4) Le rayonnement thermique contribue à une augmentation du transfert de chaleur, principalement à $Ra$ élevé.
5) Le transfert de chaleur est minimal pour $L_0 = 0$ (cavité plein) et est maximal pour une cavité vide (sans mur).

## Références

[1] A. Mezrhab et al, "Modelling of combined radiation and convection heat transfer in an enclosure with a heat generating conducting body", *International Journal of Computational Methods*, vol. 2, n° 3, (2005), pp. 431-450.
[2] A. Mezrhab et al "Computation of combined natural convection and radiation heat transfer in a partitioned cavity". Article accepté pour publication dans *Applied Energy*, (2005).
[3] H. Nakamura et Y. Asko, "Combined free convection and radiation heat transfer in rectangular cavities with a partion wall", *Heat Transfer Japanese Research*, (1986) pp. 60-81.
[4] A. Mezrhab et al "Computation of view factors for surfaces of complex shape including screening effects and using a boundary element approximation", *Engineering Computations: International Journal for Computer-Aided Engineering and Software*, Vol. 22, No. 2, (2005), 132.

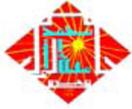 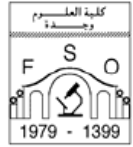 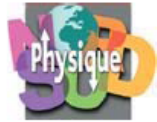 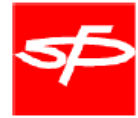

# Etude du confort thermique dans l'habitat

**Abdeladim SEDDOUKI**[1], H. AMAOUI [1], A. MEZRHAB[1], R. BELARBI[2]

[1] Faculté des sciences, Laboratoire de Mécanique & Energétique, Oujda, Maroc
[2] Université de La Rochelle, Laboratoire d'Eude des Phénomènes de Transfert Appliqués Aux Bâtiments (LEPTAB), France

**Résumé**

Protéger les bâtiments contre les variations thermiques est une nécessité qui s'est affinée au fil des années avec l'exigence de confort de leurs occupants.
La grande crise énergétique des années 70, a entraîné l'homme à une réflexion sur les économies d'énergie en respectant les conditions de confort dans l'habitation.
Les exigences de confort dans les bâtiments ont fortement évoluées durant ces dernières années, car les procédés d'isolation, de chauffage et de ventilation des bâtiments sont en profonde mutation.Tout en assurant le confort des occupants, les technologies évoluent de façon à rationaliser l'utilisation d'énergie et à s'adapter à de nouvelles sources énergétiques.
Le choix d'investigation vient dans le cadre de protéger les bâtiments contre les variations thermiques et répondre aux exigences de leurs occupants vis à vis du confort.
Avec les préoccupations grandissantes du développement durable, le secteur du bâtiment doit répondre à deux exigences primordiales: maîtriser les impacts sur l'environnement extérieur, tout en assurant des ambiances intérieures saines et confortables. Ainsi, une vision globale du confort thermique, qui tient compte de l'évolution technologique de façon à rationaliser l'utilisation de l'énergie et s'adapter à de nouvelles sources énergétiques.

# 1 Introduction

Le confort est une notion subjective. On peut dire qu'un individu est placé dans des conditions confortables lorsqu'il ne ressent aucune gène ni aucune contrainte de nature à le distraire de ses activités**.**

### 1.1 Les facteurs du confort

- La température ambiante.
- L'humidité de l'air.
- La ventilation (renouvellement, et vitesse de l'air).
- Le niveau d'activité de l'individu**.**

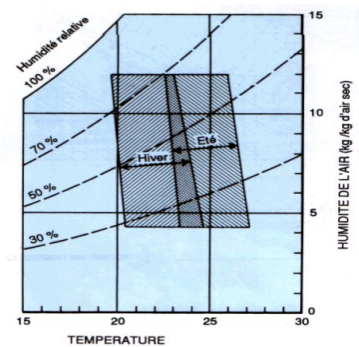

### 1.2 Principaux critères

- L'orientation du bâtiment.
- Son environnement immédiat.
- Les caractéristiques du climat local (direction des vents, précipitations…)
- Les besoins en énergie (électrique et thermique).

## 2 Stratégies thermiques

### 2.1 Stratégie du chaud (Hiver)

- Capter la chaleur du rayonnement solaire.
- La stocker dans la masse.
- La conserver par l'isolation.
- La distribuer dans le bâtiment tout en la régulant.

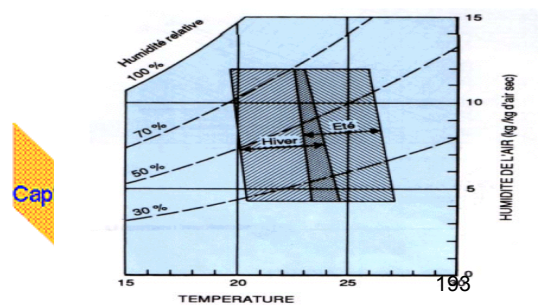

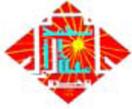 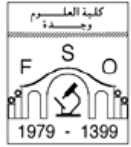 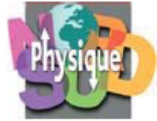 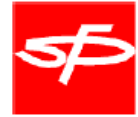

- L'intégration dans l'environnement par:
   * l'utilisation de matériaux locaux (bois, pierre…).
   * respect du site (terrassement Limité, respect de la végétation existante).

## 4   Conclusion

- Une conception architecturale finement étudiée requiert un équilibre entre la qualité globale des ambiances crées et une optimisation du coût énergétique et environnemental**.**
- Actuellement on peut assurer jusqu'à 70% des besoins énergétiques grâce à l'énergie solaire.

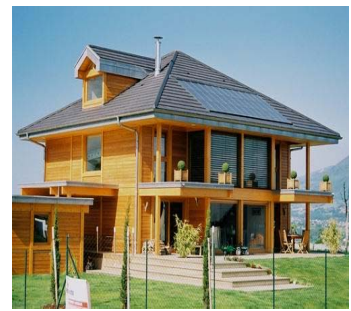

### 2.2 Stratégie du froid (Eté)

- Se protéger du rayonnement solaire.
- Minimiser les apports internes.
- Dissiper la chaleur en excès.
- Refroidir naturellement.

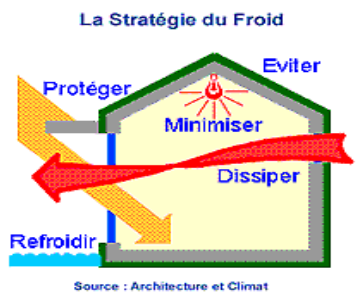

## 3-Principes généraux :

- De larges ouvertures au sud (baies vitrées, serres…).
- De petites ouvertures au nord.
- Des espaces tampons au nord (garages, débarras…).
- L'utilisation des massesthermiques comme (murs, dalles).
- La compacité des volumes.
- La circulation de l'air.

## Références

[1]  D.Hernot, G.Porcher,"Thermique Appliquée Aux Bâtiments"Edition parisienne C.F.P (Chaud Froid Plomberie) ,1984.
[2]  C.Chapon, A.Parraud,"Confort d'été et conception architecturale", Edition janvier 1993, EDF (électricité de France).